\documentclass[a4paper,12pt]{uq-thesis}

\newcommand{\bra}[1]{\left\langle{#1}\right\vert}
\newcommand{\ket}[1]{\left\vert{#1}\right\rangle}
\newcommand{\be}{\begin{equation}}
\newcommand{\bea}{\begin{eqnarray}}
\newcommand{\eea}{\end{eqnarray}}
\newcommand{\ee}{\end{equation}}
\newcommand{\tr}{\mathrm{tr}}

\newcommand{\beforepreface}{
        \pagestyle{empty}
        \titlep
        \pagenumbering{roman}
        \pagestyle{plain}}

\newcommand{\afterpreface}{
        \pagenumbering{arabic}
        \pagestyle{fancy}}

\usepackage{helvet}
\usepackage{amssymb}
\usepackage{amsmath}
\usepackage{amsthm}
\usepackage{bm}
\usepackage{array}
\usepackage{makeidx} \makeindex    
\usepackage[nottoc]{tocbibind}
\usepackage{booktabs}   
\usepackage{titlesec}
\usepackage{graphicx}    

\titleformat{\subsubsection}[hang]{\centering\slshape\bfseries}{}{0pt}{}{}

\usepackage{citeref}

\title{\Large{{Foundations and Applications of Entanglement Renormalization \vspace{.2em}\\}}}

\author{Glen Evenbly}
\prevdegrees{B.~Sc.~(Hons 1st), University of Auckland} \division{} \school{School of
Mathematics and Physics} \thesistype{Doctor of Philosophy}
\principalsupervisor{Guifre Vidal} \associatesupervisor{John Fjaerestad}

\begin{document}

\beforepreface

\chapter*{Statements of Contributions and Published Works}

\textbf{Published Works by the Author Incorporated into the Thesis}

\textit{Chapter 2}: Incorporates Ref.~\cite{evenbly08a}.

\textit{Chapter 3}: Incorporates Ref.~\cite{evenbly07a}.

\textit{Chapter 4}: Incorporates Ref.~\cite{evenbly07b}.

\textit{Chapter 5}: Incorporates Ref.~\cite{evenbly08a}.

\textit{Chapter 6}: Incorporates Ref.~\cite{pfeifer08}.

\textit{Chapter 7}: Incorporates Ref.~\cite{evenbly08b}.

\textit{Chapter 8}: Incorporates Ref.~\cite{evenbly09a}.

\chapter*{Abstract}

Understanding the collective behavior of a quantum many-body system, a system composed of a large number of interacting microscopic degrees of freedom, is a key aspect in many areas of contemporary physics. However, as a direct consequence of the difficultly of the so-called many-body problem, many exotic quantum phenomena involving extended systems, such as high temperature superconductivity, remain not well understood on a theoretical level. 

Entanglement renormalization is a recently proposed numerical method for the simulation of many-body systems which draws together ideas from the renormalization group and from the field of quantum information. By taking due care of the quantum entanglement of a system, entanglement renormalization has the potential to go beyond the limitations of previous numerical methods and to provide new insight to quantum collective phenomena. This thesis comprises a significant portion of the research development of ER following its initial proposal. This includes exploratory studies with ER in simple systems of free particles, the development of the optimisation algorithms associated to ER, and the early applications of ER in the study of quantum critical phenomena and frustrated spin systems. 

\textbf{Keywords:} Entanglement renormalization, quantum many-body systems, simulation algorithms, tensor networks, quantum entanglement, quantum information.

\tableofcontents

\clearpage
\afterpreface

\chapter{Overview}\label{chap:Introduction}

\begin{quote}
``\textit{It would indeed be remarkable if Nature fortified herself against further advances in knowledge behind the analytical difficulties of the many-body problem}."
\end{quote}%
\hspace{1cm}- Max Born (1960)

Understanding the collective behavior of quantum many-body systems, that is quantum systems composed of a large number of interacting microscopic degrees of freedom, is a central aspect in many areas of contemporary physics including particle physics and condensed matter physics. Solving a many-body problem to calculate, for instance, equilibrium properties or the outcome of a dynamic process, even while allowing for some degree of approximation, is often very difficult. Many techniques, both analytical and numeric, have been developed to address specific classes of quantum many-body problems. Techniques based upon perturbation theory have proved incredibly successful for weakly interacting systems like those encountered e.g. in quantum electro-dynamics (QED) \cite{mandl93,srednicki07}. For instance, perturbative calculations of electromagnetic fine structure constant $\alpha$ have been verified experimentally to astonishing accuracy; to within ten parts in a billion $(10^{-8})$.

In other branches of physics, notably condensed matter physics, relevant many-body systems are often no longer in the weakly interacting regime and cannot be analysed by perturbation theory. Instead, without the general purpose tool of perturbation theory (and given that few systems of interest admit analytic solutions), much of condensed matter physics has been traditionally based upon a set on approximations specific to the problem at hand. Unfortunately, satisfactory tools or approximations are not known for many models of potential interest in condensed matter theory, and these models have remained defiant to theoretical investigation. For instance the Hubbard model \cite{hubbard63}, a simple model of interacting fermions on a lattice proposed by John Hubbard in 1963, is thought to possibly describe high-temperature superconductivity. Despite intensive research efforts over many years, this possibility has neither been confirmed nor refuted, nor is even the phase diagram of the Hubbard model well understood. In general, the difficulty of solving many-body problems, such as the Hubbard model, has retarded progress in theoretical condensed matter physics. As a consequence many exotic quantum phenomena involving extended systems, such as the aforementioned high-temperature superconductor or the spin-liquid phase of matter, remain not well understood on a theoretical level. 

Given the staggering progress made in computer technology in recent decades, numerical approaches to many-body problems are becoming increasingly dominant. Unfortunately, even with vast amounts of computational power available, large quantum many-body systems still cannot be addressed exactly. For instance, one would typically like to analyse systems with a large number of particles (often many hundreds or thousands) yet, due the exponential growth of the Hilbert space with particle number, only a few tens of particles may be analyzed exactly. As of 2009, a powerful supercomputer can analyze systems of at most $N\approx 48$ spin half particles using exact diagonalisation (ED) techniques \cite{laeuchli09,richter09}. Despite this limitation, exact diagionalization (often combined with finite-size scaling techniques) remains a ubiquitous numerical tool in many areas of computational physics. 

Numerical techniques based upon Monte Carlo sampling of the multi-dimensional integrals which arise in quantum many-body problems have proved incredibly useful and versatile for studying many-body problems. These comprise a large class of algorithms collectively referred to as quantum Monte Carlo (QMC) \cite{mcmillan65,anderson75,ceperley95}. Indeed, much of the contemporary understanding of lattice models in $D=2$ or higher dimensions has come from studies with QMC. Quantum Monte Carlo allows calculation of many-body effects in the wavefunction at the cost of statistical uncertainty that can be reduced by increasing the number of samples taken. While QMC techniques have proved an invaluable tool for many-body systems they are unfortunately only viable for a certain class of Hamiltonians. Specifically, QMC techniques can only be efficiently applied to study systems for which the Feynman path integral can be evaluated as a sum over configurations with positive weights. However, there is a large class of systems for which the path integral does not have any known representation with only positive weights and the sampling time becomes exponentially large with system size; this is the famous \emph{sign problem}. Two important classes of problems in which the sign problem is encountered are frustrated models of magnetism and systems of interacting fermions. Despite intensive research efforts there remains no general solution to the sign problem (though there have been partial solutions found for particular systems, for instance \cite{henelius00,bergkvist03,corney04,alford09}) and many models of interest remain out of reach to QMC techniques.

Another leading numerical method for many body systems is the density matrix renormalization group (DMRG) \cite{white92,white93}, a method based upon ideas from the real-space renormalization group (RG) \cite{fisher98}. In the context of lattice models, the RG aims to obtain the physics of low energy states by grouping degrees of freedom and retaining only the relevant ones. Wilson's numerical renormalization group (NRG) \cite{wilson75} provided an explicit prescription to implement rescaling transformations of the lattice and was successful in solving the Kondo problem. But it was not until the advent of White's DMRG algorithm in 1991 that RG methods became (and remain today) the dominant numerical approach for 1$D$ lattice systems. Typically, the DMRG algorithm can be used to address the ground state of large 1$D$ lattice systems (hundreds of sites) with many digits of precision. Unlike QMC techniques, DMRG does not suffer from the sign problem and can be applied to study interacting fermions and frustrated spin systems. The advent of the DMRG algorithm revolutionised the study of quantum systems; by allowing highly accurate simulation of a large class of 1$D$ lattice problems DMRG has led to a deeper understanding of many types of quantum systems including fermionic systems, such as the Hubbard model, bosonic systems, problems with impurities, and quantum dots joined with quantum wires. However, in $D=2$ or higher dimensional quantum lattice systems, DMRG is limited to small system sizes ($N \approx 10 \times 10$ sites), although this is still significantly larger than what is possible with exact diagonalization alone. This limitation arises due to the failure of the matrix product state (MPS) ansatz, the tensor-network ansatz upon which DMRG is based, to properly capture the geometry of quantum correlations in higher dimensional systems.

Though QMC and DMRG are both stunningly successful techniques in their own right, certain classes of many body problem remain intractable to either method; mainly frustrated spin systems and interacting fermions on large 2$D$ lattices. These systems cannot be analyzed with QMC due to the sign problem, nor with DMRG due to the inability of DMRG to analyze large 2$D$ systems. Accordingly, there have been significant research efforts to devise new numerical many-body techniques that both $(i)$ do not possess a sign problem and $(ii)$ scale efficiently with system size in $D=2$ dimensional quantum lattice systems. Several of these proposals are based upon generalizations of the MPS tensor network ansatz, or proposals of entirely new tensor network ansatz, that properly capture the correlation structure of 2$D$ systems. Among this new breed of many-body techniques, based upon tensor network ansatz, are projected entangled pair states (PEPS) \cite{porras06,jordan08}, tensor entanglement renormalization group (TERG) \cite{gu08,gu09} and, the subject of this thesis, entanglement renormalization (ER) \cite{vidal07,vidal08}. 

We now proceed to outline the various aspects of entanglement renormalization that are analysed in this thesis.

%
%

\section{Foundations of entanglement renormalization and the MERA}
In Chapter \ref{chap:MERAintro}, we present a short introduction to entanglement renormalization and its related tensor network, the multiscale entanglement renormalization ansatz (MERA). The MERA is presented both from the perspective as a peculiar class of quantum circuit and also from the complementary perspective of the renormalization group (RG). Several realizations of the MERA in 1$D$ and 2$D$ lattices are presented. The notion of the causal cone of the MERA is introduced and the impact on computational efficiency is discussed. We also discuss how spatial symmetries can be incorporated into the MERA.  

\section{Entanglement renormalization in free fermionic systems}
Chapter \ref{chap:FreeFerm} explores the performance of entanglement renormalization in systems of free spinless fermions on 1$D$ and 2$D$ lattices. The implementation of ER considered makes use of properties of free fermions to achieve substantial reductions in computational cost, and also to simplify the subsequent analysis of results. The ability of ER to accurately coarse-grain fermionic systems in insulating, conducting and superconducting phases (whose corresponding ground states span all known forms of entropy scaling) is investigated. We examine the accuracy of the ER simulations as compared to exact results in terms of the truncation error of the correlation matrix, the accuracy of the ground energy and the accuracy of two-point correlators. The results demonstrate the ability of a coarse-graining transformation based upon ER to accurately capture the low-energy physics of a variety of free fermion systems in $D=1,2$ dimensional lattices.

\section{Entanglement renormalization in free bosonic systems}
In Chapter \ref{chap:FreeBoson}, we investigate the performance of ER in systems of free bosons, while highlighting the connection between momentum-space RG transformations and real-space RG transformations. The process of applying RG transforms to the system in momentum-space is explained and contrasted against the equivalent steps of applying real-space RG. A critical system and examples of relevant and irrelevant perturbations thereof are considered. A comparison between results obtained from exact momentum-space RG and the results from numerical real-space RG is presented. The results of this Chapter demonstrate that a real-space RG transform based upon entanglement renormalization is able to reproduce the exact results from momentum-space RG to a high accuracy in $D=1,2$ dimensional lattice systems, both for the critical and non-critical cases considered. Also demonstrated is the ability of the MERA to provide an efficient and accurate representation of the ground state of free boson systems, thus extending to the bosonic case the results of Chapter \ref{chap:FreeFerm}.

\section{Algorithms for entanglement renormalization}
In Chapter \ref{chap:MERAalg} we describe how to compute expected values of local observables and two-point correlators from a MERA, and also present optimisation algorithms that allow the MERA approximation to the ground state of an arbitrary system to be obtained. A highlight of the algorithms is their computational cost. For an inhomogeneous lattice with $N$ sites, the cost scales as $O(N)$, whereas for translation invariant systems it drops to just $O(\log N)$. Other variations of the algorithm allow us to address infinite systems, and scale invariant systems (e.g. quantum critical systems), at a cost independent of $N$. We also present benchmark calculations for different 1$D$ quantum lattice models, namely Ising, 3-level Potts, XX and Heisenberg models. We compute ground state energies, magnetizations and two-point correlators throughout the phase diagram of the Ising and Potts models, which includes a second order quantum phase transition. We find that, at the critical point of an infinite system, the error in the ground state energy decays exponentially with the refinement parameter $\chi$, whereas the two-point correlators remain accurate even at distances of millions of lattice sites. We then extract critical exponents from the order parameter and from two-point correlators. Finally, we also compute a MERA that includes the first excited state, from which the energy gap can be obtained and seen to vanish as $1/N$ at criticality. 

\section{Entanglement renormalization and quantum criticality}
Chapter \ref{chap:1DCrit} explains how to use the MERA to investigate quantum critical systems that are invariant under changes of scale. An algorithm is presented which, given a critical Hamiltonian, details how to compute a scale invariant MERA for its ground state. Then, starting from the scale invariant MERA, a procedure is described to identify the scaling operators/dimensions of the theory. A closed expression for two-point and three-point correlators is derived, and a connection between the MERA and conformal field theory (which can be used to readily identify the continuum limit of a critical lattice model) is established. Finally, benchmark calculations for the Ising and Potts models are presented. 

\section{Entanglement renormalization in two spatial dimensions}
In Chapter \ref{chap:2DMera} we build upon the MERA schemes introduced in Chapter \ref{chap:MERAintro}, and also the optimisation algorithms presented in Chapter \ref{chap:MERAalg}, to describe an implementation of the MERA that allows us to consider 2$D$ systems of arbitrary size, including infinite systems. In this way we demonstrate the scalability of entanglement renormalization in two spatial dimensions and decisively contribute to establishing the MERA as a competitive approach to systematically address 2$D$ lattice models. We also demonstrate the performance of the scheme by analysing the 2$D$ quantum Ising model, for which we obtain accurate estimates of the ground state energy and magnetizations, as well as two-point correlators (shown to scale polynomially at criticality), the energy gap, and the critical magnetic field and beta exponent. Finally, we discuss how the use of disentanglers affects the simulation costs, by comparing the MERA with a \emph{tree tensor network} (TTN).

\section{The spin-$\frac{1}{2}$ Kagome lattice Heisenberg model with entanglement renormalization}
Chapter \ref{chap:KagMera} examines a long-standing problem in condensed matter physics, the nature of the ground-state of the spin-$\frac{1}{2}$ kagome lattice Heisenberg model (KLHM), using entanglement renormalization. Progress in understanding the KLHM has been hindered, as with many other models of frustrated antiferromagnets, by the inapplicability of quantum Monte Carlo methods due to the sign problem. This investigation is the first demonstration of the utility of entanglement renormalization to study 2$D$ lattice models that are beyond the reach of quantum Monte Carlo techniques. After describing a scheme for ER on the Kagome lattice with periodic boundary conditions, we address lattices of $N=\{36,144,\infty\}$ sites. We then analyse bond energies, two-point and bond-bond correlators of the ground state on the various lattice sizes. The results give strong numerical evidence in favor of a VBC ground state for the KLHM.

\chapter{Foundations of entanglement renormalization and the MERA}
\label{chap:MERAintro}

\section{Introduction} \label{sect:MERAintro:Intro}

Entanglement renormalization (ER) \cite{vidal07} is a numerical technique based on locally reorganizing the Hilbert space of a quantum many-body system with the aim to reduce the amount of entanglement in its wave function. It was introduced to address a major computational obstacle in real space renormalization group (RG) methods \cite{wilson75,white92,white93,morningstar94,morningstar96}, responsible for limitations in their performance and range of applicability, namely the proliferation of degrees of freedom that occurs under successive applications of a RG transformation.

Entanglement renormalization is built around the assumption that, as a result of the local character of physical interactions, some of the relevant degrees of freedom in the ground state of a many-body system can be decoupled from the rest by unitarily transforming small regions of space. Accordingly, unitary transformations known as \emph{disentanglers} are applied locally to the system in order to identify and decouple such degrees of freedom, which are then safely removed and therefore do no longer appear in any subsequent coarse-grained description. This prevents the harmful accumulation of degrees of freedom and thus leads to a sustainable real space RG transformation, able to explore arbitrarily large 1$D$ and 2$D$ lattice systems, even at a quantum critical point. It also leads to the \emph{multi-scale entanglement renormalization ansatz} (MERA), a variational ansatz for many-body states \cite{vidal08}. 

The MERA, based in turn on a class of quantum circuits, is particularly successful at describing ground states at quantum criticality \cite{vidal07,vidal08,evenbly07a,evenbly07b,cincio08, giovannetti08,pfeifer08,evenbly08b} or with topological order \cite{aguado08,konig09}. From the computational viewpoint, the key property of the MERA is that it can be manipulated efficiently, due to the causal structure of the underlying quantum circuit \cite{vidal08}. As a result, it is possible to efficiently evaluate the expected value of local observables, and to efficiently optimize its tensors. Thus, well-established simulation techniques for matrix product states, such as energy minimization \cite{verstraete04,porras06} or simulation of time evolution \cite{vidal03b,vidal04}, can be readily generalized to the MERA \cite{dawson08,rizzi08}.

The goal of this Chapter is to provide a short introduction to entanglement renormalization and the MERA, establishing the notation and terminology that shall be used in the rest of this thesis. Specifically, Sect. \ref{sect:MERAintro:Quantcirc} introduces the MERA from the perspective of quantum circuits while Sect. \ref{sect:MERAintro:RG} offers a complementary interpretation of the MERA in terms of the renormalization group (RG). Several realizations of the MERA in 1$D$ and 2$D$ lattices are discussed in Sect. \ref{sect:MERAintro:ChooseMERA} and Sect. \ref{sect:MERAintro:Sym} details how spatial symmetries can be incorporated into the MERA.


\begin{figure}[!tbhp]
\begin{center}
\includegraphics[width=10cm]{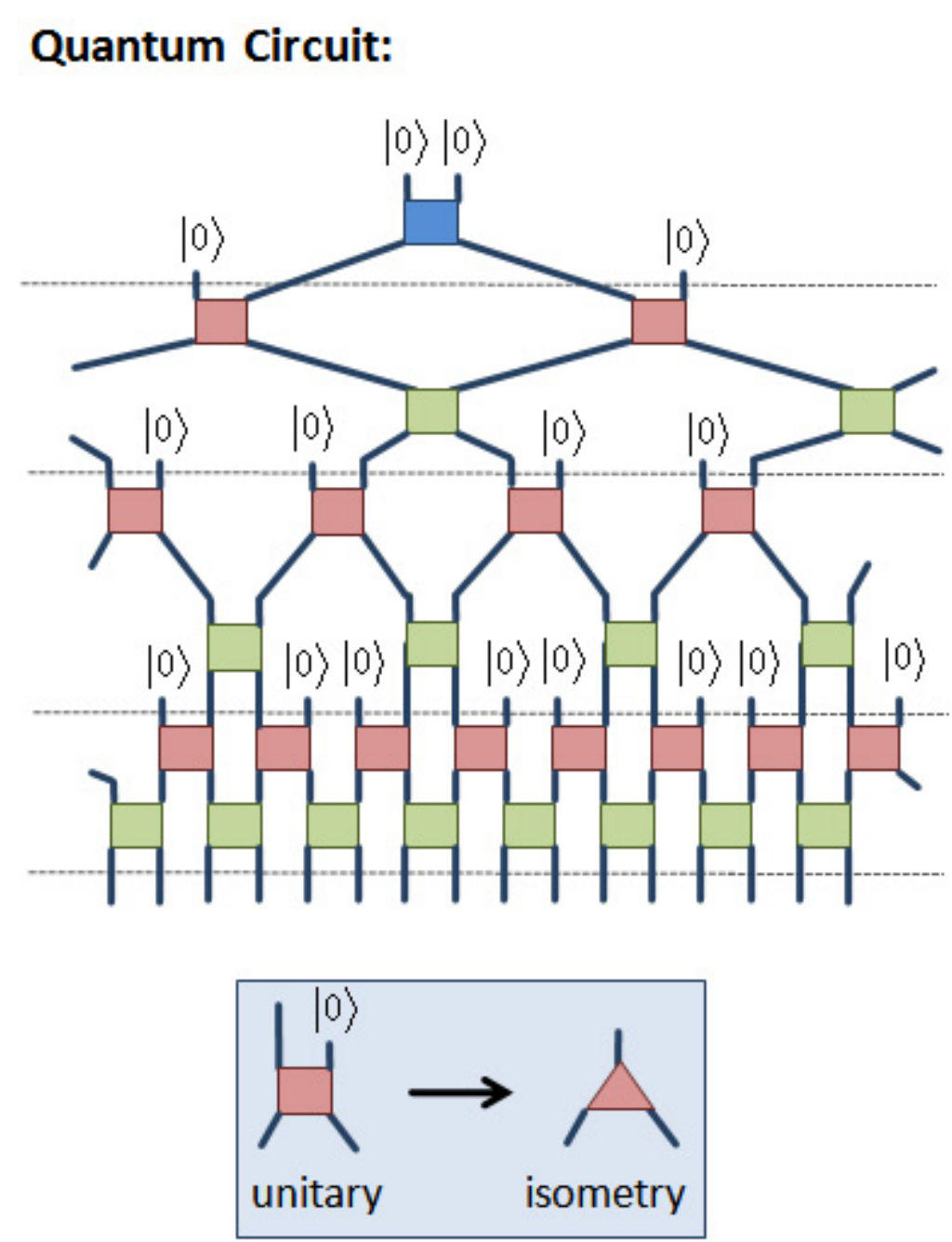}
\caption{ Quantum circuit $\mathcal{C}$ corresponding to a specific realization of the MERA, namely the binary 1$D$ MERA of Fig. \ref{fig:MERAintro:2MERA}. In this particular example, circuit $\mathcal{C}$ is made of gates involving two incoming wires and two outgoing wires, $p=p_{in}=p_{out}=2$. Some of the unitary gates in this circuit have one incoming wire in the fixed state $\ket{0}$ and can be replaced with an isometry $w$ of type (1,2). By making this replacement, we obtain the isometric circuit of Fig. \ref{fig:MERAintro:2MERA}.} 
\label{fig:MERAintro:QuantCirc}
\end{center}
\end{figure}

\section{The MERA} \label{sect:MERAintro:MERA}

Let $\mathcal{L}$ denote a $D$-dimensional lattice made of $N$ sites, where each site is described by a Hilbert space $\mathbb{V}$ of finite dimension $d$, so that $\mathbb{V}_{\mathcal{L}} \cong \mathbb{V}^{\otimes N}$. The MERA is an ansatz to describe certain pure states $\ket{\Psi}\in \mathbb{V}_{\mathcal{L}}$ of the lattice or, more generally, subspaces $\mathbb{V}_{U} \subseteq \mathbb{V}_{\mathcal{L}}$.

There are two useful ways of thinking about the MERA that can be used to motivate its specific structure as a tensor network, and also help understand its properties and how the algorithms ultimately work. One way is to regard the MERA as a quantum circuit $\mathcal{C}$ whose output wires correspond to the sites of the lattice $\mathcal{L}$ \cite{vidal08}. Alternatively, we can think of the MERA as defining a coarse-graining transformation that maps $\mathcal{L}$ into a sequence of increasingly coarser lattices, thus leading to a renormalization group transformation \cite{vidal07}. Next we briefly review these two complementary interpretations. 

\begin{figure}[!tbhp]
\begin{center}
\includegraphics[width=10cm]{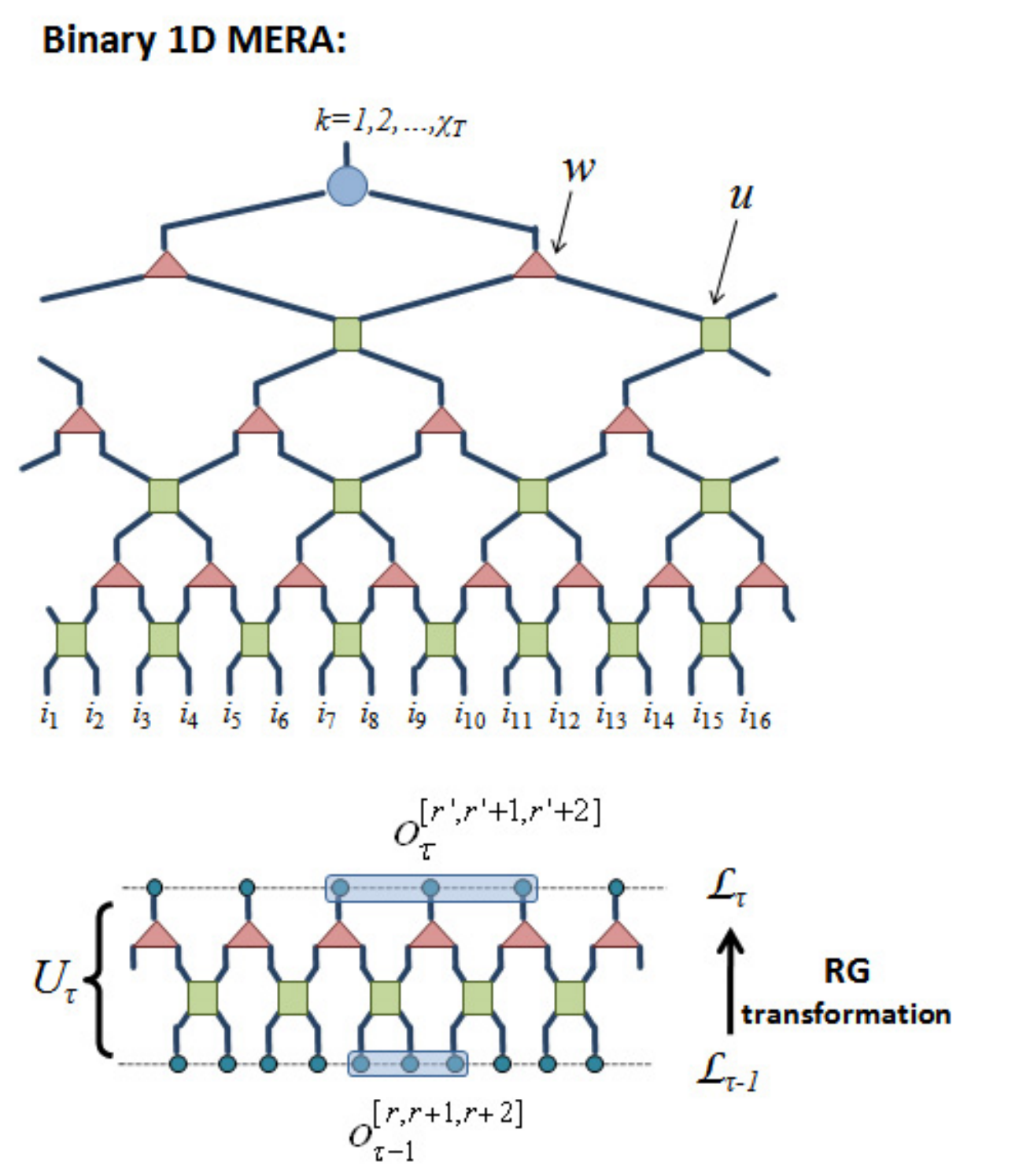}
\caption{ \emph{Top:} Example of a binary 1$D$ MERA for a lattice $\mathcal{L}$ with $N=16$ sites. It contains two types of isometric tensors, organized in $T=4$ layers. The input (output) wires of a tensor are those that enter it from the top (leave it from the bottom). The \emph{top tensor} is of type $(1,2)$ and the rank $\chi_T$ of its upper index determines the dimension of the subspace $\mathbb{V}_{U}\subseteq \mathbb{V}_{\mathcal{L}}$ represented by the MERA. The \emph{isometries} $w$ are of type $(1,2)$ and are used to replace each block of two sites with a single effective site. Finally, the \emph{disentanglers} $u$ are of type (2,2) and are used to disentangle the blocks of sites before coarse-graining. \emph{Bottom:} Under the renormalization group transformation induced by the binary 1$D$ MERA, three-site operators are mapped into three-site operators.} 
\label{fig:MERAintro:2MERA}
\end{center}
\end{figure}

\section{Quantum circuit} \label{sect:MERAintro:Quantcirc}

As a quantum circuit $\mathcal{C}$, the MERA for a pure state $\ket{\Psi} \in \mathbb{V}_{\mathcal{L}}$ is made of $N$ quantum wires, each one described by a Hilbert space $\mathbb{V}$, and unitary gates $u$ that transform the unentangled state $\ket{0}^{\otimes N}$ into $\ket{\Psi}$ (see Fig. \ref{fig:MERAintro:QuantCirc}).  

In a generic case, each unitary gate $u$ in the circuit $\mathcal{C}$ involves some small number $p$ of wires,
\begin{equation}
	u:\mathbb{V}^{\otimes p} \rightarrow\mathbb{V}^{\otimes p},~~~~~ u^\dagger u = u u^{\dagger} = \mathbb{I},
	\label{eq:MERAintro:unitary}
\end{equation}
where $\mathbb I$ is the identity operator in $\mathbb{V}^{\otimes p}$. For some gates, however, one or several of the input wires are in a fixed state $\ket{0}$. In this case we can replace the unitary gate $u$ with an isometric gate $w$
\begin{equation}
	w:\mathbb{V}_{in}\rightarrow\mathbb{V}_{out},~~~~~ w^\dagger w = \mathbb{I}_{\mathbb{V}_{in}}, ~~~
\label{eq:MERAintro:isometry}
\end{equation}
where $\mathbb{V}_{in}\cong \mathbb{V}^{\otimes p_{in}}$ is the space of the $p_{in}$ input wires that are not in a fixed state $\ket{0}$ and $\mathbb{V}_{out}\cong \mathbb{V}^{\otimes p_{out}}$ is the space of the $p_{out}=p$ output wires. We refer to $w$ as a $(p_{in}, p_{out})$ gate or tensor. 

\begin{figure}[!tbhp]
\begin{center}
\includegraphics[width=10cm]{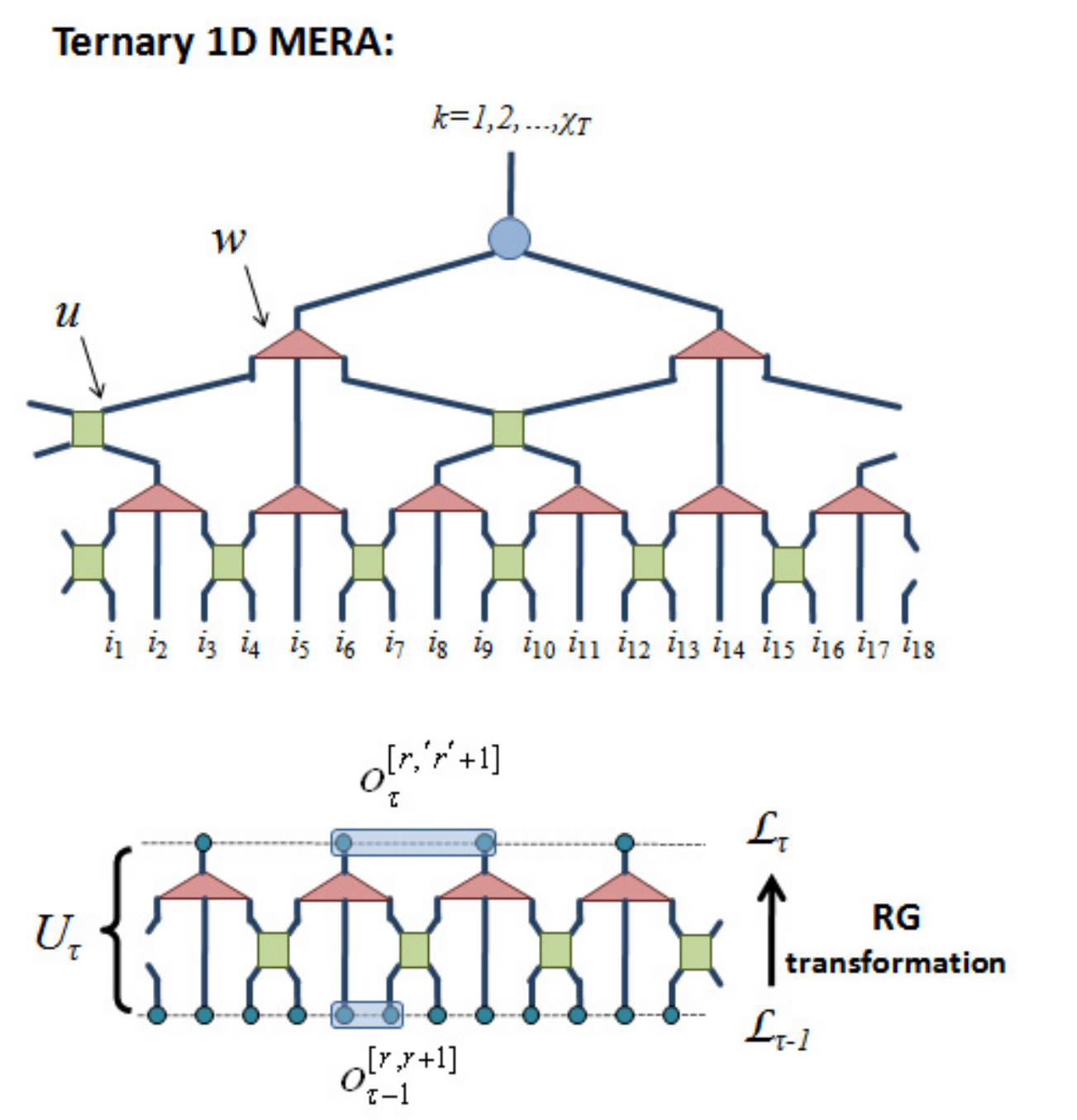}
\caption{ \emph{Top:} Example of ternary 1$D$ MERA (rank $\chi_T$, $T=3$) for a lattice of 18 sites. It differs from the binary 1$D$ MERA of Fig. \ref{fig:MERAintro:2MERA} in that the \emph{isometries} are of type $(1,3)$, so that blocks of three sites are replaced with one effective site. \emph{Bottom:} As a result, two-site operators are mapped into two-site operators during the coarse-graining.} 
\label{fig:MERAintro:3MERA}
\end{center}
\end{figure}

\begin{figure}[!tbhp]
\begin{center}
\includegraphics[width=8cm]{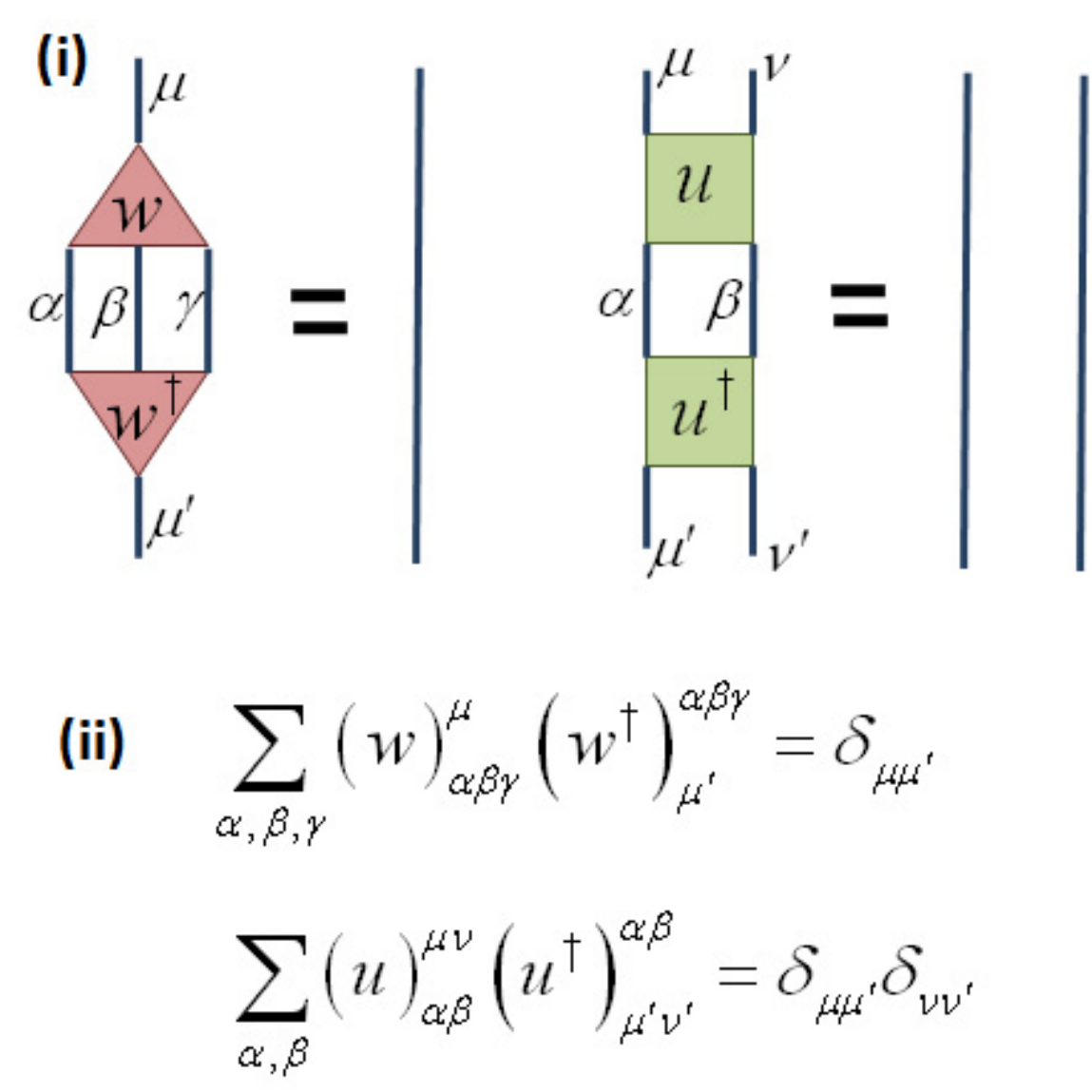}
\caption{ The tensors which comprise a MERA are constrained to be isometric, cf. Eq. \ref{eq:MERAintro:isometry}. The constraints for the isometries $w$ and disentanglers $u$ of the ternary MERA can be equivalently expressed (i) diagramatically or (ii) with equations. In this thesis we will mostly use the diagramatic notation, which remains simple for complicated tensor networks. } 
\label{fig:MERAintro:NotationCompare}
\end{center}
\end{figure}

Fig. \ref{fig:MERAintro:2MERA} shows an example of a MERA for a 1$D$ lattice $\mathcal{L}$ made of $N=16$ sites. Its tensors are of types $(1,2)$ and $(2,2)$. We call the $(1,2)$ tensors \emph{isometries} $w$ and the $(2,2)$ tensors \emph{disentanglers} $u$ for reasons that will be explained shortly, and refer to Fig. \ref{fig:MERAintro:2MERA} as a binary 1$D$ MERA, since it becomes a binary tree when we remove the disentanglers. Most of the early work for 1$D$ lattices \cite{vidal07,vidal08,evenbly07a,evenbly07b,dawson08,rizzi08} has been done using the binary 1$D$ MERA. However, there are many other possible choices. In Chapters \ref{chap:MERAalg} and \ref{chap:1DCrit}, for instance, we will mostly use the ternary 1$D$ MERA of Fig. \ref{fig:MERAintro:3MERA}, where the isometries $w$ are of type $(1,3)$ and the disentanglers $u$ remain of type $(2,2)$. Fig. \ref{fig:MERAintro:NotationCompare} makes more explicit the meaning of Eq. \ref{eq:MERAintro:isometry} for these tensors. Notice that describing tensors and their manipulations by means of diagrams is fully equivalent to using equations and often much more clear.

\begin{figure}[!tbhp]
\begin{center}
\includegraphics[width=12cm]{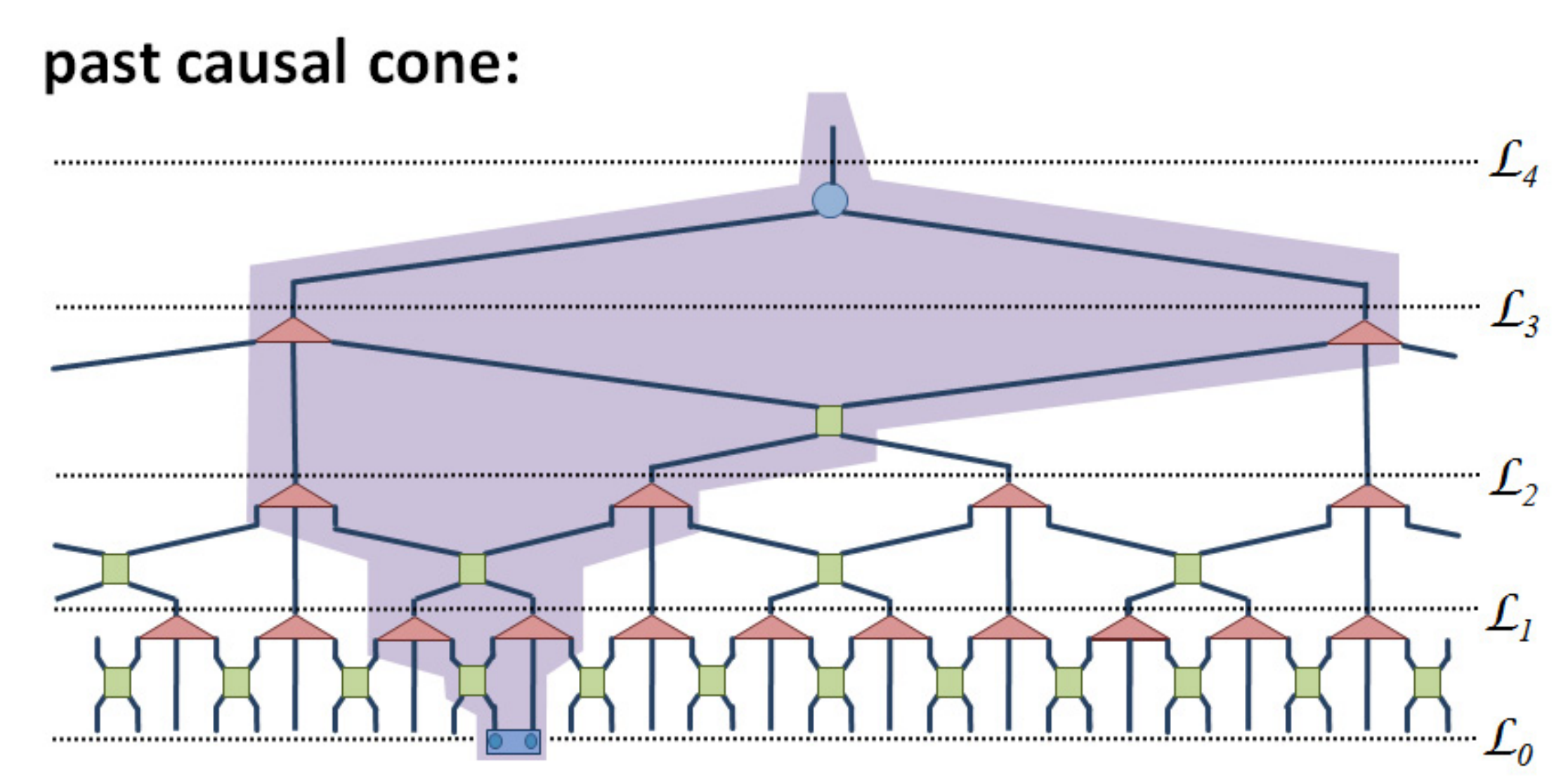}
\caption{ The past causal cone of a group of sites in $\mathcal{L}_0 \equiv \mathcal{L}$ is the subset of wires and gates that can affect the state of those sites. The example shows the causal cone of a pair of nearest neighbor sites of $\mathcal{L}_0$ for the ternary 1$D$ MERA. Notice that for each lattice $\mathcal{L}_{\tau}$, $\tau=0,1,2,3,4$, the causal cone involves at most 2 sites. This can be seen to be the case for any pair of contiguous sites of $\mathcal{L}_0$. We refer to this property by saying that the causal cones of the MERA have bounded width. } 
\label{fig:MERAintro:CausalCone}
\end{center}
\end{figure}

Eq. \ref{eq:MERAintro:isometry} encapsulates a distinctive property of the MERA as a tensor network: each of its tensors is isometric (notice that Eq. \ref{eq:MERAintro:unitary} is a particular case of Eq. \ref{eq:MERAintro:isometry}). A second key feature of the MERA refers to its causal structure. We define the past \emph{causal cone} of an outgoing wire of circuit $\mathcal{C}$ as the set of wires and gates that can affect the state on that wire. A quantum circuit $\mathcal{C}$ leads to a MERA only when the causal cone of an outgoing wire involves just a constant (that is, independent of $N$) number of wires at any fixed past time (Fig. \ref{fig:MERAintro:CausalCone}). We refer to this property by saying that the causal cone has a bounded `width'. 

The usefulness of the quantum circuit interpretation of the MERA will become clear in Chap. \ref{chap:MERAalg}, in the context of computing expected values for local observables. There, the two defining properties, namely Eq. \ref{eq:MERAintro:isometry} and the peculiar structure of the causal cones of $\mathcal{C}$, will be the key to making such calculations efficient. 

\section{Renormalization group transformation} \label{sect:MERAintro:RG}

Let us now review how the MERA defines a coarse-graining transformation for lattice systems that leads to a real-space renormalization group scheme, known as entanglement renormalization \cite{vidal07}. 

We start by grouping the tensors in the MERA into $T\approx \log N$ different layers, where each layer contains a row of isometries $w$ and a row of disentanglers $u$. We label these layers with an integer $\tau=1,2,\cdots T$, with $\tau=1$ for the lowest layer and with increasing values of $\tau$ as we climb up the tensor network, and denote by $U_{\tau}$ the isometric transformation implemented by all tensors in layer $\tau$, see Figs. \ref{fig:MERAintro:2MERA} and \ref{fig:MERAintro:3MERA}. Notice that the incoming wires of each $U_{\tau}$ define the Hilbert space of a lattice $\mathcal{L}_{\tau}$ with a number of sites $N_{\tau}$ that decreases exponentially with $\tau$ (specifically, as $N2^{-\tau}$ and $N3^{-\tau}$ for the binary and ternary 1$D$ MERA). That is, the MERA implicitly defines a sequence of lattices 
\begin{equation}
	\mathcal{L}_0 \rightarrow \mathcal{L}_1 \rightarrow \cdots \rightarrow \mathcal{L}_T,
\end{equation}
where $\mathcal{L}_0\equiv \mathcal{L}$ is the original lattice, and where we can think of lattice $\mathcal{L}_\tau$ as the result of coarse-graining lattice $\mathcal{L}_{\tau-1}$.

\begin{figure}[!tbhp]
\begin{center}
\includegraphics[width=10cm]{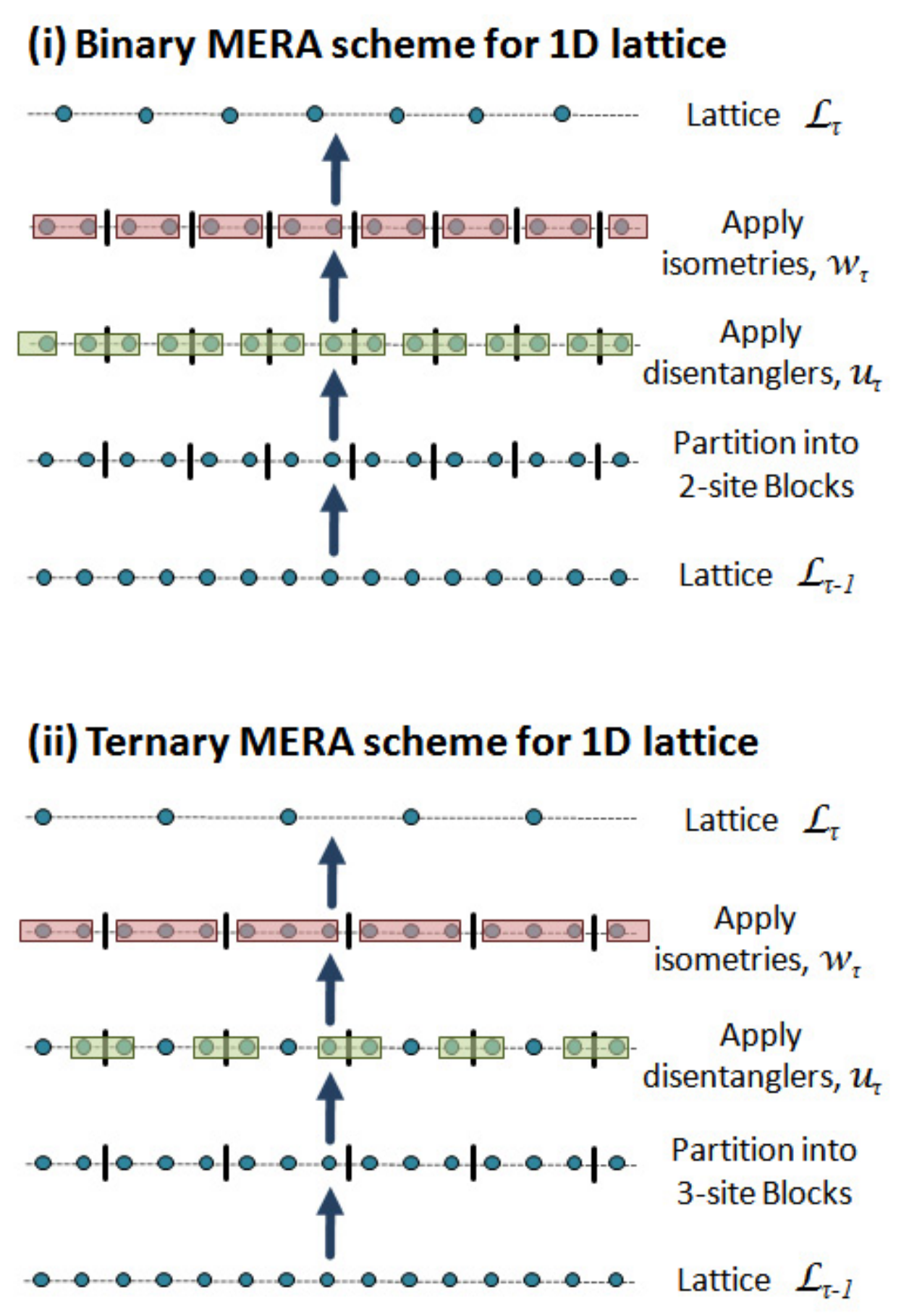}
\caption{ Detailed description of the real-space renormalization group transformation for 1$D$ lattices induced by ($i$) the binary 1$D$ MERA and ($ii$) the ternary 1$D$ MERA.} 
\label{fig:MERAintro:1DAltSchemes}
\end{center}
\end{figure}

Specifically, as illustrated in Figs. \ref{fig:MERAintro:2MERA} and \ref{fig:MERAintro:3MERA}, this coarse-graining transformation is implemented by the operator $U_{\tau}^{\dagger}$ that maps pure states of the lattice $\mathcal{L}_{\tau-1}$ into pure states of the lattice $\mathcal{L}_{\tau}$,
\begin{equation}
	U_{\tau}^{\dagger}:\mathbb{V}_{\mathcal{L}_{\tau-1}} \rightarrow \mathbb{V}_{\mathcal{L}_{\tau}},
	\label{eq:MERAintro:Utau}
\end{equation}
and that proceeds in two steps (Fig. \ref{fig:MERAintro:1DAltSchemes}). Let us partition the lattice $\mathcal{L}_{\tau-1}$ into blocks of neighboring sites. The first step consists of applying the disentanglers $u$ on the boundaries of the blocks, aiming to reduce the amount of short range entanglement in the system. Once (part of) the short-range entanglement between neighboring blocks has been removed, the isometries $w$ are used in the second step to map each block of sites of lattice $\mathcal{L}_{\tau-1}$ into a single effective site of lattice $\mathcal{L}_{\tau}$.

By composition, we obtain a sequence of increasingly coarse-grained states, 
\begin{equation}
	\ket{\Psi_0} \rightarrow \ket{\Psi_1} \rightarrow \cdots \rightarrow \ket{\Psi_{T}},
\label{eq:MERAintro:sequePsi}
\end{equation}
for the lattices $\{\mathcal{L}_0, \mathcal{L}_1, \cdots, \mathcal{L}_T\}$, where $\ket{\Psi_{\tau}} \equiv U_{\tau}^{\dagger} \ket{\Psi_{\tau-1}}$ and $\ket{\Psi_0}\equiv \ket{\Psi}$ is the original state. Overall the MERA corresponds to the transformation $U\equiv U_1 U_2 \cdots U_T$,
\begin{equation}
	U :\mathbb{V}_{\mathcal{L}_{T}}\rightarrow \mathbb{V}_{\mathcal{L}_0},
\label{eq:MERAintro:U}
\end{equation}
with $\ket{\Psi_0} = U\ket{\Psi_T}$. 

\begin{figure}[!tbhp]
\begin{center}
\includegraphics[width=7cm]{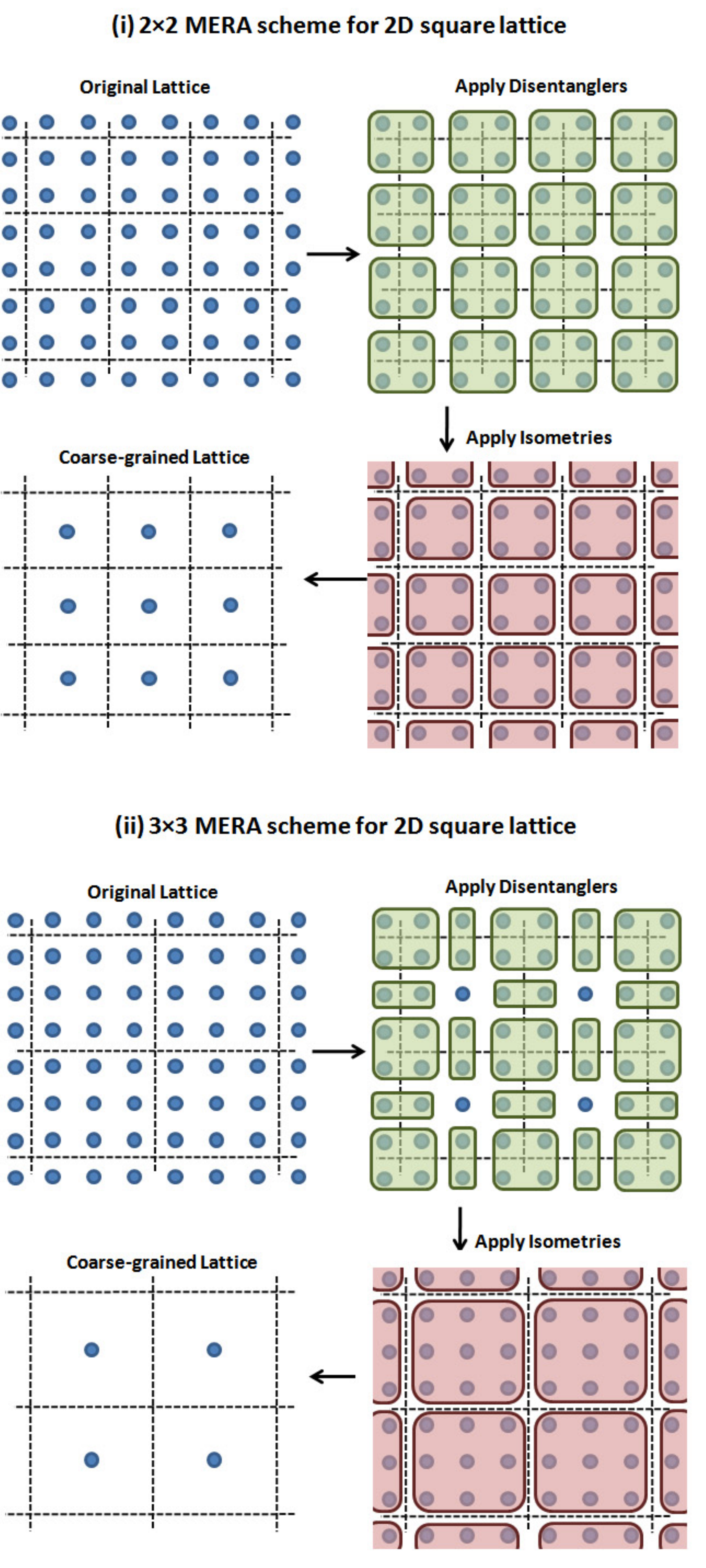}
\caption{ Detailed description of the real-space renormalization group transformation for a 2$D$ square lattice induced by two possible realizations of the MERA,  generalizing the 1$D$ schemes of Fig. \ref{fig:MERAintro:1DAltSchemes}. In the first case the isometries map a block of $2\times 2$ sites into a single site, which can be seen to imply that the natural size of a local operator, equivalently the causal width of the scheme, is $3\times 3$ sites. In the second case the isometries map a block of $3\times 3$ sites into a single site and the natural size of a local operator is $2\times 2$. As a result, the computational cost in the second scheme is much smaller than in the first scheme.} 
\label{fig:MERAintro:2DAltSchemes}
\end{center}
\end{figure}

Regarding the MERA from the perspective of the renormalization group is quite instructive. It tells us that this ansatz is likely to describe states with a specific structure of internal correlations, namely, states in which the entanglement is organized in different length scales. Let us briefly explain what we mean by this. 

We say that the state $\ket{\Psi}$ contains entanglement at a given length scale $\lambda$ if by applying a unitary operation (i.e. a disentangler) on a region $R$ of linear size $\lambda$, we are able to decouple (i.e. disentangle) some of the local degrees of freedom, that is, if we are able to convert the state $\ket{\Psi}$ into a product state $\ket{\Psi'}\otimes \ket{0}$, where $\ket{0}$ is the state of the local degrees of freedom that have been decoupled and $\ket{\Psi'}$ is the state of the rest of the system. [Here we assumed, of course, that the decoupling is not possible with a unitary operation that acts on a subregion $R'$ of the region $R$, where the size $\lambda'$ of $R'$ is smaller than the size of $R$, $\lambda' < \lambda$].

What makes the MERA useful is that the entanglement in most ground states of local Hamiltonians seems to decompose into moderate contributions corresponding to different length scales. We can identify two behaviors, depending on whether the system is in a phase characterized by symmetry-breaking order or by topological order (see \cite{levin05} and references therein). In systems with symmetry-breaking order, ground-state entanglement spans all length scales $\lambda$ smaller than the correlation length $\xi$ in the system --- and, consequently, at a quantum critical point, where the correlation length $\xi$ diverges, entanglement is present at all length scales \cite{vidal07}. In a system with topological order, instead, the ground state displays some form of (topological) entanglement affecting all length scales even when the correlation length vanishes \cite{aguado08,konig09}.

\section{Choose your MERA} \label{sect:MERAintro:ChooseMERA}
 
We have introduced the MERA as a tensor network originating in a quantum circuit. Its tensors have incoming and outgoing wires/indices according to a well-defined direction of time in the circuit. Therefore, a MERA can be regarded as a tensor network equipped with a (fictitious) time direction and with two properties:
\begin{itemize}
\item Its tensors are isometric (Eq. \ref{eq:MERAintro:isometry}). 
\item Past causal cones have bounded width (Fig. \ref{fig:MERAintro:CausalCone}). 
\end{itemize}

From a computational perspective, these are the only properties that we need to retain. In particular, there is no need to keep the vector space dimension of the quantum wires (equivalently, of the sites in the coarse-grained lattice) constant throughout the tensor network. Accordingly, we will consider a MERA where the vector space dimension of a site of lattice $\mathcal{L}_{\tau}$, denoted $\chi_{\tau}$, may depend on the layer $\tau$ (this dimension could also be different for each individual site of layer $\tau$, but for simplicity we will not consider this case here). Notice that $\chi_0=d$ corresponds to the sites of the original lattice $\mathcal{L}$. 

\textbf{Bond dimension.---} Often, however, the sites in most layers will have the same vector space dimension (except, for instance, the sites of the original lattice $\mathcal{L}$, with $\chi_0=d$, or the single site of the top lattice $\mathcal{L}_{T}$, with dimension $\chi_T$). In this case we denote the dominant dimension simply by $\chi$, and we refer to the MERA as having \emph{bond dimension} $\chi$. The computational cost of the algorithms described in subsequent sections is often expressed as a power of the bond dimension $\chi$.

\textbf{Rank.---} We refer to the dimension $\chi_T$ of the space $\mathbb{V}_{\mathcal{L}_{T}}$ (corresponding to the single site of the uppermost lattice $\mathcal{L}_T$) as the rank of the MERA. For $\chi_T=1$, the MERA represents a pure state $\ket{\Psi}\in\mathbb{V}_{\mathcal{L}}$. More generally, a rank $\chi_T$ MERA encodes a $\chi_T$-dimensional subspace $\mathbb{V}_{U} \subseteq \mathbb{V}_{\mathcal{L}}$. For instance, given a Hamiltonian $H$ on the lattice $\mathcal{L}$, we could use a rank $\chi_T$ MERA to describe the ground subspace of $H$ (assuming it had dimension $\chi_T$); or the ground state of $H$ (if it was not degenerate) and its $\chi_T-1$ excitations with lowest energy. The isometric transformation $U$ in Eq. \ref{eq:MERAintro:U} can be used to build a projector $P \equiv UU^{\dagger}$,
\begin{equation}
	P :\mathbb{V}_{\mathcal{L}} \rightarrow \mathbb{V}_{\mathcal{L}},~~~~~~~~P^2 = P,~~~ \tr(P) = \chi_T,
	\label{eq:MERAintro:P}
\end{equation}
onto the subspace $\mathbb{V}_{U} \subseteq \mathbb{V}_{\mathcal{L}}$.

Given the above definition of the MERA, many different realizations are possible depending on what kind of isometric tensors are used and how they are interconnected. We have already met two examples for a 1$D$ lattice, based on a binary and ternary underlying tree. Fig. \ref{fig:MERAintro:2DAltSchemes} shows two schemes for a 2$D$ square lattice. It is natural to ask, given a lattice geometry, what realization of the MERA is the most convenient from a computational point of view. A definitive answer to this question does not seem simple. An important factor, however, is given by the fixed-point size of the support of local observables under successive RG transformations---which corresponds to the width of the past causal cones. 

\textbf{Support of local observables.---} In each MERA scheme, under successive coarse-graining transformations a local operator eventually becomes supported in a characteristic number of sites. This is the result of two competing effects: disentanglers $u$ tend to extend the support of the local observable (by adding new sites at its boundary), whereas the isometries $w$ tend to reduce it (by transforming blocks of sites into single sites). For instance, in the binary 1$D$ MERA, local observables end up supported in three contiguous sites (Fig. \ref{fig:MERAintro:2MERA}), whereas in the ternary 1$D$ MERA local observables become supported in two contiguous sites (Fig. \ref{fig:MERAintro:3MERA}).

Therefore, an important difference between the binary and ternary 1$D$ schemes is in the natural support of local observables. This can be seen to imply that the cost of a computation scales as a larger power of the bond dimension $\chi$ for the binary scheme than for the ternary scheme, namely as $O(\chi^9)$ compared to $O(\chi^8)$. However, it turns out that the binary scheme is more effective at removing entanglement, and as a result a smaller $\chi$ is already sufficient in order to achieve the same degree of accuracy in the computation of, say, a ground state energy. In the end, we find that for the 1$D$ systems analyzed in Sect. \ref{sect:MERAalg:benchmark}, the two effects compensate and the cost required in both schemes in order to achieve the same accuracy is comparable. On the other hand, in the ternary 1$D$ MERA, two-point correlators between selected sites can be computed at a cost $O(\chi^8)$, whereas analogous calculations in the binary 1$D$ MERA are much more expensive. Therefore in any context where the calculation of two-point correlators is important, the ternary 1$D$ MERA is a better choice.

The number of possible realizations of the MERA for 2$D$ lattices is greater than for 1$D$ lattices. For a square lattice, the two schemes of Fig. \ref{fig:MERAintro:2DAltSchemes} are obvious generalizations of the above ones for 1$D$ lattices. The first scheme, proposed in \cite{evenbly07a} (see also \cite{cincio08}), involves isometries of type $(1,4)$ and the natural support of local observables is a block of $3\times 3$ sites. The second scheme involves isometries of type $(1,9)$ and local observables end up supported in blocks of $2\times 2$ sites. Here, the much narrower causal cones of the second scheme leads to a much better scaling of the computational cost with $\chi$, only $O(\chi^{16})$ compared to $O(\chi^{28})$ for the first scheme.

Another remark in relation to possible realizations concerns the type of tensors we use. So far we have insisted in distinguishing between disentanglers $u$ (unitary tensors of type $p\rightarrow p$) and isometries $w$ (isometric tensors of type $1\rightarrow p'$). We will continue to use this terminology throughout this paper, but we emphasize that a more general form of isometric tensor, e.g. of type $(2,4)$, that both disentangles the system and coarse-grains sites, is possible and is actually used in some realizations \cite{evenbly08b}.

\begin{figure}[!tbhp]
\begin{center}
\includegraphics[width=12cm]{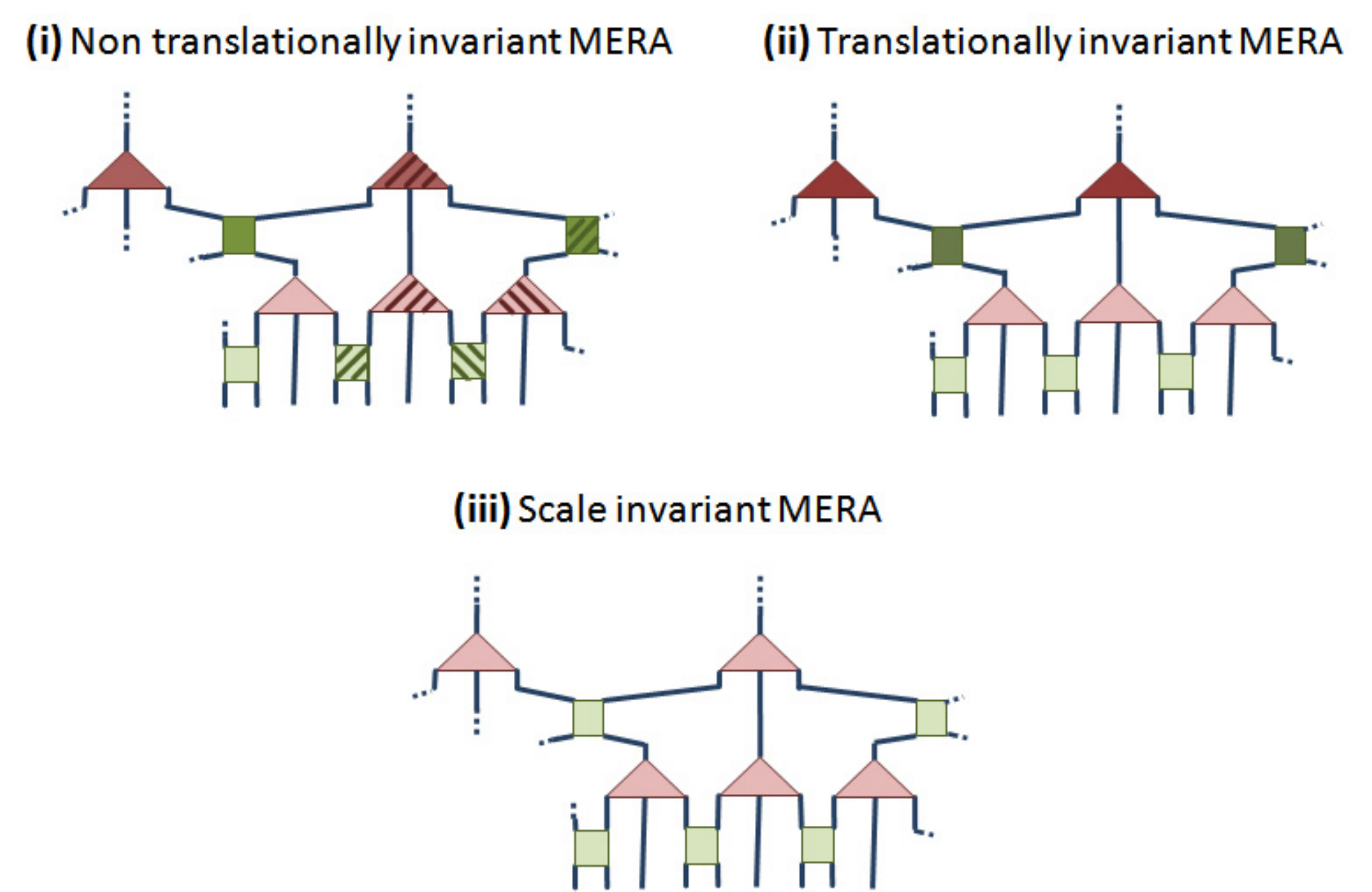}
\caption{Ternary 1$D$ MERA in the presence of space symmetries. (i) In order to represent an inhomogeneous state/subspace, all disentanglers $u$ and isometries $w$ are different (denoted by different colouring). Notice that there are $N/3$ disentanglers (isometries) in the first layer, $N/9$ in the second, and more generally $N/3^{\tau}$ in layer $\tau$, so that the total number of tensors is $2N \sum_{\tau=1}^{\log N} 1/3^{\tau} < 2N$. Therefore the total number of parameters required to specify the MERA is proportional to the size $N$ of the lattice $\mathcal{L}$. (ii) In order to represent a state/subspace that is invariant under translations, we choose all disentanglers and isometries on a given layer of the MERA to be the same. In this case the MERA is completely specified by $O(\log N)$ disentanglers and isometries. (iii) In a scale invariant MERA, the same disentangler and isometry is in addition used in all layers.} 
\label{fig:MERAintro:MERAtypes}
\end{center}
\end{figure}

\section{Exploiting symmetries} \label{sect:MERAintro:Sym}

Symmetries have a direct impact on the efficiency of computations, because they can be used to drastically reduce the number of parameters in the MERA. Important examples are given by space symmetries, such as translation and scale invariance, see Fig. \ref{fig:MERAintro:MERAtypes}.

The MERA is made of $O(N)$ disentanglers and isometries. In order to describe an inhomogeneous state $\ket{\Psi}\in \mathbb{V}_{\mathcal{L}}$ or subspace $\mathbb{V}_{U} \subseteq \mathbb{V}_{\mathcal{L}}$, all these tensors are chosen to be different. Therefore, for fixed $\chi$ the number of parameters in the MERA scales linearly in $N$.

However, in the presence of translation invariance, one can use a \emph{translation invariant} MERA, where we choose all the disentanglers $u$ and isometries $w$ of any given layer $\tau$ to be the same, thus reducing the number of parameters to $O(\log N)$ (if there are $T\approx \log N$ layers). We emphasize that a translation invariant MERA, as just defined, does not necessarily represent a translation invariant state $\ket{\Psi}\in \mathbb{V}_{\mathcal{L}}$ or subspace $\mathbb{V}_{U} \subseteq \mathbb{V}_{\mathcal{L}}$. The reason is that different sites of $\mathcal{L}$ are placed in inequivalent positions with respect to the MERA. As a result, often the MERA can only approximately reproduce translation invariant states/subspaces, although the departure from translation invariance is seen to typically decrease fast with increasing $\chi$. In order to further mitigate inhomogeneities, we often consider an average of local observables/reduced density matrices over all possible sites, as will be discussed in Chap. \ref{chap:MERAalg}.

In systems that are invariant under changes of scale, we will use a \emph{scale invariant} MERA, where all the disentanglers and isometries can be chosen to be the same and we only need to store a constant number of parameters. The scale invariant MERA is useful to represent the ground state of some quantum critical systems \cite{vidal07} and the ground subspace of systems with topological order at the infrared limit of the RG flow \cite{aguado08,konig09}.

A reduction in parameters (as a function of $\chi$) is also possible in the presence of internal symmetries, such as $U(1)$ (e.g. particle conservation) or $SU(2)$ (e.g. spin isotropy). Exploitation of internal symmetries is discussed in Ref. \cite{sukhi09} and shall not be further considered in this thesis.

\section{Conclusion} \label{sect:MERAintro:Conclusion}

In this Chapter, the MERA has been introduced both from the perspective of a perculiar class of quantum circuit and from the perspective of an RG transformation of the lattice. Several implementations of the MERA for $D=1$ and $D=2$ systems have been described and we have briefly discussed how the presence of symmetries can be exploited to reduce the computational cost of the MERA. The binary MERA of Fig. \ref{fig:MERAintro:2MERA} and the 4-to-1 MERA scheme of Fig. \ref{fig:MERAintro:2DAltSchemes} shall be used in Chapters \ref{chap:FreeFerm} and \ref{chap:FreeBoson} to explore the performance of the MERA in $D=1,2$ dimensional lattices of free fermions and free bosons.

\chapter{Entanglement renormalization in free fermionic systems}
\label{chap:FreeFerm}

\section{Introduction}
The \emph{renormalization group} (RG), concerned with the change of physics with the observation scale, is among the main ideas underlying the theoretical structure of statistical mechanics and quantum field theory, and is of central importance in the modern formulation of critical phenomena and phase transitions \cite{fisher98}. Its influence extends well beyond the conceptual domain: RG transformations are also the basis of numerical approaches to the study of strongly correlated many-body systems.

In a lattice model, a real-space RG transformation produces a coarse-grained system by first joining the lattice sites into blocks and then replacing each block with an effective site \cite{kadanoff67}. Two very natural requirements for a RG transformation are: ($i$) it should preserve the long-distance physics of the system; ($ii$) when this physics is invariant under changes of scale, the system should be a fixed point of the RG transformation. 

For the important case of a quantum system at zero temperature, the first requirement is fulfilled if, as determined by White in his density matrix renormalization group (DMRG) \cite{white92,white93}, the vector space of the effective site retains the local support of the ground state. \emph{Entanglement renormalization} (ER) has been proposed in order to simultaneously meet the second requirement. By using \emph{disentanglers}, ER aims to produce a coarse-grained lattice locally identical to the original one, in the sense that their sites have the same vector space dimension. When this is accomplished, the original system and its coarse-grained version can be meaningfully compared, e.g. through their Hamiltonians or ground state properties, leading to a proper real-space RG flow. 

In this Chapter we explore the performance of ER in systems of free spinless fermions on 1$D$ and 2$D$ lattices using the binary MERA and 4-to-1 MERA schemes described in Figs. \ref{fig:MERAintro:1DAltSchemes} and \ref{fig:MERAintro:2DAltSchemes} of Chap. \ref{chap:MERAintro}. Specifically, we consider systems specified by the quadratic Hamiltonian
\begin{equation}
\hat{H} = \sum_{\langle rs \rangle} [a_r^{\dagger}a_s + a_s^{\dagger}a_r - \gamma(a_{r}^{\dagger}a_{s}^{\dagger} + a_{s}a_{r})] - 2\lambda \sum_r a_{r}^{\dagger} a_r,
\label{eq:ferm:Ham} 
\end{equation}
where $\lambda$ and $\gamma$ are the chemical and pairing potentials and the first sum involves only nearest neighbors. In spite of its simplicity, Hamiltonian $\hat{H}$ contains a rich phase diagram as a function of $\lambda$ and $\gamma$, including insulating, conducting and superconducting phases \cite{li06}. Importantly, the corresponding ground states span all known forms of entropy scaling \cite{li06, barthel06}. 
In addition, $\hat{H}$ can be diagonalized through linear (Fourier and Bogoliubov) transformations of the fermion operators $\hat a$ and $\hat a^\dag$ while, by Wick's theorem, all properties of its \emph{Gaussian} ground state $\ket{\Psi_{\mbox{\tiny GS}}}$ can be extracted from the two-point correlators $\left\langle {\hat a_r^\dag  \hat a_s} \right\rangle$ and $\left\langle {\hat a_r  \hat a_s} \right\rangle$. Then, provided that our RG transformation also maps fermion modes linearly, the entire analysis can be conducted in the space of two-point correlators and quadratic Hamiltonians of $N$ fermionic modes, as represented by $N\times N$ matrices. Hence quadratic fermionic models such as Eq. \ref{eq:ferm:Ham}) offer an appealing testing ground for ER, one where computational costs have been greatly simplified (e.g. $\hat{H}$ can be diagonalized exactly with just $O(N^3)$ operations) while keeping a rich variety of non-trivial ground state structures. 

\section{ER applied to free fermions}
We start by rephrasing, in the language of correlation matrices, the process of coarse-graining a $D$-dimensional (hypercubic) lattice. We assume that the system is in the ground state $\ket{\Psi_{\mbox{\tiny GS}}}$ of $\hat{H}$, which we compute using standard analytic techniques (see e.g. \cite{li06}). 
It is convenient to redraw the hypercubic lattice so that each site contains $P\equiv p^D$ fermion modes for some integer $p$. Then a hypercube of $2^D$ sites defines a \emph{block} that contains $P2^{D}$ modes. The goal of the RG transformation is to replace this block with just one effective site made of $P'$ modes, with $P'<P2^{D}$. We would like to have $P'=P$, so that the sites of the coarse-grained and original lattices are identical and we can compare the corresponding Hamiltonians or ground-state reduced density matrices. However, in the coarse-graining step only modes of the block that are disentangled from the rest of the system can be removed. As a result, $P'$ often must be larger than $P$.

For the sake of simplicity, we continue the analysis for the case of a 1$D$ lattice. Let us temporarily replace the $N$ spinless fermion operators $\hat a$ in Eq. \ref{eq:ferm:Ham} with $2N$ (self-adjoint) Majorana fermion operators $\check c$,   
\begin{equation}
\check c_{2r - 1} \equiv \hat a_r + \hat a_r^{\dagger},~~~~~~~~~\check c_{2r} \equiv \frac{\hat a_r - \hat a_r^{\dagger}}{i}. \label{eq:ferm:Majorana}
\end{equation}
The ground state $\ket{\Psi_{\mbox{\tiny GS}}}$ is then completely specified by
\begin{equation}
\left\langle {\check c_r \check c_s } \right\rangle  = \delta _{rs}  + i \Gamma_{rs}, \label{eq:ferm:Gamma} 
\end{equation}
where $\Gamma$, henceforth referred to as the \emph{correlation matrix}, is real and antisymmetric. Similarly, the reduced density matrix $\rho_{\mbox{\tiny GS}}$ for a block made of 2 sites, that is with $L=2P$ spinless modes (equivalently, $2L$ Majorana modes) is described by a $2L\times 2L$ submatrix $\Gamma_{L}$ of $\Gamma$. This matrix is brought into (block) diagonal form by a special orthogonal transformation $V$,
\begin{equation}
 V\Gamma _L V^\dag = \bigoplus \limits_{r = 1}^L \left[ {\begin{array}{cc}
   0 & {v_r }  \\
   { - v_r } & 0  \\
\end{array}} \right], ~~~~  V \in \textrm{SO}(2L),\label{eq:ferm:GammaL} 
\end{equation}
where $0 \le v_r \le 1$ are the eigenvalues of $\Gamma_L$, each one associated with a pair of Majorana fermions. These pairs recombine into $L$ spinless fermions in a product state \cite{vidal03,latorre04} 
\begin{equation}
	\rho_{\mbox{\tiny GS}} = \bigotimes_{r=1}^L\varrho_{r} = \bigotimes_{r=1}^L \left( {\begin{array}{cc}
   \frac{1+v_r}{2} & 0  \\
   0 & \frac{1-v_r}{2}  \\
\end{array}} \right), \label{eq:ferm:rho}
\end{equation}
where $\varrho_r$, the state of a spinless fermion mode, is \emph{mixed} if $v_r<1$ and \emph{pure} if $v_r=1$. Notice that since the ground state $\ket{\Psi_{\mbox{\tiny GS}}}$ is a pure state, a mode in a mixed state must be \emph{entangled} with modes outside the block, whereas a mode in a pure state is \emph{unentangled} from the rest of the system. We build an effective site by removing from the block, or \emph{projecting out} from $\Gamma_L$, all the modes that are unentangled (pure), and just keeping those $P'$ modes that are entangled (mixed). In this way, the coarse-grained lattice retains the ground state properties.

\vspace{0.6cm}
\begin{center}
  \fbox{
    \begin{minipage}{15 cm}
 \begin{center}
 \vspace{0.2cm}
 {\bf The equivalence of Hilbert space truncation and the removal of fermionic modes}
 \end{center}
Here we describe the process of coarse-graining a lattice by replacing blocks of sites with effective sites. We show that the truncation of the Hilbert space of a given block can be implemented by eliminating some of the modes in that block. We consider a fermionic lattice system in its (gaussian) ground state $\ket{\Psi_{\mbox{\tiny GS}}}$. Let $\mathbb{V}$ be the vector space of a block containing $L$ modes and let $\rho_{\mbox{\tiny GS}}$ denote the reduced density matrix of $\ket{\Psi}_{\mbox{\tiny GS}}$ on the block. We assume that the support of $\ket{\Psi}_{\mbox{\tiny GS}}$ is concentrated in a subpace $\mathbb{V}^{\rho} \subset \mathbb{V}$. Then, following White \cite{white92,white93}, the optimal coarse-graining of the block is obtained by defining an effective site $s'$ with vector space $\mathbb{V}^{s'} = \mathbb{V}^{\rho}$. In our case, $\rho_{\mbox{\tiny GS}}$ is the tensor product of density matrices $\varrho_r$ for individual modes \cite{vidal03,latorre04},
\begin{equation}
	\rho_{\mbox{\tiny GS}} = \bigotimes_{r=1}^L\varrho_{r} = \bigotimes_{r=1}^L \left( {\begin{array}{cc}
   \frac{1+v_r}{2} & 0  \\
   0 & \frac{1-v_r}{2}  \\
\end{array}} \right). \label{eq:ferm:rho2}
\end{equation}
Suppose that the first $P'$ modes are in a mixed state and the remaining $L-P'$ modes are in a pure state. Then we can write
\begin{equation}
	\rho_{\mbox{\tiny GS}} = (\bigotimes_{r=1}^{P'}\varrho_{r}) \otimes (\bigotimes_{r=P'+1}^{L}\varrho_{r}) \equiv \sigma \otimes \pi,
\end{equation}
where $\sigma$ is a mixed state with rank $2^{P'}$ whereas $\pi$ is a projector with rank 1. Let $\mathbb{V}= \mathbb{V}^{\sigma}\otimes \mathbb{V}^{\pi}$ be a tensor factorization of $\mathbb{V}$ such that $\sigma = \tr_{\mathbb{V}^{\pi}}(\rho)$. The key observation is that $\mathbb{V}^{\sigma} \cong \mathbb{V}^{\rho}$, and that $\rho$ and $\sigma$ have the same none-vanishing eigenvalues.
Therefore, we have two equivalent ways of constructing the space $\mathbb{V}^{s'}$ for the effective site $s'$ while preserving the support of the ground state density matrix $\rho_{\mbox{\tiny GS}}$. On the one hand, $\mathbb{V}^{s'}$ can be obtained by projecting $\mathbb{V}$ on the support $\mathbb{V}^{\rho}$ of $\rho$. On the other, $\mathbb{V}$ can also be build by factorizing the space $\mathbb{V}$ into two factor spaces $\mathbb{V}^{\sigma}$ and $\mathbb{V}^{\pi}$, and by then tracing out the second factor, corresponding to modes in a pure state. Both constructions lead to an equivalent effective lattice.
Finally, tracing out the factor space of mode $r$ corresponds, in the language of correlation matrices $\Gamma_L$, to removing the $r$th row and a $r$th column of $V\Gamma_L V^{\dagger}$ in Eq. \ref{eq:ferm:GammaL}, process to which we referred to as \emph{projecting out} the mode.
\vspace{0.2cm}
    \end{minipage}
  }
\end{center}
\newpage

\begin{figure}[hptb]
\begin{center}
\includegraphics[width=10cm]{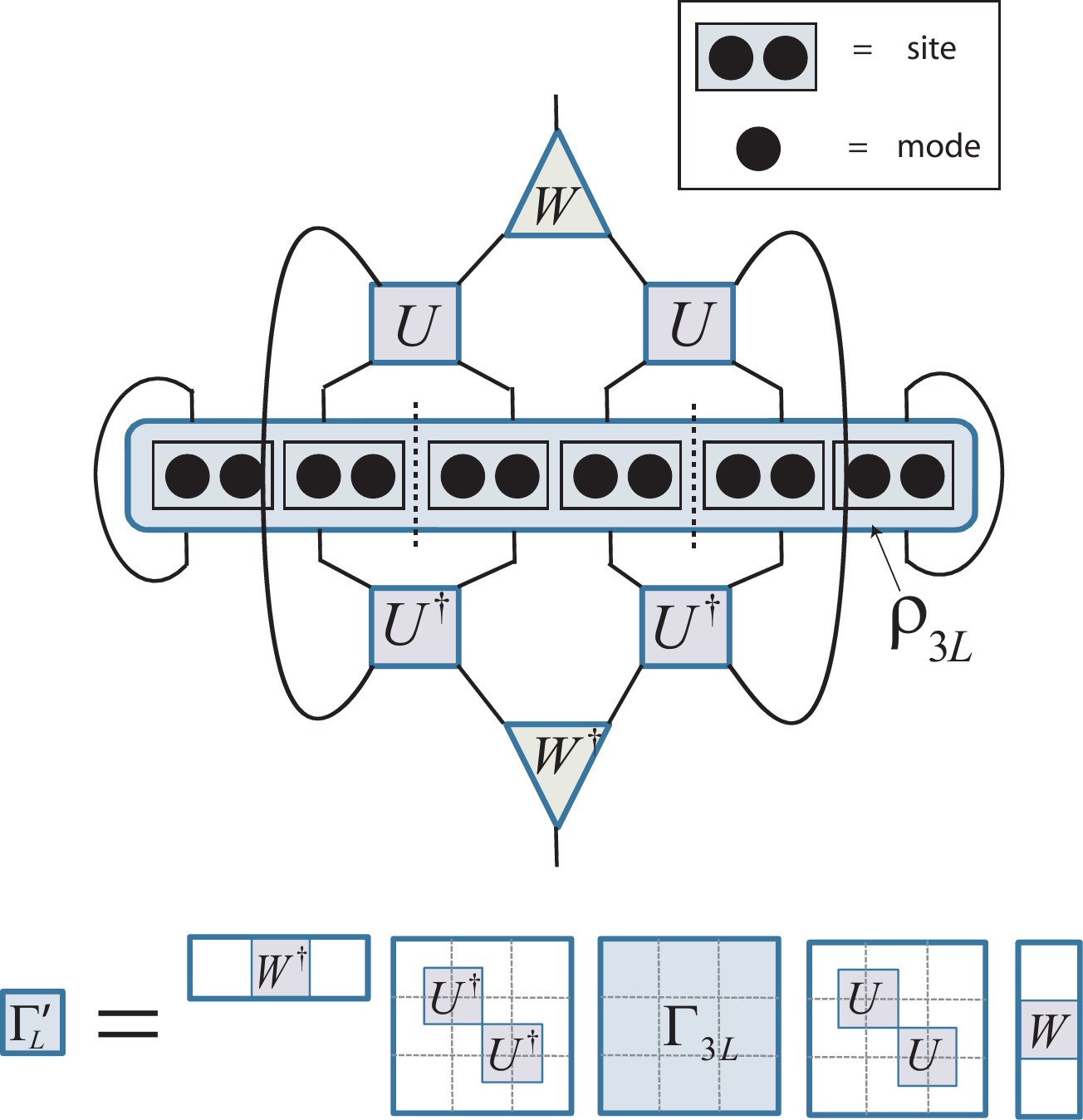}
\caption{\emph{Top:} A block of two sites (four modes) is coarse-grained into an effective site by first applying disentanglers $U$ across the boundary of the block and then using isometry $W$ to project out two modes. \emph{Bottom:} Same RG transformation written in the language of correlation matrices, Eq. \ref{eq:ferm:truncated}.} \label{fig:ferm:disentanglers}
\end{center}
\end{figure}

The key idea of ER, see Fig. \ref{fig:ferm:disentanglers}, is to use disentangling unitary transformations, or \emph{disentanglers}, to diminish $P'$ by increasing the number of modes in the block that are unentangled from the rest of the system. A disentangler is implemented through a special orthogonal matrix $U\in \textrm{SO}(2L)$ that acts on two neigboring sites across the boundary of the block,  wheareas the coarse-graining is implemented by an isometry $W = RY_{P'}$ that selects the $P'$ spinless fermion modes to be kept in the effective site, where $R \in \textrm{SO}(2L)$ and
\begin{equation}
Y_{P'} \equiv \mathop  \bigoplus \limits_{r = 1}^L \left[ {\begin{array}{*{20}c}
   0 & {g_r }  \\
   { - g_r } & 0  \\
\end{array}} \right],\quad g_r  = \left\{ {\begin{array}{*{20}c}
   1 & {r \le P'}  \\
   0 & {r > P'}  \\
\end{array}}. \right.\label{eq:ferm:P}
\end{equation}
Let $\Gamma_{L3}$ describe three consecutive blocks. Then the correlation matrix $\Gamma'_L$ for the effective site reads (Fig. \ref{fig:ferm:disentanglers})
\begin{equation}
\Gamma'_L = W^{\dagger}\left( {U \oplus U} \right)^{\dagger}\Gamma_{3L} \left( {U   \oplus U  } \right)W. \label{eq:ferm:truncated} 
\end{equation} 
Similarly, the correlation matrix $\bar{\Gamma}'_L$  for the modes to be removed is
\begin{eqnarray}
&&\bar{\Gamma}'_L  = \bar{W}^{\dagger}\left( {U \oplus U} \right)^{\dagger}\Gamma_{3L} \left( {U   \oplus U  } \right)\bar{W}, \label{eq:ferm:gone}\\ 
&&\bar{W} \equiv R(Y_L - Y_{P'}), ~~~~~~Y_L\equiv \oplus_{r=1}^L \left[ {\begin{array}{*{20}c}
   0 & 1  \\
   - 1 & 0  \\
\end{array}} \right]
\end{eqnarray} 

Our goal is to maximize the \emph{purity} of the modes to be projected out, so that they become as unentangled as possible. The sum of their purities, $\sum_{r=P'+1}^L v_r$, is half of the antisymmetric trace of $\tilde{\Gamma}'_L$, $\tr ( \tilde{\Gamma}'_L Y_L^{\dagger})$. Consequently, $U$ and $W$ are obtained from the optimization
\begin{equation}
 \max_{U,R\in \textrm{SO}(2N)} ~~\tr \left( \tilde{\Gamma}'_L Y_L^{\dagger}\right), \label{eq:ferm:optimization}
\end{equation}
that we address through a sequence of alternating optimizations for $U$ and $R$ \cite{evenbly08a}. 

\begin{figure}[!tb]
\begin{center}
\includegraphics[width=10cm]{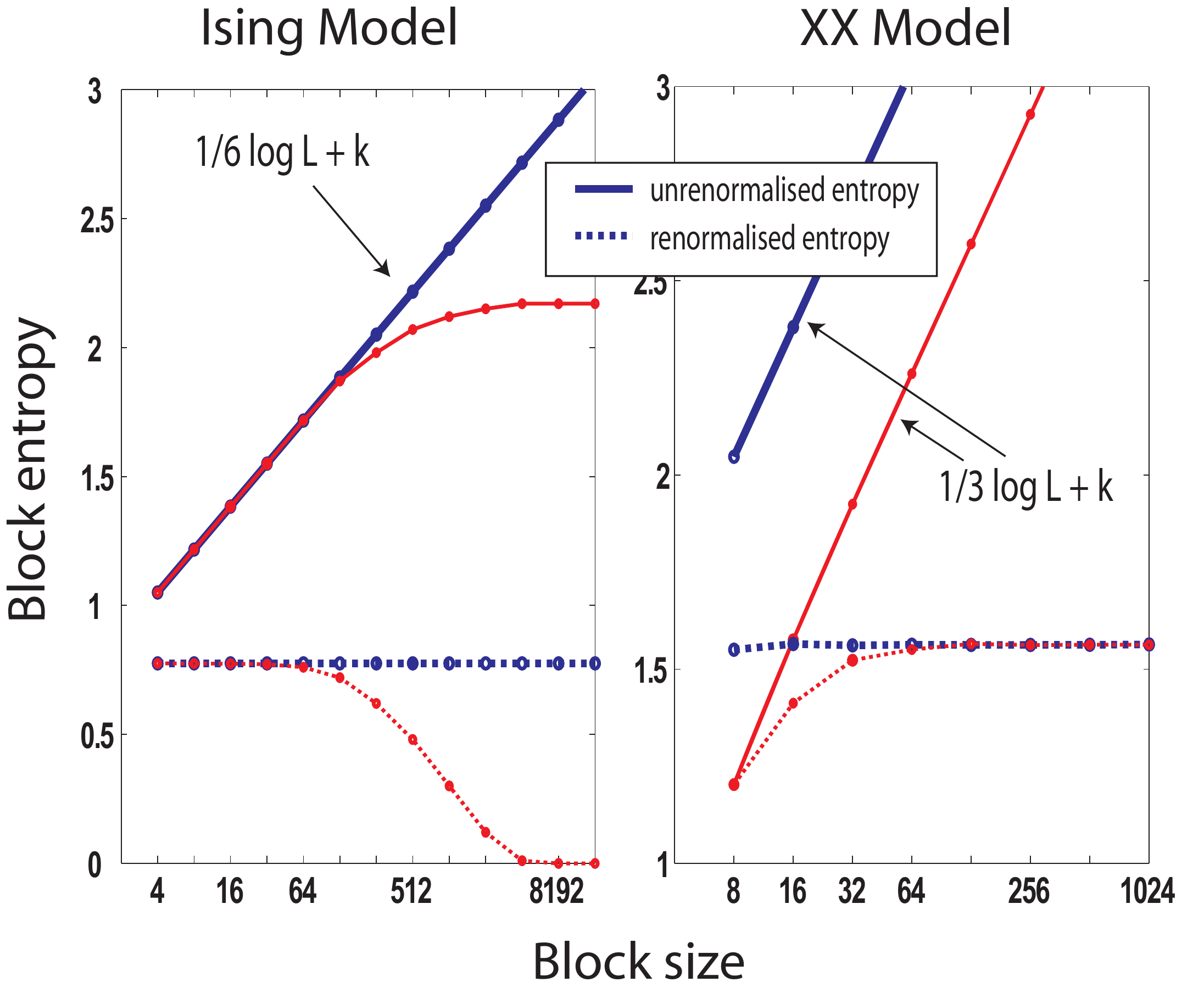}
\caption{ Scaling of the entanglement entropy $S_L$ \cite{vidal03,latorre04} in 1$D$ systems. \emph{Left:} Quantum Ising model, $\gamma=1$. Bold (solid/dotted) lines represent entanglement at criticality, $\lambda=1$. The system is an entangled fixed point of our RG transformation: the correlation matrices $\{\Gamma^{(1)}, \Gamma^{(2)}, \cdots \}$ quickly converge to a fixed $\Gamma_{\mbox{\tiny{Ising}}}^{*}$. In particular, the renormalized entanglement of a block is constant. Thin lines correspond to a non-critical system, $\lambda=1.001$, which the RG flow, as generated by ER, eventually brings a product (unentangled) ground state. \emph{Right:} Quantum XX model, $\gamma=0$. Bold/thin lines represent two critical cases, $\lambda=0$ and $\lambda=\textrm{cos} (15\pi/16)$. They belong to the same universality class and are found to indeed converge to the same correlation matrix $\Gamma_{\mbox{\tiny{XX}}}^{*}$, (with $\Gamma_{\mbox{\tiny{XX}}}^{*} \neq \Gamma_{\mbox{\tiny{Ising}}}^{*}$) and in particular to the same renormalized entropy.}\label{fig:ferm:1D}
\end{center} 
\end{figure}

Then, given the correlation matrix $\Gamma$ for $\ket{\Psi_{\mbox{\tiny GS}}}$, the RG transformation is implemented in three steps: ($i$) first a submatrix $\Gamma_{3L}$ for three consecutive blocks is extracted from $\Gamma$; ($ii$) then disentangler $U$ and isometry $W$ are computed using the optimization (\ref{eq:ferm:optimization}) while keeping $P'=P$ modes in the effective site; ($iii$) finally, $U$ and $W$ are used to transform the original $N$-mode system into a coarse-grained system with just $N/2$ modes and effective correlation matrix $\Gamma^{(1)}$. Some of the modes that are removed are still slightly mixed. Their mixness $\epsilon_r \equiv 1-v_r$ quantifies the errors introduced. Iteration of the RG transformation produces a sequence of increasingly coarse-grained lattices, described by correlation matrices $\{\Gamma^{(1)}, \Gamma^{(2)}, \cdots \}$. The corresponding disentanglers $\{U^{(1)}, U^{(2)}, \cdots\}$ and isometries $\{W^{(1)}, W^{(2)}, \cdots\}$ constitute the \emph{multi-scale entanglement renormalization ansatz} (MERA) \cite{vidal08} for the ground state $\ket{\Psi_{\mbox{\tiny GS}}}$. 

\begin{figure}[tb]
\begin{center}
\includegraphics[width=10cm]{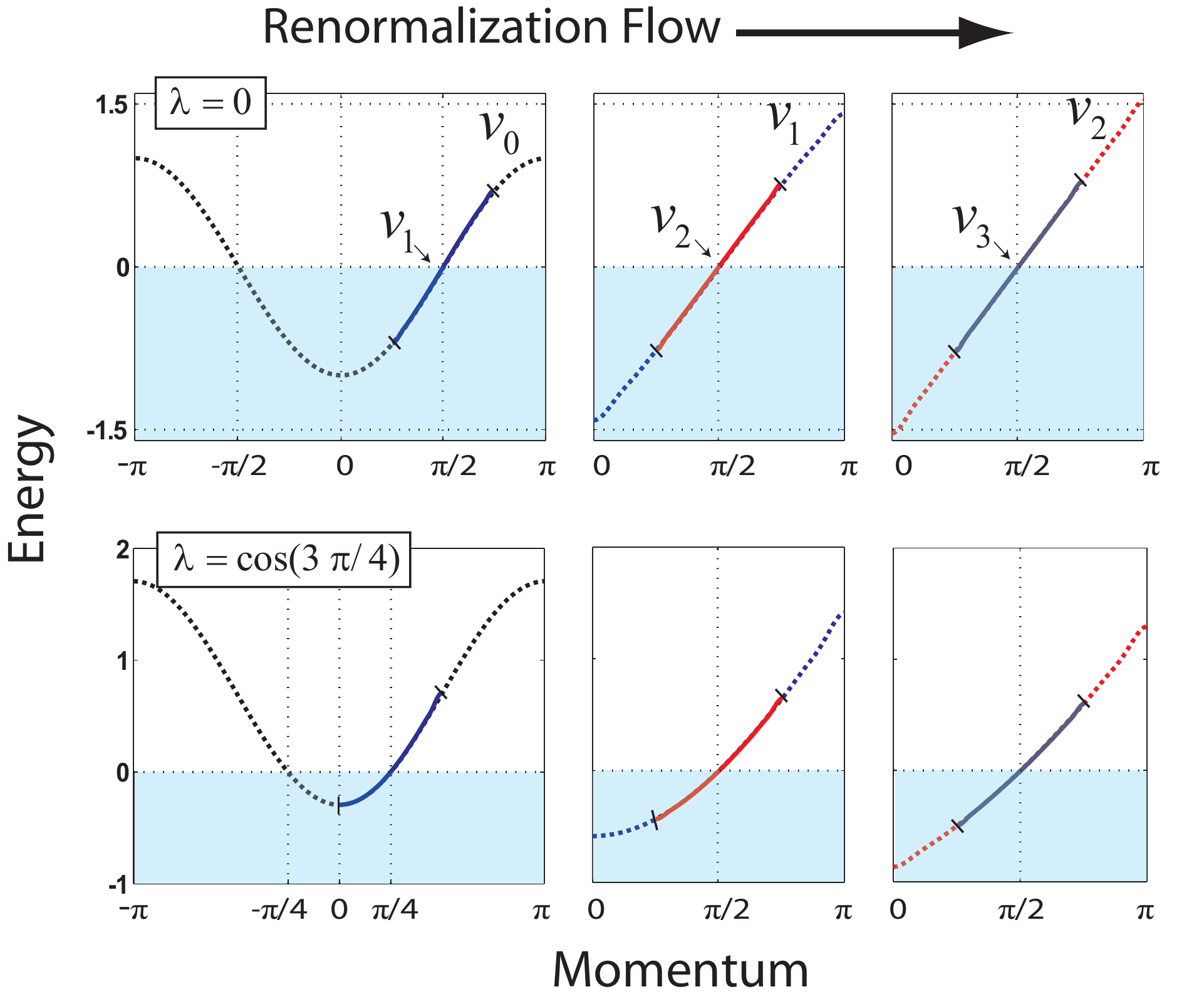}
\caption{Dispersion relation of Hamiltonian (\ref{eq:ferm:Ham}) in 1$D$ with $\gamma=0$, the quantum spin XX model, under successive RG transformations. Shading indicates the Fermi sea. A sequence of local, coarse-grained Hamiltonians is obtained $\{H^{(1)}, H^{(2)}, \cdots \}$ with their corresponding dispersion relations $\{\nu_1,\nu_2, \cdots\}$ converging to a straight line, a fixed point of the RG flow. Convergence is achieved very quickly at half filling ($\lambda=0$) and slower for $\lambda = \textrm{cos}(3\pi/4)$. These results have been obtained by minimizing the energy while keeping $8$ modes in each effective site.} \label{fig:ferm:Ham}
\end{center} 
\end{figure}

\begin{figure}[!tb]
\begin{center}
\includegraphics[width=10cm]{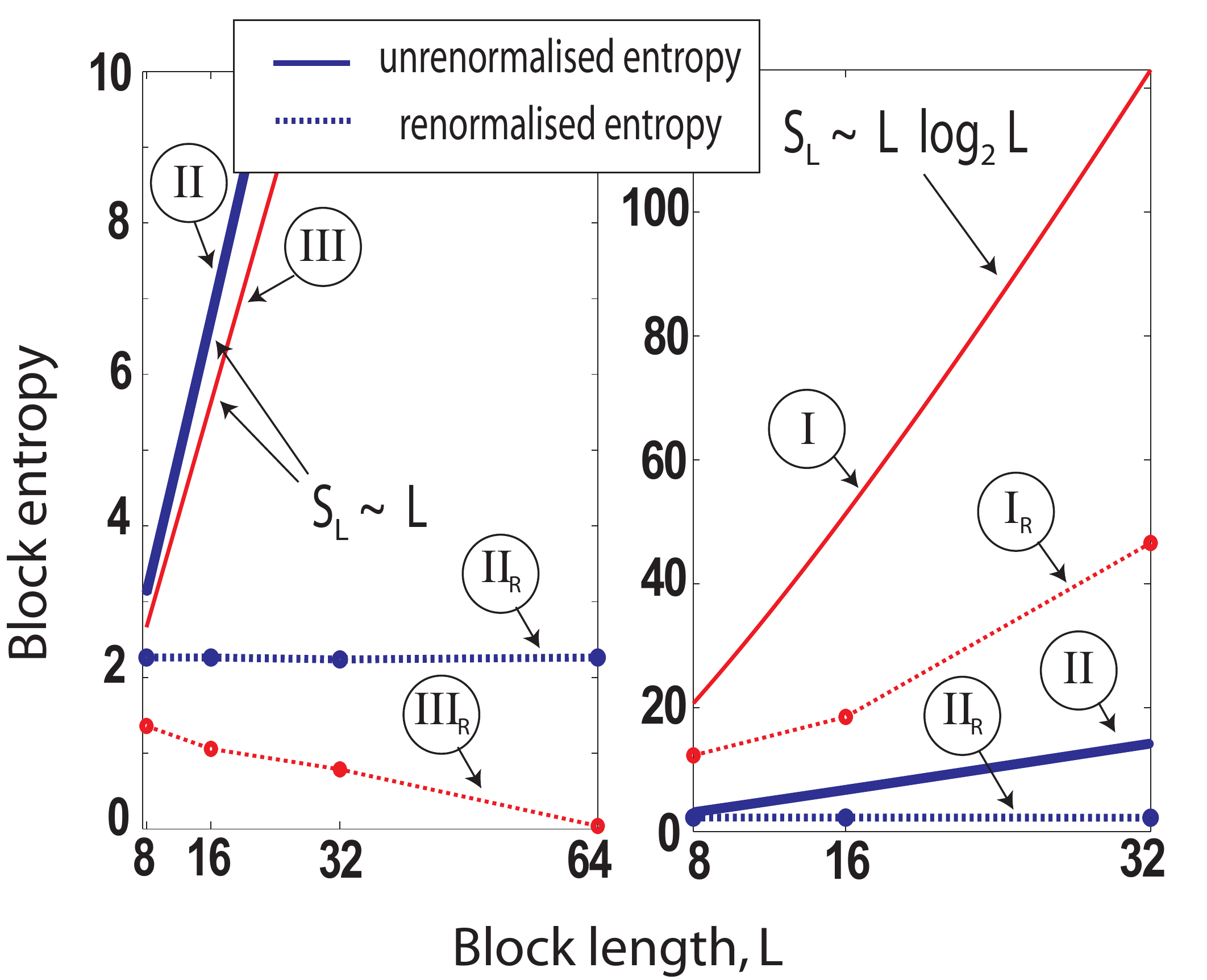}
\caption{Entanglement entropy $S_{L}$ of a block of $L\times L$ modes in 2$D$ models. \emph{Left:} in the critical phase II and the non-critical phase III (bold/fine lines respectively) the entanglement entropy grows linearly with the size $L$ of the boundary of the block, $S_L \sim L$ (boundary law). As in 1$D$, the renormalized entanglement is constant for the critical model and it eventually vanishes for the non-critical model. We have considered $\gamma=1$ and $\lambda=2, \lambda=2.05$ for the critical/non-critical case. \emph{Right:} The critical phase II system $(\gamma,\lambda)=(1,2)$ is replotted for comparison against critical phase I, $(\gamma,\lambda)=(0,0)$, where the system has a 1$D$ Fermi surface and the entanglement entropy has a logarithmic correction, $S_L\sim L\log L$. Here disentanglers are not able to reduce the renormalized entanglement down to a constant.} \label{fig:ferm:2D}
\end{center} 
\end{figure}

\section{Results and discussion}
We have applied the present RG approach to Hamiltonian (\ref{eq:ferm:Ham}) in the thermodynamic limit $N\rightarrow \infty$. First we consider 1$D$ systems, where the whole $(\gamma, \lambda)$ plane can be mapped into the quantum spin XY model using a Jordan-Wigner transformation. ($i$) For the line $\gamma=1$ [equivalent to the quantum spin Ising model] we consider $P=2$ modes per site and apply $13$ iterations of the RG transformation, so that a final effective site (with just $P=2$ modes) corresponds to $2\times 2^{13} = 16384$ modes of the original system. At the critical point $\lambda=1$, which is the most demanding, the mixedness of the removed modes is at most $\epsilon_r = 1.2\times 10^{-4}$. The effect on local observables, even after the 13 iterations, is remarkably small: the error in the critical ground state energy is less than $10^{-7}$, while the two-point correlators $\left\langle {\hat a_r^\dag  \hat a_s} \right\rangle$, reconstructed from the MERA, accumulate a relative error that ranges from $10^{-7}$ for nearest neighbors to $10\%$ for $|r-s| \approx 4,000$. Had we not used disentanglers, the error in the energy would be $10^{-3}$ after only a single RG transformation and an error of $10\%$ in the two-point correlators is already achieved for $|r-s| = 42$. 
($ii$) The line $\gamma=0$ [equivalent to the quantum spin XX model] is critical for $|\lambda| < 1$. Here we consider $P=4$ modes per site and apply again 13 iterations of the RG transformation, reaching sizes of $4\times 2^{13} = 32768$ modes. The errors in energy and correlators are similar to those in the line $\gamma=1$. In both cases, an analysis of the RG flow and its fixed points in terms of entanglement is quite insightful, see Fig. \ref{fig:ferm:1D}. ER can also be used to generate a RG transformation in the space of Hamiltonians, by replacing Eq. \ref{eq:ferm:optimization} with a minimization of the energy. Fig. \ref{fig:ferm:Ham} shows that critical systems are also fixed points of this alternative approach, that preserves the low energy spectrum.

In 2$D$ the model has three phases, denoted I, II and III in Ref. \cite{li06}, where the distinct forms of entanglement scaling were characterized. In phases II (critical, with a Fermi surface consisting of a finite number of points) and III (non-critical, with a gap in the energy spectrum) we are once more able to coarse-grain the system in a quasi-exact, sustainable manner. This is remarkable. The entropy of a square block made of $L^2$ modes grows as the size of its boundary, $S_L \sim L$ \cite{li06}. This implies that the number of modes we should keep in an effective site grows \emph{exponentially} with the number of iterations of the RG transformation, which is precisely why DMRG does not work for large 2$D$ systems. Instead, disentanglers bring this number again down to just a \emph{constant}. As a result one can, in principle, explore systems of arbitrary sizes. In particular, by considering $P = 4^2$ modes per site we apply $\tau =4$ iterations of the RG transformation, with a final block effectively spanning $P\times 4^{\tau+1}= 16384$ modes, whilst maintaining truncation errors of the same scale as the 1$D$ models analyzed, $\epsilon_r = 1.1\times 10^{-4}$. As in the 1$D$ case, the structure of fixed points of the RG flow can be understood in terms of the renormalized entanglement, see Fig. \ref{fig:ferm:2D}. On the other hand, Phase I (critical, with a one-dimensional Fermi surface) is so entangled that ER is no longer able to prevent the growth in the number $P'$ of modes that need to be kept per site. The system displays a logarithmic correction to the entropy, $S_L \sim L\log L$ \cite{li06,barthel06,wolf06,gioev06}, while the MERA can only reproduce a linear scaling $S_L\sim L$  \cite{vidal08} if just a constant number of modes are kept per site, $P'=P$.

\section{Conclusions}

We have presented, in the simplified context of fermion models with quadratic Hamiltonian, unambiguous evidence of the validity of the ER approach in 1$D$ and 2$D$ systems. Similar derivations can be also conducted for bosonic lattice systems with quadratic Hamiltonians, as is investigated in Chap. \ref{chap:FreeBoson}. These results show that the MERA \cite{vidal08} is an efficient description of certain 2$D$ ground states. A number of examples also confirm that ($i$) ER produces a quasi-exact, real-space RG transformation where the coarse-grained lattice is locally equivalent to the original one, enabling the study of RG flow both in the space of ground states and Hamiltonians; ($ii$) non-critical systems end up in a stable fixed point of this RG flow, where the corresponding ground state is a product (i.e. fully disentangled) state, whereas scale invariant critical systems end up in an unstable fixed point, with an entangled ground state. Moreover, ER sheds new light into the ground state structure of two-dimensional systems with a one-dimensional Fermi surface. There, the presence of logarithmic corrections in the entropy $S_L$ of a large block \cite{li06,barthel06,wolf06,gioev06} cannot be accounted for with the MERA with constant number of modes $P$ per level of coarse-graining, hinting for the need for a generalized MERA in order to properly describe such systems.

\chapter[Entanglement renormalization in free bosonic systems]{Entanglement renormalization in free bosonic systems: real-space versus momentum-space renormalization group transforms}
\label{chap:FreeBoson}

\section{Introduction} \label{Sec:Boson:Intro}

The renormalization group (RG) is a set of tools and ideas used to investigate how the physics of an extended system changes with the scale of observation \cite{kadanoff67,wilson75,fisher98,cardy96,shirkov99,delamotte04,shankar94,white92,white93,scholl05}. The RG plays a prominent role in the conceptual foundation of several areas of physics concerned with systems that consists of many interacting degrees of freedom, as is the case of quantum field theory, statistical mechanics and condensed matter theory \cite{kadanoff67,fisher98,cardy96,shirkov99,delamotte04,shankar94}. In addition it also provides the basis for important numerical approaches to study such systems \cite{wilson75,white92,white93,scholl05,morningstar94,morningstar96}.

\begin{figure}[b]
  \begin{center}
    \includegraphics[width=10cm]{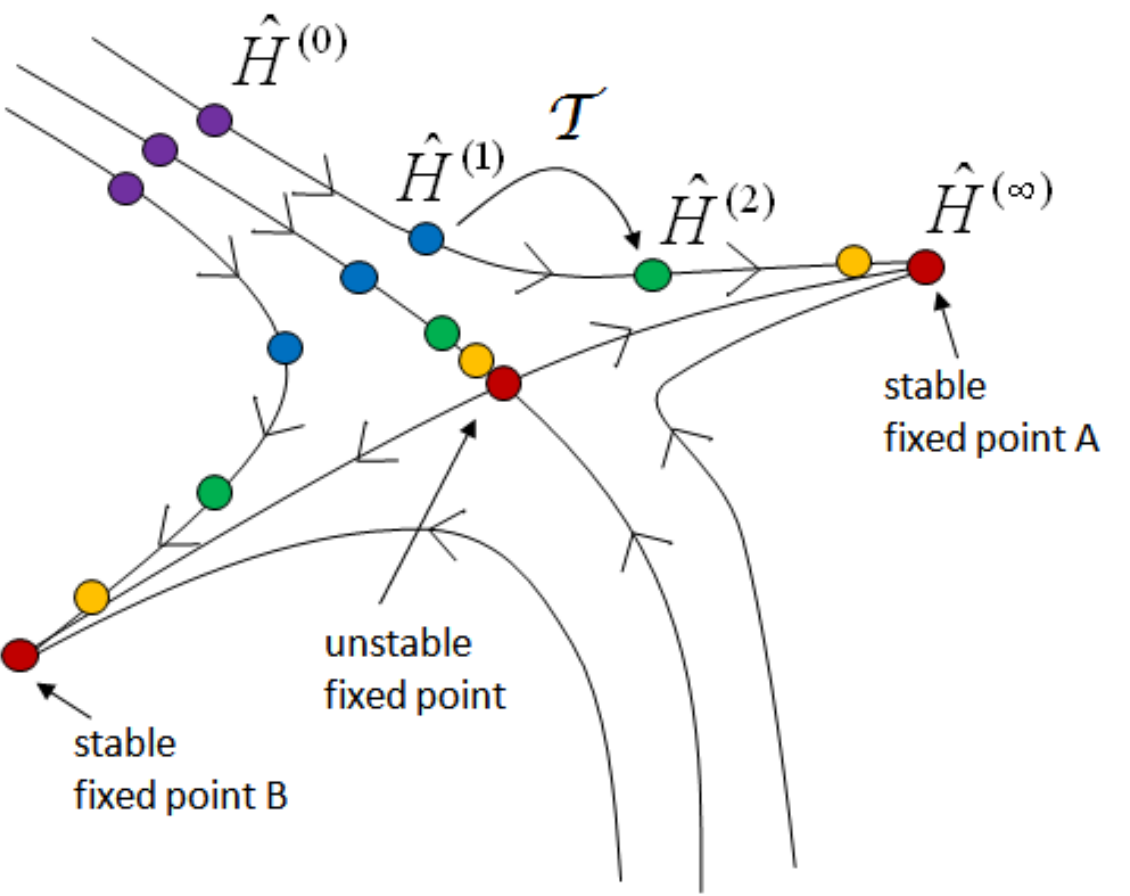}
  \caption{Given a description of an extended system in terms of a Hamiltonian $\hat H^{(0)}$ one may use successive RG transforms $\mathcal T$ to obtain a sequence of Hamiltonians $\left( \hat H^{(0)} , \hat H^{(1)} , \hat H^{(2)} , \ldots \right)$ each an effective theory describing the original system at successively lower energy scales (or longer distances). Characterising the \emph{fixed points} of the RG flow may help provide an understanding e.g. of the low-energy behavior of the system in the thermodynamic limit or of the stability of the system under various perturbations.}
  \label{fig:Boson:RGflow}
 \end{center}
\end{figure}

Given a microscopic description of an extended system in terms of its basic degrees of freedom and their interactions, RG methods aim to obtain an \emph{effective theory}, one that retains only some of these degrees of freedom but is nevertheless still able to reproduce its low energy (or long distance) physics. The effective theory is obtained through coarse-graining transformations that remove those degrees of freedom deemed to be \emph{frozen} at the observation scale of interest. For instance, given a Hamiltonian $\hat H^{(0)}$ for an extended system one may aim to use successive RG transforms to obtain a sequence of coarse-grained Hamiltonians $\left( \hat H^{(0)} , \hat H^{(1)} , \hat H^{(2)} , \ldots \right)$ each an effective theory describing the original system at successively lower energy scales, c.f. Fig. \ref{fig:Boson:RGflow}. Broadly speaking, RG techniques fall into two categories depending on how the coarse-graining is implemented, namely momentum-space RG \cite{delamotte04} and real-space RG \cite{white92,white93,scholl05,morningstar94,morningstar96}. 

Momentum-space RG is applied to theories that are expressed in Fourier space. It works by integrating out high-momentum modes of a field and it is often associated to perturbative approaches. Instead, real-space RG is applied directly to theories that are written in terms of local degrees of freedom, say spins in the case of a spin system defined on a lattice. It is not linked to perturbation theory and can in particular be applied to strongly interacting systems. As proposed by Kadanov \cite{kadanoff67}, the coarse-graining transformation is implemented by replacing a block of spins with a single effective spin, a procedure refined by Wilson \cite{wilson75} and subsequently turned by White into the density matrix renormalization group (DMRG) algorithm \cite{white92,white93, scholl05}, an impressively precise numerical tool to study one-dimensional systems.

A major difficulty of momentum-space RG comes precisely from the fact that it requires, as a starting point, a description of the system in Fourier space. Such description is not always available and might not be obtained easily. Consider for instance a system of interacting spins, as specified by some generic spin-spin interaction. There, obtaining a momentum-space representation might be as difficult as solving the whole theory. In this and many other cases, a RG approach must be performed in real space.

In spite of its indisputable success, the DMRG algorithm \cite{white92,white93, scholl05} suffers from a shortcoming that has important implications. Because of the accumulation of short-ranged entanglement near the boundary of a spin block, the dimension of the Hilbert space used to effectively describe the block must grow with each iteration of the RG transformation. As a result, for instance, unstable fixed points of the RG flow (scale invariant critical systems) cannot be fixed points of the DMRG algorithm. Another, more practical consequence of this growth is that it limits the size of $1D$ critical systems that can be analyzed and, most importantly, it severely limits the success of DMRG computations in higher spatial dimensions. 

Entanglement renormalization (ER) is a real-space RG method proposed in order to overcome the above difficulties \cite{vidal07}. The main feature of ER is the use of \emph{disentanglers}. These are unitary transformations, locally applied near the boundary of a spin block, that remove short-ranged entanglement before the system is coarse-grained. As a result, the effective dimension of the Hilbert space for a spin block can be kept constant under successive RG transformations, so that the approach can be applied to arbitrarily large systems. In this Chapter we explore the ability of entanglement renormalization to produce a sensible RG flow, one with the expected structure of fixed points and flow directions according to momentum-space RG, in $D=1,2$ dimensional harmonic lattice systems. Such systems are an ideal testing-ground for ER. On the one hand, they have well studied properties \cite{plenio05, audenaert02, skrovseth05, cramer06} and can be fully characterized in terms of correlation matrices, a fact that simplifies the analysis and conveniently reduces the computational complexity of ER calculations. On the other hand, an RG analysis of free-particle theories can be conducted simply and without approximations in momentum-space allowing for a comparison between the numerical results obtained using ER and the exact solution. The setting of harmonic lattices also allows for ER to be formulated in the language of bosonic modes, a formalism familiar to researchers in the areas of condensed matter physics and quantum field theory.


The results of this Chapter, presented in Sect. \ref{Sec:Boson:Results}, demonstrate that a real-space RG transform based upon entanglement renormalization is able to reproduce the exact results from momentum-space RG to a high accuracy in $D=1,2$ dimensional lattice systems, both for the critical and non-critical cases considered. Also demonstrated is the ability of the MERA to provide an efficient and accurate representation of the ground state of free boson systems, thus extending to the bosonic case the results of Ref. \cite{evenbly07a} and those of the previous Chapter. These results provide strong evidence that the ER approach, which can be implemented without making use of the special properties of free-particle systems, could be used to investigate low-energy properties of strongly interacting systems not tractable with momentum-space RG approaches \cite{dawson08, rizzi08, evenbly08a}.

\begin{figure}[!tbhp]
  \begin{center}
    \includegraphics[width=12cm]{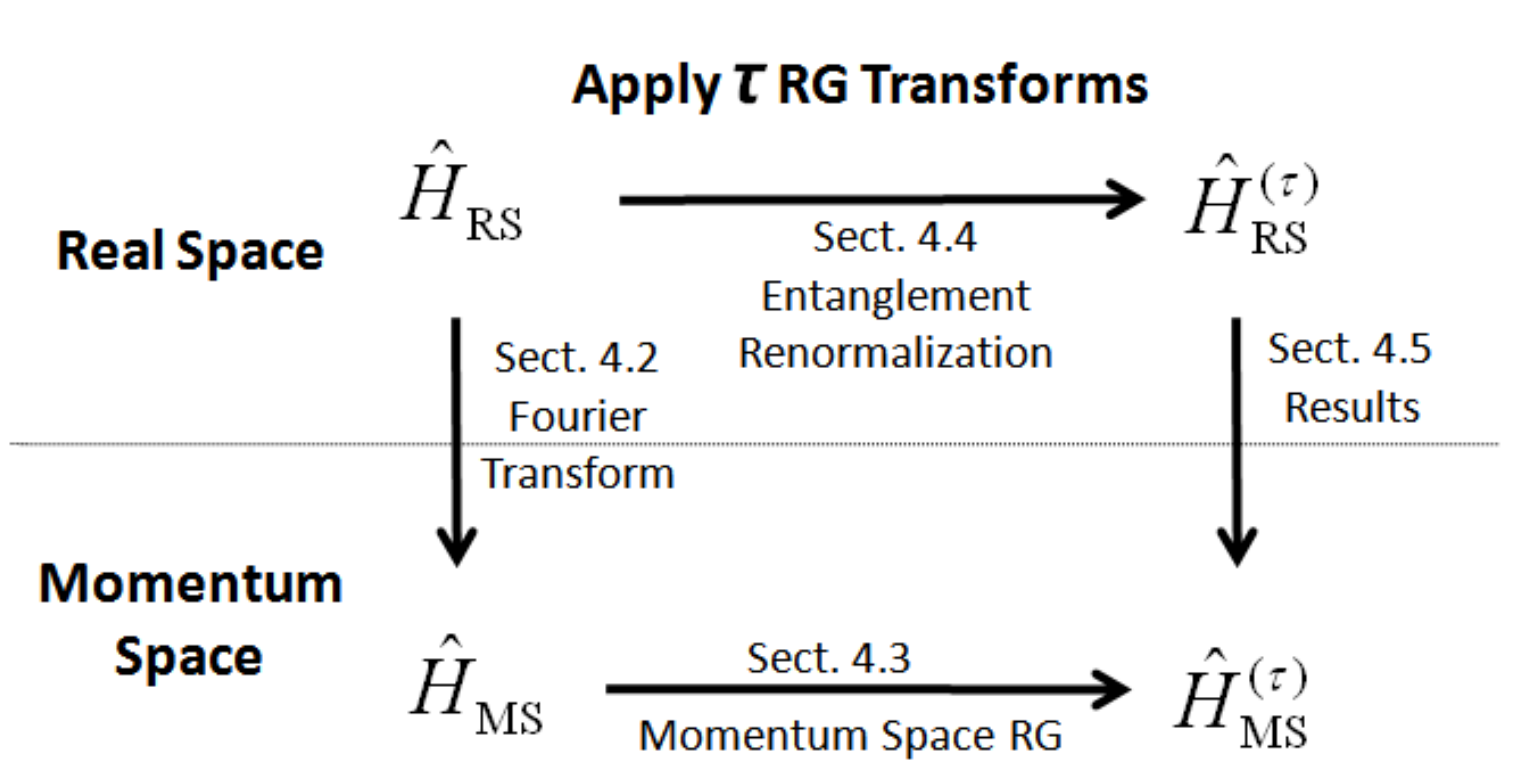}
  \caption{An outline of this Chapter. The harmonic lattice Hamiltonian $\hat H_{RS}$, defined in terms of interaction between local degrees of freedom, may be coarse-grained directly with a numeric implementation of a real-space RG transform, such as entanglement renormalization (ER), as outlined in Sect. \ref{Sec:Boson:RSRG}. Iterating the RG transform $\tau$ times we get the $\tau^\textrm{th}$ effective Hamiltonian $\hat H_{RS}^{(\tau)}$. An effective Hamiltonian may also be obtained by first transforming the Hamiltonian to a momentum-space representation $\hat H_\textrm{MS}$, via Fourier transform of the canonical coordinates, as described in Sect. \ref{Sec:Boson:Harm}. Momentum-space RG transforms may be applied (analytically) to $\hat H_\textrm{MS}$ as described Sect. \ref{Sec:Boson:MRG}. The dispersion relations from the real-space Hamiltonians $\hat H_{RS}^{(\tau)}$ are compared to those of the corresponding momentum-space Hamiltonians, $\hat H_\textrm{MS}^{(\tau)}$ in the results of Sect. \ref{Sec:Boson:Results}.}
  \label{fig:Boson:ER-MSschematic}
 \end{center}
\end{figure}

The Chapter is organized in sections as follows (see also Fig. \ref{fig:Boson:ER-MSschematic}). Sect. \ref{Sec:Boson:Harm} introduces the harmonic systems under consideration. In Sect. \ref{Sec:Boson:MRG} the process of renormalizing the system in momentum-space is explained, highlighting conceptual features of the RG. The critical system and examples of relevant and irrelevant perturbations are considered. Sect. \ref{Sec:Boson:RSRG} explains the details of the real-space RG implementation, both in terms of renormalizing the Hamiltonian and in terms of renormalizing the ground state directly. In Sect. \ref{Sec:Boson:Results} a comparison between results obtained from exact momentum-space RG and the results from numerical real-space RG is presented. 

\begin{figure}[!tbhp]
  \begin{center}
    \includegraphics[width=12cm]{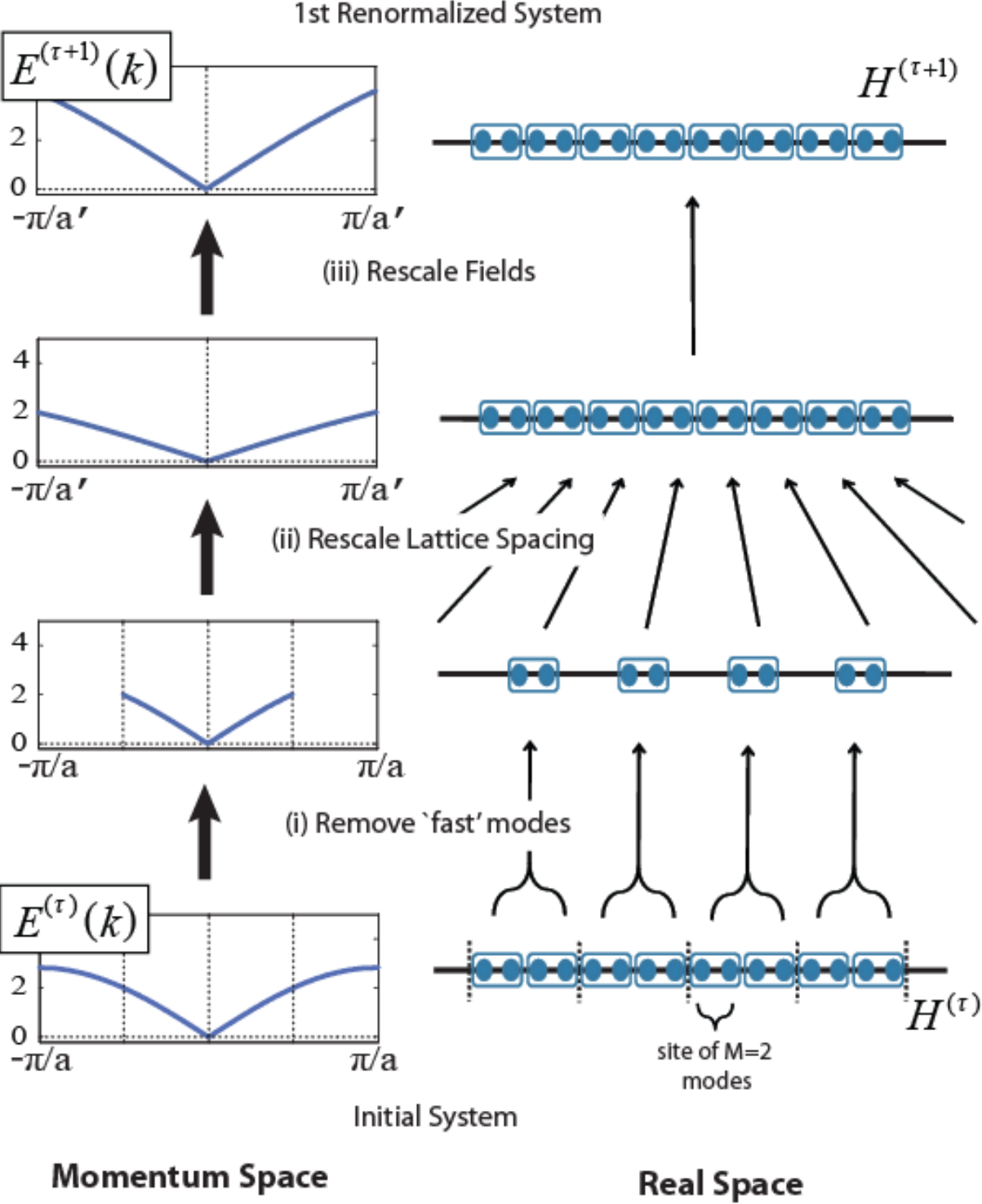}
  \caption{A comparison of an RG iteration for a $D=1$ dimensional system (left) in terms of a dispersion relation, $E(k)$, in momentum-space and (right) in terms of a lattice in real-space. The RG transformation maps the $\tau^\textrm{th}$ effective theory $\hat H^{(\tau)}$ into a new effective theory $\hat H^{(\tau+1)}$ while preserving the low-energy (or long distance) physics of the original theory. The three steps which comprise the iteration are described in detail in Sect. \ref{Sec:Boson:MRG} for momentum-space RG and Sect. \ref{Sec:Boson:HamRG} for the real-space RG. There are many possible ways the real-space coarse-graining step (i) can be implemented, Fig. \ref{fig:Boson:ProjMera} describes two non-equivalent implementations.}
  \label{fig:Boson:ERcompare}
 \end{center}
\end{figure}

\section{Coupled harmonic oscillators} \label{Sec:Boson:Harm}
In this Chapter the low-energy subspaces of harmonic lattices in $D=1,2$ spatial dimensions are to be analysed using both real-space and momentum-space RG transformations. We begin with a brief introduction to harmonic lattices detailing equivalent representations in real-space and momentum-space coordinates, and the Fourier transform that shifts between such descriptions. For clarity the following derivations shall only be presented for the $1D$ system as the generalization to $2D$, or higher dimensional systems, is straight forward. The Hamiltonian for a chain of $N$ harmonic oscillators each with mass $m$, angular frequency $\omega$, and coupled with nearest neighbors via `spring constant' $K$, is written
\begin{align}
 \hat H &= c_0 \sum\limits_{r = 1}^N \left( {\frac{1}{{2m}}\hat p_r^2  + \frac{{m\omega ^2 }}{2}\hat q_r^2  + K\left( {\hat q_{r + 1}  - \hat q_r } \right)^2 }\right)   \nonumber\\ 
  &= \sum\limits_{r = 1}^N \left( {\hat p_r^2  + m^2 \omega ^2 \hat q_r^2  + 2\tilde K\left( {\hat q_{r + 1}  - \hat q_r } \right)^2 }\right) \label{eq:Boson:s1e1}   
\end{align}
where in the second line we have chosen $c_0=2m$ and defined $\tilde K=mK$ for convenience. Note that periodic boundary conditions are assumed. The operators $\hat p_i$ and $\hat q_i$ are the usual canonical coordinates with commutation $\left[ {\hat p_k ,\hat q_l } \right] = i\hbar\delta_{kl}$. In our present considerations it is convenient to focus on the critical (massless) Hamiltonian
\begin{equation}
\hat H_0  = \sum\limits_{r = 1}^N \left( {\hat p_r^2  + 2\tilde K\left( {\hat q_{r + 1}  - \hat q_r } \right)^2 } \right), \label{eq:Boson:s1e2}
\end{equation}
though the non-zero mass case will later be reintroduced in Sect. \ref{Sec:Boson:Rel}. As a preliminary to the momentum-space RG, the Hamiltonian $\hat H_0$ shall be recast into momentum-space variables via Fourier transform of the canonical coordinates. The Fourier-space coordinates $\check p$ and $\check q$ are defined
\begin{align}
 \check p_\kappa   = \frac{1}{{\sqrt N }}\sum\limits_{r = 1}^N {\hat p_r e^{ - 2\pi ir\kappa /N} }  \nonumber\\ 
 \check q_\kappa   = \frac{1}{{\sqrt N }}\sum\limits_{r = 1}^N {\hat q_r e^{ - 2\pi ir\kappa /N} } .  \label{eq:Boson:s1e2b}
\end{align}
Substitution of the Fourier-space coordinates brings the Hamiltonian of Eq. \ref{eq:Boson:s1e2} into diagonal form, here a set of $N$ uncoupled oscillators
\begin{equation}
\hat H_0  = \sum\limits_{\kappa  =  - (N - 1)/2}^{(N - 1)/2} \left( {\check p_\kappa ^2  + 8\tilde K\sin ^2 \left( \frac{\pi \kappa}{ N} \right)\check q_\kappa ^2 }\right) .\label{eq:Boson:s1e3} 
\end{equation}
Defining $k=2\pi\kappa/(aN)$, with constant `$a$' representative of the lattice spacing, the thermodynamic limit ($N\rightarrow \infty$) can be taken
\begin{equation}
\hat H_0  = \int\limits_{k =  - \pi/a }^{\pi/a}  \left( {\check p\left( k \right)^2  + 8\tilde K\sin ^2 \left( \frac{ka}{2} \right)\check q\left( k \right)^2 }\right)dk . \label{eq:Boson:s1e4}
\end{equation}
The theory has a natural momentum cut-off $\Lambda=\pm \pi/a$ originating from the finite lattice spacing; taking the lattice spacing `$a$' to zero recovers the continuum limit with corresponds to the field theory of a real, massless scalar field. Two equivalent representations of the harmonic chain have been obtained; that of Eq. \ref{eq:Boson:s1e2} written in terms of spatial modes (amenable to numeric, real-space RG) and that of Eq. \ref{eq:Boson:s1e4} written in terms of momentum modes (amenable to analytic, momentum-space RG). The \emph{dispersion relation} $E_0(k)$ of the system, which describes the energy of momentum mode $k$ and is known from the solution to a single oscillator, is given by
\begin{equation}
E_0 \left( k \right) = 2\sqrt 2 \tilde K\left| {\sin \left( {ka/2} \right)} \right|. \label{eq:Boson:s1e5}
\end{equation}
The RG transformations of the Hamiltonian shall be chosen such that the resulting \emph{effective theory} preserves the low-energy structure of the original theory, or equivalently, preserves the small $k$ part of the dispersion relation. As the low-energy part of the dispersion is gapless and linear we would expect that, under the RG flow, the renormalized Hamiltonians would tend to a fixed point that has a linear, gapless dispersion.

\section{Momentum-space RG} \label{Sec:Boson:MRG}
In this section the harmonic system of Eq. \ref{eq:Boson:s1e4}, which has been recast in a Fourier basis, is analysed using momentum-space RG. Although the RG analysis of interacting systems is more complicated that the free particle analysis undertaken here, and often involves perturbation theory, the basic procedure is the same. It is useful to consider the RG transformation as occuring in three steps. (i) Firstly the momentum cut-off is reduced, $\Lambda\mapsto \Lambda'=\Lambda/2$, and modes greater than the cut-off are integrated out of the theory. As there is no interaction between the momentum modes of Eq. \ref{eq:Boson:s1e4}, this step is presently very simple; the cut-off is reduced $\Lambda\mapsto\Lambda'=\Lambda/2$ whilst leaving the form of the Hamiltonian for modes with momentum $k<\Lambda'$ unchanged
\begin{equation}
\hat H'_0  = \int\limits_{k =  - \pi /2a}^{\pi /2a} \left( {\check p\left( k \right)^2  + 8\tilde K\sin ^2 \left( \frac{ka}{2} \right)\check q\left( k \right)^2 }\right) dk. \label{eq:Boson:s2e1}
\end{equation}
(ii) Next the length associated to the system is changed, this can be implemented as a scaling of the lattice spacing\footnote{The approach of rescaling the lattice spacing is common in the context of condensed matter problems; equivalently we could have rescaled the momentum of the theory, $k\mapsto k'=2k$, as is the approach most often used for the RG in a quantum field theory setting.}, $a\mapsto a'=2a$, which gives
\begin{equation}
\hat H''_0  = \int\limits_{k =  - \pi/a' }^{\pi/a'} \left( {\check p\left( {k} \right)^2  + 8\tilde K\sin ^2 \left( \frac{ka'}{4} \right)\check q\left( {k} \right)^2 }\right) dk. \label{eq:Boson:s2e2}
\end{equation}
A change has been made to the observation scale of the system in terms of \emph{length}. Next a change in the observation scale is made in terms of \emph{energy}. Indeed, in the final step (iii) the fields are rescaled
\begin{align}
 \check p\left( {k} \right) \mapsto \check p'\left( {k} \right) &= \frac{1}{\sqrt{2 }}\check p\left( {k} \right) \nonumber\\ 
 \check q\left( {k} \right) \mapsto \check q'\left( {k} \right) &= \frac{1}{\sqrt{2 }}\check q\left( {k} \right) \label{eq:Boson:s2e3},
\end{align}
so that the new field operators have a modified commutation relation $[p'(k),q'(k)]=i\hbar/2$, in accordance with the desired change of energy scale. In principle the RG transformation is complete, however in this instance a further transform is required to recast the critical Hamiltonian into a manifestly invariant form. The field operators are rescaled once more 
\begin{align}
 \hat p' \mapsto \hat p'' &= \sqrt{2} \hat p' \nonumber\\ 
 \hat q' \mapsto \hat q'' &= \frac{1}{{\sqrt{2}}}\hat q'.  \label{eq:Boson:s2e4}
\end{align}
In contrast to the previous transform of Eq. \ref{eq:Boson:s2e3}, this transform is \emph{commutation preserving} hence does not affect the physics of the system; namely the dispersion relation remains unchanged. Implementing the third step, together with the auxiliary transform of Eq. \ref{eq:Boson:s2e4}, the first renormalized Hamiltonian $\hat H_0^{(1)}$ is given
\begin{equation}
\hat H_0^{(1)}  = \int\limits_{k =  - \pi/a' }^{\pi/a'}  \left( {\check p''\left( {k} \right)^2  + 32\tilde K\sin ^2 \left( \frac{ka'}{4} \right)\check q''\left( {k} \right)^2 }\right) dk. \label{eq:Boson:s2e5}
\end{equation}
The RG transformation is summarized: (i) the degrees of freedom that are not relevant to the low energy physics are removed, followed by a changes of observation scale in terms of (ii) \emph{length} ($a\mapsto a'=2a$) and (iii) \emph{energy} ($\hbar\mapsto \hbar'=\hbar/2$). Starting from a theory with a natural length scale, the lattice spacing $a$, the RG transform is thus used to derive an effective theory with the new length scale $a'$. The scale factors chosen in steps (ii) and (iii) may depend on the implementation of the RG as well as the problem to which it is being applied. Iterating the RG transform (dropping the `primes' from notation), the $\tau^\textrm{th}$ renormalized Hamiltonian is 
\begin{equation}
 \hat H_0^{(\tau)}  = \int\limits_{k =  - \pi/a }^{\pi/a}  \left( {\check p\left( k \right)^2  + 2^{2\tau + 3} \tilde K\sin ^2 \left( \frac{ka}{2^{\tau + 1} } \right)\check q\left( k \right)^2 }\right) dk,  \label{eq:Boson:s2e6a}
\end{equation}
with corresponding dispersion relation
\begin{equation}
E_0^{(\tau)} \left( k \right) = 2^{\tau + \frac{3}{2}} \sqrt{\tilde K}\left| {\sin \left( {ka/2^{\tau + 1} } \right)} \right| .\label{eq:Boson:s2e6b}
\end{equation}
In the limit of infinitely many transforms, $\tau\rightarrow\infty$, we get the Hamiltonian $H_0^{(\infty)}$ at the fixed point of the RG flow 
\begin{equation}
H_0^{(\infty)} = \int\limits_{k =  - \pi/a }^{\pi/a}  \left( \check p\left( k \right)^2  + \tilde K \left( k^2 a^2 \right)\check q\left( k \right)^2 \right) dk \label{eq:Boson:s2e6}. 
\end{equation}
The fixed point Hamiltonian has a purely linear, gapless dispersion
\begin{equation}
  E_0^{(\infty)}\left( k \right)= a\sqrt {2 \tilde K} \left| k \right|  .\label{eq:Boson:s2e7} 
\end{equation}
as anticipated.

\subsection{Relevant perturbation} \label{Sec:Boson:Rel}
In the previous section the massless harmonic lattice of Eq. \ref{eq:Boson:s1e4}, a critical system, was shown to be a (non-trivial) fixed point of the RG flow. We now consider the stability of this Hamiltonian under the addition of perturbations. Perturbations to the critical theory can be classified as being relevant or irrelevant depending on whether the deviations from the fixed point, induced by the perturbations, grow or diminish under the RG flow. In order to study relevent perturbations a mass term $\hat H_{{\rm{rel}}}$ is reintroduced to the critical system of Eq. \ref{eq:Boson:s1e4}; as shall be shown shortly this term \emph{grows} under the RG flow. Hence, it significantly modifies the low-energy physics from that of the unperturbed system. The mass term is diagonal in both real-space and momentum-space representations
\begin{equation}
 \hat H_{{\rm{rel}}} \equiv \sum\limits_{r} {\hat q_r^2 } = \int\limits_{k =  - \pi/a }^{\pi/a}  { \check q\left( k \right)^2 dk}. \label{eq:Boson:s3e1}  
\end{equation}
The perturbed Hamiltonian $\hat H_m = \hat H_0 + m^2 \hat H_\textrm{rel}$ is equal to the original Hamiltonian of Eq. \ref{eq:Boson:s1e1} describing coupled harmonic oscillators of mass $m$. Since this perturbation term does not reorder mode energies ($E(k)$ is still an increasing function of $|k|$) the same RG transformations may be performed on $\hat H_\textrm{rel}$ as on the unperturbed system $\hat H_0$. This analysis shows that the $\tau^\textrm{th}$ renormalized perturbation term $\hat H_\textrm{rel}^{(\tau)}$ grows exponentially with the RG iteration $\tau$
\begin{equation}
\hat H_\textrm{rel}^{(\tau)}  = 2^{2\tau} \hat H_\textrm{rel},\label{eq:Boson:s3e2} 
\end{equation}
Thus even the addition of even a small mass $m$ to the critical system leads to a large difference between the perturbed $\hat H_m$ and original $\hat H_0$ Hamiltonians after only a few RG iterations; that is to say, the perturbed system tends to a different fixed point of the RG flow. This is further evidenced by consideration of the dispersion relation $E_m^{(\tau)}$ of the perturbed system $\hat H_m$
\begin{align}
 E_m^{(\tau)} \left( k \right) &= 2^\tau \sqrt {m^2  + 8\tilde K\sin ^2 \left( {ka/2^{\tau + 1} } \right)} \nonumber \\ 
  &= 2^\tau m + \frac{{a^2\tilde K}}{{2^\tau m}}k^2 + O\left( {\frac{a^2 \tilde K}{{m^3  2^{3\tau -3} }}} \right).\label{eq:Boson:s3e3}
\end{align}
In contrast to the linear, gapless dispersion $E_0^{(\tau)}$ of the massless theory described by Eq. \ref{eq:Boson:s2e7}, we now see a quadratic dependence of the energy on the momentum $k$, together with an energy gap that grows \emph{exponentially} with the number $\tau$ of RG iterations.

\subsection{Irrelevant perturbation} \label{Sec:Boson:Irrel}
Perturbations that become smaller along the RG flow are termed \emph{irrelevant perturbations}, as these do not affect the asymptotic, low-energy behavior of the system. In this section we construct and analyze an example of such a perturbation. The irrelevant perturbation $\hat H_{{\rm{irrel}}}$ is constructed from neighboring and next-to-nearest neighboring quadratic couplings
\begin{equation}
 \hat H_{{\rm{irrel}}}  \equiv \sum\limits_{r = 1}^N \left( { - \hat q_r^2  + \left( {\hat q_{r + 1}  - \hat q_r } \right)^2  + \frac{1}{4}} \left( {\hat q_{r + 2}  + \hat q_r } \right)^2\right). \label{eq:Boson:s4e1}
\end{equation}
In the Fourier basis $\hat H_{{\rm{irrel}}}$ has representation
\begin{equation}
\hat H_{{\rm{irrel}}} = 4 \int\limits_{k =  - \pi/a }^{\pi/a} {\sin ^4}\left( \frac{ ka}{2}\right) \check q\left( k \right)^2 dk. \label{eq:Boson:s4e1b}
\end{equation} 
We consider the perturbed system $\hat H_\alpha= \hat H_0 +\alpha \hat H_\textrm{irrel}$ with the perturbation strength chosen $\alpha>0$. The $\tau^\textrm{th}$ renormalized perturbation $\hat H_{{\rm{irrel}}}^{(\tau)}$ may be obtained through the same sequence of RG transforms as applied to the unperturbed system
\begin{align}
\hat H_\textrm{irrel}^{(\tau)}  &= \int\limits_{k =  - \pi/a }^{\pi/a}  {2^{2\tau + 2} \sin ^4 \left( \frac{ka}{2^{\tau + 1}} \right)\hat q\left( k \right)^2 dk}  \nonumber\\ 
  &= 2^{ - 2\tau - 2} \int\limits_{k =  - \pi/a }^{\pi/a}  {\left(a^4 k^4+ O\left(\frac{a^6 k^6}{2^{2\tau + 2}}\right) \right) \hat q\left( k \right)^2 dk},  \label{eq:Boson:s4e2} 
\end{align}
The perturbation $\hat H_\textrm{irrel}$ is thus exponentially suppressed under the RG flow. Equivalently, the addition of this term to the critical system has an effect on the low-energy physics that diminishes with each RG iteration, as is seen directly from the dispersion $E_\alpha^{(\tau)}$ of the perturbed system $\hat H_\alpha$
\begin{align}
 E_{\alpha}^{(\tau)} (k) &= 2^{\tau + 1} \left| {\sin \left( \frac{ka}{2^{\tau + 1} } \right)} \right|\sqrt {2\tilde K + \alpha \sin ^2 \left( \frac{ka}{2^{\tau + 1} } \right)} \nonumber\\ 
  &= a\sqrt {2\tilde K} \left| k \right| + O\left( {2^{ - 2\tau} } \right),  \label{eq:Boson:s4e3}
\end{align}
which converges to the same linear, gapless fixed point as did the unperturbed system of Eq. \ref{eq:Boson:s2e7}. 

\begin{figure}[!tb]
  \begin{center}
    \includegraphics[width=10cm]{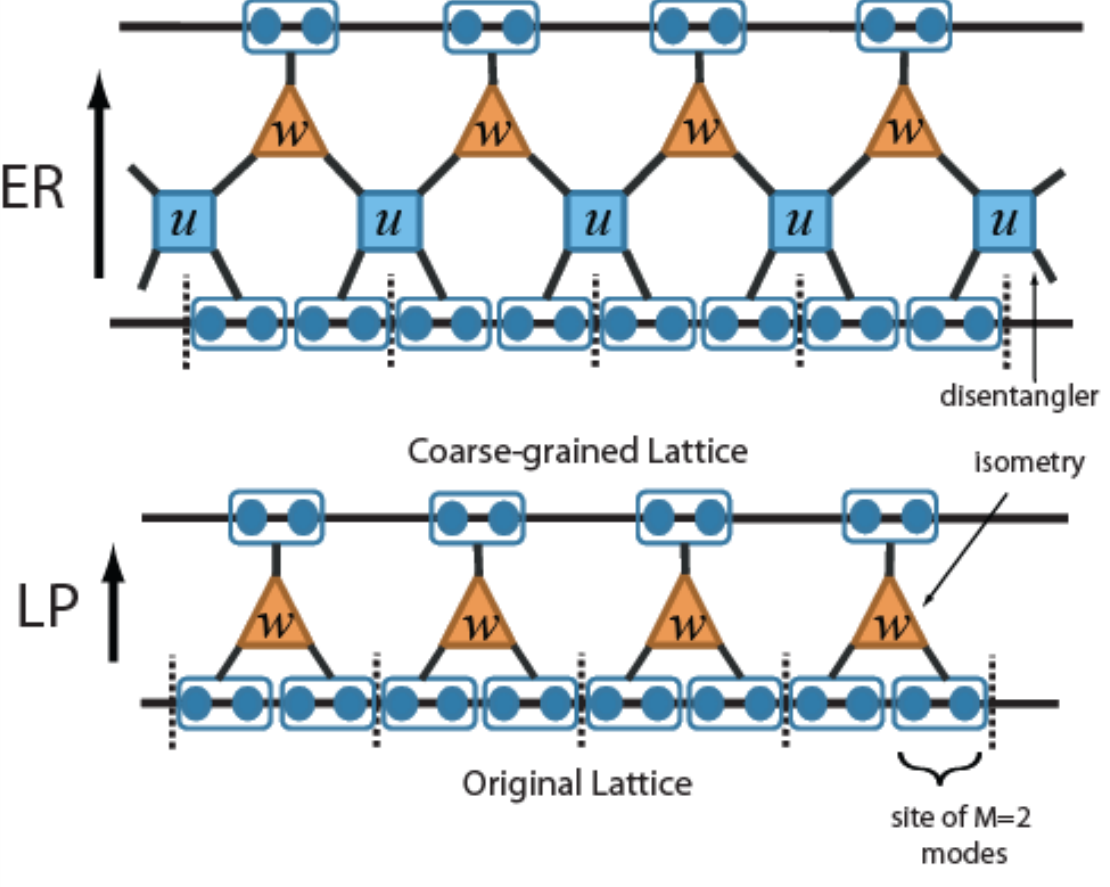}
  \caption{The coarse-graining step of the real-space RG, in which a block of two sites (each composed of $M$ bosonic modes) of the original lattice $\mathcal L$ is mapped to a single site of the coarse-grained lattice $\mathcal L '$, can be accomplished in different ways. (Bottom) The simplest method involves applying a series of local projections, realized by isometric tensors $w$, whose function is to select and retain relevant block degrees of freedom. This shall subsequently be referred to as the local projection (LP) method. (Top) Entanglement Renormalization (ER) differs from the LP coarse-graining by the inclusion unitary disentanglers $u$, enacted across block boundaries, before the truncation with isometries. Here we follow the binary coarse grainging scheme introduced in Chap. \ref{chap:MERAintro}.}
  \label{fig:Boson:ProjMera}
 \end{center}
\end{figure}

\section{Real-space RG} \label{Sec:Boson:RSRG}
In this section an implementation of real-space RG based upon coarse-graining transformations of the lattice is described. Following the seminal works of Migdal, Kadanoff, and Wilson in real-space RG \cite{kadanoff67, wilson75}, the coarse-graining transformation maps a \emph{block} of sites from the original lattice $\mathcal L$ into a single site of a coarser lattice $\mathcal{L}'$. Let us consider a $1D$ lattice $\mathcal{L}$ of $N$ sites, each site described by a vector space $\mathbb V$. We divide $\mathcal L$ into blocks of two sites and, following Wilson, implement a coarse-graining transformation by means of an isometry $w$
\begin{equation}
w: \mathbb V' \mapsto \mathbb V^{\otimes 2},\ \ w^{\dag} w =I_\mathbb{V'} \label{eq:Boson:s5e1}
\end{equation}
where $\mathbb V^{\otimes 2}$ is the vector space of two sites, $\mathbb V'$ is the vector space of a site in the coarser lattice $\mathcal L'$ of $N'=N/2$ sites and $I_\mathbb{V'}$ is the identity in $\mathbb V'$. This coarse-graining transformation, which we shall refer to as a \emph{local projection} (LP) transformation, is depicted graphically in Fig. \ref{fig:Boson:ProjMera}. From an initial Hamiltonian $\hat H$ defined on lattice $\mathcal L$ we can obtain an effective Hamiltonian $\hat H'$ on lattice $\mathcal L'$ via the transformation
\begin{equation}
\hat H'=W^\dag \hat H W,\ \  W = w^{\otimes N/2}. \label{eq:Boson:s5e2}
\end{equation}
The LP transformation has the property of preserving locality of operators; for instance if the original Hamiltonian was a sum of two-body interactions, $\hat H = \sum\nolimits_{i = 1}^N {h_{i,i + 1} }$, then the effective Hamiltonian would remain a sum of two-body interactions, $\hat H' = \sum\nolimits_{i = 1}^{N'} {h'_{i,i + 1} }$. An alternative coarse-graining transformation, known as entanglement renormalization (ER), follows as a modification of the LP scheme. As with the LP scheme, we map two sites of $\mathcal L$ into a single effective site of $\mathcal L'$ via an isometry $w$, however in ER one first enacts unitary \emph{disentanglers} $u$
\begin{equation}
u: \mathbb V^{\otimes 2} \mapsto \mathbb V^{\otimes 2},\ \  u^{\dag} u = u u^\dag =I_{\mathbb{V}^{\otimes 2}} \label{eq:Boson:s5e3}
\end{equation}
across the boundaries of adjacent blocks, as show Fig. \ref{fig:Boson:ProjMera}. Thus the effective Hamiltonian $\hat H'$, as given by a transformation with entanglement renormalization, is  
\begin{equation}
\hat H'=\left( W^\dag U^\dag \right) \hat H \left( U W \right) ,\ \  U = u^{\otimes N/2}. \label{eq:Boson:s5e4}
\end{equation}
Inclusion of the disentanglers in the RG scheme, although having the unfortunate effect of increasing the computational cost of numeric implementation, has profound implications regarding the ability of the method to produce a sensible real-space RG flow, as shall be demonstrated in the results of Sect. \ref{Sec:Boson:Results}. 

Starting from the original Hamiltonian $( \mathcal{L}^{(0)} ,\hat H^{(0)} ) \equiv ( \mathcal{L} ,\hat H )$ one could, by iteration of the RG map, generate a sequence of effective Hamiltonians defined on successively coarser lattices  
\begin{equation}
\left( \mathcal{L}^{(0)}, \hat H^{(0)} \right) \stackrel{\mathcal{T}_0}{\rightarrow} \left( \mathcal{L}^{(1)}, \hat H^{(1)} \right) \stackrel{\mathcal{T}_1}{\rightarrow} \left( \mathcal{L}^{(2)}, \hat H^{(2)} \right) \stackrel{\mathcal{T}_2}{\rightarrow} \ldots \label{eq:Boson:s5e5},
\end{equation}
with the transformation $\mathcal T$ representing either a LP or ER transformation. The LP transformation $\mathcal T_\textrm{LP}^{(\tau)}$ is characterised by the corresponding isometry $w^{(\tau)}$, while the ER transformation $\mathcal T_\textrm{ER}^{(\tau)}$ is characterised jointly by an isometry and a disentangler, $(w^{(\tau)},u^{(\tau)})$. Equation \ref{eq:Boson:s5e5} can be used to directly investigate how $\hat H$ changes under scale transformations allowing e.g. one to characterise the stability of $\hat H$ under perturbations or investigate properties of the system in the thermodynamic limit directly. 

In order to make meaningful comparisons between effective Hamiltonians defined at different length scales, say between $\hat H^{(\tau)}$ and $\hat H^{(\tau+1)}$ defined on lattices $\mathcal L^{(\tau)}$ and $\mathcal L^{(\tau+1)}$ respectively, it is required that the dimension of the Hilbert space of a site in $\mathcal L^{(\tau+1)}$ remains the same as that of $\mathcal L^{(\tau)}$, or equivalently that $\mathbb V' = \mathbb V$ in Eq. \ref{eq:Boson:s5e1}. This ensures that, for instance, two-body operators defined on $\mathcal L^{(\tau)}$ and $\mathcal L^{(\tau+1)}$ exist in the same parameter space, allowing a direct comparison between the coefficients that describe the operators. Keeping the dimension of effective lattice sites in lattice $\mathcal L^{(\tau)}$ constant between RG transforms is also desirable for computational purposes. Indeed, if the dimension of effective sites were to grow with each RG iteration then the computational cost of implementing the transformations would also grow, limiting the number of transforms which may be performed. An investigation of whether the LP and ER transformations can accurately coarse-grain Hamiltonians over repeated RG iterations while keeping a fixed local dimension is a focus of this Chapter.

Note that, in general, the disentanglers and isometries $(u,w)$ that best preserve the low-energy space of a Hamiltonian $\hat H^{(0)}$ under coarse-graining will depend on the specific Hamiltonian $\hat H^{(0)}$ itself; in practice optimisation techniques such as \cite{dawson08, rizzi08, evenbly08a}, or the techniques presented in Chap. \ref{chap:MERAalg}, are required to compute the tensors $(u,w)$. There are however some systems where analytic expressions for these tensors may be obtained \cite{aguado08, konig09}.

\subsection{RG of the harmonic lattice} \label{Sec:Boson:HamRG}
We now turn our attention to the analysis of the harmonic lattice system with ER\footnote{The LP coarse-graining may be viewed as a simplification of ER in which the disentanglers $u$ are set to identity transforms. Therefore only implementation of ER need be explicitly described.}. Following the ideas discussed in the previous section, several possible algorithms \cite{dawson08, rizzi08, evenbly08a} could be applied to study the low-energy subspace of the harmonic systems directly. Application of these algorithms requires only that the Hamiltonian in question is composed of a sum of local interaction terms, as is the case with the systems we consider. However, as this study is focused on Hamiltonians containing only \emph{quadratic couplings}, this property can be exploited in order to significantly reduce the cost of the numerical RG as well as simplify the analysis of the results. The Hamiltonian of Eq. \ref{eq:Boson:s1e1}, describing a $1D$ harmonic chain of $N$ modes, can be written in a concise quadratic form
\begin{equation}
 \hat H = \sum\limits_{i,j = 1}^{2N} { R_i^T \mathcal{H}_{ij}  R_j }\label{eq:Boson:s6e1}   
\end{equation}
by defining a quadrature vector $\vec R \equiv \left( {\vec p,\vec q} \right)$ with
\begin{equation}
\vec p \equiv \left( {\begin{array}{*{20}c}
   {\hat p_1 }  \\
   {\hat p_2 }  \\
    \vdots   \\
   {\hat p_N }  \\
\end{array}} \right),\vec q \equiv \left( {\begin{array}{*{20}c}
   {\hat q_1 }  \\
   {\hat q_2 }  \\
    \vdots   \\
   {\hat q_N }  \\
\end{array}} \right)\label{eq:Boson:s6e1b}
\end{equation}
where $\mathcal H$, henceforth referred to as the Hamiltonian matrix, is a $2N\times 2N$ Hermitian matrix. The coarse-graining transformations shall be chosen such that the effective Hamiltonians also only contain quadratic couplings, described by some new Hamiltonian matrix $\mathcal H'$. Thus the RG analysis may be performed in the space of Hamiltonian matrices $\mathcal H$, as opposed to the (much larger) space of full-fledged Hamiltonians $\hat H$ and a more efficient realization of ER is possible. Retaining the quadratic form of the Hamiltonian entails limiting the disentanglers $u$ and isometries $w$ which comprise the ER map to \emph{cannonical} transformations, namely transformations that preserve commutation relations. Consider a transformation of the Hamiltonian matrix by a $2N \times 2N$ matrix $S$
\begin{equation}
\mathcal{H} \mapsto \mathcal{H}' = S^T \mathcal{H}S.\label{eq:Boson:s6e2}
\end{equation}
In order for the transformation to be commutation preserving it is required that the transform $S$ be a symplectic matrix, $S\in \textrm{Sp}(2N,\mathbb{R})$. A symplectic transform can be characterized as leaving the symplectic matrix $\Sigma$ invariant under conjugation, $S^T \Sigma S = \Sigma$. Given our convention of grouping the quadrature vectors in Eq. \ref{eq:Boson:s6e1b} the symplectic matrix takes the form 
\begin{equation}
\Sigma  \equiv \left( {\begin{array}{*{20}c}
   0 & {{I}_N }  \\
   {{I}_N } & 0  \\
\end{array}} \right)\label{eq:Boson:s6e2b}
\end{equation}
with $I_N$ as the $N\times N$ identity. Additionally, the Hamiltonians under consideration take even simpler form than Eq. \ref{eq:Boson:s6e1}; as there is no coupling between $\hat p$ and $\hat q$ quadrature degrees of freedom in the harmonic chain, the Hamiltonian may be expressed as
\begin{equation}
\hat H = \vec p^T \vec p + \vec q^T \mathcal{H}_q \vec q. \label{eq:Boson:s6e3} 
\end{equation}
It is thus convenient to restrict symplectic transforms $S$ to those which preserve the form of Eq. \ref{eq:Boson:s6e3}; we only consider transforms of the type $S =V  \oplus V$, with $V$ a \emph{special orthogonal} transformation, $V\in \textrm{SO}(N)$. It can be easily checked that $V  \oplus V$ is a symplectic transform. In fact, this is an element of the maximal compact subgroup of Sp$(2N,\mathbb R)$. The $\hat p$-quadrature part of the Hamiltonian in Eq. \ref{eq:Boson:s6e3} remains trivial under these transformations, allowing us to focus on the $\hat q$-quadrature part $\mathcal H_q$ of the Hamiltonian matrix, which transforms as
\begin{equation}
{\mathcal{H}_q}' = V^T \mathcal{H}_q V.  \label{eq:Boson:s6e4}
\end{equation}
Let us group a number $M$ of contiguous bosonic modes of the $1D$ harmonic chain together; each group of $M$ modes shall henceforth be referred to as a \emph{site} of the original lattice $\mathcal{L}$. The disentanglers $u$, which act on two sites (that is, on $2M$ modes), are chosen as special orthogonal transforms $u\in \textrm{SO}(2M)$. Isometries $w$ are realized as a composition of a special orthogonal transform followed by a projection, $w = w_0 w_\textrm{proj.}$, with
\begin{equation}
w_0 \in \textrm{SO}(2M),\ \  w_\textrm{proj.} = (0_M  \oplus I_M).  \label{eq:Boson:s6e5}
\end{equation}
The transformation of the entire lattice is achieved by first constructing the direct sum of the local operators
\begin{equation}
W =  {\mathop  \bigoplus \limits_{i = 1}^{N/2M} w},\ \   U =  {\mathop  \bigoplus \limits_{i = 1}^{N/2M} u}.  \label{eq:Boson:s6e6}
\end{equation}
Given the disentangler and isometry $(u,w)$, the Hamiltonian $\mathcal H_q$ is coarse-grained into a new Hamiltonian $\mathcal H_q '$ defined as
\begin{equation}
{{\mathcal{H}}_q} '= W^\textrm{T} U^\textrm{T} \left( \mathcal{H}_q \right) U W.\label{eq:Boson:s6e7}
\end{equation}
By the definition of the isometry in Eq. \ref{eq:Boson:s6e5} it is ensured that if the initial lattice $\mathcal L$ has $M$ bosonic modes per lattice sites then the coarser lattices $\mathcal L ^{(\tau)}$ also have $M$ modes per lattice site. As discussed earlier in Sect. \ref{Sec:Boson:RSRG}, keeping the number of degrees of freedom per lattice site constant between RG maps is necessary to allow meaningful comparison of operators at different length scales. Note that the number of modes per site $M$ plays the role of a refinement parameter; choice of a larger $M$ retains more parameters in the description of the effective theory, yielding more accurate results, at the cost of greater computational expense. The simplest choice of a one-to-one correspondence between bosonic modes and lattice sites, i.e. setting $M=1$, does not give sufficiently accurate numerics, hence the need for grouping $M>1$ modes into each lattice site. 

\begin{figure}[!tbhp]
  \begin{center}
    \includegraphics[width=12cm]{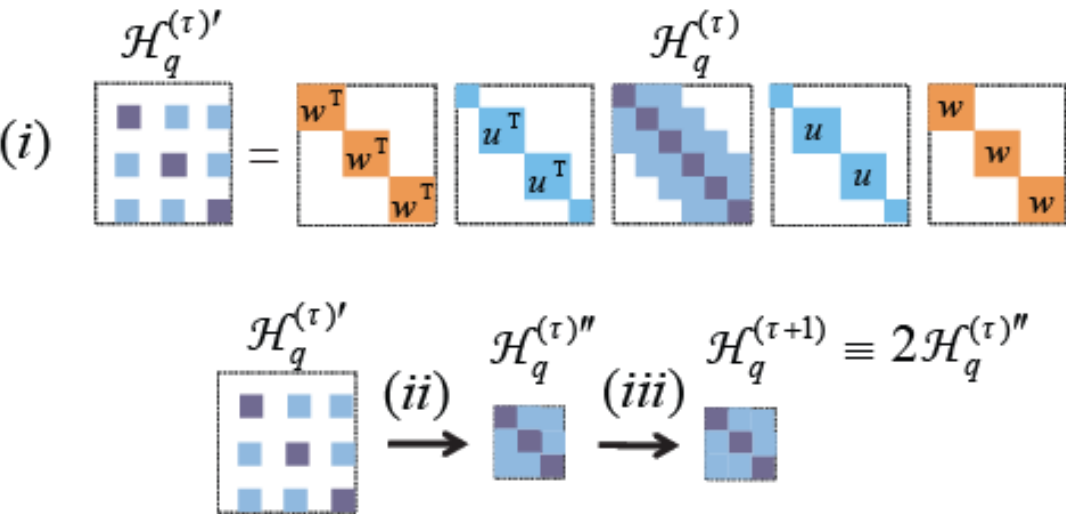}
  \caption{An iteration of entanglement renormalization, broken into three steps, is depicted in terms of the direct sum structure of the Hamiltonian matrix $\mathcal H_q$. Dark shaded squares in $\mathcal H_q$ represent couplings within the site of $M$ modes, light shaded squares are the couplings between sites (at most next-to-nearest neighbor). Step(i), removing `fast' modes is realized by transforming $\mathcal H_q$ by conjugation with the disentanglers $u$ and isometries $w$. Step(ii), rescaling the momentum, is achieved by removing the zero rows/columns from $\mathcal H_q'$. Step(iii), rescaling the fields, is achieved by directly scaling $\mathcal H_q''$ by a factor of 2. These three steps combined take the $\tau^\textrm{th}$ renormalized Hamiltonian matrix $\mathcal H_q^{(\tau)}$ to the $(\tau+1)^\textrm{th}$ renormalized Hamiltonian matrix $\mathcal H_q^{(\tau+1)}$. }
  \label{fig:Boson:RScovariance}
 \end{center}
\end{figure}

It is only for a proper choice of disentangler and isometry $(u,w)$ that $\mathcal H_q '$ of Eq. \ref{eq:Boson:s6e7} retains the low-energy subspace of the original $\mathcal H_q$. This proper choice is found by optimisation over all possible $(u,w)$. It is desired that the RG transform be optimized to project onto the \emph{minimum} energy subspace of the original Hamiltonian; a matrix equation which describes the minimization can be written
\begin{equation}
\mathop {\min }\limits_{u,w} \left( \textrm{tr} \{ {H_q '} \} \right),  \label{eq:Boson:s6e8}
\end{equation}
with the effective Hamiltonian $\mathcal H_q '$ as Eq. \ref{eq:Boson:s6e7}. The matrix $\mathcal H_q '$ describes a Hamiltonian that is translation invariant between blocks of $2M$ modes, hence the trace of $\mathcal H_q '$ (which describes the entire system) may be minimised by minimising the trace of an individual block of $\mathcal H_q '$. An optimisation method, based upon alternating updates for isometries $w$ and disentanglers $u$, can be used to find suitable $(u,w)$ that minimise Eq. \ref{eq:Boson:s6e8} and best preserve the low-energy space of the original Hamiltonian $\mathcal H_q$. The optimisation method to be used here is similar to the general algorithm described in Sect. IV of Ref. \cite{evenbly08a}.

Assuming that suitable $(u,w)$ have been obtained, the full real-space RG transformation of the Hamiltonian may be achieved in three steps, as illustrated Fig. \ref{fig:Boson:RScovariance}, analogous to the three steps of momentum-space procedure described Sect. \ref{Sec:Boson:MRG}. Firstly, (i) the Hamiltonian matrix $\mathcal H_q$ is transformed with direct sums of the disentanglers and isometries as Eq. \ref{eq:Boson:s6e7}. Recall from Eq. \ref{eq:Boson:s6e5} that an isometry $w$, which acts upon a block of $2M$ modes, consists of a special orthogonal transform followed by a projection, $w = w_0 w_\textrm{proj}$. The projection has form $w_\textrm{proj}=(0_M\oplus I_M)$ with the trivial (zero) part $0_M$ describing the $M$ modes to be removed from the system and the identity part $I_M$ describing the $M$ modes retained in the effective description. In the next step, (ii) the zero rows/columns, those which were acted upon in the previous step by $0_M$, are removed from Hamiltonian matrix to form a new matrix ${\mathcal{H}_q^{(0)}} ''$. (iii) The final step of rescaling the fields is realized by scaling the Hamiltonian matrix by a factor, the same rescaling as Eqs. \ref{eq:Boson:s2e3} and \ref{eq:Boson:s2e4} for momentum-space RG is realized by defining $\mathcal{H}_q^{(1)}\equiv 2 {\mathcal{H}_q^{(0)}} ''$, with ${\mathcal{H}_q^{(1)}}$ as the first renormalized Hamiltonian. Iterating this procedure $\tau$ times gives the $\tau^\textrm{th}$ renormalized Hamiltonian ${\mathcal{H}_q^{(\tau)}}$. In the thermodynamic limit, $N\rightarrow \infty$, the transformation can be iterated an arbitrary number of times to obtain a sequence of Hamiltonian matrices
\begin{equation}
\left( {\mathcal H^{(0)} } \right) \stackrel{(u^{(1)} ,w^{(1)} )}{\longrightarrow} \left( {\mathcal H^{(1)} } \right) \stackrel{(u^{(2)} ,w^{(2)} )}{\longrightarrow} \left( {\mathcal H^{(2)} } \right)\stackrel{(u^{(3} ,w^{(3)} )}{\longrightarrow}  \ldots  \label{eq:Boson:s6e9}
\end{equation}
each describing a theory with quadratic interactions between at most next-nearest-neighbor sites, and each defined on an identical lattice $\mathcal L$ that has $M$ bosonic modes per lattice site. 

\subsection{Ground state RG} \label{Sec:Boson:GroundRG}
In the previous section an implementation of real-space RG that could be used to coarse-grain harmonic lattice Hamiltonians was described. We now detail a similar procedure which allows \emph{ground-states} of the harmonic lattices to be coarse-grained directly. Since we are dealing with systems of free-particles, the covariance matrix $\gamma$ gives a complete description of the (Gaussian) ground state  $\left| {\psi _{{\rm{GS}}} } \right\rangle$. In the case of a Hamiltonian as in Eq. \ref{eq:Boson:s6e3}, the convariance matrix is of the form $\gamma  = \gamma _p  \oplus \gamma _q $, where $\gamma _p$  and $\gamma _q$ are defined as
\begin{align}
 \left( {\gamma _p } \right)_{ij}  &\equiv 2\left\langle {\psi _{{\rm{GS}}} } \right|\hat p_i \hat p_j \left| {\psi _{{\rm{GS}}} } \right\rangle  \nonumber\\ 
 \left( {\gamma _q } \right)_{ij}  &\equiv 2\left\langle {\psi _{{\rm{GS}}} } \right|\hat q_i \hat q_j \left| {\psi _{{\rm{GS}}} } \right\rangle.\label{eq:Boson:s7e1}   
\end{align}
The derivation of analytic expressions for $( \gamma_p$, $\gamma_q )$ is a standard calculation and can be found e.g. Ref. \cite{audenaert02}, but is also presented in Appendix \ref{chap:BosonGround} for completeness. Again, we choose the disentanglers $u$ and isometries $w$ to be canonical transformations. We obtain a sequence of increasingly coarse-grained states each defined by covariance matrix $\gamma^{(\tau)}$ 
\begin{equation}
\left( {\gamma ^{(0)} } \right) \stackrel{(u^{(1)} ,w^{(1)} )}{\longrightarrow} \left( {\gamma ^{(1)} } \right)\stackrel{(u^{(2)} ,w^{(2)} )}{\longrightarrow} \left( {\gamma ^{(2)} } \right)\stackrel{(u^{(3)} ,w^{(3)} )}{\longrightarrow} \ldots \label{eq:Boson:s7e2}
\end{equation}
with $\gamma^{(0)} \equiv \gamma$ as the original ground state. The coarse-graining transformations of the ground state shall be realized by symplectic transforms $S$ acting in the space of the covariance matrix, $\gamma  \mapsto \gamma ' = S^T \gamma S$. As there is no correlation between $\hat p$ and $\hat q$ quadrature coordinates, i.e. $\gamma  = \gamma _p  \oplus \gamma _q $, we may restrict consideration to symplectic transforms $S\in\textrm{Sp}(2N,\mathbb R)$ that are only of the form $S = A \oplus (A^{ - 1} )^\textrm{T}$ with $A$ a real, invertible $N\times N$ invertible matrix. The covariance matrix $\gamma$ transforms under conjugation by matrix $S$, which implies that
\begin{align}
 \gamma _p  &\mapsto \gamma '_p  = A^T \left( \gamma _p \right) A \nonumber\\ 
 \gamma _q  &\mapsto \gamma '_q  = A^{ - 1} \left( \gamma _q \right) (A^{ - 1} )^T, \label{eq:Boson:s7e3}  
\end{align}
yeilding a new state $\gamma ' = \gamma_p ' \oplus \gamma_q '$. The disentanglers $u$, which act on two contiguous sites each of $M$ modes, are thus realized as real, invertible $2M\times 2M$ matrices that transform the covariance matrix as per Eq. \ref{eq:Boson:s7e3}. Isometries $w$, which act on two contiguous sites within a block, are realised as $w = w_0 w_\textrm{proj}$ with $w_0$ as a real, invertible $2M\times 2M$ matrix and $w_\textrm{proj} = (0_M  \oplus I_M)$ a projection onto $M$ modes of the block. The direct-sum of the local operators $(u,w)$ is constructed
\begin{equation}
 W_\textrm{proj}  =  {\mathop  \bigoplus \limits_{i = 1}^{N/2L} w_\textrm{proj} } ,\ \ W_0  = {\mathop  \bigoplus \limits_{i = 1}^{N/2L} w_0 }, \label{eq:Boson:s7e4}    
\end{equation}
\begin{equation}
 U =  {\mathop  \bigoplus \limits_{i = 1}^{N/2L} u} . \label{eq:Boson:s7e5} 
\end{equation}
in order to coarse-grain the entire lattice. Starting from the covariance matrix $\gamma = \gamma_p \oplus \gamma_q$ describing the ground state of the original system and, for any choice of $(u,w)$, a new state $\gamma' = \gamma_p ' \oplus \gamma_q '$ can be obtained on a coarser lattice $\mathcal L '$ through an ER transform
\begin{align}
 \gamma _p '  &= \left( W_{{\rm{proj}}} W_0^\textrm{T} U^\textrm{T} \right) \gamma _p \left( U W_0 W_{{\rm{proj}}} \right),  \nonumber \\
 \gamma _q '  &= \left( W_{{\rm{proj}}} W_0^{ - 1} U^{-1} \right)           \gamma _q \left( (U^{-1})^\textrm{T} (W_0^{ - 1})^\textrm{T} W_{{\rm{proj}}} \right). \label{eq:Boson:s7e6}   
\end{align}
It is only for proper choice of tensors $(u,w)$ that the above transformation correctly preserves the properties of the original state and produces a meaningful coarse-grained state. We now address the issue of how the proper tensors $(u,w)$ can be found. In the application of the RG to the system Hamiltonian, both in momentum-space and real-space formulation, the modes truncated at each iteration were chosen as \emph{high-energy modes} in order to leave the low-energy structure of the original theory intact. For the ground state RG a different criteria is required to judge which modes should be truncated from the system. The proper truncation criteria in order to preserve the ground state properties, proposed by White as part of his DMRG algorithm \cite{white92,white93}, requires that the truncation of a block should be chosen to keep the support of the density matrix for the block. In the present formulation of bosonic modes, in which the representation of the state is given by a covariance matrix as opposed to a density matrix, this rule imposes that only modes in a block that can be identified as being in a \emph{product state} with the rest of the system can be truncated and safely removed from the description of the state. Thus in comparison with Hamiltonian RG, which was optimised to truncate modes such that effective theory had minimal energy, here we optimise for $(u,w)$ so that the truncated modes have minimal entanglement with the rest of the system.

The entanglement of a block with the rest of the system is known to be related to the symplectic eigenvalues of $\gamma$ for the block \cite{plenio05,audenaert02}. Let the covariance matrix $(\gamma) |_{\mathcal B} = ({\gamma _p}) |_\mathcal{B} \oplus ({\gamma_q}) |_\mathcal B $ describe the correlations within a block $\mathcal B$ of two $M$-mode sites. The $2M$ symplectic eigenvalues $\lambda_i$ of the block $\mathcal B$ are the eigenvalues of the matrix formed by taking the product of matrices $({\gamma _p}) |_\mathcal{B}$ and $({\gamma_q}) |_\mathcal B$
\begin{equation}
\lambda _{i}  = {\rm{Spect}}\left\{ ({\gamma _p}) |_\mathcal{B} ({\gamma_q}) |_\mathcal B \right\}.\label{eq:Boson:s7e7}
\end{equation}
The Heisenberg position-momentum uncertainty relation, which here may be simplified as $\left\langle {\hat p^2 } \right\rangle \left\langle {\hat q^2 } \right\rangle  \ge 1/4$, enforces that all symplectic eigenvalues are positive and have magnitude greater than unity, $\lambda_i\ge 1$ for all $i$. The entanglement entropy $S_i$ of a mode `$i$' with symplectic eigenvalue $\lambda_i$ is
\begin{equation}
S_{i}  = \left[ {f\left( {\frac{{\sqrt {\lambda _i }  - 1}}{2}} \right) - f\left( {\frac{{\sqrt {\lambda _i }  + 1}}{2}} \right)} \right] \label{eq:Boson:s7e8}
\end{equation}
with $f(x) =  - x\log x$. It is seen that the entropy $S_i$ is zero when mode $i$ is in a minimum uncertainty state, $\lambda_i=1$, and an also that $S_i$ is \emph{increasing} function of $\lambda_i$. It follows that, if a mode with eigenvalue $\lambda_i=1$ can be identified within a block $\mathcal B$, then we are assured that the mode is in a product state with the rest of the system and may be safely truncated. Hence the tensors $(u,w)$ should be optimized to minimize the eigenvalues $\lambda_i$, and thus the entanglement entropy, of the modes to be projected out. The projection $W_\textrm{proj}$ in Eq. \ref{eq:Boson:s7e4} was defined to select the modes to be retained; we now construct the complimentary projector $\tilde W_\textrm{proj}$
\begin{equation}
\tilde W_\textrm{proj}  =  {\mathop  \bigoplus \limits_{i = 1}^{N/2M} \left( {I_{2M}  - w_\textrm{proj} } \right)} \label{eq:Boson:s7e9}
\end{equation}
which projects onto the space of the modes to be truncated. The part $\tilde \gamma$ of the covariance matrix $\gamma$ that is to be projected out during the RG iteration may be written $\tilde \gamma = \tilde \gamma_p \oplus \tilde \gamma_q$ with
\begin{align}
 \tilde \gamma _p  &= \left( \tilde W_{\textrm{proj}} W_0^\textrm{T} U^\textrm{T} \right) \gamma _p \left( U W_0 \tilde W_{\textrm{proj}} \right) \nonumber \\ 
 \tilde \gamma _q  &= \left( \tilde W_{{\rm{proj}}}  {W_0^{ - 1} }   U^{-1} \right)       \gamma _q \left( (U^{-1})^\textrm{T} (W_0^{ - 1})^\textrm{T} \tilde W_{{\rm{proj}}}\right). \label{eq:Boson:s7e10}
\end{align}
Given that the modes projected out of block $\mathcal B$ are to have minimum entropy, the transforms $(u,w)$ should be chosen to minimise the matrix equation
\begin{equation}
\mathop {\min }\limits_{u,w} \left( {\textrm{tr}}\left\{ {(\tilde \gamma _q ) }|_\mathcal{B}  {(\tilde \gamma _p )  }| _\mathcal{B}  \right\} \right). \label{eq:Boson:s7e11}
\end{equation}
As with the energy minimization described by Eq. \ref{eq:Boson:s6e8} for the Hamiltonian RG , this equation is optimised variationally to find good disentanglers $u$ and isometries $w$.

\begin{figure}[!tbhp]
  \begin{center}
    \includegraphics[width=9cm]{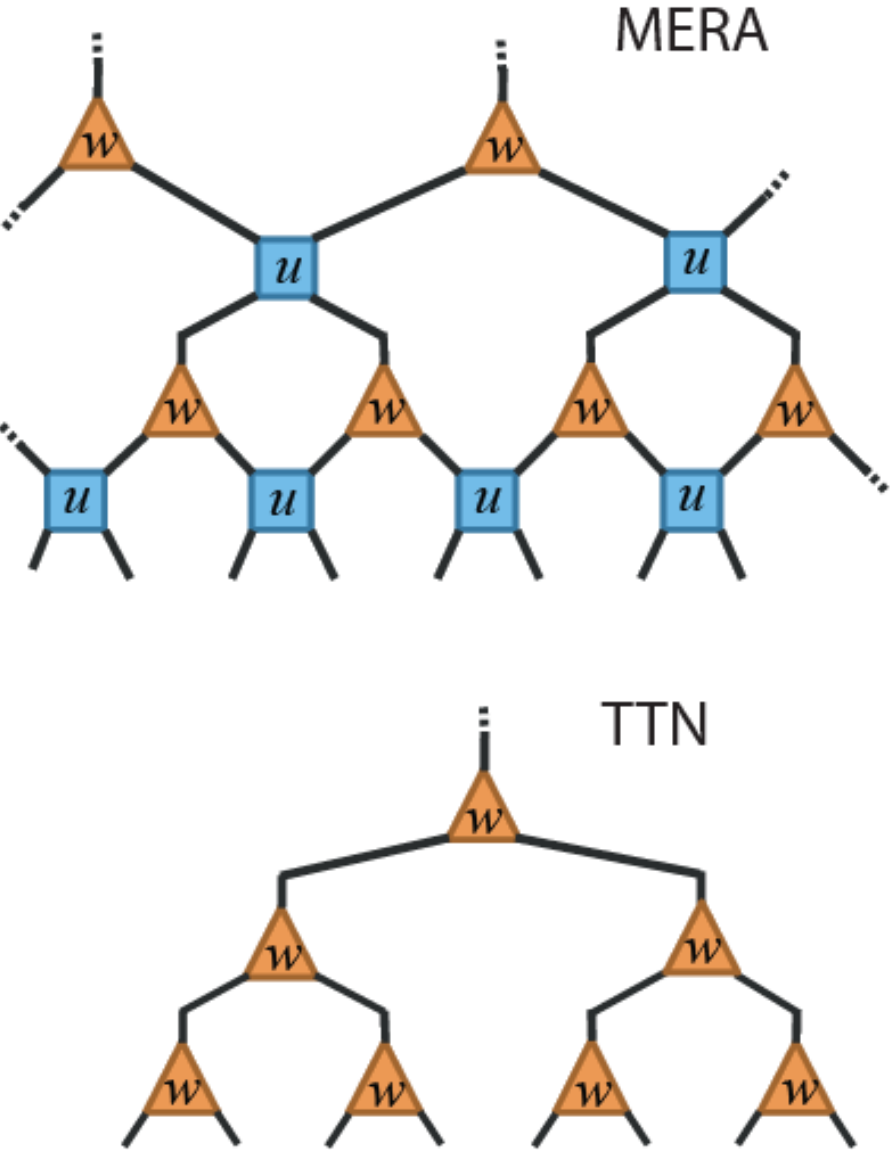}
  \caption{(Top) Applying successive RG maps based upon ER, as depicted in Fig. \ref{fig:Boson:ProjMera}, to the ground state leads to an approximate representation of the ground state in terms of a set of disentanglers and isometries $(u^{(\tau)}, w^{(\tau)})$ connected in a MERA tensor network. Here we utilise the binary MERA scheme introduced in Chap. \ref{chap:MERAintro}. (bottom) Implementing the RG based upon an LP coarse-graining gives an approximation to the state in terms of a \emph{tree-tensor network} (TTN) \cite{shi06, tagliacozzo09}, a different class of ansatz for many-body states on a lattice.}  
  \label{fig:Boson:MERAvsTTN}
 \end{center}
\end{figure}

The application of RG transformations to the Hamiltonian could be interpreted as producing effective theories for the low-energy subspace of the original system. Is there a similar interpretation for the coarse-graining of the ground state? To address this question we assume that the ER map has been iterated $\tau$ successive times on the original ground state $\gamma^{(0)}$ to get the coarse-grained state $\gamma^{(\tau)}$. Each of the $\tau$ coarse-grainings is characterised by a disentangler $u$ and isometry $w$, thus the set of these transforms $\left( u^{(j)}, w^{(j)} \right)$ for $j = 1,2,\ldots,\tau$ characterise the sequence of RG maps. If the modes truncated during each iteration were in an exact product state, equivalently the eigenvalues of every mode `$i$' truncated was $\lambda_i = 1$, then the sequence of transformations could be inverted as follows. Starting from $\gamma^{(\tau)}$, truncated modes (which were in a product state) are replaced back, and each of the transforms of Eq. \ref{eq:Boson:s7e6} is inverted
\begin{equation}
\left( {\gamma ^{(\tau)} } \right) \stackrel{(u^{(\tau)} ,w^{(\tau)} )}{\longrightarrow} \left( {\gamma ^{(\tau-1)} } \right)\stackrel{(u^{(\tau-1)} ,w^{(\tau-1)} )}{\longrightarrow} \ldots \stackrel{(u^{(1)} ,w^{(1)} )}{\longrightarrow} \left( {\gamma ^{(0)} } \right), \label{eq:Boson:s7e12}
\end{equation}
as to recover the exact original state $\gamma^{(0)}$. The coarse-graining of the ground state can be interpreted as \emph{storing} information about the short range properties of the state into the tensors $\left( u^{(i)}, w^{(i)} \right)$ while preserving the long range information about the state; the set of tensors $\left( u^{(i)}, w^{(i)} \right)$ together with state $\gamma^{(\tau)}$ thus serve as a representation of the original state $\gamma^{(0)}$. The set $\gamma^{(\tau)}$ and $\left( u^{(i)}, w^{(i)} \right)$ form the \emph{multi-scale entanglement renormalization ansatz} (MERA) \cite{vidal08}, a variational ansatz for states on the lattice, c.f. Fig \ref{fig:Boson:MERAvsTTN}. [If the ground state RG is performed with an LP coarse-graining, as opposed to one based upon ER, a \emph{tree tensor network} (TTN) \cite{shi06, tagliacozzo09} approximation to the state is obtained in terms of the isometries $ w^{(i)} $]. For the harmonic lattices we consider, as with most non-trivial models, the coarse-graining cannot be performed exactly and a MERA will be an approximate, rather than exact, representation of the true ground state. In the present case, irrespective of how the transforms $\left( u, w\right)$ are chosen, the modes to be truncated will still be slightly entangled as manifested in their symplectic eigenvalues, which will fulfill $\lambda_i > 1$. This entanglement will be ignored during the coarse-graining thus limiting the precision with which the original ground state $\gamma^{(0)}$ may be recovered.

In the present setting of free bosons, in which the exact ground state $\gamma^{(0)}$ is already known, we compute the MERA and TTN approximations to the ground state via coarse-graining the ground state covariance matrix. 

The point of this exercise is that it allows us to assess the accuracy with which a MERA or a TTN can represent the many-body state. It also provides the opportunity to study the RG flow of the ground state $\gamma^{(\tau)}$ under scale transformations. However, had our goal been to investigate properties of the unknown ground state of a system $\hat H$ with the help of real-space RG, it would have been absurd to assume knowledge of the exact ground state $\left| {\psi _{{\rm{GS}}} } \right\rangle$ from the beginning. In that case, an approximation $\left| {\tilde \psi _{{\rm{GS}}} } \right\rangle$ to the ground state may be found through e.g. variational minimization of energy $\left\langle {\tilde \psi _{GS} } \right|\hat H\left| {\tilde \psi _{GS} } \right\rangle$ as per Ref. \cite{evenbly08a} or through an alternative method \cite{dawson08, rizzi08}.

\begin{figure}[tb]
  \begin{center}
    \includegraphics[width=10cm]{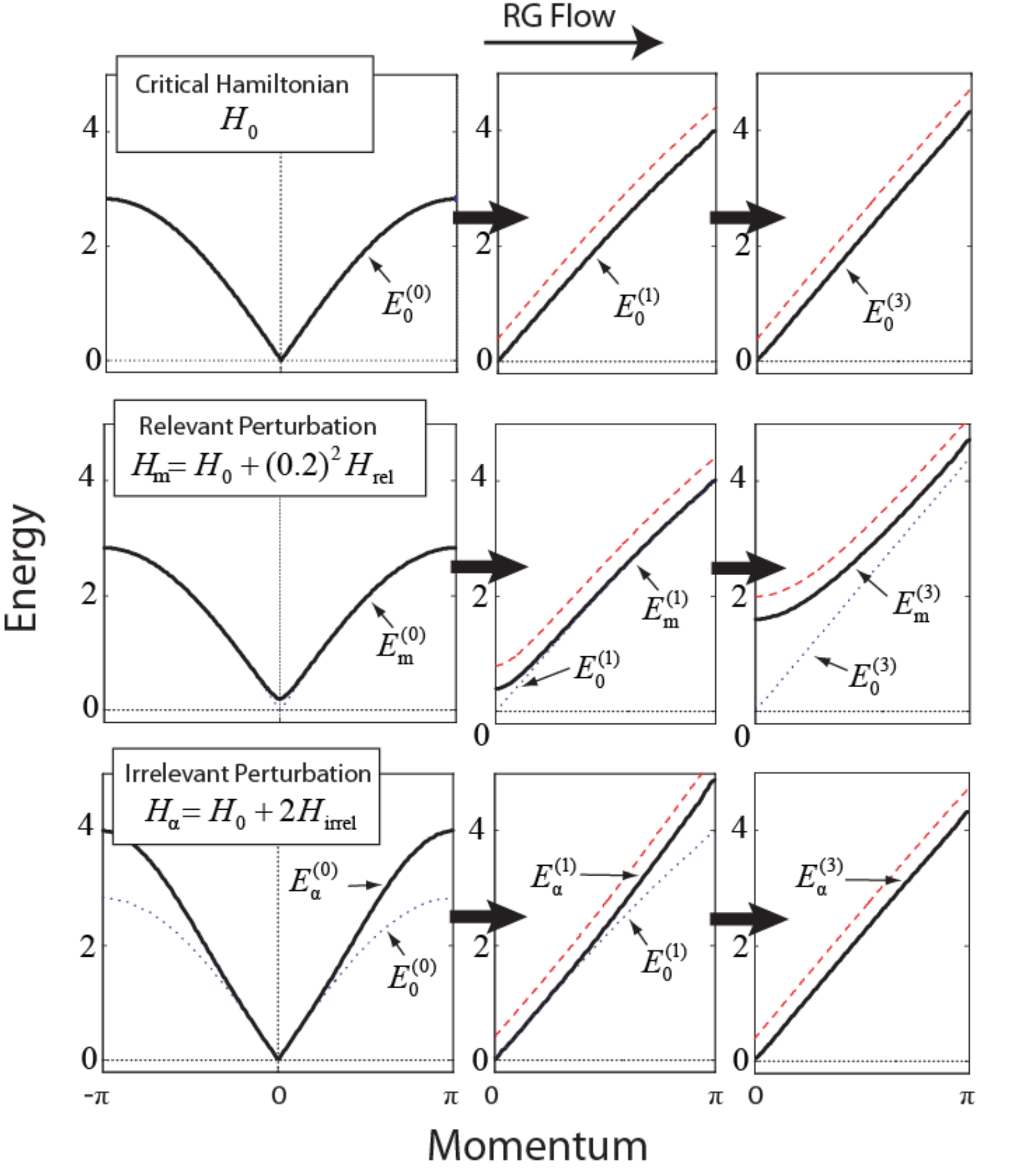}
  \caption{Sequences of dispersion relations of Hamiltonians and after $\tau=0,1,3$ RG transforms, comparing the real-space ER results (bold) with the exact momentum-space results (dashed, offset by 0.2). (Top Series) The critical Hamiltonian $\hat H_0$ tends to a linear dispersion under the RG map with ER, in good agreement with momentum-space results. (Middle Series) The addition of even small mass, $m=0.2$, to the critical system gives a marked difference in the dispersion relation after $\tau=3\ $ RG transforms. (Bottom Series) The effect of an irrelevant perturbation quickly diminishes under RG flow; by $\tau=3$ iterations the original and the perturbed dispersions are virtually identical, $E_{{\alpha}}^{(3)} \approx E_0^{(3)}$.}
  \label{fig:Boson:BoseHamRenorm}
 \end{center}
\end{figure}

\begin{figure}[tb]
  \begin{center}
    \includegraphics[width=10cm]{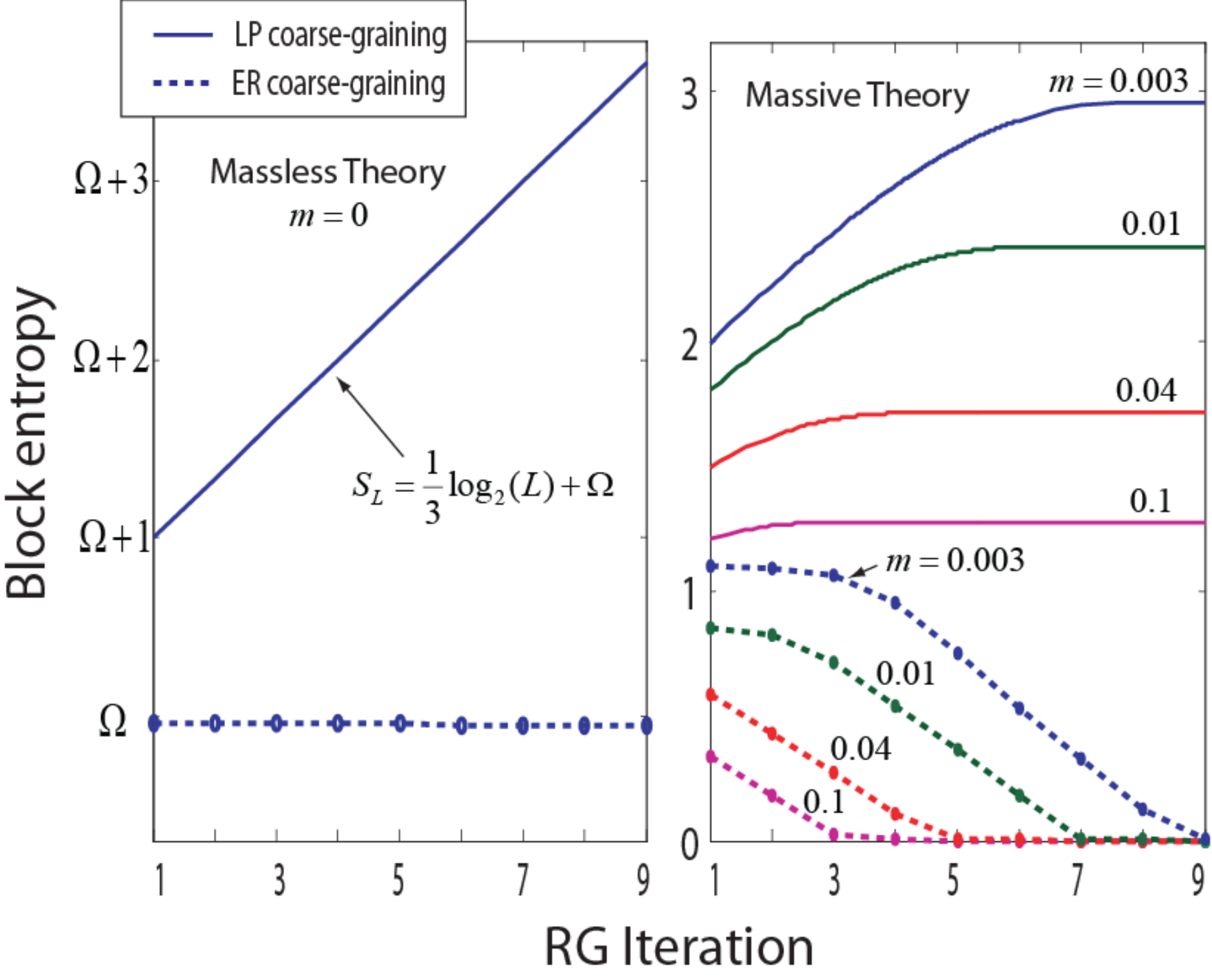}
  \caption{The entanglement entropy $S$ of a site in the $\tau^\textrm{th}$ coarse-grained ground state $\gamma^{(\tau)}$ of the $1D$ harmonic chain. (Left) In the critical (massless) regime the entropy of a site is infinite, as realized by the infinite constant $\Omega$; however the change in entropy along the RG flow can be computed via a limiting process (see Appendix \ref{chap:BosonGround}). The entropy of the state renormalized with an LP coarse-graining increases by a constant with each RG iteration, reproducing the logarithmic growth law, $S_L = (1/3)\log_2 L + c$, as expected from conformal field theory \cite{vidal03, calabrese04}, whilst the entropy of the system renormalized with ER remains constant along the RG flow. Furthermore the sequence of renormalized ground-states $\{ {\gamma ^{(1)} ,\gamma ^{(2)} , \ldots } \}$ rapidly converge to a fixed state $\gamma^*$ under the ER transforms. (Right) For several values of finite mass, the entropy of the states renormalized with the LP scheme saturate at a length scale governed by correlation length, whilst the theories renormalized with ER factorize into a product state (zero entropy) at the approximately same length. }
  \label{fig:Boson:BoseEntPlotFlat}
 \end{center}
\end{figure}

\begin{figure}[!tb]
  \begin{center}
    \includegraphics[width=10cm]{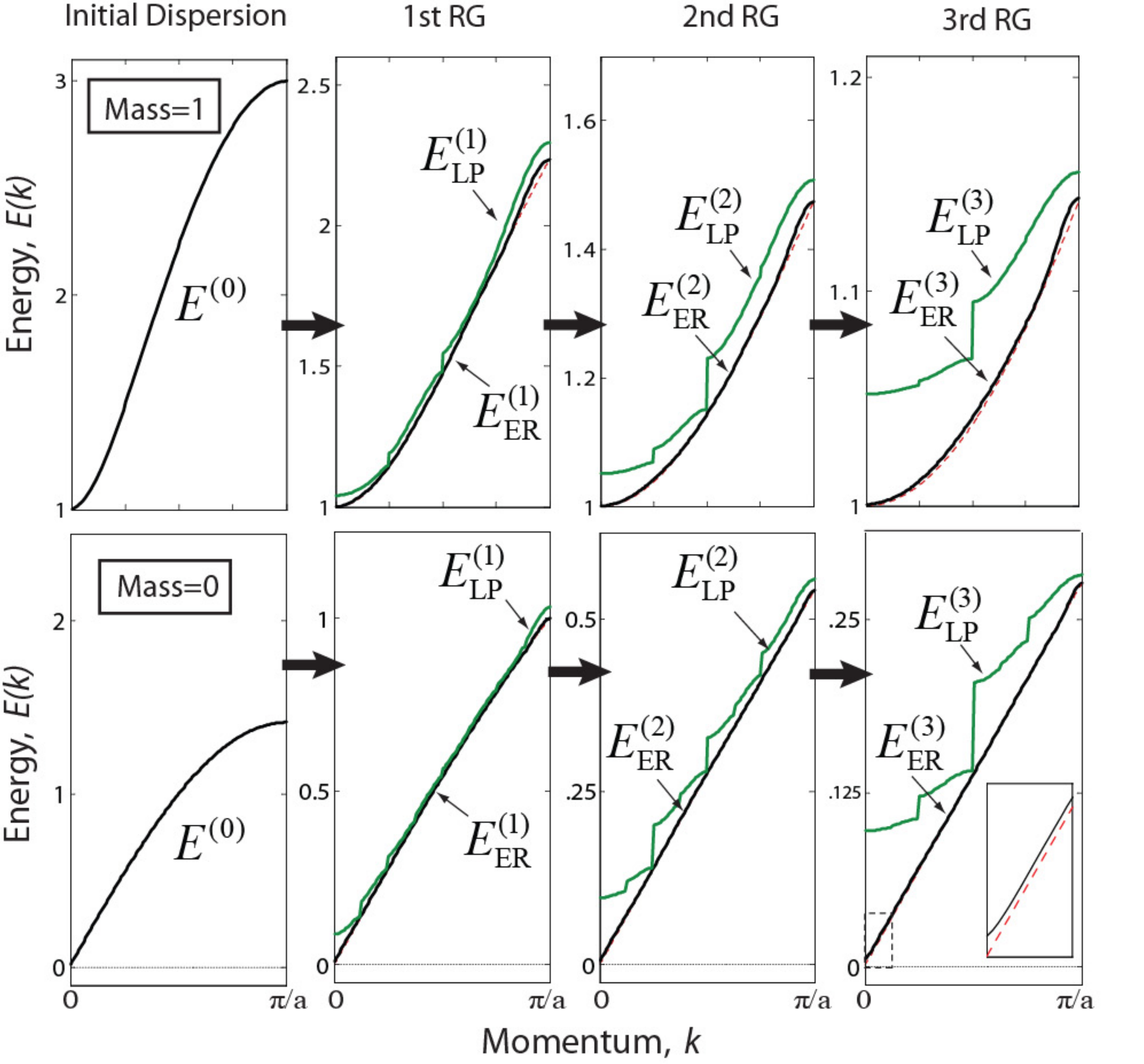}
  \caption{(Left) Sequences of dispersion relations (in non-rescaled energy units) for the (top) gapped and (bottom) critical $1D$ harmonic systems after $\tau=0,1,2,3$ real-space RG transforms, comparing the results from the local projection (LP) method, $E_\textrm{LP}^{(\tau)}$, to those from the entanglement renormalization (ER) coarse-graining, $E_\textrm{ER}^{(\tau)}$. The numeric dispersions produced by ER agree with the exact results obtained from momentum-space RG (dashed) as to be almost visually indistinguishable; though small errors are noticeable near the momentum cut-off, $k=\pi/a$. The dispersion relations given by coarse-graining performed with the LP scheme, whilst reasonably accurate after the $1^\textrm{st}$ iteration, rapidly diverge from the exact result with successive RG iterations. In all cases the numeric RG was performed keeping $M=4$ modes per site. }
  \label{fig:Boson:BoseHamError}
 \end{center}
\end{figure}

\begin{figure}[!tb]
  \begin{center}
    \includegraphics[width=10cm]{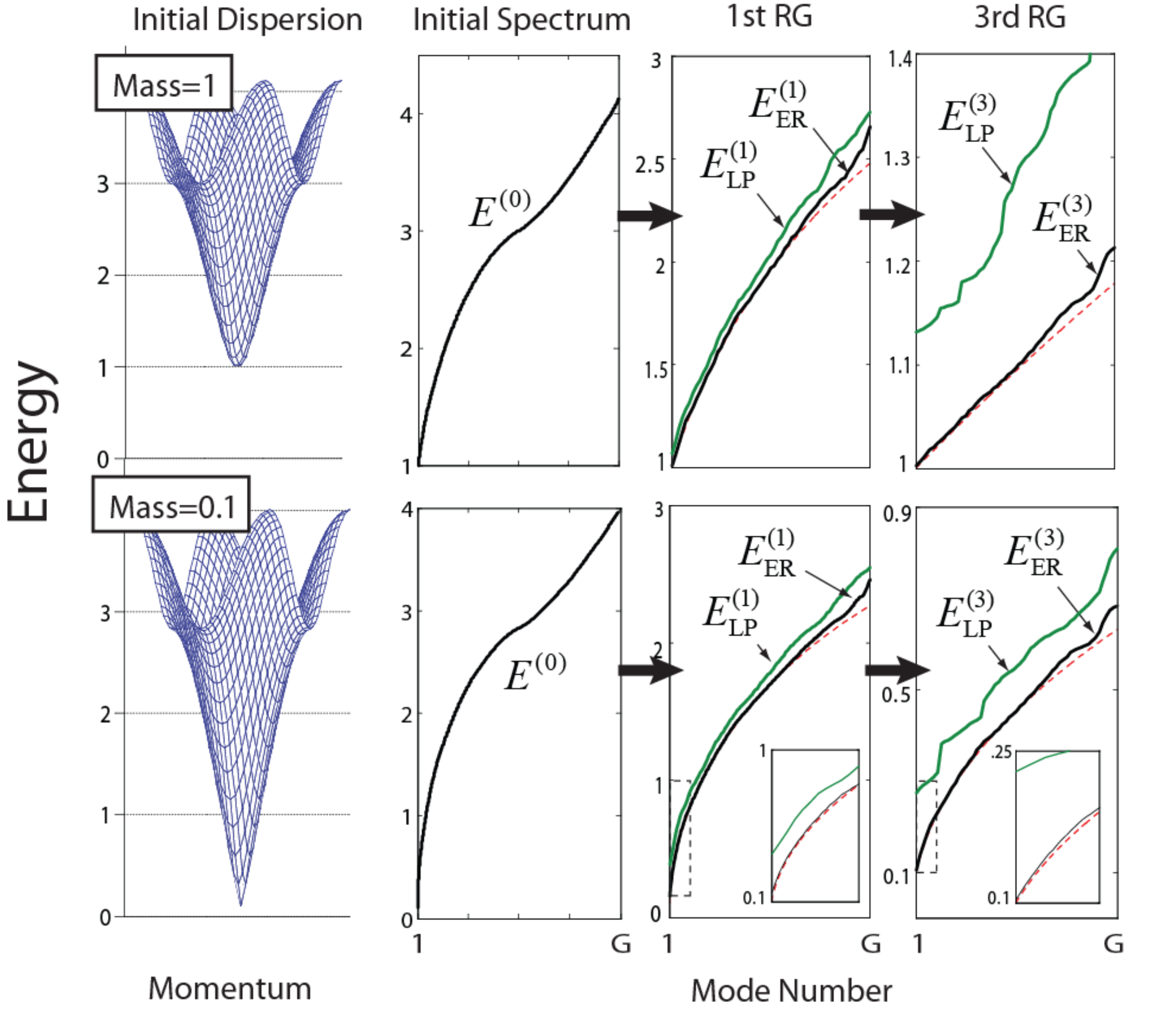}
  \caption{Sequences of energy spectra (in non-rescaled energy units) for the (top) gapped and (bottom) near-critical $2D$ harmonic lattice system after $\tau=0,1,3$ real-space RG transforms. The performance of the numeric real-space methods, local projection (LP) and entanglement renormalization (ER), are benchmarked against the exact solutions from momentum-space RG (dashed). The energy spectra are obtained by sampling the dispersion relation $E(k_1,k_2)$ on a finite grid of $G$ points, with $G$ chosen very large, and then ordering the values, $\{E_1\le E_2\le \cdots \le E_G\}$. The spectra of the systems renormalized with entanglement renormalization, $E_\textrm{ER}^{(\tau)}$, retain a high level of accuracy through all $\tau=3$ RG iterations, though do lose precision near the high energy cut-off of the effective theory. Energy spectra $E_\textrm{LP}^{(\tau)}$ of the effective theories obtained with an LP coarse-graining have diverged considerably from the exact result by $\tau=3$ RG iterations. The numeric RG was performed keeping $M=9$ modes per site.  }
  \label{fig:Boson:BoseError2D}
 \end{center}
\end{figure}

\section{Results and discussion} \label{Sec:Boson:Results}
\subsection{RG flow of Hamiltonians} \label{Sec:Boson:HamResult}

In Sect. \ref{Sec:Boson:MRG} the low energy subspace of harmonic chains were analysed exactly with momentum-space RG. This analysis yielded sequences of dispersion relations describing the RG flow towards a critical fixed point as well as the RG flow resulting from adding relevant and irrelevant perturbation terms, as given by Eqs. \ref{eq:Boson:s2e7}, \ref{eq:Boson:s3e3} and \ref{eq:Boson:s4e3} respectively. The same harmonic systems can also be analysed numerically by applying successive real-space RG transformations $\mathcal{T}$, based upon ER, following the variational method described in Sect. \ref{Sec:Boson:HamRG}. This produces a sequence of Hamiltonian matrices $\left( \mathcal{H}^{(0)}, \mathcal{H}^{(1)}, \mathcal{H}^{(2)}, \ldots \right)$, each an effective theory describing successively lower energy subspace of the original system. By construction, $\mathcal{H}^{(\tau)}$ only contains quadratic couplings between nearest and next-to-nearest neighboring sites. The dispersion relations of the effective Hamiltonians $\mathcal H^{(\tau)}$ are found by Fourier transform in a similar manner as presented for the original Hamiltonian in Sect. \ref{Sec:Boson:Harm}. Figure \ref{fig:Boson:BoseHamRenorm} shows the comparison of the analytic (momentum-space) and numeric (real-space) dispersion relations.

The dispersion relations produced from numeric real-space RG with ER are seen to approximate the exact results to a high degree of accuracy for the three of the cases considered: a critical system, and relevant and irrelevant perturbations on the critical system. Furthermore, the critical Hamiltonian $\mathcal H_0^{(0)}$ converges to a fixed point of the RG flow, $\mathcal{H}_0^{(\tau)} = \mathcal{H}^{*}$, after $\tau \ge 3$ RG transforms. This fixed point Hamiltonian matrix $\mathcal{H}^{*}$ is manifestly invariant under further transformations, $\mathcal{T} (\mathcal {H}^{*}) = \mathcal {H}^{*}$, to within small numerical errors. Consequently the isometries $w$ and disentanglers $u$, which comprise the ER transform $\mathcal{T}$, also converged to fixed points $w^{(\tau)} = w^*$ and $u^{(\tau)} = u^*$ for $\tau \ge 3$. The Hamiltonian matrix $\mathcal{H}_\alpha^{(0)}$ of the critical system with the addition of an irrelevant perturbation, $\hat H_\alpha^{(0)} = \hat H_0^{(0)} + \alpha \hat H_\textrm{irrel}^{(0)}$, converges to the same fixed point as the unperturbed Hamiltonian $\mathcal{H}_\alpha^{(\tau)} = \mathcal{H}_0^{(\tau)} = \mathcal{H}^{*}$ for $\tau \ge 3$ transforms.

\subsection{RG flow of ground states} \label{Sec:Boson:GroundResult}

On the other hand, in Sect. \ref{Sec:Boson:GroundRG} we describe a real-space RG to coarse-grain the ground state of the $1D$ harmonic chain. The coarse-graining transformations can be based upon either the LP or ER schemes of Fig. \ref{fig:Boson:ProjMera}. The real-space RG produces a sequence of increasingly coarse-grained ground-states $\left( \gamma^{(0)}, \gamma^{(1)}, \gamma^{(2)}, \ldots \right)$, where $\gamma^{(0)}$ is the ground state covariance matrix of a harmonic chain with mass $m$. Fig. \ref{fig:Boson:BoseEntPlotFlat} displays the entanglement entropy $S$, as defined in Eq. \ref{eq:Boson:s7e8}, of a site in the coarse-grained state $\gamma^{(\tau)}$ as a function of the RG iteration $\tau$.

The entropy of the ground state $\gamma_0^{(\tau)}$ of the critical chain, $m=0$, when coarse-grained with the LP scheme, increases by a constant with each RG iteration. More precisely, if we recall that a site in lattice $\mathcal L ^{(\tau)}$ corresponds to a block of $L = 2^\tau$ sites of the original lattice, this growth of entropy reproduces the expected logarithmic scaling, $S_L = (1/3) \log_2 L + c$, for $1D$ critical systems \cite{vidal03, calabrese04}. This growth demonstrates that it is impossible for a critical ground state to be a fixed point of the LP scheme.

The ground state of a system with finite mass $m$ can also be analysed with the LP method. Initially, the ground state displays the same logarithmic growth of entropy as in the critical case, but it saturates approximately after $\tau = \log_2 \zeta$ iterations of the RG transformation, where $\zeta$ is the correlation length.

Turning to the ER scheme, if disentanglers are included in the coarse-graining step then the entropy per site of state $\gamma_0^{(\tau)}$ remains constant under RG transforms. This is made possible by the disentanglers, which remove short-range entanglement at each iteration. Moreover, for the critical case, the sequence of successively coarse-grained ground-states $\gamma_0^{(\tau)}$ explicitly converges to a fixed point, $\gamma_0^{(\tau)}=\gamma^*$ for $\tau \ge 3$. The non-critical ground states $\gamma^{(\tau)}$ of a harmonic chain with finite mass $m$ converged to a trivial fixed point, a product state $\gamma^{**}$, under RG iteration. This occurs at the length scale of the correlation length $\zeta$. The numeric results have been obtained by keeping $M=4$ bosonic modes per lattice site. The MERA approximation to the ground state, obtained from successive coarse-grainings with ER, proves remarkably accurate. For the critical system, which is the hardest to analyse from a computational point of view, the MERA obtained from $\tau =9$ RG iterations can reproduce the exact correlators of the ground state to error bounded by $1\times 10^{-4}$.

\begin{figure}[!tb]
  \begin{center}
    \includegraphics[width=10cm]{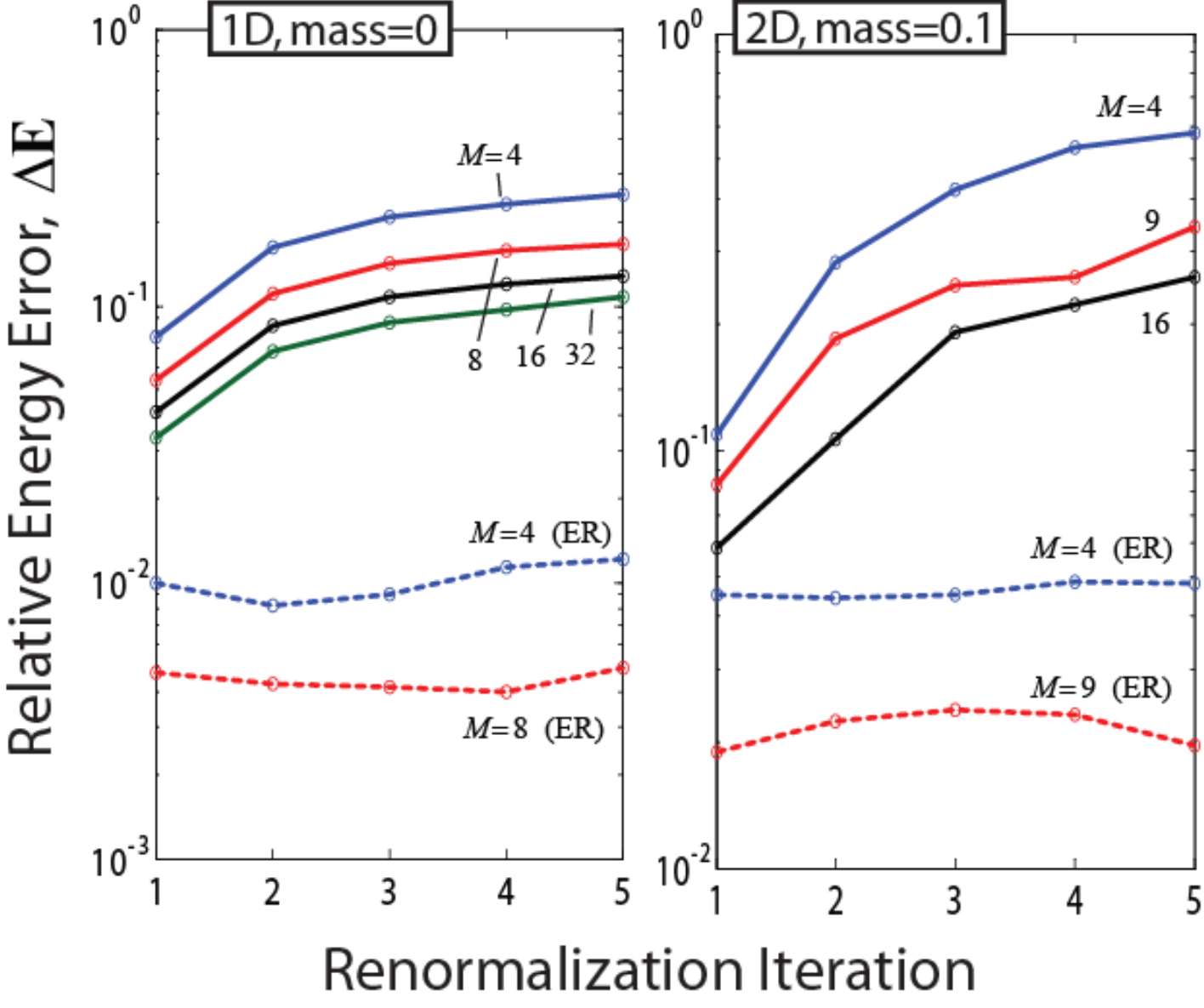}
  \caption{The mean energy $\bar{E}_\textrm{RS}$, as defined Eq. \ref{eq:Boson:s8e1}, of the effective theories obtained through real-space RG, based upon either an LP (solid) or ER (dashed) coarse-graining, are compared against exact momentum-space results $\bar{E}_\textrm{MS}$ in terms of the relative error $\Delta {E} = (\bar{E}_\textrm{RS} - \bar{E}_\textrm{MS}) / \bar{E}_\textrm{MS}$. Performing the real-space coarse-graining with a larger number of modes per site $M$, which allows for more parameters to be kept in the description of the effective theories, is shown to give more accurate results for the real-space RG. (left) For the $1D$ critical harmonic chain the mean of the energy spectrum obtained from renormalizing with ER, while keeping $M=4$ modes per site, remains approximately $1\%$ greater than exact value throughout the RG iterations. The spectra obtained from the LP method, even when using significantly larger $M$, only gave at best $10\%$ accuracy after the same number of RG iterations. (right) For the $2D$ near-critical harmonic lattice, ER gives accuracies after 5 iterations of no less than $4\%$ and $2\%$ for $M=4$ and $M=9$ respectively; this is in contrast to the LP method which is only accurate to within $60\%$ for $M=4$ and $30\%$ for $M=9$. Barring small fluctuations, the accuracy of the coarse-graining with ER remains relatively stable with RG iteration. }
  \label{fig:Boson:1D2Derrraw}
 \end{center}
\end{figure}

\subsection{Local projection vs entanglement renormalization} \label{Sec:Boson:LPvsER}

Figs. \ref{fig:Boson:BoseHamError} and \ref{fig:Boson:BoseError2D} present numeric dispersion relations for the harmonic lattices in $D=1$ and $D=2$ spatial dimensions respectively, and compare results produced by coarse-graining with LP to those obtained by coarse-graining with ER. The dispersion relations for the effective theories produced from the LP method are reasonably accurate after a small number of RG iterations. However, they rapidly diverge from the exact results in subsequent iterations. On the other hand, coarse-graining with ER is shown to keep significantly better support of the low energy subspace and, most importantly, to maintain accuracy over repeated RG iterations. The numeric spectra obtained with ER for the $2D$ lattices of Fig. \ref{fig:Boson:BoseError2D} displays a loss of accuracy towards the high-energy cut-off, which indicates difficulty in keeping a sharp cut-off numerically. However, this is of little concern as the primary interest lies in the low-energy physics and not the high-energy cut-off of the effective theory.

The average mode energy $\bar E$ of an effective theory, defined in terms of its dispersion relation $E(k)$, is
\begin{equation}
\bar{E} \equiv \frac{1}{{2\Lambda }}\int_{ - \Lambda }^\Lambda  {E(k)dk} \label{eq:Boson:s8e1}
\end{equation}
for a $1D$ system and similar for $2D$. The average mode energy is used to qualitatively analyze the accuracy of the effective theories obtained with real-space RG, as presented in Fig. \ref{fig:Boson:1D2Derrraw}. Keeping more modes-per-site $M$, hence more parameters in the description of the effective theory, increases the accuracy with which the numeric RG may be performed. However, even with the choice of a very large $M$, very large the LP method shows significant increase in error along the RG flow for both the $1D$ critical and $2D$ near-critical harmonic systems. This figure also confirms what was established visually in Figs. \ref{fig:Boson:BoseHamError} and \ref{fig:Boson:BoseError2D}; that coarse-graining with ER not only produces a more accurate low-energy theory than with LP, but also maintains constant accuracy over successive RG maps.  

\section{Conclusions} \label{Sec:Boson:Conclude}
Real-space RG techniques, primarily in the form of the DMRG algorithm \cite{white92,white93, scholl05}, have proved an invaluable tool for the numeric analysis of low-energy properties of extended quantum systems. However, this class of real-space RG methods (including the LP coarse-graining described here) suffer from an important deficiency: namely they do not reproduce proper RG flows. As demonstrated by Fig. \ref{fig:Boson:BoseEntPlotFlat}, a $1D$ critical system could not possibly be a fixed point of the LP map due to the growth of entropy along the RG flow, which is caused by the accumulation of short-range degrees of freedom. This deficiency in turn limits the size of $1D$ critical and $2D$ lattices that can be analysed with such methods.      

In this Chapter we have demonstrated, by comparison against exact results from momentum-space RG, that a coarse-graining based upon entanglement renormalization can reproduce proper RG flows. Demonstrations included the analysis of a critical Hamiltonian $\mathcal H_0$ and its ground state $\gamma_0$, which were shown to rapidly converge to non-trivial fixed points of the RG flow, $\mathcal H_0^*$ and $\gamma_0^*$ respectively, and analysis of several non-critical systems which converged to trivial fixed-points of the RG flow. By addition of an irrelevant perturbation to the critical Hamiltonian $\mathcal H_0$, a new Hamiltonian $\mathcal H_\alpha$ was obtained that described the same phase as $\mathcal H_0$, but differed in the local interaction. As is required of a proper RG flow, $\mathcal H_\alpha$ and $\mathcal H_0$ converged to the same fixed point $\mathcal H_0^*$ under the RG map defined by ER. It is important to note that the tensors $(u,w)$, which comprised the real-space RG transform, were not chosen based on heuristic arguments or the desire for a particular outcome. Instead they were found through optimisation based upon either energy minimization, as in Eq. \ref{eq:Boson:s6e8}, or attempting to retain the support of the ground state, as in Eq. \ref{eq:Boson:s7e11}.

In addition, the coarse-graining transformation based upon ER was shown to induce a \emph{sustainable} RG map; one that could be applied arbitrarily many times without significant loss of accuracy and without the need to increase the local dimension $M$ of the effective theories (so that the computational cost is also kept constant). The sustainability of the ER based RG map allows investigation of low-energy properties in arbitrarily large or infinite $1D$ and $2D$ lattices, as has also been demonstrated in recent studies in which local observables are evaluated directly in the thermodynamic limit \cite{evenbly08a, evenbly08b, evenbly09a} and critical exponents computed \cite{pfeifer08,montangero09} without the need for finite size scaling techniques. 

The implementation of entanglement renormalization presented in this Chapter exploited properties of free-particle systems in order to reduce the computational cost. More general algorithms exist, as shall be described in the following Chapter, which allow implementation without making use of the special properties of free particle systems. Thus it is possible to use real-space RG, based upon ER, as a means of investigating the low-energy properties of strongly-correlated systems where perturbative approaches are not valid. However, the implementation of ER in the interacting case is computationally expensive- especially so for $2D$ lattices. The application of ER to interacting quantum systems on $2D$ lattices is investigated in Chap. \ref{chap:2DMera} and Chap. \ref{chap:KagMera}.

\chapter{Algorithms for entanglement renormalization} \label{chap:MERAalg}

\section{Introduction} \label{sect:MERAalg:Intro}

Entanglement renormalization provides a conceptual framework for implementing real-space RG transforms on lattice systems. As was described Chapter \ref{chap:MERAintro}, ER is naturally related to a class of quantum many-body states, the so-called multiscale entanglement renormalization ansatz (MERA). Perhaps the most important application of ER, and certainly the application that is the focus of this thesis, is as a numerical tool for investigation of low-energy subspaces of quantum many-body systems. However, the usefulness of ER extends beyond that of solely as a numerical tool \cite{aguado08,konig09,swingle09}. Chapters \ref{chap:FreeFerm} and \ref{chap:FreeBoson} achieved the first step in realizing ER as a numeric tool for quantum many-body systems by demonstrating the ability of the MERA to accurately represent a variety of ground state and low energy wavefunctions in $D=1,2$ dimensional systems of free fermions and free bosons. Nevertheless, the variational methods used in these Chapters to optimise the MERA was specific to the free-particle systems under consideration. In this Chapter we present opimisations algorithms that allow one to compute the MERA for the ground-state or low-energy subspace of a generic local Hamiltonian. We also describe how expected values may be computed from the MERA and provide benchmark results for a variety of $1D$ spin models.  

The Chapter is organised as follows. Sect. \ref{sect:MERAalg:local} explains how to compute expected values of local observables and two-point correlators. Central to this discussion is the \emph{past causal cone} of a small block of lattice sites and the \emph{ascending} and \emph{descending} superoperators, which can be used to move local observables and density matrices up and down the causal cone. Sect. \ref{sect:MERAalg:optim} considers how to optimize a single tensor of the MERA during an energy minimization. This optimization involves linearizing a quadratic cost function for the (isometric) tensor, and computing its \emph{environment}. In Sect. \ref{sect:MERAalg:algorithm} we describe algorithms to minimize the energy of the state/subspace represented by a MERA. A highlight of the algorithms is their computational cost. For an inhomogeneous lattice with $N$ sites, the cost scales as $O(N)$, whereas for translation invariant systems it drops to just $O(\log N)$. Other variations of the algorithm allow us to address infinite systems, and scale invariant systems (e.g. quantum critical systems), at a cost independent of $N$. Sect. \ref{sect:MERAalg:benchmark} presents benchmark calculations for different 1$D$ quantum lattice models, namely Ising, 3-level Potts, XX and Heisenberg models. We compute ground state energies, magnetizations and two-point correlators throughout the phase diagram of the Ising and Potts models, which includes a second order quantum phase transition. We find that, at the critical point of an infinite system, the error in the ground state energy decays exponentially with the refinement parameter $\chi$, whereas the two-point correlators remain accurate even at distances of millions of lattice sites. We then extract critical exponents from the order parameter and from two-point correlators. Finally, we also compute a MERA that includes the first excited state, from which the energy gap can be obtained and seen to vanish as $1/N$ at criticality.

\section{Computation of expected values of local observables and correlators} \label{sect:MERAalg:local}

We begin this section by briefly recalling the MERA formalism presented in Sects. \ref{sect:MERAintro:MERA}, \ref{sect:MERAintro:Quantcirc} and \ref{sect:MERAintro:RG} of Chap. \ref{chap:MERAintro}. Let $\mathcal{L}$ denote a $D$-dimensional lattice made of $N$ sites, where each site is described by a Hilbert space $\mathbb{V}$ of finite dimension $d$, so that $\mathbb{V}_{\mathcal{L}} \cong \mathbb{V}^{\otimes N}$. The MERA is an ansatz to describe certain pure states $\ket{\Psi}\in \mathbb{V}_{\mathcal{L}}$ of the lattice or, more generally, to describe a $\chi_T$ dimensional subspace $\mathbb{V}_{U} \subseteq \mathbb{V}_{\mathcal{L}}$. The MERA for the $N$ site lattice may be grouped into $T\approx \log N$ different layers, where each layer contains a row of isometries $w$ and a row of disentanglers $u$. We label these layers with an integer $\tau=1,2,\cdots T$, with $\tau=1$ for the lowest layer and with increasing values of $\tau$ as we climb up the tensor network. The MERA implicitly defines a sequence of lattices 
\begin{equation}
	\mathcal{L}_0 \rightarrow \mathcal{L}_1 \rightarrow \cdots \rightarrow \mathcal{L}_T,
\end{equation}
where $\mathcal{L}_0\equiv \mathcal{L}$ is the original lattice, and where we can think of lattice $\mathcal{L}_\tau$ as the result of coarse-graining lattice $\mathcal{L}_{\tau-1}$.

Let $o^{[r,r+1]}$ denote a local observable defined on two contiguous sites $[r,r\!+\!1]$ of $\mathcal{L}$. In this section we explain how to compute the expected value 
\begin{equation}
	\langle o^{[r,r+1]} \rangle_{\mathbb{V}_{U}} \equiv \tr (o^{[r,r+1]}P).
\label{eq:MERAalg:ev_o}
\end{equation}
Here $P$ is a projector (see Eq. \ref{eq:MERAintro:P}) onto the $\chi_T$-dimensional subspace $\mathbb{V}_{U}\subseteq \mathbb{V}_{\mathcal{L}}$ represented by the MERA. For a rank $\chi_T=1$ MERA, representing a pure state $\ket{\Psi}\in  \mathbb{V}_{\mathcal{L}}$, the above expression reduces to 
\begin{equation}
	\langle o^{[r,r+1]} \rangle_{\Psi} \equiv \bra{\Psi}o^{[r,r+1]}\ket{\Psi}. 
\end{equation}
Evaluating Eq. \ref{eq:MERAalg:ev_o} is necessary in order to extract physically relevant information from the MERA, such as e.g. the energy and magnetization in a spin system. In addition, the manipulations involved are also required as a central part of the optimization algorithms described in Sects. \ref{sect:MERAalg:optim} and \ref{sect:MERAalg:algorithm}. The results of this section remain relevant even in cases where no optimization algorithm is required (for instance when an exact expression of the MERA is known \cite{aguado08,konig09}).

As explained below, the expected value of Eq. \ref{eq:MERAalg:ev_o} can be computed in a number of ways:
\begin{itemize}
\item By repeated use of the \emph{ascending superoperator} $\mathcal{A}$, the local operator $o^{[r,r+1]}$ is mapped onto a coarse-grained operator $o_T$ on lattice $\mathcal{L}_T$. Eq. \ref{eq:MERAalg:ev_o} can then be evaluated as the trace of the coarse-grained operator $o_T$, $\tr (o^{[r,r+1]}P) = \tr(o_T)$.
\item Alternatively, by repeated use of the \emph{descending superoperator} $\mathcal{D}$, a two-site reduced density matrix $\rho^{[r,r+1]}$ for lattice $\mathcal{L}$ is obtained. Eq. \ref{eq:MERAalg:ev_o} can then evaluated as $\tr (o^{[r,r+1]}P) = \tr (o^{[r,r+1]}\rho^{[r,r+1]})$.
\item More generally, the ascending and descending superoperators $\mathcal{A}$ and $\mathcal{D}$ can be used to compute an operator $o^{[r',r'+1]}_{\tau}$ and density matrix $\rho^{[r',r'+1]}_{\tau}$ for the coarse-grained lattice $\mathcal{L}_{\tau}$. Eq. \ref{eq:MERAalg:ev_o} can then be evaluated as  $\tr (o^{[r,r+1]}P) = \tr (o^{[r',r'+1]}_{\tau}\rho^{[r',r'+1]}_{\tau})$. 
\end{itemize}
First we introduce the ascending and descending superoperators $\mathcal{A}$ and $\mathcal{D}$ and explain in detail how to perform the computation of the expected value of Eq. \ref{eq:MERAalg:ev_o}. Then we address also the computation of the expected value
\begin{equation}
		\langle O \rangle_{\mathbb{V}_{U}} \equiv \tr (OP),~~~~~~~~	O \equiv \sum_{r} o^{[r,r+1]},
\label{eq:MERAalg:ev_O}
\end{equation}
where $O$ is an operator that decomposes as a sum of local operators in $\mathcal{L}$;
as well as the computation of two-point correlators. Finally, we revisit these tasks in the presence of translation invariance and scale invariance.

The ascending and descending superoperators are an essential part of the MERA formalism that was introduced in Ref. \cite{vidal08} (see e.g. Fig. 4 of Ref. \cite{vidal08} for an explicit representation of the descending superoperator $\mathcal{D}$). These superoperators have been also called MERA quantum channel/MERA transfer matrix in Ref. \cite{giovannetti08}.

\subsection{Ascending and descending superoperators}

\begin{figure}[!tbhp]
\begin{center}
\includegraphics[width=10cm]{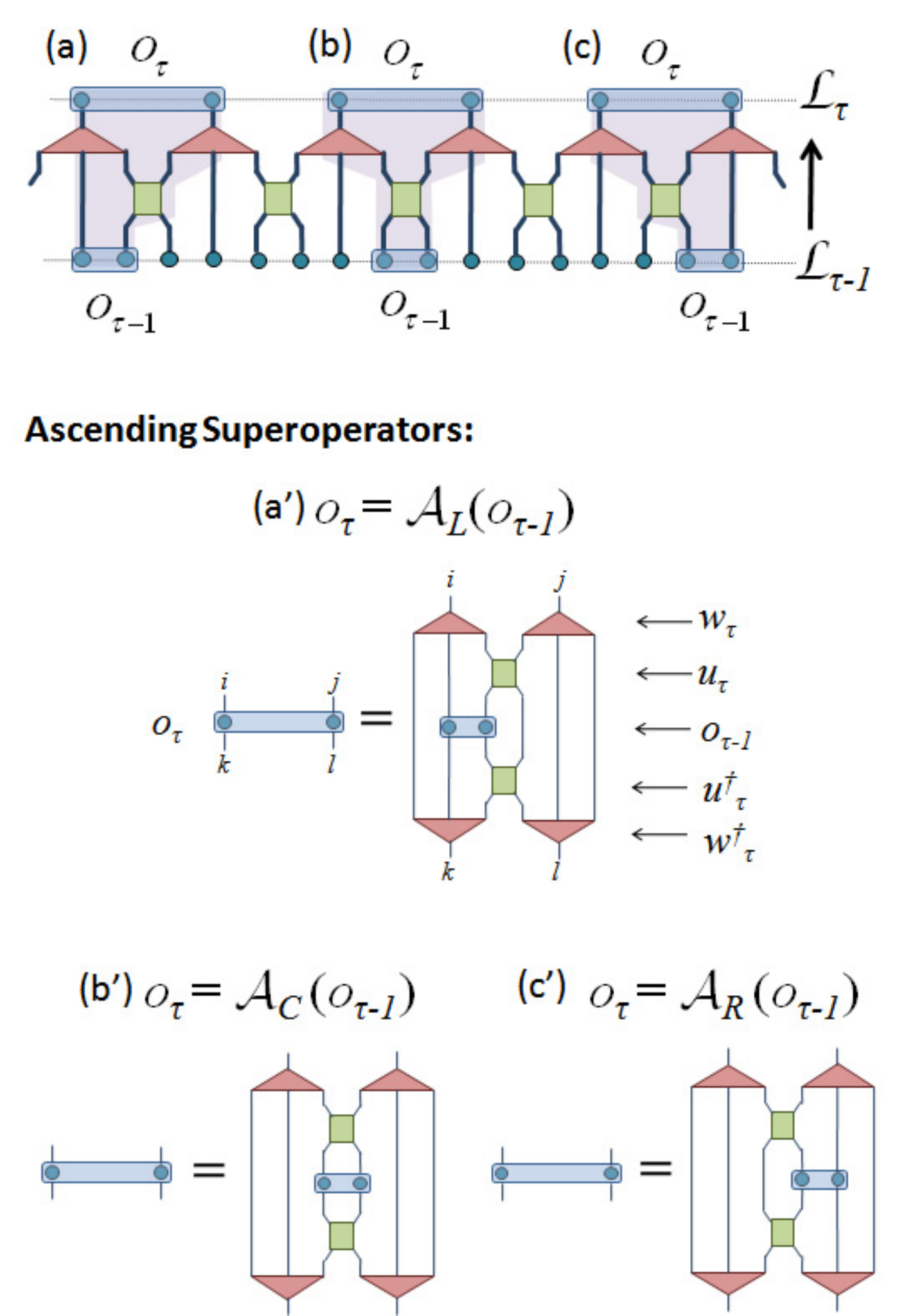}
\caption{ The ascending superoperator $\mathcal{A}$ transforms a local operator $o_{\tau-1}$ of lattice $\mathcal{L}_{\tau-1}$ into a local operator $o_{\tau}$ of lattice $\mathcal{L}_{\tau}$ (for simplicity we omit the label $[r,r+1]$ that specifies the sites on which $o_{\tau-1}$ and $o_{\tau}$ are supported). Depending on the relative position between the support of $o_{\tau-1}$ and the closest disentangler, the operator can be lifted to lattice $\mathcal{L}_{\tau}$ in three different ways, indicated in the figure as (a), (b) and (c). Correspondingly, there are three structurally different forms of the ascending superoperator $\mathcal{A}$, namely left $\mathcal{A}_L$, center $\mathcal{A}_C$ and right $\mathcal{A}_R$, indicated as (a'), (b') and (c'). Notice that the figure completely specifies the tensor network representation of the superoperator, which is written in terms of the relevant disentanglers and isometries (and their Hermitian conjugates). An explicit form for the \emph{average ascending superoperator} $\bar{\mathcal{A}}$ of Eq. \ref{eq:MERAalg:avAsc} is obtained by averaging the above three tensor networks.}
\label{fig:MERAalg:AscendSuper}
\end{center}
\end{figure}

In the previous section we have seen that the MERA defines a sequence of increasingly coarser lattices $\{\mathcal{L}_0, \mathcal{L}_1, \cdots, \mathcal{L}_T\}$. Under the coarse-graining transformation $U_{\tau}^\dagger$ of Eq. \ref{eq:MERAintro:Utau}, a local operator $o^{[r,r+1]}_{\tau-1}$, supported on two consecutive sites $[r,r\!+\!1]$ of lattice $\mathcal{L}_{\tau-1}$, is mapped onto another local operator $o^{[r',r'+1]}_{\tau}$ supported on two consecutive sites $[r',r'\!+\!1]$ of lattice $\mathcal{L}_{\tau}$ (Fig. \ref{fig:MERAintro:3MERA}). This is so because in $U_{\tau}^\dagger o^{[r,r+1]}_{\tau-1} U_{\tau}$ most disentanglers and isometries of $U_{\tau}$ and $U_{\tau}^{\dagger}$ are annihilated in pairs according to Eq. \ref{eq:MERAintro:isometry}. The resulting transformation is implemented by means of the \emph{ascending} superoperator $\mathcal{A}$ described in Fig. \ref{fig:MERAalg:AscendSuper}, 
\begin{equation}
 o^{[r',r'+1]}_{\tau} = \mathcal{A}(o^{[r,r+1]}_{\tau-1}).
\label{eq:MERAalg:Ascending}
\end{equation}
In order to keep our notation simple, we do not specify on which lattice/sites the superoperator $\mathcal{A}$ is applied, even though $\mathcal{A}$ actually depends on $\tau$, $r$ and $r'$. Instead, when necessary we will simply indicate which of its three structurally different forms (namely, left $\mathcal{A}_L$, center $\mathcal{A}_C$ or right $\mathcal{A}_R$ in Fig. \ref{fig:MERAalg:AscendSuper}) is being used.

\begin{figure}[!tbhp]
\begin{center}
\includegraphics[width=10cm]{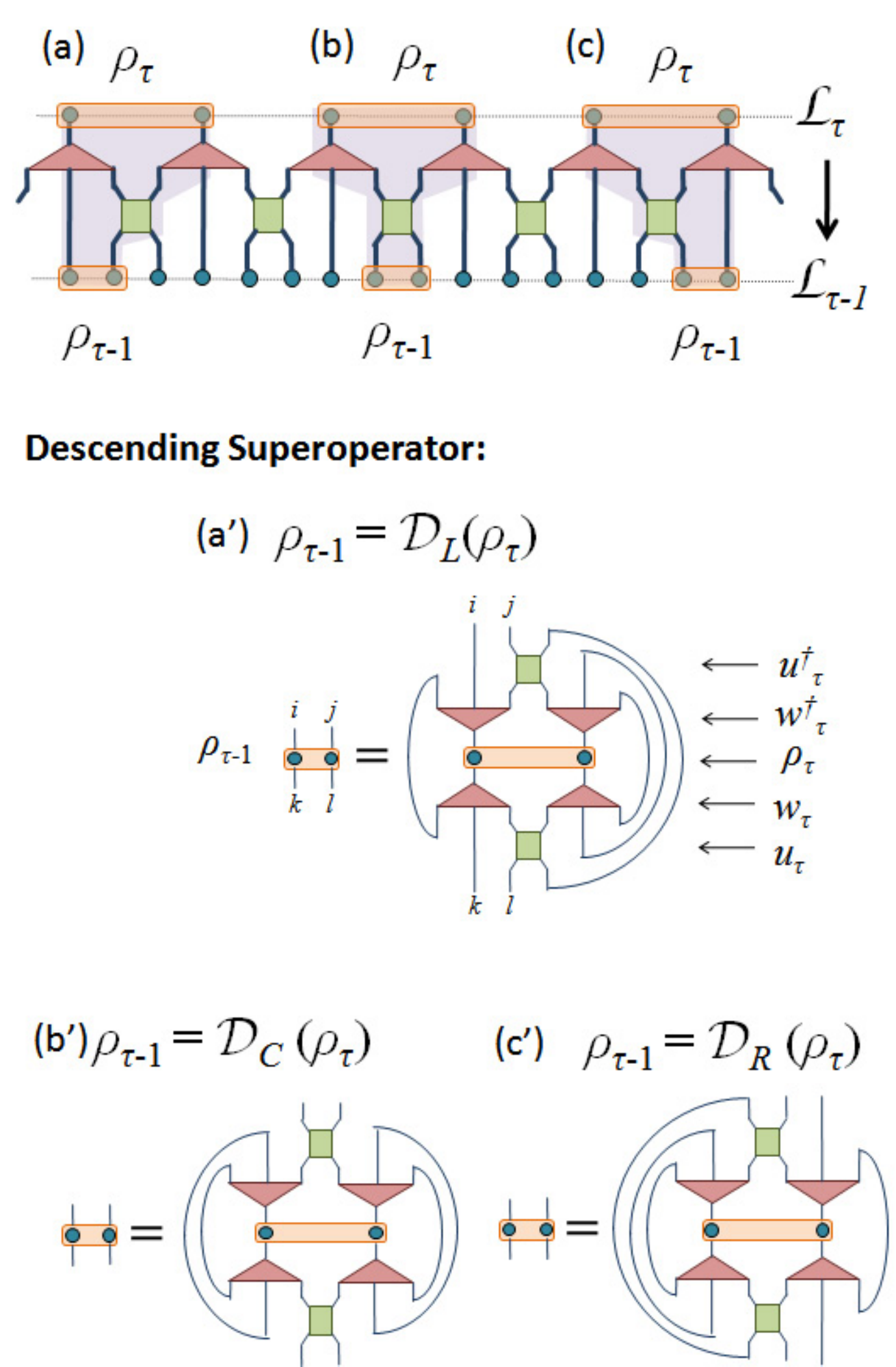}
\caption{ The descending superoperator $\mathcal{D}$ transforms a local density matrix $\rho_{\tau}$ of lattice $\mathcal{L}_{\tau}$ into a local density matrix $\rho_{\tau-1}$ of lattice $\mathcal{L}_{\tau-1}$. Depending on the relative position between the support of $\rho_{\tau-1}$ and the closest disentangler, the density matrix $\rho_{\tau}$ canbe lowered to lattice $\mathcal{L}_{\tau-1}$ in three different ways, indicated in the figure as (a), (b) and (c). Correspondingly, there are three structurally different forms of the descending superoperator $\mathcal{D}$, namely left $\mathcal{D}_L$, center $\mathcal{D}_C$ and right $\mathcal{D}_R$, indicated as (a'), (b') and (c'). An explicit form for the \emph{average descending superoperator} $\bar{\mathcal{D}}$ of Eq. \ref{eq:MERAalg:avDesc} is obtained by averaging the above three tensor networks.}
\label{fig:MERAalg:DescendSuper}
\end{center}
\end{figure}

\begin{figure}[!tbhp]
\begin{center}
\includegraphics[width=10cm]{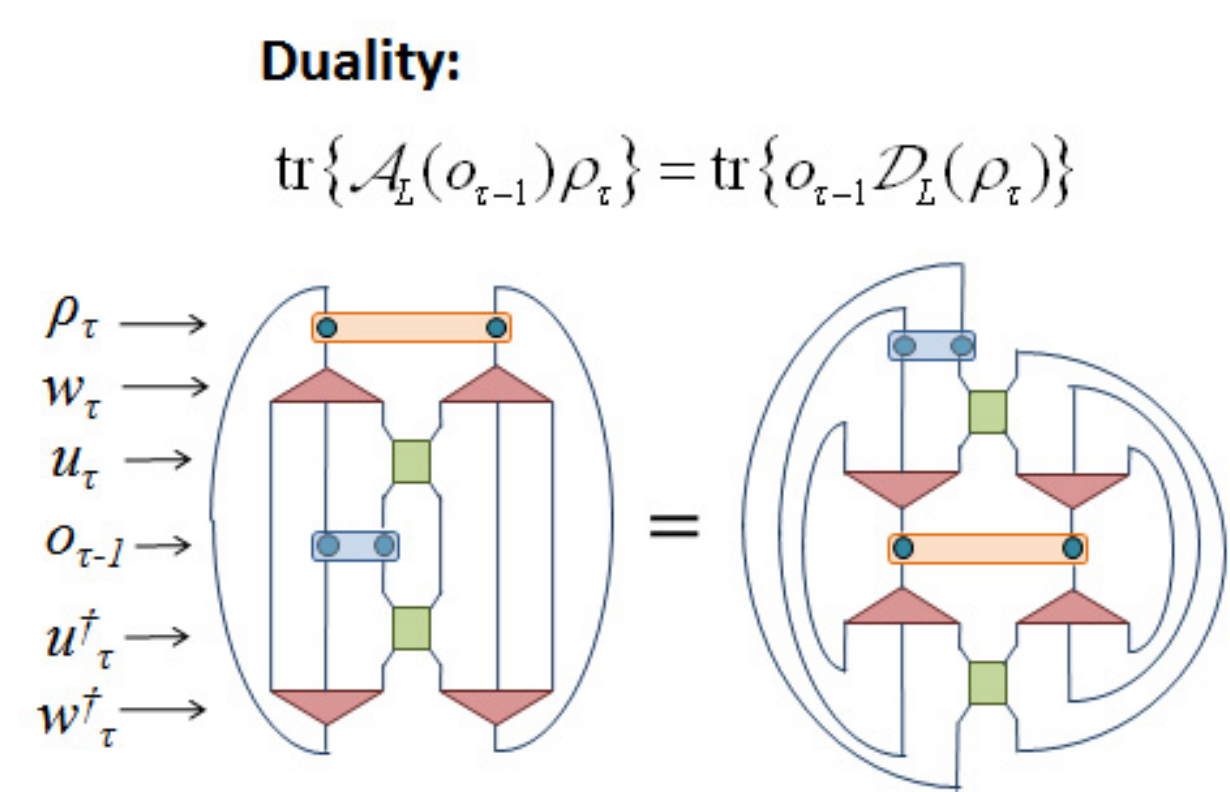}
\caption{ The ascending and descending superoperators, $\mathcal{A}$ and $\mathcal{D}$, are dual to each other, see Eq. \ref{eq:MERAalg:Duality}. This becomes evident by inspecting the above figure, where the superoperators are explicitly decomposed in terms of disentanglers and isometries.} 
\label{fig:MERAalg:Duality}
\end{center}
\end{figure}

As above, let $[r,r\!+\!1]$ denote two consecutive sites of lattice $\mathcal{L}_{\tau-1}$ and let $[r',r'\!+\!1]$ denote two consecutive sites of lattice $\mathcal{L}_{\tau}$ that lay inside the past causal cone of $[r,r\!+\!1] \in \mathcal{L}_{\tau-1}$. Given a density matrix $\rho_{\tau}^{[r',r'+1]}$ in $\mathcal{L}_{\tau}$, the \emph{descending superoperator} $\mathcal{D}$ of Fig. \ref{fig:MERAalg:DescendSuper} produces a density matrix $\rho_{\tau-1}^{[r,r+1]}$ in $\mathcal{L}_{\tau-1}$,
\begin{equation}
 \rho^{[r,r+1]}_{\tau-1} = \mathcal{D}(\rho^{[r',r'+1]}_{\tau}).
\end{equation}
Notice that the descending superoperator $\mathcal{D}$ (which depends on $\tau$, $r$ and $r'$) is the dual of the ascending superoperator $\mathcal{A}$, $\mathcal{D} = \mathcal{A}^{\star}$. Indeed, as can be checked in Fig. \ref{fig:MERAalg:Duality}, by construction we have that, for any $o^{[r,r+1]}_{\tau-1}$ and $\rho^{[r',r'+1]}_{\tau}$,
\begin{equation}
	\tr \left( o^{[r,r+1]}_{\tau-1}\mathcal{D}(\rho^{[r',r'+1]}_{\tau}) \right) 
	= \tr \left( \mathcal{A}(o^{[r,r+1]}_{\tau-1}) \rho^{[r',r'+1]}_{\tau} \right).
	\label{eq:MERAalg:Duality}
\end{equation}
Correspondingly, there are also three structurally different forms of the descending superoperators, namely left $\mathcal{D}_L$, center $\mathcal{D}_C$ and right $\mathcal{D}_R$ in Fig. \ref{fig:MERAalg:DescendSuper}.

\begin{figure}[!tbhp]
\begin{center}
\includegraphics[width=10cm]{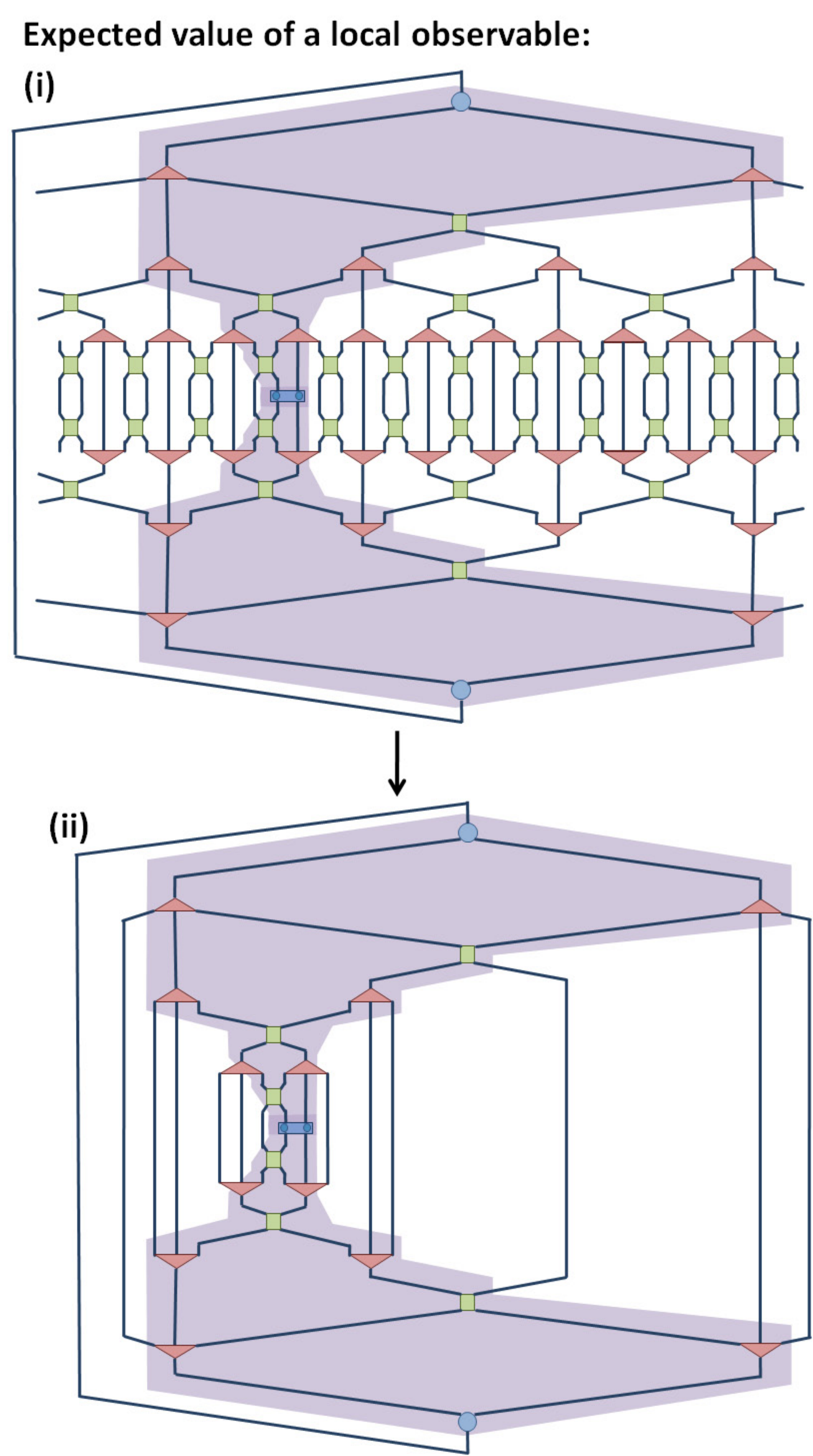}
\caption{ (i) Tensor network corresponding to the expected value $\tr(o^{[r,r+1]}P)$ of Eq. \ref{eq:MERAalg:ev_o}. The two-site operator $o^{[r,r+1]}$ is represented by a four-legged rectangle in the middle of the tensor network. The shaded region represents the past causal cone of sites $r,r+1 \in \mathcal{L}$. (ii) All isometric tensors that lay outside the past causal cone of sites $r,r+1 \in \mathcal{L}$ annihilate and we are left with a simpler tensor network.} 
\label{fig:MERAalg:LocalObs1}
\end{center}
\end{figure}

\begin{figure}[!tbhp]
\begin{center}
\includegraphics[width=10cm]{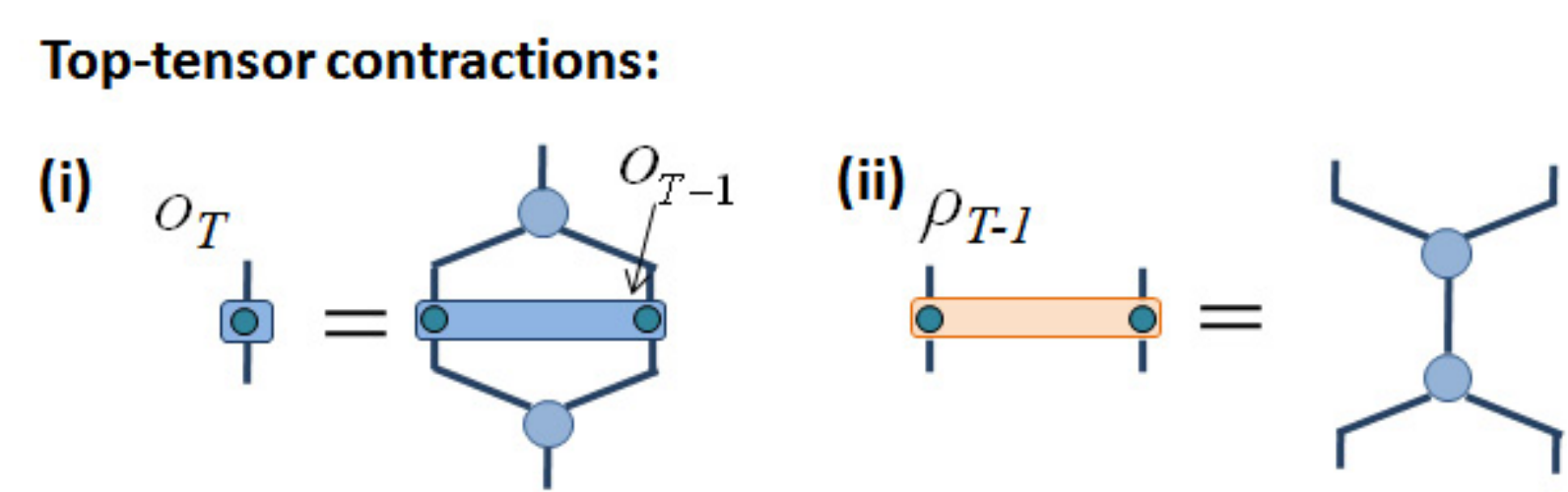}
\caption{ (i) The \emph{top tensor} transforms a two-site operator $o_{T-1}$ defined on lattice $\mathcal{L}_{T-1}$ into a one-site operator (a $\chi_T \times \chi_T$ matrix) $o_{T}$ on the top of the MERA. (ii) The two-site density matrix $\rho_{T-1}$ on lattice $\mathcal{L}_{T-1}$ is obtained through contraction of the top tensor with its conjugate. Notice that $\rho_{T-1}$, as well as all $\rho_\tau$, are normalized to have trace $\tr(\rho_{\tau}) = \chi_T$.}
\label{fig:MERAalg:TopTensor}
\end{center}
\end{figure}

\subsection{Evaluation of a two-site operator}

We can now proceed to compute the expected value $\tr (o^{[r,r+1]}P)$ of Eq. \ref{eq:MERAalg:ev_o} from the MERA. This computation corresponds to contracting the tensor network depicted in the upper half of Fig. \ref{fig:MERAalg:LocalObs1}.

In a key first step, the contraction of the tensor network for $\tr (o^{[r,r+1]}P)$ is significantly simplified by the fact that, by virtue of Eq. \ref{eq:MERAintro:isometry}, each isometric tensor outside the past causal cone of sites $[r,r\!+\!1] \in \mathcal{L}$ is annihilated by its Hermitian conjugate. As a result, we are left with a new tensor network that contains only (two copies of) the tensors in the causal cone, as represented in the second half of Fig. \ref{fig:MERAalg:LocalObs1}. Because the past causal cones in the MERA have a bounded width, this tensor network can now be contracted with a computational effort that grows with $N$ just as O$(\log N)$. One can proceed in several ways:

\begin{figure}[!tbhp]
\begin{center}
\includegraphics[width=10cm]{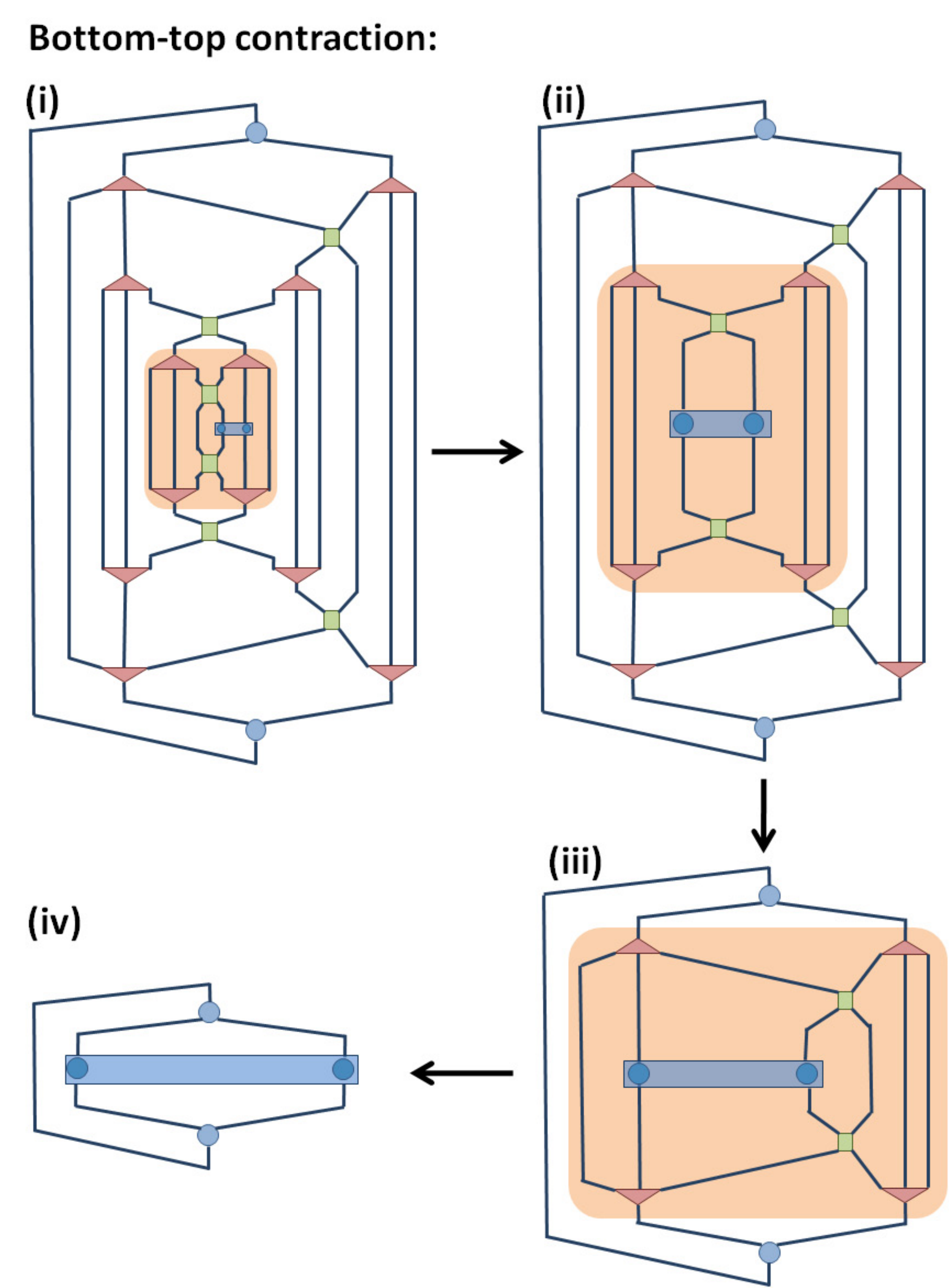}
\caption{ The contraction of the tensor network in the lower half of Fig. \ref{fig:MERAalg:LocalObs1} using the bottom-top approach corresponds to employing the ascending super operator $\mathcal{A}$ a number of times. In this particular case, we first use (i) $\mathcal{A}_R$, then (ii) $\mathcal{A}_C$ and then (iii) $\mathcal{A}_L$, to bring the tensor network into a simple form whose contraction gives a complex number: the expected value of Eq. \ref{eq:MERAalg:ev_o}. } 
\label{fig:MERAalg:BottomTop}
\end{center}
\end{figure}

\begin{figure}[!tbhp]
\begin{center}
\includegraphics[width=10cm]{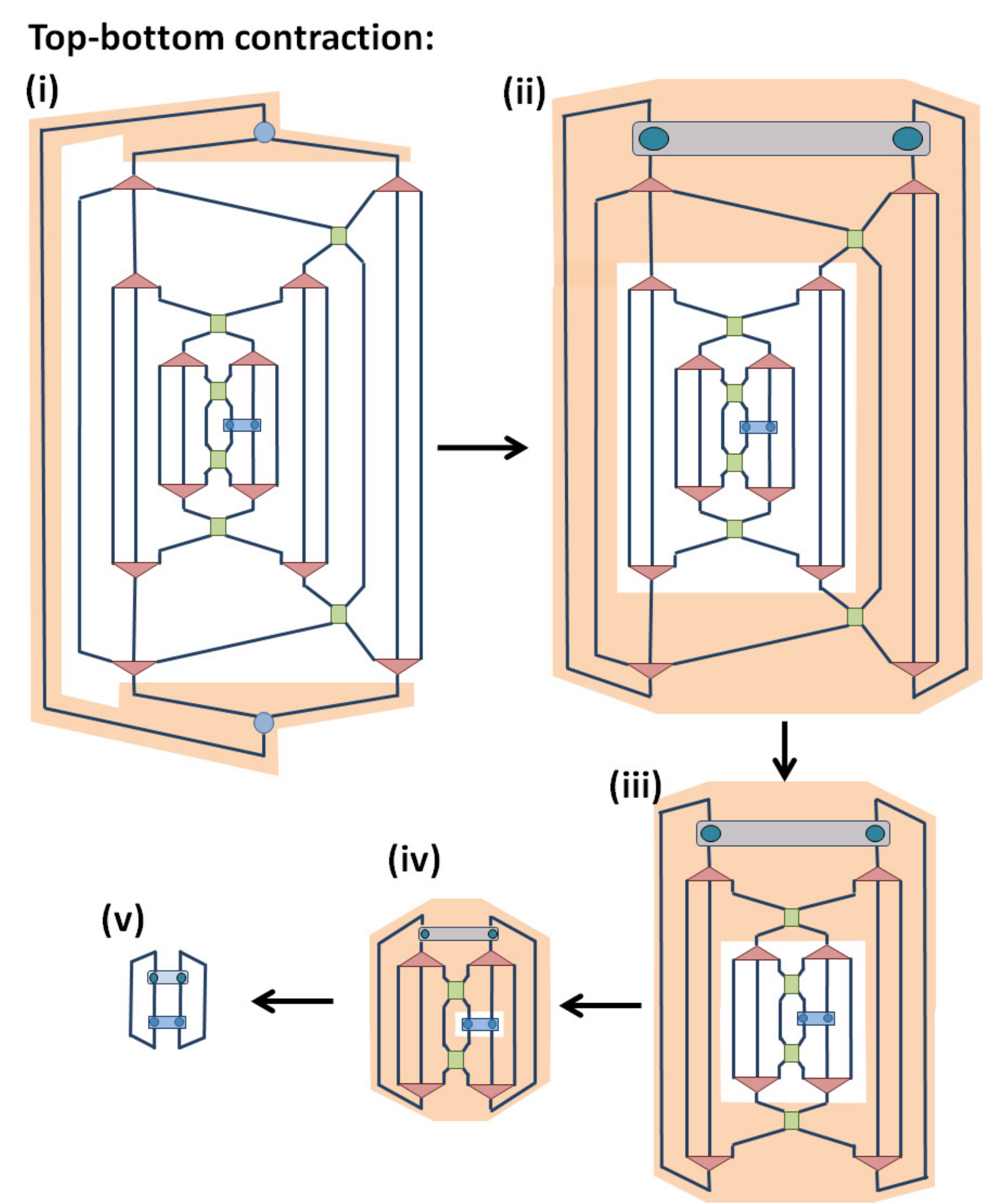}
\caption{ The contraction of the tensor network in the lower half of Fig. \ref{fig:MERAalg:LocalObs1} using the top-bottom approach corresponds to first implementing (i) a \emph{top tensor} contraction followed by repeated application of the descending super operator $\mathcal{D}$. Specifically, here we first use (ii) $\mathcal{D}_L$, then (iii) $\mathcal{D}_C$ and then (iv) $\mathcal{D}_R$, in order to compute the appropriate density matrix $\rho^{[r,r+1]}$ for two sites $[r,r+1] \in \mathcal{L}$. With the density matrix $\rho^{[r,r+1]}$ we can finally compute (v) the expectation value $\tr (o^{[r,r+1]}P) = \tr(o^{[r,r+1]} \rho^{[r,r+1]})$.} 
\label{fig:MERAalg:TopBottom}
\end{center}
\end{figure}
 
\textbf{Bottom-top.---} In the bottom-top approach, we would start by contracting the indices of $o^{[r,r+1]}$ and the disentanglers and isometries of the first layer ($\tau=1$) of the causal cone; then we would contract the indices of disentanglers and isometries of the second layer ($\tau = 2$); and so on (Fig. \ref{fig:MERAalg:BottomTop}). However, this corresponds to repeatedly applying the ascending superoperator $\mathcal{A}$ on $o_0^{[r,r+1]} \equiv o^{[r,r+1]}$. Therefore this is precisely how we proceed, obtaining a sequence of increasingly coarse-grained operators
\begin{equation}
	o^{[r,r+1]}_0 ~ \stackrel{\mathcal{A}}{\rightarrow} ~ o^{[r_1,r_1+1]}_1 ~ \stackrel{\mathcal{A}}{\rightarrow} ~ o^{[r_2,r_2+1]}_2 ~ \stackrel{\mathcal{A}}{\rightarrow}~ \cdots ~ o_T
\end{equation}
supported on lattices $\mathcal{L}_0$, $\mathcal{L}_1$, $\mathcal{L}_2$, $\cdots$, and $\mathcal{L}_{T}$ respectively.
Here, the $\chi_T\times \chi_T$ matrix $o_T$ at the top of the MERA is obtained according to Fig. \ref{fig:MERAalg:TopTensor} and the expected value of Eq. \ref{eq:MERAalg:ev_o} corresponds to its trace,
\begin{equation}
	\tr(o^{[r,r+1]}P) = \tr(o_T).
\end{equation}
 
\textbf{Top-bottom.---} In the top-bottom approach, we would instead start by contracting the indices of the tensors in the top layer ($\tau=T$) of the causal cone; then we would contract the indices of the tensors in the layer right below ($\tau=T-1$); and so on (Fig. \ref{fig:MERAalg:TopBottom}). However, that corresponds to first computing a density matrix $\rho_{T-1}$ for the two sites of $\mathcal{L}_{T-1}$ according to Fig. \ref{fig:MERAalg:TopTensor}  and then repeatedly applying the descending superoperator $\mathcal{D}$. Therefore this is how we proceed, producing a sequence of two-site density matrices 
\begin{equation}
	\rho_{T-1} ~ \stackrel{\mathcal{D}}{\rightarrow} ~ \cdots ~ \rho^{[r_2,r_2+1]}_2 ~ \stackrel{\mathcal{D}} {\rightarrow} ~\rho^{[r_1,r_1+1]}_1 ~ \stackrel{\mathcal{D}} {\rightarrow} ~ \rho_0^{[r,r+1]}
	\label{eq:MERAalg:sequenceDM}
\end{equation}
supported on lattices $\mathcal{L}_{T-1}$, $\cdots$, $\mathcal{L}_2$, $\mathcal{L}_1$ and $\mathcal{L}_{0}$ respectively \footnote{Each operator $\rho_{\tau}$ in Eq. {\ref{eq:MERAalg:sequenceDM}} is both Hermitian ($\rho_{\tau}^{\dagger} = \rho_{\tau}$) and non-negative ($\bra{\phi}\rho_{\tau}\ket{\phi} \geq 0, \forall \phi$) but its trace is $\tr(\rho_{\tau})=\chi_T$. For simplicity, we call $\rho_{\tau}$ a density matrix for any $\chi_T \geq 1$, even though it is only a proper density matrix for $\chi_T=1$.}. The last density matrix $\rho^{[r,r+1]} \equiv \rho^{[r,r+1]}_{0}$ describes the state of the two sites of $\mathcal{L}$ on which the local operator $o^{[r,r+1]}$ is supported. Therefore we can evaluate the expected value of $o^{[r,r+1]}$,
\begin{equation}
	\tr (o^{[r,r+1]}P) = \tr(o^{[r,r+1]} \rho^{[r,r+1]}).
\end{equation}

\textbf{Middle ground.---} More generally, one can also evaluate the expected value of Eq. \ref{eq:MERAalg:ev_o} through a mixed strategy where the ascending and descending superoperators are used to compute the operator $o^{[r_{\tau}, r_{\tau}+1]}_{\tau}$ and density matrix $\rho^{[r_{\tau}, r_{\tau}+1]}_{\tau}$ supported on lattice $\mathcal{L}_{\tau}$, which fulfill 
\begin{equation}
	\tr (o^{[r,r+1]}P) = \tr(o^{[r_{\tau}, r_{\tau}+1]}_{\tau} \rho^{[r_{\tau}, r_{\tau}+1]}_{\tau}).
\end{equation}

In all the cases above, one needs to use the ascending/descending superoperators about $T\approx \log N$ times, at a cost O$(\chi^8)$, so that the total computational cost is O$(\chi^8\log N)$.

\subsection{Evaluation of a sum of two-site operators}

In order to compute the expected value 
\begin{equation}
	\langle O \rangle_{\mathbb{V}_{U}} \equiv \tr (OP),~~~~~~~~~~O \equiv \sum_{r} o^{[r,r+1]}
\label{eq:MERAalg:ev_O1}
\end{equation}
of an operator $O$ on $\mathcal{L}$ that decomposes as the sum of two-site operators, we can write
\begin{equation}
\tr( O P) = \sum_{r} \tr( o^{[r,r+1]} P)
\label{eq:MERAalg:ev_O2}
\end{equation}
and individually evaluate each contribution $\tr( o^{[r,r+1]} P)$ by using e.g. the bottom-top strategy of the previous subsection, with a cost O$(\chi^8N\log N)$. However, by properly organizing the calculation, the cost of computing $\tr( O P)$ can be reduced to O$(\chi^8 N)$. We next describe how this is achieved. The strategy is closely related to the computation of expected values in the presence of translation invariance, as discussed later in this section. Again, there are several possible approaches:

\textbf{Bottom-top.---} We consider the sequence of operators
\begin{equation}
	O_0 ~ \stackrel{U_{1}^{\dagger}}{\rightarrow} ~ O_1 ~ \stackrel{U_{2}^{\dagger}}{\rightarrow} ~ O_2 ~ \stackrel{U_{3}^{\dagger}}{\rightarrow}~ \cdots ~ O_T, ~~~~~~~~O_0 \equiv O,
\label{eq:MERAalg:Os}
\end{equation}
where the operator $O_{\tau}$ is the sum of $N/3^{\tau}$ local operators,
\begin{equation}
	O_{\tau} = \sum_{r=1}^{N/3^{\tau}} o_{\tau}^{[r,r+1]}.
	\label{eq:MERAalg:O2}
\end{equation}
$O_{\tau-1}$ is obtained from $O_{\tau-1}$ by coarse-graining, $O_{\tau} = U_{\tau}^{\dagger} O_{\tau-1} U_{\tau}$. Each local operator $o^{[r,r+1]}_{\tau}$ in $O_{\tau}$ is the sum of three local operators from $O_{\tau-1}$ (see (a),(b) and (c) in Fig. \ref{fig:MERAalg:AscendSuper}), which are lifted to $\mathcal{L}_{\tau}$ by the three different forms of the ascending superoperator, $\mathcal{A}_L$, $\mathcal{A}_C$ and $\mathcal{A}_R$. Since $O_{\tau-1}$ has $N/3^{\tau-1}$ local operators, $O_{\tau}$ is obtained from $O_{\tau-1}$ by using the ascending superoperator $\mathcal{A}$ only $N/3^{\tau-1}$ times. Then, since $\sum_{\tau=0}^{T} 3^{-\tau} < 2$, this means that the entire sequence of Eq. \ref{eq:MERAalg:Os} requires using $\mathcal{A}$ only O$(N)$ times. Once $O_T$ is obtained, the expected value of $O$ follows from
\begin{equation}
	\tr(OP) = \tr(O_T).
\end{equation}

\textbf{Top-bottom.---} Here we consider instead the sequence of ensembles of density matrices
\begin{equation}
	E_{T-1} ~ \stackrel{U_{T-1}}{\rightarrow} ~ \cdots ~ E_2 ~ \stackrel{U_2} {\rightarrow} ~E_1 ~ \stackrel{U_1} {\rightarrow} ~ E_0,
	\label{eq:MERAalg:sequenceE}
\end{equation}
where $E_{\tau}$ is an ensemble of the $N/3^{\tau}$ two-site density matrices $\rho_{\tau}^{[r,r+1]}$ supported on nearest neighbor sites of $\mathcal{L}_{\tau}$,
\begin{equation}
	E_{\tau} \equiv \left\{ \rho_{\tau}^{[1,2]},  \rho_{\tau}^{[2,3]}, \cdots, \rho_{\tau}^{[N/3^{\tau},1]} \right\}
\end{equation}
From each density matrix in the ensemble $E_{\tau}$ we can generate three density matrices in the ensemble $E_{\tau-1}$ by applying the three different forms of the descending superoperator,  $\mathcal{D}_L$, $\mathcal{D}_C$ and $\mathcal{D}_R$ (see (a),(b) and (c) in Fig. \ref{fig:MERAalg:DescendSuper}). All the $N/3^{\tau-1}$ density matrices of ensemble $E_{\tau-1}$ can be obtained from density matrices of $E_{\tau}$ in this way. Since $\sum_{\tau=0}^{T} 3^{-\tau} < 2$, we see that by using the descending superoperator $\mathcal{D}$ only O$(N)$ times, we are able to compute all the density matrices in the sequence of ensembles of Eq. \ref{eq:MERAalg:sequenceE}. Once the ensemble $E_0\equiv E$ has been obtained, 
\begin{equation}
	E = \left\{ \rho^{[1,2]}, \rho^{[2,3]}, \cdots, \rho^{[N,1]} \right\},
\end{equation}
the expected value of $O$ follows from
\begin{equation}
	\tr(OP) = \sum_r \tr(o^{[r,r+1]}\rho^{[r,r+1]}).
\end{equation}

\textbf{Middle ground.---} More generally, we could build operator $O_\tau$ as well as ensemble $E_{\tau}$ and evaluate the expected value of $O$ from the equality
\begin{equation}
	\tr(OP) = \sum_{r} \tr(o_{\tau}^{[r,r+1]}\rho_{\tau}^{[r,r+1]}).
\end{equation}

Each of the strategies above require the use of the ascending/descending superoperators O$(N)$ times and therefore can indeed be accomplished with cost O$(\chi^8 N)$.

\begin{figure}[!tbhp]
\begin{center}
\includegraphics[width=10cm]{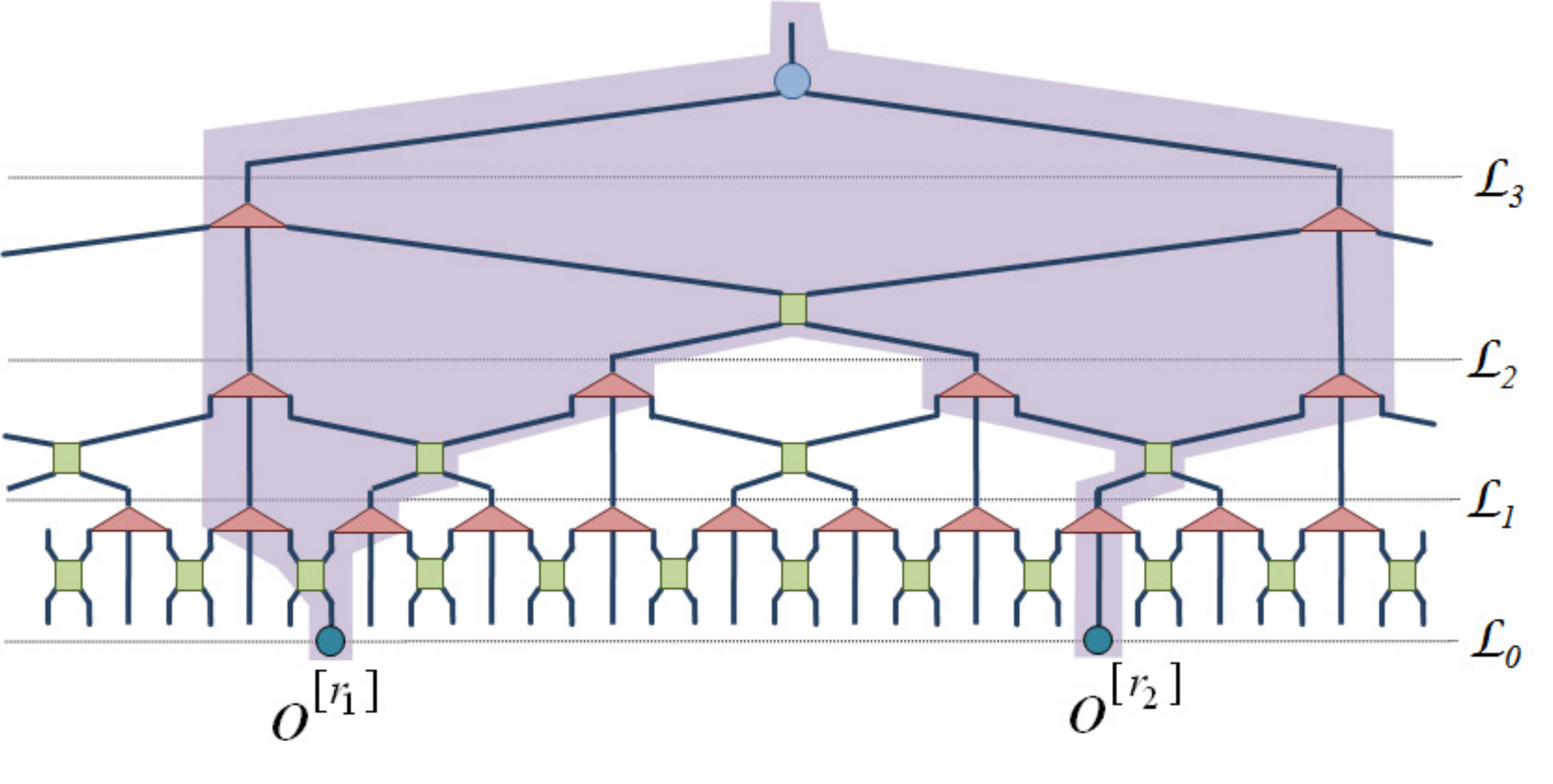}
\caption{ In order to compute a two-point correlator $C_2(r_1,r_2)$ we need to consider the union of the past causal cones of sites $r_1$ and $r_2$. Notice that, in contrast with the case of a single local operator, the joint causal cone of two distant sites typically involves more than two contiguous sites of some lattice $\mathcal{L}_{\tau}$. This makes the computational cost scale as a power of $\chi$ larger than $\chi^8$.}
\label{fig:MERAalg:CorrDifficult}
\end{center}
\end{figure}

\begin{figure}[!tbhp]
\begin{center}
\includegraphics[width=10cm]{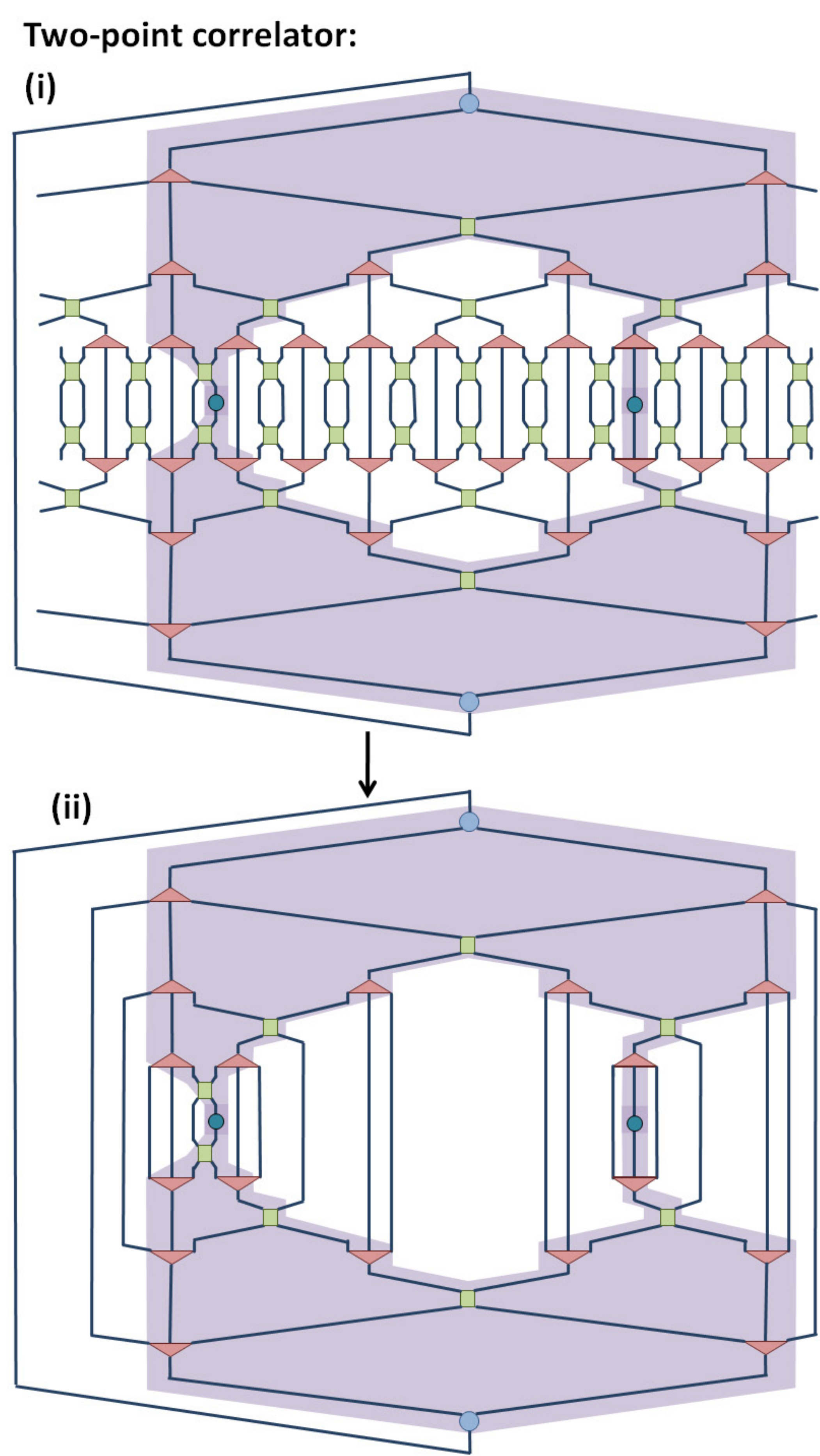}
\caption{ (i) Tensor network to be contracted in order to evaluate a two-point correlator $C_2(r_1,r_2)$.
Similarly to the case of a local observable Fig. \ref{fig:MERAintro:3MERA}, the tensors outside of the casual cone annihilate in pairs due to their isometric character, Eq. \ref{eq:MERAintro:isometry}. The resulting tensor network (ii) is much simpler network. However, for a generic pair of sites $r_1,r_2\in\mathcal{L}$, the joint past causal cone will contain more than just two sites per layer, resulting in a computational cost that scales with $\chi$ as a power larger than $\chi^8$.} 
\label{fig:MERAalg:CorrDifficultFull}
\end{center}
\end{figure}

\subsection{Evaluation of two-point correlators}
 
Let us now consider the computation of a two-point correlator of the form
\begin{equation}
C_2(r_1,r_2) \equiv	\bra{\Psi}o^{[r_1]}\otimes o^{[r_2]}\ket{\Psi},
\label{eq:MERAalg:two-point}
\end{equation}
where $o^{[r]}$ and $o^{[s]}$ denote two one-site operators applied on two arbitrary sites $r$ and $s$ of $\mathcal{L}$, see Fig. \ref{fig:MERAalg:CorrDifficult}. Fig. \ref{fig:MERAalg:CorrDifficultFull} shows the tensor network to be contracted. Again, we can use Eq. \ref{eq:MERAintro:isometry} to remove all disentanglers and isometries that lay outside the joint past causal cone for sites $r$ and $s$. Then, we can proceed to contract the resulting tensor network, for instance through a bottom-top or top-bottom approach, with the help of the ascending and descending superoperators (and generalizations thereof). Notice that since at intermediate layers the two legs of the causal cone may contain two sites each one, in general we will need to compute operators/density matrices that span more than just two sites, and the cost of their computation will be larger than O$(\chi^8)$. 

\begin{figure}[!tbhp]
\begin{center}
\includegraphics[width=10cm]{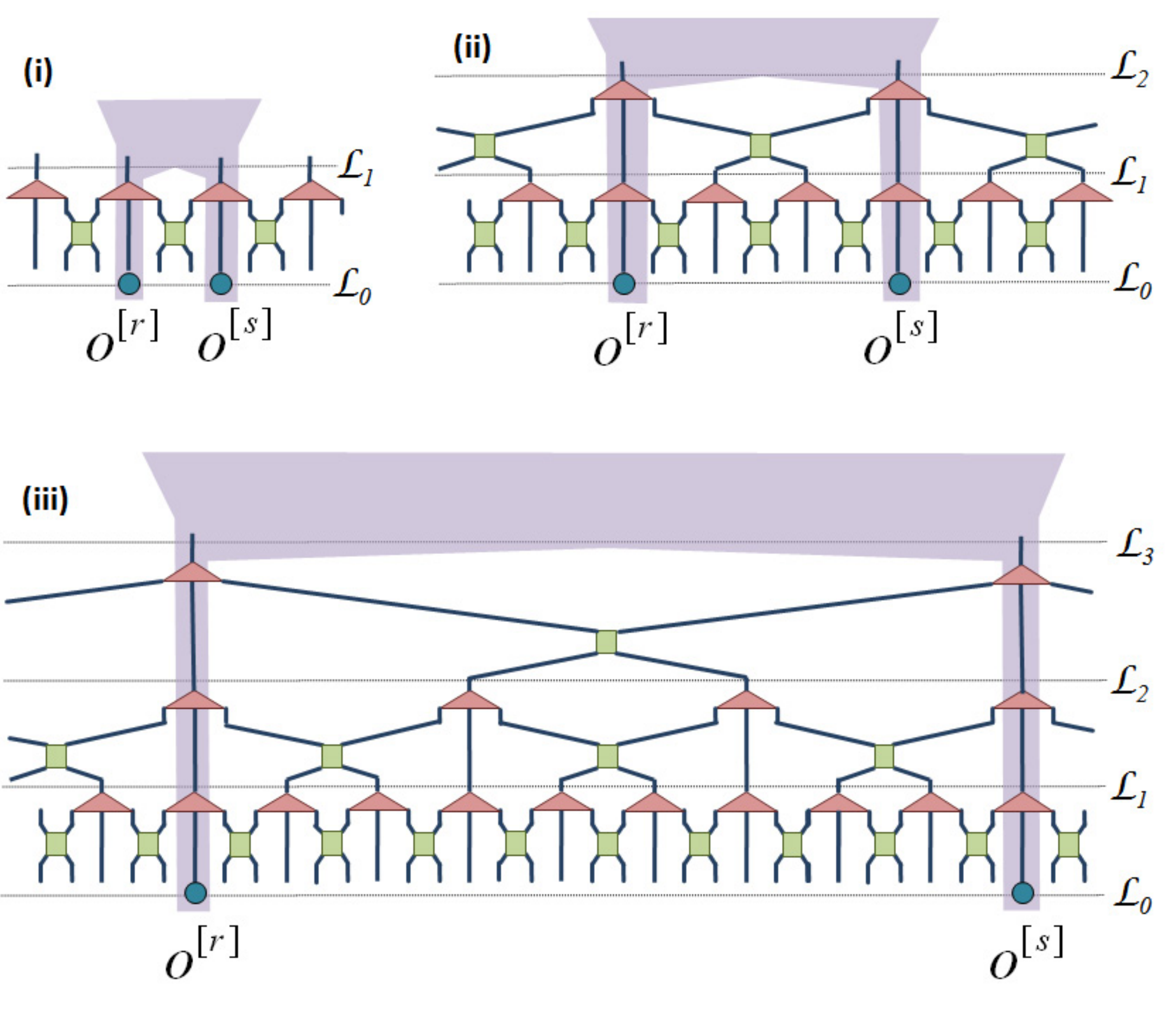}
\caption{ Two-point correlators for specific pairs of sites $r,s$ [at distances of $3^q$ sites for $q=1,2,3...$] can be computed with cost $O(\chi^8\log N)$. This is due to the fact that the causal cones for each of $r,s$ contains only one site until they meet--- (i) at $\mathcal L_1$, (ii) at $\mathcal L_2$ or (iii) at $\mathcal L_3$.}
\label{fig:MERAalg:CorrEasy}
\end{center}
\end{figure}

\begin{figure}[!tbhp]
\begin{center}
\includegraphics[width=10cm]{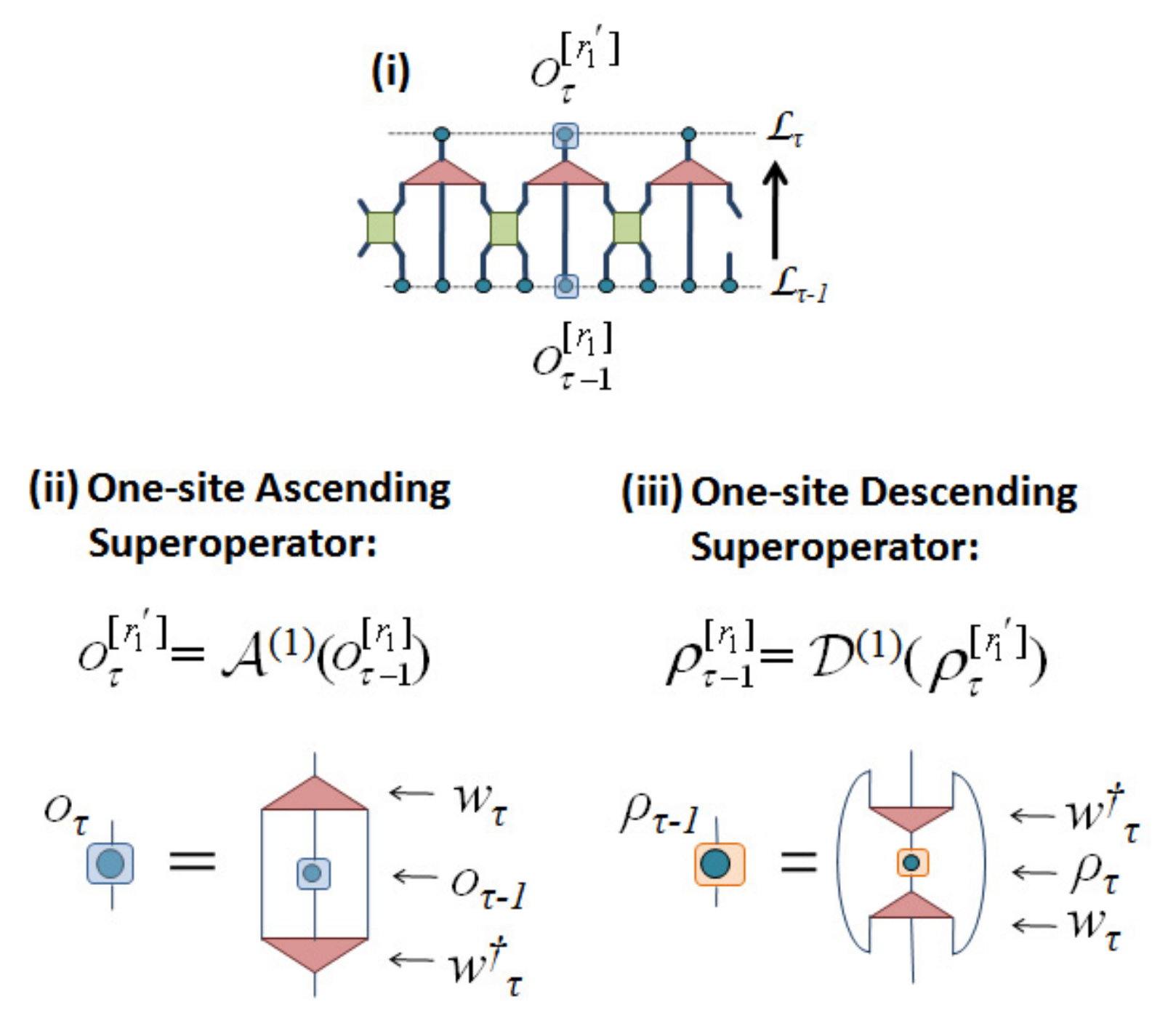}
\caption{ A one-site operator $o_{\tau-1}$ supported on certain sites of $\mathcal{L}_{\tau-1}$ (corresponding to the central wire of an isometry $w_{\tau}$) is mapped onto a single-site operator on $\mathcal{L}_{\tau}$. In this case the (i) ascending and (ii) descending superoperators $\mathcal{A}^{(1)}$ and $\mathcal{D}^{(1)}$ have a very simple form.} 
\label{fig:MERAalg:OneSiteSuper}
\end{center}
\end{figure}

However, for specific choices of sites $r,s\in \mathcal{L}$, we are still able to compute $C_2(r,s)$ with overall cost $O(\chi^8\log N)$, as illustrated in Fig. \ref{fig:MERAalg:CorrEasy}. We emphasize that this was not possible in the binary 1$D$ MERA and is one of the main reasons to work with the ternary 1$D$ MERA. For such choices of sites $r$ and $s$, each of the two legs of the joint past causal cone contains just one site until, at some layer $\tau_0$, they fuse into a single two-site leg. We can introduce one-site ascending and descending superoperators $\mathcal{A}^{(1)}$ and $\mathcal{D}^{(1)}$ (Fig. \ref{fig:MERAalg:OneSiteSuper}), in terms of which we can express, for $\tau \leq \tau_0$, the transformation of a product operator $o_{\tau-1}^{[r]}\otimes o^{[s]}_{\tau-1}$ into a product operator
\begin{equation}
	o_{\tau}^{[r']}\otimes o^{[s']}_{\tau} = \mathcal{A}^{(1)}(o_{\tau-1}^{[r]})\otimes \mathcal{A}^{(1)}(o^{[s]}_{\tau-1}),
\end{equation}
or of a density matrix $\rho_{\tau}^{[r',s']}$ into a density matrix
\begin{equation}
	\rho_{\tau-1}^{[r,s]} = (\mathcal{D}^{(1)} \otimes \mathcal{D}^{(1)})(\rho_{\tau}^{[r',s']}),
\end{equation}
where $r,s \in \mathcal{L}_{\tau-1}$ and $r',s' \in \mathcal{L}_{\tau}$ are sites corresponding to single-site legs of the causal cone. In, say, the bottom-top approach we can compute the correlator of Eq. \ref{eq:MERAalg:two-point} by using the single-site ascending superoperator $\mathcal{A}^{(1)}$ for layers $\tau\leq\tau_0$ and then the two-site ascending super-operator $\mathcal{A}$ for layer $\tau > \tau_0$.

\subsection{Translation invariance} 

The computation of the expected value $\tr(o^{[r,r+1]}P)$ of a single local operator $o^{[r,r+1]}$ in the case of a translation invariant MERA (see Fig. \ref{fig:MERAintro:MERAtypes}) can proceed as explained earlier in this section. In the present case one would expect the result to be independent of the sites $[r,r+1]\in\mathcal{L}$ on which the operator is supported, but a finite bond dimension $\chi$ typically introduces small space inhomogeneities in the reduced density matrix $\rho^{[r,r+1]}$ and therefore also in $\tr(o^{[r,r+1]}P) = \tr(o^{[r,r+1]}\rho^{[r,r+1]})$. 

Given a two-site operator $o$, an expected value that is independent of $[r,r+1]$ can be obtained by computing an average over sites,
\begin{eqnarray}
	\tr(o^{[r,r+1]}P) ~~\rightarrow && \frac{1}{N} \sum_r \tr(o^{[r,r+1]}P) \label{eq:MERAalg:av_ev1}\\ 
	&=& \frac{1}{N} \sum_r \tr(o^{[r,r+1]}\rho^{[r,r+1]}),
	\label{eq:MERAalg:av_ev2}
\end{eqnarray}
where the terms $o^{[r,r+1]}$ denote translations of the same operator $o$. This average can be computed e.g. by obtaining the $N$ density matrices $\rho^{[r,r+1]}$ individually and then adding them together, with an overall cost O$(\chi^8 N)$. However, with a better organization of the calculation the cost can be reduced to O$(\chi^8 \log N)$.

We first need to introduce average versions of the ascending and descending superoperators. Given a two-site operator $o_{\tau-1}$ in lattice $\mathcal{L}_{\tau-1}$, we can build a two-site operator $o_{\tau}$ by using an average of the three two-site operators resulting from lifting $o_{\tau-1}$ to lattice $\mathcal{L}_{\tau}$, namely $\mathcal{A}_{L}(o_{\tau-1})$, $\mathcal{A}_{L}(o_{\tau-1})$ and $\mathcal{A}_{L}(o_{\tau-1})$. In terms of the \emph{average ascending superoperator} $\bar{\mathcal{A}}$,
\begin{equation}
	\bar{\mathcal{A}} \equiv \frac{1}{3}(\mathcal{A}_L + \mathcal{A}_C + \mathcal{A}_R),
	\label{eq:MERAalg:avAsc}
\end{equation}
this transformation reads
\begin{equation}
	o_{\tau} = \bar{\mathcal{A}}(o_{\tau-1}).	
	\label{eq:MERAalg:avAsc2}
\end{equation}
Importantly, if we coarse-grain the translation invariant operator 
\begin{equation}
 \frac{1}{N_{\tau\!-\!1}} \sum_r o_{\tau\!-\!1}^{[r,r+1]},~~~~~~~~~~N_x\equiv N/3^{x},
 \label{eq:MERAalg:TIo}
\end{equation}
where $N_{\tau\!-\!1}$ is the number of sites of $\mathcal{L}_{\tau\!-\!1}$ and the terms $o_{\tau\!-\!1}^{[r,r+1]}$ denote translations of $o_{\tau\!-\!1}$, the resulting operator can be written as
\begin{equation}
 \frac{1}{N_{\tau}} \sum_r o_{\tau}^{[r,r+1]},
\end{equation}
where the terms $o_{\tau-1}^{[r,r+1]}$ denote translations of $o_{\tau}$ and where $o_{\tau}$ and $o_{\tau-1}$ are related through Eq. \ref{eq:MERAalg:avAsc2}. In other words, the average ascending superoperator $\bar{\mathcal{A}}$ can also be used to characterize the coarse-graining, in the translation invariant case, of operators of the form of Eq. \ref{eq:MERAalg:ev_O1}.

Let $\bar{\rho}_{\tau}$ denote the two-site density matrix obtained by averaging over all density matrices $\rho^{[r,r+1]}_{\tau}$ on different pairs $[r,r+1]$ of two contiguous sites of $\mathcal{L}_{\tau}$, 
\begin{equation}
	\bar{\rho}_{\tau} \equiv \frac{1}{N_{\tau}} \sum_r \rho_{\tau}^{[r,r+1]},
\end{equation}
and similarly for lattice $\mathcal{L}_{\tau-1}$,
\begin{equation}
	\bar{\rho}_{\tau-1} \equiv \frac{1}{N_{\tau-1}} \sum_r \rho_{\tau-1}^{[r,r+1]}.
\end{equation}
Recall that each density matrix $\rho^{[r,r+1]}_{\tau}$ on lattice $\mathcal{L}_{\tau}$ gives rise to three density matrices in $\mathcal{L}_{\tau-1}$ according to the three versions of the descending superoperator, namely $\mathcal{D}_{L}$, $\mathcal{D}_{C}$ and $\mathcal{D}_{R}$. It follows that the density matrix $\bar{\rho}_{\tau-1}$ can be obtained from the density matrix $\bar{\rho}_{\tau}$ by using the \emph{average descending superoperator},
\begin{equation}
	\bar{\mathcal{D}}  \equiv \frac{1}{3}(\mathcal{D}_L + \mathcal{D}_C + \mathcal{D}_R),
	\label{eq:MERAalg:avDesc}
\end{equation}
that is
\begin{equation}
	\bar{\rho}_{\tau-1} = \bar{\mathcal{A}}(\bar{\rho}_{\tau}).	
	\label{eq:MERAalg:avDesc2}
\end{equation}

We can now proceed to compute the average expected value of Eqs. \ref{eq:MERAalg:av_ev1}-\ref{eq:MERAalg:av_ev2}. This can be accomplished in several alternative ways.

\textbf{Bottom-top.---} Given a two-site operator $o$, we compute a sequence of increasingly coarse-grained operators 
\begin{equation}
	o_0 ~ \stackrel{\bar{\mathcal{A}}}{\rightarrow} ~ o_1 ~ \stackrel{\bar{\mathcal{A}}}{\rightarrow} ~ o_2 ~ \stackrel{\bar{\mathcal{A}}}{\rightarrow}~ \cdots ~ o_T,~~~~~~~o_0\equiv o,
\end{equation}
where $o_{\tau}$ is obtained from $o_{\tau-1}$ by means of the average ascending superoperator $\bar{\mathcal{A}}$. Then we simply have
\begin{equation}
	\frac{1}{N} \sum_r \tr(o^{[r,r+1]}P) = \tr(o_T).
\end{equation}

\textbf{Top-bottom.---} Alternatively, we can compute the sequence of average density matrices
\begin{equation}
	\bar{\rho}_{T-1} ~ \stackrel{\bar{\mathcal{D}}}{\rightarrow} ~ \cdots ~ \bar{\rho}_2 ~ \stackrel{\bar{\mathcal{D}}} {\rightarrow} ~\bar{\rho}_1 ~ \stackrel{\bar{\mathcal{D}}} {\rightarrow} ~ \bar{\rho}_0,
\end{equation}
where $\bar{\rho}_{\tau-1}$ is obtained from $\bar{\rho}_{\tau}$ by means of the average descending superoperator $\bar{\mathcal{D}}$ and where $\bar{\rho} \equiv \bar{\rho}_{0}$ corresponds to
the average density matrix on lattice $\mathcal{L}$, 
\begin{equation}
	\bar{\rho} \equiv \frac{1}{N} \sum_r \rho^{[r,r+1]},
\end{equation}
in terms of which we can express the average expected value as
\begin{equation}
	\frac{1}{N} \sum_r \tr(o^{[r,r+1]}P) = \tr(o\bar{\rho}).
\end{equation}

\textbf{Middle ground.---} As costumary, we can also use both $\bar{\mathcal{A}}$ and $\bar{\mathcal{D}}$ to compute $o_{\tau}$ and $\bar{\rho}_{\tau}$, and evaluate the average expected value as
\begin{equation}
	\frac{1}{N} \sum_r \tr(o^{[r,r+1]}P) = \tr(o_\tau\bar{\rho}_{\tau}).
\end{equation}

In all the above strategies the average ascending and descending superoperators $\bar{\mathcal{A}}$ and $\bar{\mathcal{D}}$ are used O$(\log(N))$ times and therefore the computational cost scales as O$(\chi^8 \log N)$.

To summarize, with a translation invariant MERA we can coarse-grain a single two-site operator $o$ (with a transformation that involves averaging over all possible causal cones) or compute the average density matrix $\bar{\rho}$ by using the average ascending/descending superoperators. This leads to a sequence of operators $o_{\tau}$ and density matrices $\bar{\rho}_{\tau}$,
\begin{eqnarray}
	&&o_0  \stackrel{\bar{\mathcal{A}}}{\rightarrow}  o_1 \stackrel{\bar{\mathcal{A}}}{\rightarrow} ~\cdots~ \stackrel{\bar{\mathcal{A}}}{\rightarrow} o_T, ~~~~~~~~~ o_0 \equiv o, \label{eq:MERAalg:ATI}\\
	&&\bar{\rho}_0  \stackrel{\bar{\mathcal{D}}}{\leftarrow}  \bar{\rho}_1 \stackrel{\bar{\mathcal{D}}}{\leftarrow} ~\cdots~ \stackrel{\bar{\mathcal{D}}}{\leftarrow} \bar{\rho}_T, ~~~~~~~~~ \bar{\rho}_0 \equiv \bar{\rho}, \label{eq:MERAalg:DTI}
\end{eqnarray}
from which the expected value of $o$ is obtained as $\tr(o\bar{\rho})$, as $\tr(o_T)$ or, more generally, as $\tr(o_{\tau}\bar{\rho}_{\tau})$.

\subsection{Scale invariance}

In the case of a translation invariant MERA that is also scale invariant (see Fig. \ref{fig:MERAintro:MERAtypes}), the average ascending superoperators $\bar{\mathcal{A}}$ is identical on each layer $\tau$, since it is always made of the same disentangler $u$ and isometry $w$. We then refer to it as the \emph{scaling superoperator} $\mathcal{S}$ \cite{pfeifer08}. Its dual $\mathcal{S}^{*}$ corresponds to the descending superoperator $\bar{\mathcal{D}}$. 

As derived in Ref. \cite{giovannetti08}, the expected value of a local observable $o$ in the thermodynamic limit can be obtained by analyzing the spectral decomposition of the scaling superoperator $\mathcal{S}$,
\begin{equation}
	\mathcal{S}(\bullet) = \sum_{\alpha} \lambda_{\alpha} \phi_{\alpha} \tr(\hat{\phi}_{\alpha} \bullet),~~~~~~\tr(\hat{\phi}_{\alpha} \phi_{\beta}) = \delta_{\alpha\beta}.
	\label{eq:MERAalg:spectral}
\end{equation}
We refer to the eigenoperators $\phi_{\alpha}$ of $\mathcal{S}$, 
\begin{equation}
	\mathcal{S}(\phi_{\alpha}) = \lambda_{\alpha} \phi_{\alpha},
\end{equation}
as the \emph{scaling operators}. Notice that the operators $\hat{\phi}_{\alpha}$, which are bi-orthonormal to the operators $\phi_{\alpha}$, are eigenoperators of $\mathcal{S}^{*}$,
\begin{equation}
	\mathcal{S}^{*}(\hat{\phi}_{\alpha}) = \lambda_{\alpha} \hat{\phi}_{\alpha}.
\end{equation}

We recall that the scaling operator $\mathcal{S}$ is made of isometric tensors (cf. Eq. \ref{eq:MERAintro:isometry}) and therefore the identity operator $\mathbb{I}$ is an eigenoperator of $\mathcal{S}$ with eigenvalue 1 (that is, $\mathcal{S}$ is unital),
\begin{equation}
	\mathcal{S}(\mathbb{I}) = \mathbb{I}.
\end{equation}
On the other hand, since the MERA is built as a quantum circuit ---and descending through the causal cone corresponds to advancing in the time of a quantum evolution--- it is obvious that the descending superoperator $\mathcal{D}$ is a quantum channel, and so are $\bar{\mathcal{D}}$ and $\mathcal{S}^{*}$ (see also \cite{giovannetti08}). In particular, $\mathcal{S}^{*}$ is a contractive superoperator \cite{bratteli79}, which means that the eigenvalues $\lambda_{\alpha}$ in Eq. \ref{eq:MERAalg:spectral} are constrained to fulfill  $|\lambda_{\alpha}| \leq 1$. In practical simulations \cite{pfeifer08} one finds that the identity operator $\mathbb{I}$ is the only eigenoperator of $\mathcal{S}$ with eigenvalue one, $\lambda_{\mathbb{I}} = 1$, and that $|\lambda_{\alpha}| < 1$ for $\alpha \neq \mathbb{I}$. Let $\hat{\rho}$ denote the corresponding unique fixed point of $\mathcal{S}^{*}$, 
\begin{equation}
	\mathcal{S}^{*}(\hat{\rho}) = \hat{\rho}.
	\label{eq:MERAalg:hat_rho}
\end{equation}
In an infinite system, the local density matrix of any lattice $\mathcal{L}_{\tau}$ (with finite $\tau$) results from applying $\mathcal{S}^{*}$ on $\rho^{T}$ an infinite number of times, and it is therefore equal to the fixed point $\hat{\rho}$ \cite{giovannetti08}. Consequently, Eqs. \ref{eq:MERAalg:ATI} and \ref{eq:MERAalg:DTI} are then replaced with
\begin{eqnarray}
	&&o_0  \stackrel{\mathcal{S}}{\rightarrow}  o_1 \stackrel{\mathcal{S}}{\rightarrow} o_2 \stackrel{\mathcal{S}}{\rightarrow} o_3 ~\cdots , ~~~~~~~~~ o_0 \equiv o, \label{eq:MERAalg:ASI}\\
	&&\hat{\rho}~  \stackrel{~\mathcal{S}^{*}}{\leftarrow}  \hat{\rho} \stackrel{~\mathcal{S}^{*}}{\leftarrow}  \hat{\rho} \stackrel{~\mathcal{S}^{*}}{\leftarrow}  \hat{\rho} ~~\cdots, ~~~~~~~~~  \label{eq:MERAalg:DSI}
\end{eqnarray}
where in addition, by decomposing $o$ in terms of the scaling operators $\phi_{\alpha}$,
\begin{equation}
	o = \sum_{\alpha} c_{\alpha} \phi_{\alpha}, ~~~c_{\alpha} \equiv \tr(\hat{\phi}_{\alpha} o),
\label{eq:MERAalg:decompose_o}
\end{equation}
we can explicitly compute $o_{\tau}$:
\begin{equation}
	o_{\tau} = \big(\underbrace{\mathcal{S}\circ \cdots \circ \mathcal{S}}_{\tau \mbox{ \scriptsize{times}}}\big)   (o) = \sum_{\alpha} c_{\alpha} (\lambda_{\alpha})^{\tau}\phi_{\alpha}.
\end{equation}
This expression shows that, unless $c_{\mathbb{I}} \neq 0$, the operator $o_{\tau}$ decreases exponentially with $\tau$ (recall that $|\lambda_{\alpha}| < 1$ for $\alpha \neq \mathbb{I}$) and its expected value must vanish. The average expected value of $o$ then reads:
\begin{equation}
	\lim_{N\rightarrow \infty} \frac{1}{N} \sum_r \tr(o^{[r,r+1]}P) = \tr(o\hat{\rho}).
\label{eq:MERAalg:SI_o}
\end{equation}

Two-point correlators for selected positions can also be expressed in a simple way, by considering the \emph{one-site scaling superoperator} $\mathcal{S}^{(1)}$, which is how we refer to the superoperator $\mathcal{A}^{(1)}$ of Fig. \ref{fig:MERAalg:OneSiteSuper} in the case of a scale invariant MERA. Its spectral decomposition, 
\begin{equation}
	\mathcal{S}^{(1)}(\bullet) = \sum_{\alpha} \mu_{\alpha} \psi_{\alpha} \tr(\hat{\psi}_{\alpha} \bullet),~~~~~~\tr(\hat{\psi}_{\alpha} \psi_{\beta}) = \delta_{\alpha\beta},
	\label{eq:MERAalg:spectraltwo}
\end{equation}
provides us with a new set of (one-site) scaling operators $\psi_{\alpha}$. Given two arbitrary one-site operators $o$ and $o'$, we can always decompose them in terms of these $\psi_{\alpha}$ (similarly as in Eq. \ref{eq:MERAalg:decompose_o}). Thus we can focus directly on a correlator of the form	$\langle \psi_{\alpha}^{[r]} \psi_{\beta}^{[s]}\rangle$. Here $r$ and $s$ are restricted to selected positions as in Fig. \ref{fig:MERAalg:CorrEasy}. Then we have
\begin{equation}
	\langle \psi_{\alpha}^{[r]} \psi_{\beta}^{[s]}\rangle = \frac{C_{\alpha\beta}} {|r-s|^{\Delta_{\alpha}+\Delta_{\beta}}},
\label{eq:MERAalg:SICorr}
\end{equation}
where $\Delta_{\alpha} \equiv \log_{3} \mu_{\alpha}$ is the \emph{scaling dimension} of the scaling operator $\psi_{\alpha}$, whereas $C_{\alpha\beta}$ is given by
\begin{equation}
	C_{\alpha\beta} \equiv \tr\left((\psi_{\alpha}\otimes \psi_{\beta})\hat{\rho}\right),
\label{eq:MERAalg:SICorr2}
\end{equation}
with $\hat{\rho}$ from Eq. \ref{eq:MERAalg:hat_rho}. 

In deriving Eq. \ref{eq:MERAalg:SICorr} we have used that, by construction,  $|r-s| = 3^q$ for some $q=1,2,3,\cdots$. Coarse-graining $\psi_{\alpha}^{[r]} \psi_{\beta}^{[s]}$ a number $q$ of times produces a multiplicative factor $(\mu_{\alpha}\mu_{\beta})^q$ and the residual two-site operator $\psi_{\alpha}^{[0]} \psi_{\beta}^{[1]}$, whose expected value gives $C_{\alpha\beta}$. On the other hand, by noting that $\mu^{q} = \mu^{\log_3 |r-s|} = |r-s|^{\log_3 \mu} = |r-s|^{\log_3 \Delta}$, we arrive at $(\mu_{\alpha}\mu_{\beta})^q = |r-s|^{\Delta_{\alpha}+\Delta_{\beta}}$, which explains the denominator in Eq. \ref{eq:MERAalg:SICorr}. 

In conclusion, from the scale invariant MERA we can characterize the expected value of local observables and two-point correlators, as expressed in Eqs. \ref{eq:MERAalg:SI_o} and \ref{eq:MERAalg:SICorr}-\ref{eq:MERAalg:SICorr2}. All critical exponents of the theory can be extracted from the scaling dimensions $\Delta_{\alpha}$. The scale invariant MERA is further explored in Chapter \ref{chap:1DCrit}, including a description of an algorithm to optimise for a scale invariant MERA and benchmark computations of scaling dimensions $\Delta_{\alpha}$ for several different spin models.

The required manipulations include computing $\bar{\rho}$ from $\mathcal{S}$ (using sparse diagonalization techniques) and diagonalizing $\mathcal{S}^{(1)}$, all of which can be accomplished with the ternary 1$D$ MERA with cost O$(\chi^8)$.

\section{Optimization of a disentangler/isometry} \label{sect:MERAalg:optim}

In preparation for the algorithms to be described in the next section, here we explain how to optimize a single tensor of the MERA.

Let $H$ be a Hamiltonian made of nearest neighbor, two-site interactions $h^{[r,r+1]}$,
\begin{equation}
	H = \sum_{r} h^{[r,r+1]}.
	\label{eq:MERAalg:H}
\end{equation}
For purposes of the optimization below, we choose each term $h^{[r,r+1]}$ so that it has no positive eigenvalues, $h^{[r,r+1]} \leq 0$. [This can be achieved with the simple replacement $h^{[r,r+1]} \rightarrow h^{[r,r+1]} - \lambda_{\max} I$, where $\lambda_{\max}$ is the largest eigenvalue of $h^{[r,r+1]}$.]

\begin{figure}[!tbhp]
\begin{center}
\includegraphics[width=10cm]{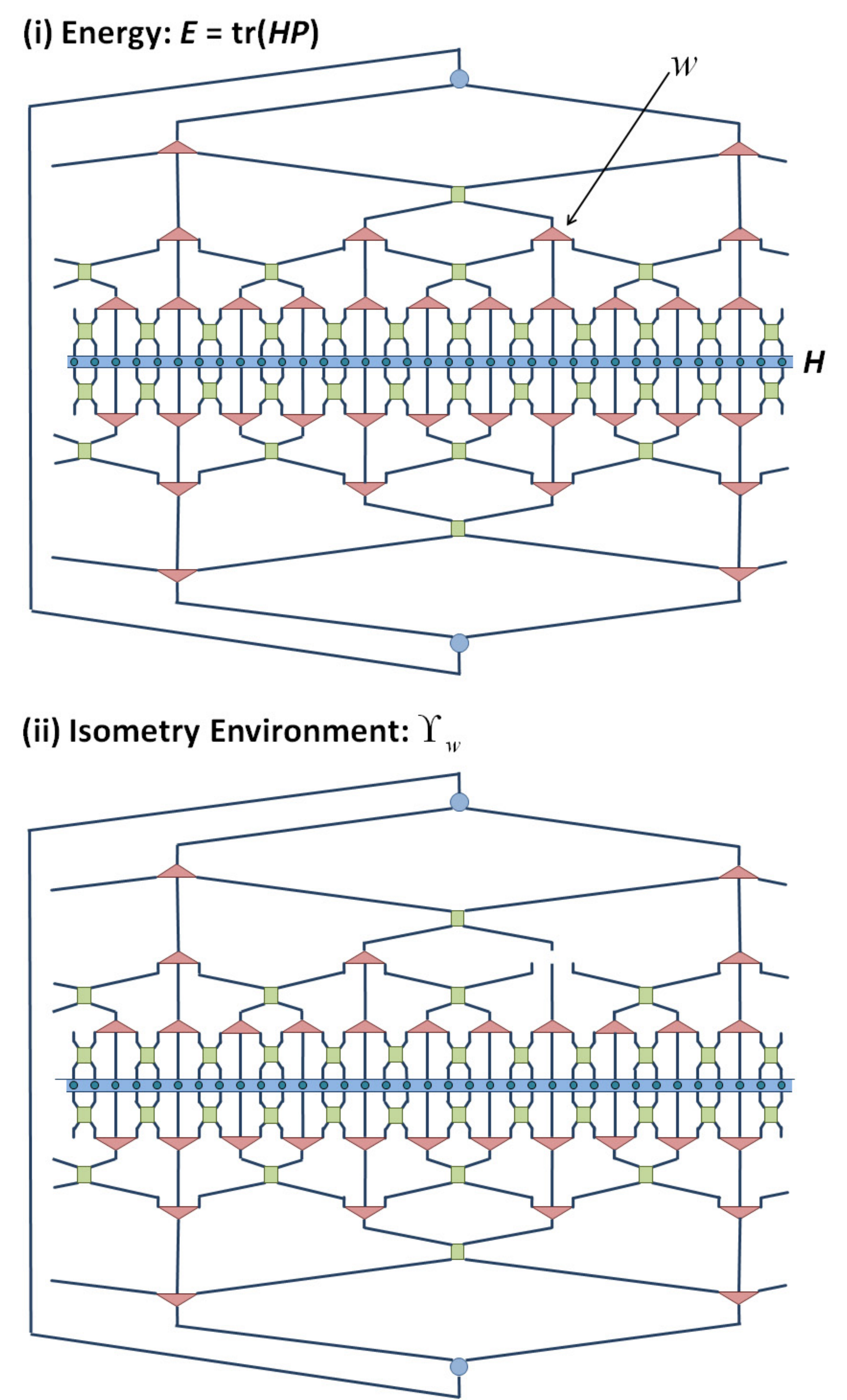}
\caption{ (i) The energy of a MERA, defined $E \equiv \tr(HP)$, is represented explicitly in terms of a tensor network. The removal of an isometry $w$ from this network gives (ii) the environment $\Upsilon_w$ for $w$ (and similarly for disentanglers $u$). By construction we have that $E = \textrm{tr} (w \Upsilon_w) $.}
\label{fig:MERAalg:IsoLargeEnviro}
\end{center}
\end{figure}

Our goal for the time being will be to minimize the energy (Fig. \ref{fig:MERAalg:IsoLargeEnviro}.i)
\begin{equation}
	E \equiv \tr(HP),
\label{eq:MERAalg:E}
\end{equation}
where $P$ is a projector onto the $\chi_T$-dimensional subspace $\mathbb{V}_{U} \in \mathbb{V}_{\mathcal{L}}$ by modifying only one of the tensors of the MERA. The optimization of a disentangler $u$ is very similar to that of an isometry $w$, and we can focus on describing the latter in more detail. 

Suppose then that, given a MERA, we want to optimize an isometry $w$ while keeping the rest of the tensors fixed. The cost function $E$ is quadratic in $w$ (more specifically, it depends bi-linearly on $w$ and $w^\dagger$),
\begin{equation}
	E(w) = \tr (\sum_s w M_s w^{\dagger} N_s) + c_1,
\end{equation}
where $M_s$ and $N_s$ are two sets of matrices and $c_1$ is a constant (that originates in all the Hamiltonian terms of Eq. \ref{eq:MERAalg:H} outside the future causal cone of $w$). Unfortunately there is no known algorithm to solve a quadratic problem subject to the additional isometric constraint of Eq. \ref{eq:MERAintro:isometry}. One can, however, attempt several approximate strategies. Here we describe an iterative approach based on linearizing the cost function $E(w)$. 
 
In this approach, we temporarily regard $w$ and $w^{\dagger}$ as independent tensors, and optimize $w$ while keeping $w^{\dagger}$ fixed. The cost function reads, up to the irrelevant constant, simply
\begin{equation}
	E^{\star}(w) \equiv \tr( w \Upsilon_w ), ~~~~~~\Upsilon_w \equiv \sum_s M_s ~w^{\dagger}N_s,
\end{equation}
where we call the matrix $\Upsilon_w$ the \emph{environment} of the isometry $w$ and we treat it as if it was indepedent of $w$. $E^{\star}(w)$ is then minimized by the choice $w = -WV^{\dagger}$, where $V$ and $W$ are the unitary transformations in the singular value decomposition of the environment, $\Upsilon_w = VSW^{\dagger}$,
\begin{equation}
	\min_w E^{\star}(w) = \min_w (w VSW^{\dagger}) =  -\tr(S) = -\sum_\alpha s_\alpha
\end{equation}
(here $s_{\alpha}\geq 0$ are the singular values of $\Upsilon_w$). 

\begin{figure}[!tbhp]
\begin{center}
\includegraphics[width=10cm]{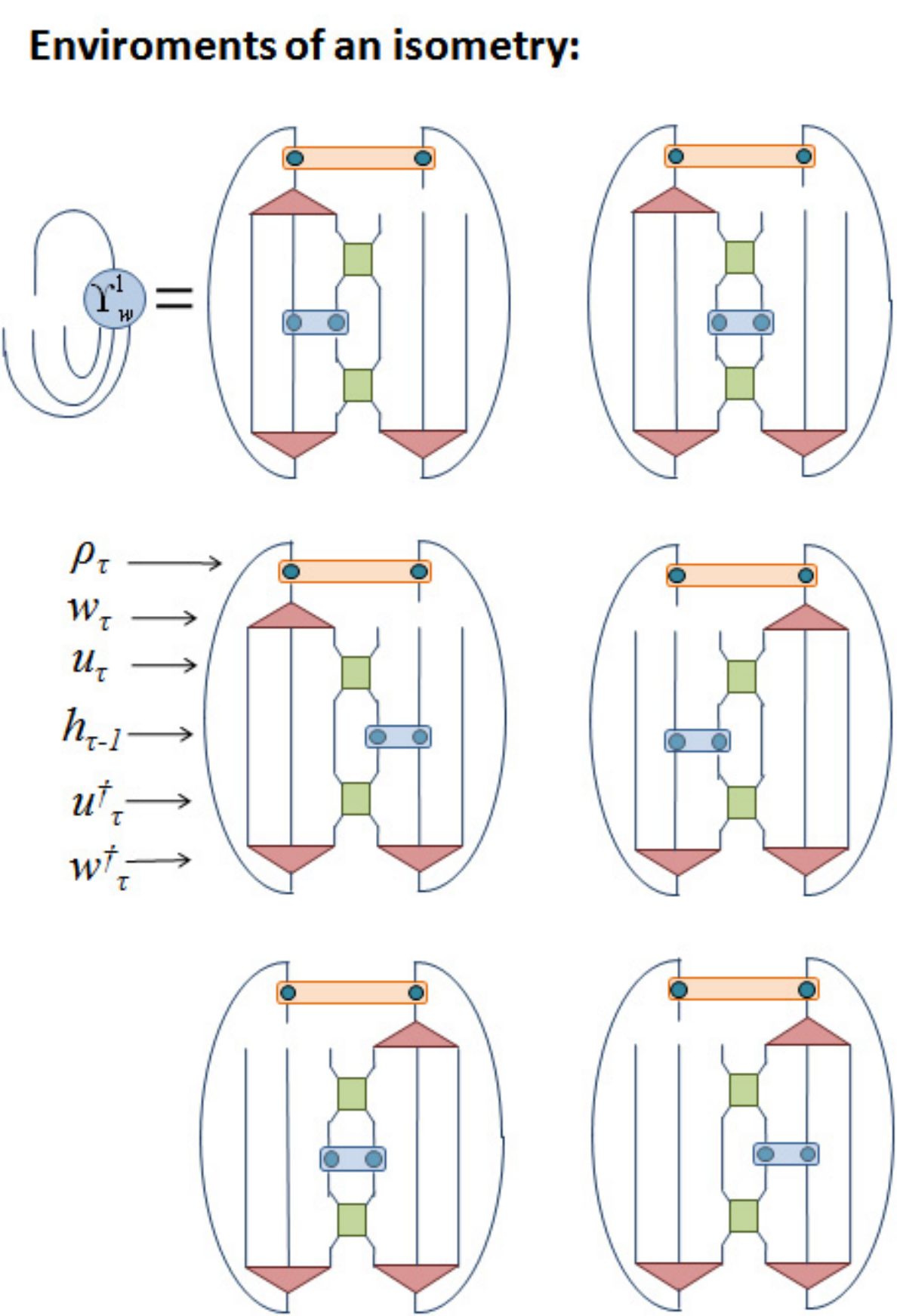}
\caption{ Tensor network corresponding to the 6 different contributions $\Upsilon_{w}^{i}$ to the environment $\Upsilon_w= \sum\nolimits_{i=1}^6 \Upsilon_{w}^{i}$ of the isometry $w$. Notice that at each iteration of L1-L4 we need to recompute each $\Upsilon_{w}^i$ since it depends on the updated $w^{\dagger}$. Nevertheless, the Hamiltonian term and density matrix that appears in $\Upsilon_u$ remain the same throughout the optimization and only need to be computed once.}
\label{fig:MERAalg:IsoEnviro}
\end{center}
\end{figure}

Accordingly, given an initial isometry $w$, the optimization is performed by iterating the following four steps $q_{\mbox{\tiny{one}}}$ times:

\newcounter{Lcount} 
\begin{list}{L\arabic{Lcount}.}  {\usecounter{Lcount}}\setcounter{Lcount}{0} 
	\item Compute the environment $\Upsilon_w$ with the newest version of $w^{\dagger}$ (as explained below, see also Fig \ref{fig:MERAalg:IsoEnviro}).
	\item Compute the singular value decomposition $\Upsilon_w = VSW^{\dagger}$.
	\item Compute the new isometry $w' = -WV^{\dagger}$.
	\item Replace $w^{\dagger}$ with $w'^{\dagger}$.
\end{list}

The environment $\Upsilon_w$ of an isometry $w$ (at layer $\tau$) can be decomposed as the sum of 6 contributions $\Upsilon_w^i$ ($i=1,\cdots,6$), each one expressed as a tensor network that involves neighboring isometric tensors of the same layer $\tau$ (disentanglers and isometries) as well as one Hamiltonian term $h^{[r,r+1]}_{\tau-1}$ and one density matrix $\rho_{\tau}^{[r',r'+1]}$, see Fig. \ref{fig:MERAalg:IsoEnviro}. The two-site Hamiltonian term $h^{[r,r+1]}_{\tau-1}$ collects contributions from all the Hamiltonian terms in Eq. \ref{eq:MERAalg:H} included in the future causal cone of the sites $r,r+1$ of $\mathcal{L}_{\tau-1}$ and is computed with the help of the ascending superoperator $\mathcal{A}$. Similarly, the two-site density matrix $\rho_{\tau}^{[r',r'+1]}$ is computed with the help of the descending superoperator $\mathcal{D}$. The computation of $h^{[r,r+1]}_{\tau-1}$ and $\rho_{\tau}^{[r',r'+1]}$, which only needs to be performed once during the optimization of $w$, has a cost $O(\chi^8 \log N)$. 

On the other hand, once we have $h^{[r,r+1]}_{\tau-1}$ and $\rho_{\tau}^{[r',r'+1]}$, computing $\Upsilon_{\omega}$ has a cost $O(\chi^8)$ and needs to be repeated at each iteration of the steps L1-L4, with a total cost $O(\chi^8 q_{\mbox{\tiny{one}}})$. In actual MERA simulations we find that the cost function $E_{w}$ typically drops very close to the eventual minimum already after a small number of iterations $q_{\mbox{\tiny{one}}}$ of the order of 10.

\begin{figure}[!tbhp]
\begin{center}
\includegraphics[width=10cm]{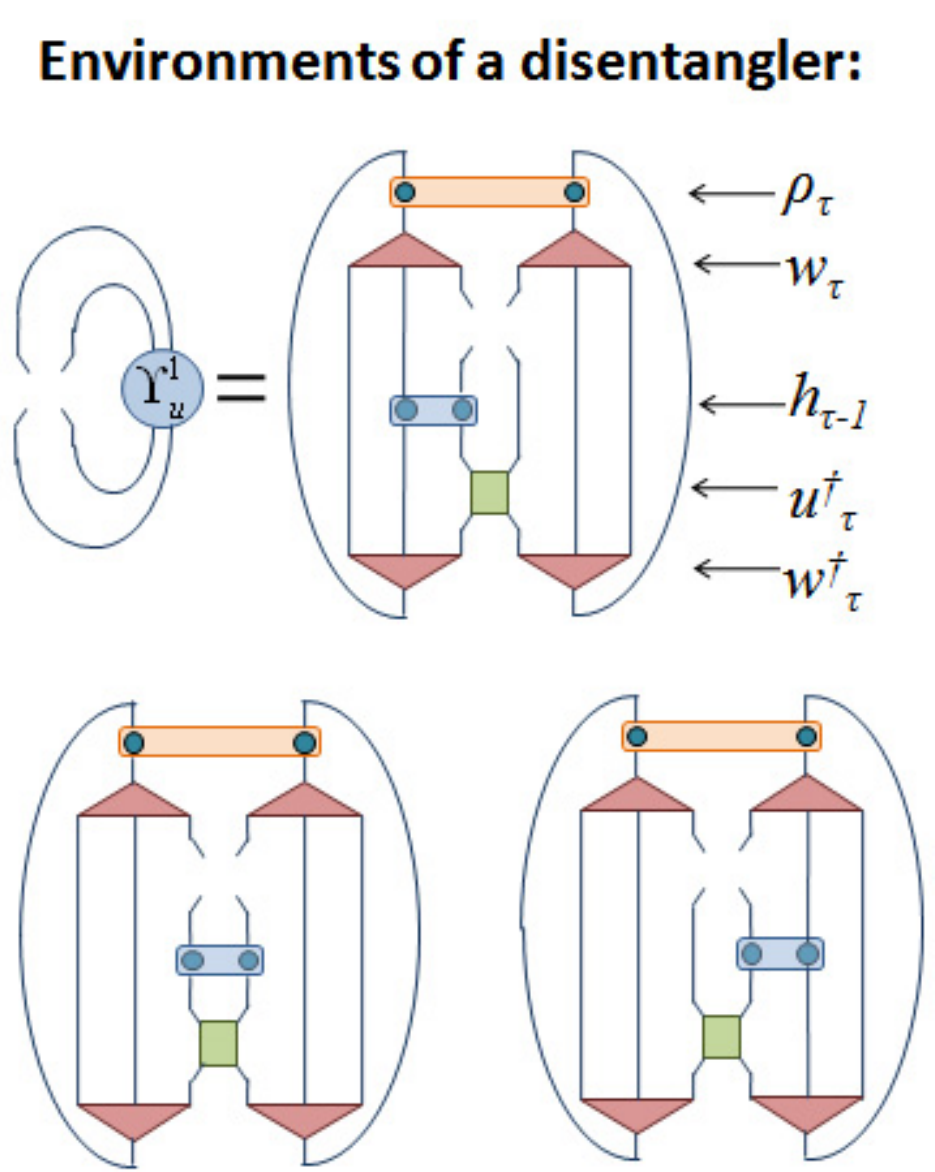}
\caption{ Tensor networks corresponding to the 3 different contributions $\Upsilon_{u}^{i}$ to the environment $\Upsilon_u= \sum\nolimits_{i=1}^3 \Upsilon_{u}^{i}$ of the disentangler $u$. Notice that at each iteration of L1-L4 we need to recompute each $\Upsilon_{u}^i$ since it depends on the updated $u^{\dagger}$. Nevertheless, the Hamiltonian term and density matrix that appears in $\Upsilon_u$ remain the same throughout the optimization and only need to be computed once.}
\label{fig:MERAalg:DisEnviro}
\end{center}
\end{figure}

The optimization of a disentangler $u$ is achieved analogously, but in this case the environment $\Upsilon_u$ decomposes into three contributions $\Upsilon_u^i$ ($i=1, 2, 3$), see Fig. \ref{fig:MERAalg:DisEnviro}. The required Hamiltonian terms and density matrices can be computed at a cost $O(\chi^8 \log N)$, while the optimization of $u$ following steps L1-L4 has a cost $O(\chi^8 q_{one})$.

\section{Optimization of the MERA} \label{sect:MERAalg:algorithm}

In this section we explain a simple algorithm to optimize the MERA so that it minimizes the energy of a local Hamiltonian of the form Eq. \ref{eq:MERAalg:H}. We first describe the algorithm for a generic system, and then discuss a number of specialized variations. These are directed to exploit translation invariance, scale invariance and to simulate systems where there is a finite range of correlations.

\subsection{The algorithm}

The basic idea of the algorithm is to attempt to minimize the cost function of Eq. \ref{eq:MERAalg:E} by sequentially optimizing individual tensors of the MERA, where each tensor is optimized as explained in the previous section.

By choosing to sweep the MERA in an organized way, we are able to update all its $O(N)$ tensors once with cost $O(\chi^8 N)$. Here we describe a bottom-top approach  where the MERA is updated layer by layer, starting with the bottom layer $\tau=1$ and progressing upwards all the way to the top layer (top-bottom and combined approaches are also possible). 

Given a starting MERA and the Hamiltonian of Eq. \ref{eq:MERAalg:H}, a bottom-top sweep is organized as follows:
 
\begin{list}{A\arabic{Lcount}.}  {\usecounter{Lcount}}\setcounter{Lcount}{0} 
\item Compute all two-site density matrices $\rho^{[r,r+1]}_{\tau}$ for all layers $\tau$ and sites $r \in \mathcal{L}_{\tau}$.
\end{list}
Starting from the lowest layer and for growing values of $\tau=1,2,\cdots, T-1$, repeat the following two steps:
\begin{list}{A\arabic{Lcount}.}  {\usecounter{Lcount}}\setcounter{Lcount}{1} 
	\item Update all disentanglers $u$ and isometries $w$ of layer $\tau$.
	\item Compute all two-site Hamiltonian terms $h^{[r,r+1]}_{\tau}$ for layer $\tau$.
\end{list}
Then, finally,
\begin{list}{A\arabic{Lcount}.}  {\usecounter{Lcount}}\setcounter{Lcount}{3} 
\item Update the top tensor of the MERA.
\end{list}

In step A1, we compute all nearest neighbor reduced density matrices $\rho^{[r,r+1]}_{\tau}$ for all the lattices $\mathcal{L}_{\tau}$, so that they can be used in step A2. We first compute the density matrix for the two sites of $\mathcal{L}_{T-1}$ as explained in Fig. \ref{fig:MERAalg:TopTensor}. Then we use the descending superoperator $\mathcal{D}$ to compute the 6 possible nearest neighbor, two-site density matrices of $\mathcal{L}_{T-2}$. More generally, given all the relevant density matrices of $\mathcal{L}_{\tau}$, we use $\mathcal{D}$ to obtain all the relevant density matrices of $\mathcal{L}_{\tau-1}$. In this way, the number of operations is proportional to the number of computed density matrices, namely $O(N)$, and the total cost is $O(\chi^8 N)$.

Step A2 breaks into a sequence of single-tensor optimizations that sweeps a given layer $\tau$ of the MERA. Each individual optimization in that layer is performed as explained in the previous section. Note that in order to optimize, say, an isometry $w$, we build its environment $\Upsilon_w$ by using (i) the density matrices computed in step A1; (ii) the Hamiltonian terms that were either given at the start for $\tau=1$ or have been computed in A3 for $\tau>1$; (iii) the neighboring disentanglers and isometries within the layer $\tau$. We can proceed, for instance, by updating all disentanglers of the layer from left to right, and then update all the isometries. This can be repeated a number $q_{\mbox{\tiny{lay}}}$ of times until the cost function does not change significantly.

In step A3, the new disentanglers and isometries of layer $\tau$ are used to build the ascending superoperator $\mathcal{A}$, which we then apply to the Hamiltonian terms of layer $\tau-1$ to compute the Hamiltonian terms for layer $\tau$. As explained after Eq. \ref{eq:MERAalg:O2}, each Hamiltonian term in layer $\tau$ is built from three contributions from layer $\tau-1$.

In step A4, the optimized top tensor corresponds to the $\chi_T$ eigenvectors with smaller energy eigenvalues of the Hamiltonian of the two-site lattice $\mathcal{L}_{T-1}$, obtained by exact diagonalization.

The overall optimization of the MERA consists of iterating steps A1-A4 until some pre-established degree of convergence in the energy $E$ is achieved. Suppose this occurs after $q_{\mbox{\tiny{iter}}}$ iterations. Then the cost of the optimization scales as $O(\chi^8 N q_{\mbox{\tiny{one}}}q_{\mbox{\tiny{lay}}}q_{\mbox{\tiny{iter}}})$. We observe that it is often convenient to keep $q_{\mbox{\tiny{one}}}$ and $q_{\mbox{\tiny{lay}}}$ relatively small (say between 1 and 5), since it is not worth spending much effort optimizing a single tensor/layer that will have to be optimized again later on with a modified cost function.

\subsection{Translation invariant systems}
\label{sect:MERAalg:trans}

When the Hamiltonian $H$ is invariant under translations, we can use a translation invariant MERA \footnote{A translation invariant MERA, characterized by one disentangler and one isometry at each layer, need not represent a translation invariant state $\ket{\Psi}$. Hence the need to consider the average density matrix $\bar{\rho}$}.

In this case, each layer $\tau$ is characterized by a disentangler $u_\tau$ and an isometry $w_{\tau}$. In addition, on each lattice $\mathcal{L}_{\tau}$ we have one two-site hamiltonian $h_\tau$ and one average density matrix $\bar{\rho}_{\tau}$. Then a bottom-top sweep of the MERA breaks into the steps A1-A4 for the inhomogeneous case above, but with the following simplifications:

In step A1, we compute $\bar{\rho}_{\tau-1}$ from $\bar{\rho}_{\tau}$ using the average descending superoperator $\bar{\mathcal{D}}$ of Eq. \ref{eq:MERAalg:avDesc},
\begin{equation}
	\bar{\rho}_{\tau} \rightarrow \bar{\rho}_{\tau-1} = \bar{\mathcal{D}}(\bar{\rho}_{\tau}).
\end{equation}
Then the whole sequence $\{\bar{\rho}_{T-1}, \cdots, \bar{\rho}_1, \bar{\rho}_0 \}$, with $T\approx \log N$, is computed with cost $O(\chi^8 \log N)$.

In step A2, the minimization of the energy $E$ by optimizing, say, the isometry $w_{\tau}$ is no longer a quadratic problem (since a larger power of $w_{\tau}$ appears now in the cost function). Nevertheless, we still linearize the cost function $E$ and optimize $w_{\tau}$ according to the steps L1-L4 of the previous section. Namely, we build the environment $\Upsilon_{w}$ (which now contains copies of $w_{\tau}$ and $w_{\tau}^{\dagger}$, all of them treated as frozen), compute its singular value decomposition to build the optimal $w'_{\tau}$, and then replace $w_{\tau}$ and $w_{\tau}^{\dagger}$ with $w'_{\tau}$ and $w'^{\dagger}_{\tau}$ in the tensor network for $\Upsilon_{w}$ before starting the next iteration. 

In step A3, the new hamiltonian term $h_{\tau}$ is obtained from $h_{\tau-1}$ using the average ascending superoperator $\bar{\mathcal{A}}$,
\begin{equation}
	h_{\tau-1} \rightarrow h_{\tau} = \bar{\mathcal{A}}(h_{\tau-1}).
\end{equation}

Step A4 proceeds as in the inhomogeneous case.

The overall cost of optimizing the MERA is in this case  $O(\chi^8 \log (N) q_{\mbox{\tiny{one}}}q_{\mbox{\tiny{lay}}}q_{\mbox{\tiny{iter}}})$.


\subsection{Scale invariant systems}
\label{sect:MERAalg:critical}

Given the Hamiltonian $H$ for an infinite lattice at a quantum critical point, where we expect the system to be invariant under rescaling, we can use a scale invariant MERA to represent its ground state \cite{vidal07,vidal08,evenbly07a,evenbly07b,giovannetti08,pfeifer08}. A scale invariant MERA is also relevant in the context of topological order in the infra-red limit of the RG flow \cite{aguado08,konig09}, both for finite and infinite systems; we will not consider such systems here.

Let us assume that all disentanglers and isometries are copies of a unique pair $(u,w)$. Then, as explained in Ref. \cite{pfeifer08}, the optimization algorithm can be specialized to take advantage of scale invariance as follows:

In step A1, we apply sparse diagonalization techniques to compute the fixed point density matrix $\hat{\rho}$ from the superoperator $\mathcal{S}^{*}$. This amount to applying $\mathcal{S}^{*}$ a number of times and therefore can be accomplished with cost $O(\chi^8)$. 

In step A2, the environment for e.g. the isometry $w$, $\Upsilon_w$, is computed as a weighted sum of environments for different layers $\tau=1,2,\cdots$. In a translation invariant MERA the environment for layer $\tau$ is a function $f(u_{\tau},w_{\tau},\rho_\tau,h_{\tau-1})$ of the pair $(u_{\tau},w_{\tau})$, the density matrix $\rho_{\tau}$ and the Hamiltonian term $h_{\tau-1}$ (specifically, $f$ is the sum of the diagrams in Fig. \ref{fig:MERAalg:IsoEnviro}). A scale invariant MERA corresponds to the replacements
\begin{equation}
	(u_{\tau},w_{\tau}) \rightarrow (u,w),~~~~~~ \rho_{\tau} \rightarrow \hat{\rho},
\end{equation}
so that only $h_{\tau-1}$ retains dependence on $\tau$. We then choose the average environment
\begin{equation}
	\Upsilon_{w} \equiv \sum_{\tau=1}^{\infty} \frac{1}{3^{\tau}} f(u,w,\hat{\rho},h_{\tau-1}),
\end{equation}
where the weight $1/3^{\tau}$ reflects the fact that for each isometry at layer $\tau$ there are $3$ isometries at layer $\tau-1$. Using linearity of the $f$ in its fourth argument we arrive at
\begin{equation}
	\Upsilon_{w} = f(u,w,\hat{\rho},\bar{h}), ~~~~~\bar{h} \equiv \sum_{\tau=1}^{\infty} \frac{1}{3^{\tau}} h_{\tau-1}.
\end{equation}
In practice, only a few terms of the expansion of $\bar{h}$ (say $\tau=1,2,3,4$) seem to be necessary. Given $\Upsilon_{w}$, the optimization proceeds as usual with a singular value decomposition.

Steps A3 and A4 are not necessary. 

That is, the algorithm to minimize the expected value of $H$ consists simply in iterating the following two steps:

\begin{list}{ScInv\arabic{Lcount}.}  {\usecounter{Lcount}}\setcounter{Lcount}{0} 
	\item Given a pair ($u,w$), compute a pair $(\hat{\rho},\bar{h})$.
	\item Given the pairs ($u,w$) and $(\hat{\rho},\bar{h})$, update the pair ($u,w$).
\end{list}

In practical simulations it is convenient to include a few (say one or two) transitional layers at the bottom of the MERA, each one characterized by a different pair ($u_{\tau}$, $w_{\tau}$). In this way the bond dimension $\chi$ of the MERA can be made independent of the dimension $d$ of the sites of $\mathcal{L}$. These transitional layers are optimized using the algorithm for translation invariant systems.


\begin{figure}[!tbhp]
\begin{center}
\includegraphics[width=10cm]{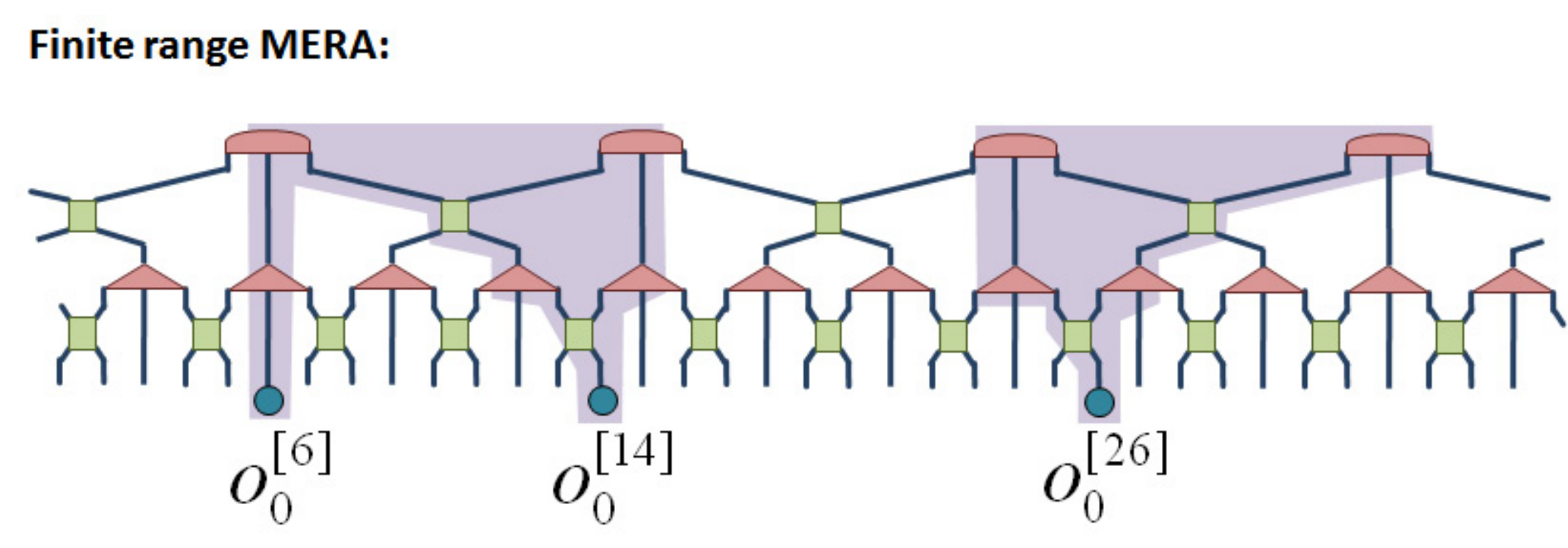}
\caption{ A \emph{finite correlation range} MERA with $T'=2$ layers is used to represent a state of $N=36$ sites. Since it lacks the uppermost layers, only sites within a finite distance or range $\zeta \approx 3^{T'}$ of each other may be correlated. More precisely, only pairs of sites $(r_1,r_2)$ whose past casual cones intersect can be correlated, as it is the case with the pair of sites $(6,14)$ but not with $(14,26)$, for which we have $\langle {o^{[14]} o^{[26]} } \rangle  = \langle {o^{[14]} } \rangle \langle {o^{[26]} } \rangle$.}
\label{fig:MERAalg:FiniteCorr}
\end{center}
\end{figure}

\subsection{Finite range of correlations}
\label{sect:MERAalg:constant}

A third variation of the basic algorithm consists in setting the number $T'$ of layers in the MERA to a value smaller than its usual one $T \approx \log_3 N$ (in such a way that the number $N_{T'}$ of sites on the top lattice $\mathcal{L}_{T'}$ may still be quite large) and to consider a state $\ket{\Psi}$ of the lattice $\mathcal{L}$ such that after $T'$ coarse-graining transformations it has become a product state,
\begin{equation}
	\ket{\Psi_0} \rightarrow \ket{\Psi_1} \rightarrow \cdots \rightarrow \ket{\Psi_{T'}},
\end{equation}
where $\ket{\Psi_{T'}} = \ket{0}^{\otimes N_{T'}}$.
For instance, Fig. \ref{fig:MERAalg:FiniteCorr} shows a MERA for $N=36$ and $T'=2$. The four top tensors of this MERA are of type $(0,3)$, where the lack of upper index indicates that the top lattice $\mathcal{L}_{T'}$ is in a product state of its $N_{T'} = 4$ sites.

We refer to this ansatz as the \emph{finite range} MERA, since it is such that correlations in $\ket{\Psi}$ are restricted to a finite range $\zeta$, roughly $\zeta\approx 3^T$ sites, in the sense that regions separated by more than $\zeta$ sites display no correlations. This is due to the fact that the past causal cones of distance regions of $\mathcal{L}$ have zero intersection, see Fig. \ref{fig:MERAalg:FiniteCorr}.

Given a ground state $\ket{\Psi}$ with a finite correlation length $\xi$, the finite range MERA with $\zeta \approx\xi$ turns out to be a better option to represent $\ket{\Psi}$ than the standard MERA with $T \approx \log_3 N$ layers, in that it offers a more compact description and the cost of the simulations is also lower since there are less tensors to be optimized. The algorithm is adapted in a straightforward way. The only significant difference is that the top isometries, being of type $(0,3)$, do not require any density matrix in their optimization (their environment is only a function of neighboring disentanglers and isometries, and of Hamiltonian terms).

A clear advantage of the finite range MERA is in a translation invariant system, where the cost of a simulation with range $\zeta = 3^{T'}$ is O$(\chi^8 \log_3 \zeta)$, that is, independent of $N$. This allows us to take the limit of an infinite system. We find that, given a translation invariant Hamiltonian $H = \sum_{r=1}^N h^{[r,r+1]}$, where $h^{[r,r+1]}$ is the same for all $r\in\mathcal{L}$, the optimization of a finite range MERA will lead to the same collection of optimal disentanglers and isometries $\{(u_1,w_1),(u_2,w_2), \cdots, (u_{T'},w_{T'})\}$, for different lattice sizes $N, N',N''\cdots$ larger than $\zeta$. This is due to the existence of disconnected causal cones, which imply that the cost functions  for the optimization are not sensitive to the total system size provided it is larger than $\zeta$. As a result, $\{(u_1,w_1),(u_2,w_2), \cdots, (u_{T'},w_{T'})\}$ can be used to define not just one but a whole collection of states $\ket{\Psi(N)}$, $\ket{\Psi(N')}$, $\ket{\Psi(N'')}$, $\cdots$, for lattices of different sizes $N,N',N'', \cdots$, such that they all have the same two-site density matrix $\rho$ and therefore also the same expected value of the energy per link,
\begin{equation}
	\bra{\Psi(N)}h\ket{\Psi(N)} = \bra{\Psi(N')}h\ket{\Psi(N')} = \cdots
\end{equation}
In particular, we can use the finite range MERA algorithm to obtain an upper bond for the ground state energy of an infinite system, even though only $T'$ pairs $(u_{\tau},w_{\tau})$ are optimized.

\begin{figure}[!tbhp]
\begin{center}
\includegraphics[width=10cm]{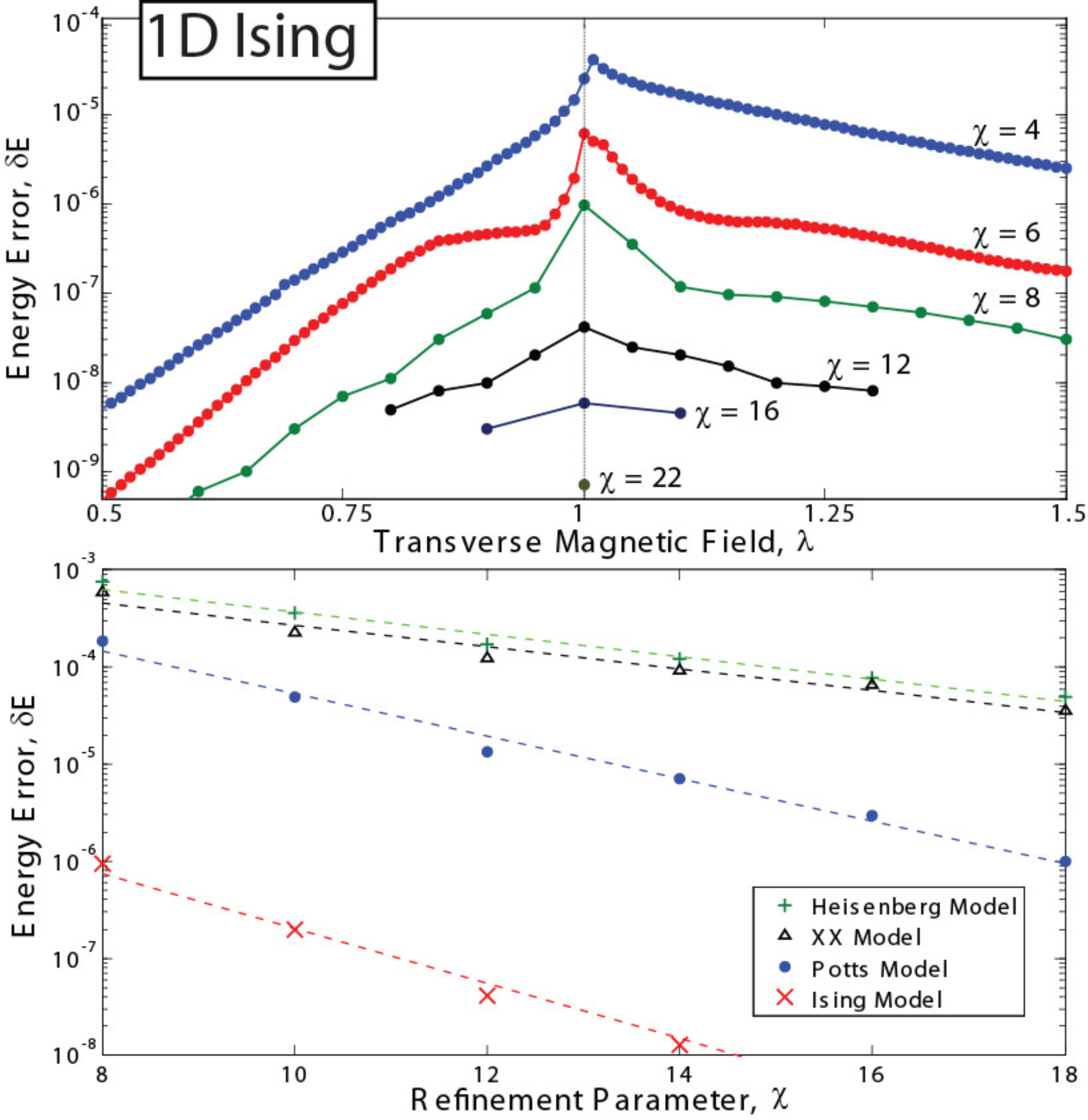}
\caption{ \emph{Top:} The energy error of the MERA approximations to the ground-state of the infinite Ising model, as compared against exact analytic values, is plotted both for different transverse magnetic field strengths and different values of the MERA refinement parameter $\chi$. The finite correlation range algorithm (with at most $T=5$ levels) was used for non-critical ground states, whilst the scale invariant MERA was used for simulations at the critical point. It is seen that representing the ground-state is most computationally demanding at the critical point, although even at criticality the MERA approximates the ground-state to between 5 digits of accuracy ($\chi=4$) and 10 digits of accuracy ($\chi=22$). \emph{Bottom:} Scale-invariant MERA are used to compute the ground-states of infinite, critical, 1$D$ spin chains of Eqs. \ref{eq:MERAalg:hamising}-\ref{eq:MERAalg:hams} for several values of $\chi$. In all instances one observes a roughly exponential convergence in energy over a wide range of values for $\chi$ as indicated by trend lines (dashed). Energy errors for Ising, XX and Heisenberg models are taken relative to the analytic values for ground energy whilst energy errors presented for the Potts model are taken relative to the energy of a $\chi=22$ simulation.} \label{fig:MERAalg:IsingPottsEnergyError}
\end{center}
\end{figure}

\section{Benchmark calculations for 1D systems}
\label{sect:MERAalg:benchmark}

In order to benchmark the algorithms of the previous section, we have analyzed zero temperature, low energy properties of a number of 1$D$ quantum spin systems. Specifically, we have considered the Ising model \cite{pfeuty70,burkhardt85}, the 3-state Potts model \cite{solyom81}, the XX model \cite{lieb61} and the Heisenberg models \cite{baxter82}, with Hamiltonians
\begin{eqnarray}
H_{{\rm{Ising}}}  &=& \sum_r \left(\lambda \sigma^{[r]}_z  + \sigma^{[r]}_x \sigma^{[r+1]}_x \right) \label{eq:MERAalg:hamising}\\
H_{{\rm{Potts}}}  &=&  \sum_r \left(\lambda M^{[r]}_{z}  + \sum_{a=1,2} M^{[r]}_{x,a} M^{[r+1]}_{x,3-a} \right)\label{hampotts}
\end{eqnarray}
\begin{eqnarray}
H_{{\rm{XX}}}   &=&  \sum_r \left(\sigma^{[r]}_x \sigma^{[r+1]}_x + \sigma^{[r]}_y \sigma^{[r+1]}_y \right) \\
H_{{\rm{Heisenberg}}}  &=&  \sum_r \left(\sigma^{[r]}_x \sigma^{[r+1]}_x + \sigma^{[r]}_y \sigma^{[r+1]}_y + \sigma^{[r]}_z \sigma^{[r+1]}_z \right)\nonumber\\ 
 \label{eq:MERAalg:hams}
\end{eqnarray}
where $\sigma_x$, $\sigma_y$ and $\sigma_z$ are the spin $1/2$ Pauli matrices and $M_{x,1}$, $M_{x,2}$ and $M_z$ are the matrices
\begin{eqnarray}
M_z  &\equiv& \left( {\begin{array}{*{20}c}
   2 & 0 & 0  \\
   0 & { - 1} & 0  \\
   0 & 0 & { - 1}  \\
\end{array}} \right),\\
M_{x,1}  &\equiv& \left( {\begin{array}{*{20}c}
   0 & 1 & 0  \\
   0 & 0 & 1  \\
   1 & 0 & 0  \\
\end{array}} \right),\; M_{x,2}  \equiv \left( {\begin{array}{*{20}c}
   0 & 0 & 1  \\
   1 & 0 & 0  \\
   0 & 1 & 0  \\
\end{array}} \right).
\end{eqnarray}
We assume periodic boundary conditions in all instances and use a translation invariant MERA to represent an approximation to the ground state and, in some models, also the first excited state. For Ising and Potts models the parameter $\lambda$ is the strength of the transverse magnetic field applied along the $z$-axis, with $\lambda_c=1$ corresponding to a quantum phase transition. Both the XX model and Heisenberg model are quantum critical as written. 

\begin{figure}[!tbhp]
\begin{center}
\includegraphics[width=10cm]{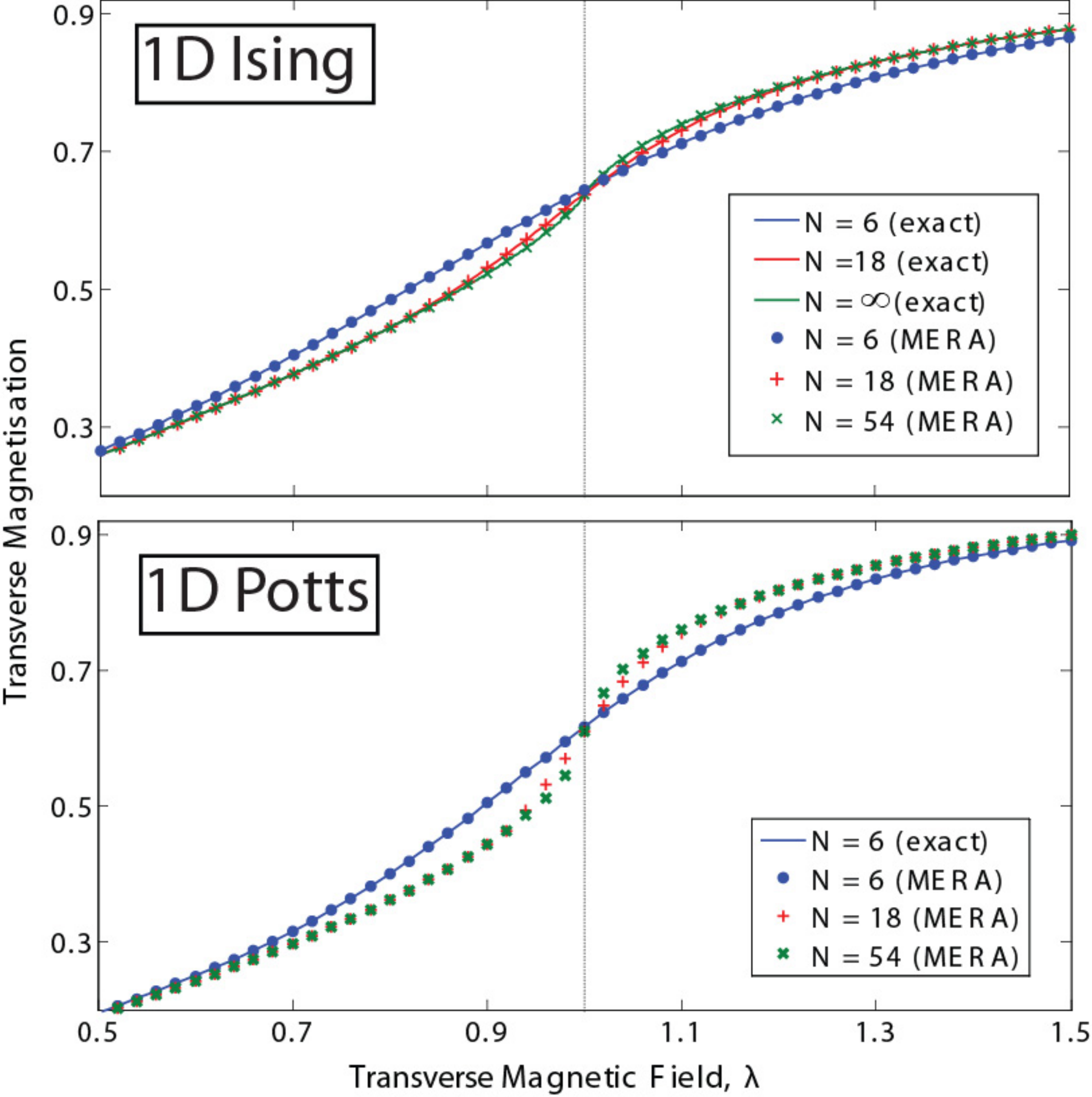}
\caption{ The transverse magnetization $\left\langle {\sigma _z } \right\rangle$ for Ising and $\frac{1}{2}\left\langle {M _z } \right\rangle$ for Potts, is plotted for translation invariant chains of several sizes $N$. \emph{Top:} For the Ising model, the magnetization given from $\chi=8$ MERA matches those from exact diagonalization for small system sizes ($N=6,18$), whilst the magnetisation from the $N=54$ MERA approximates that from the thermodynamic limit (known analytically). \emph{Bottom:} Equivalent magnetisations for the Potts model, here computed with a $\chi=12$ MERA. Simulations with larger $N$ systems show little change from the $N=54$ data, again indicating that $N=54$ is already close to the thermodynamic limit.} \label{fig:MERAalg:IsingPottsExpectZ}
\end{center}
\end{figure}

Fig.~\ref{fig:MERAalg:IsingPottsEnergyError} shows the accuracy obtained for ground-state energies of the above models in the limit of an infinite chain, as a function of the refinement parameter $\chi$. Simulations were performed with either the finite correlation range algorithm (for the non-critical Ising) or the scale invariant algorithm (for critical systems). In all cases one observes roughly exponential convergence to the exact energy with increasing $\chi$. For any fixed value of $\chi$, the MERA consistently yields more accuracy for some models than for others. For the Ising model, the cheapest simulation considered ($\chi=4$) produced 5 digits of accuracy, whilst the most computationally expensive simulation ($\chi=22$) produced 10 digits of accuracy. The time taken for the MERA to converge, running on a 3GHz dual-core desktop PC with 8Gb of RAM, is approximately a few minutes/hours/days/weeks for $\chi=4,8,16,22$ respectively. We stress that these simulations were performed on single desktop computers; a parallel implementation of the code running on a computer cluster might bring significantly larger values of $\chi$ within computational reach. 

\begin{figure}[!tbhp]
\begin{center}
\includegraphics[width=10cm]{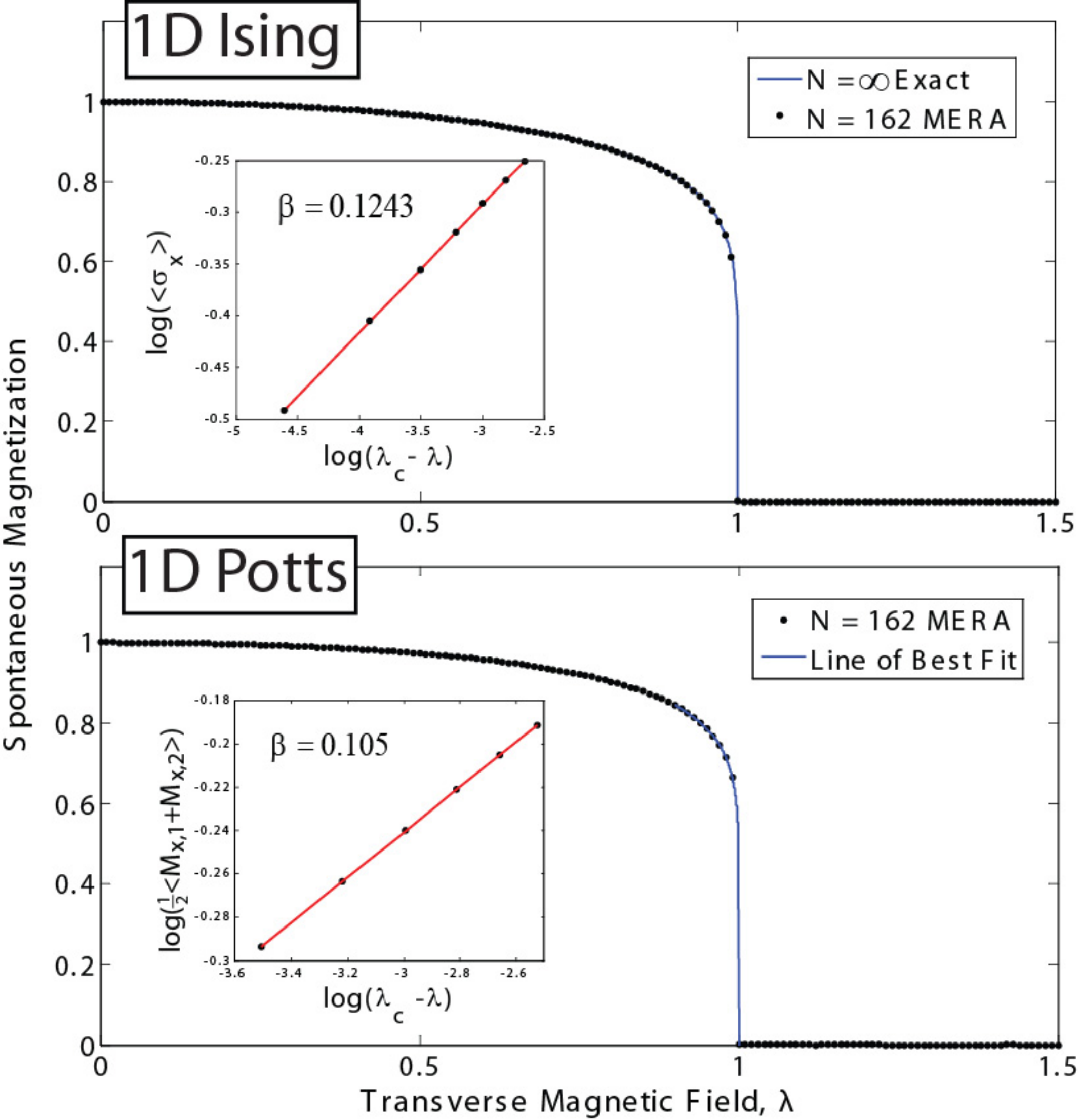}
\caption{ \emph{Top:} Spontaneous magnetization $\left\langle {\sigma _x } \right\rangle$ computed with a $\chi=8$ MERA for a periodic Ising system of $N=162$ sites. The results closely approximate the analytic values of magnetization known for the thermodynamic limit. A fit of the data near the critical point yeilds a critical exponent $\beta_\textrm{MERA} = 0.1243$, with the exact exponent known as $\beta_\textrm{ex} = 1/8$. \emph{Bottom:} An equivalent phase portrait of the Potts model, here with spontaneous magnetization $\frac{1}{2}\left\langle {M_{x,1}  + M_{x,2} } \right\rangle$, is computed with a $\chi=14$ MERA and is plotted with a fit of the data near the critical point. The fit yields a critical exponent $\beta_\textrm{MERA} = 0.105$ with the exact exponent known to be $\beta_\textrm{ex} = 1/9$.} \label{fig:MERAalg:IsingPottsSpont}
\end{center}
\end{figure}

Fig. ~\ref{fig:MERAalg:IsingPottsExpectZ} demonstrates the ability of the MERA to reproduce finite size effects. It shows the transverse magnetization as a function of the transverse magnetic field for several system sizes. The results smoothly interpolate between those for small system sizes and those for an infinite chain, and match the available exact solutions. On the other hand, the MERA can also be used to explore the phase diagram of a system. Fig.~\ref{fig:MERAalg:IsingPottsSpont} shows the spontaneous magnetization, which is the system's order parameter, for a $1D$ chain of $N=162$ sites for both Ising and Potts models, where $N$ has been chosen large enough that the results under consideration do not change singnificantly with the system size (thermodynamic limit). A fit for the critical exponent of the Ising model gives $\beta_\textrm{MERA}=0.1243$ whilst the fit for the Potts model produces $\beta_\textrm{MERA}=0.105$. These values are within less than $1\%$ and $6\%$ of the exact exponents $\beta=1/8$ and $\beta=1/9$ for the Ising and Potts models respectively. Obtaining an accurate value for this critical exponent through a fit of the data near the critical point is difficult due to the steepness of the curve near the critical point. Through an alternative method involving the scaling super-operator $\mathcal{S}$ (Sect. \ref{sect:MERAalg:local}), more accurate critical exponents can be obtained, as shall be demonstarted in Chapter \ref{chap:1DCrit}. 

\begin{figure}[!tbhp]
\begin{center}
\includegraphics[width=10cm]{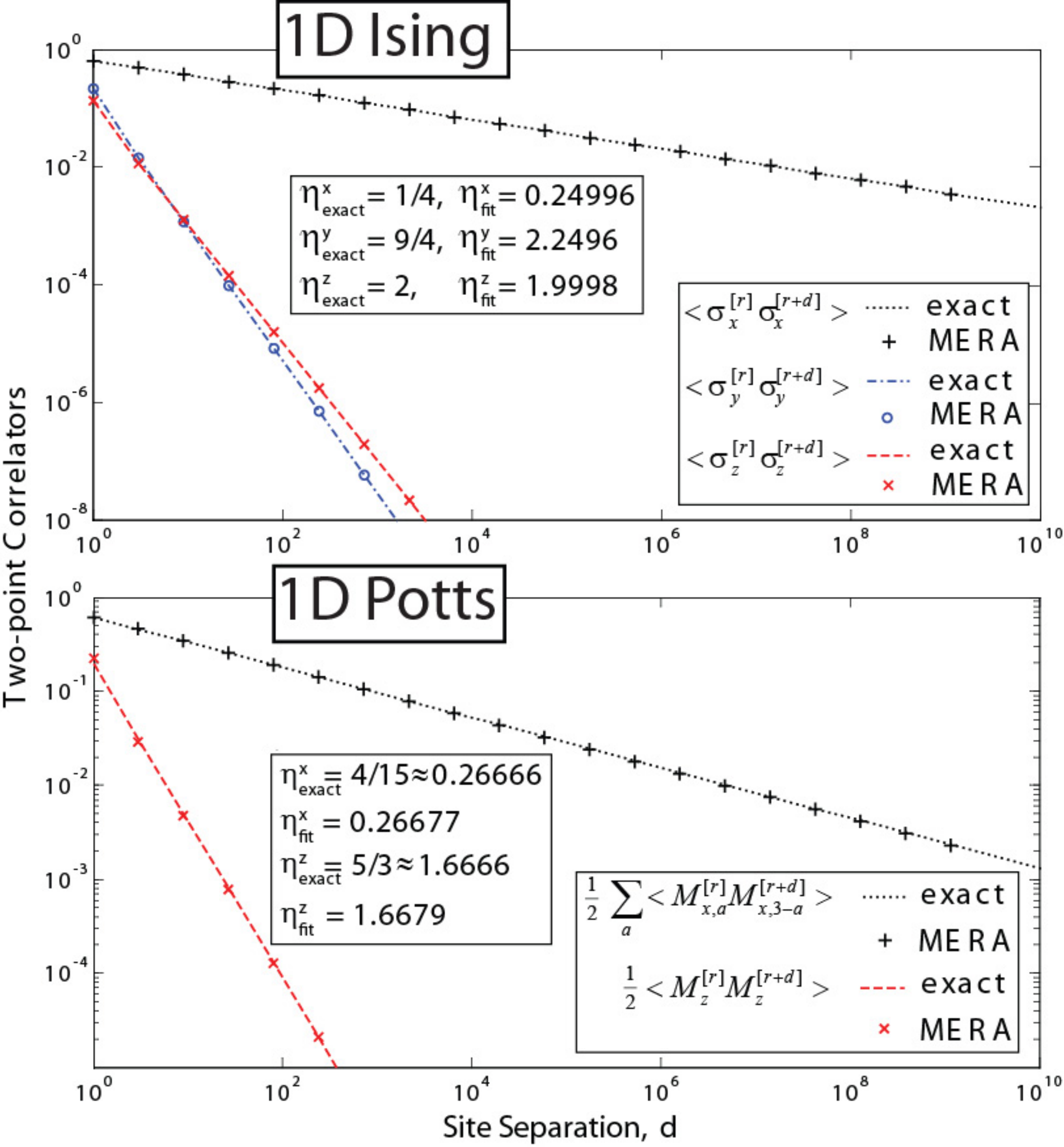}
\caption{ Two-point correlators for infinite 1$D$ Ising and Potts chains at criticality ($\lambda=1$), as computed with $\chi=22$ scale-invariant MERA. Correlators for the Ising model are compared against analytic solutions \cite{pfeuty70,burkhardt85} whilst those for the Potts model are plotted against the polynomial decay predicted from CFT \cite{francesco97}. A scale-invariant MERA produces polynomial decay of correlators at all length scales; a fit of the form $\langle{\sigma _x^{[r]} \sigma _x^{[r + d]} } \rangle  \propto d^{ - \eta ^x }$ for Ising correlators generated by the MERA gives the decay exponent $\eta ^x=0.24996$, close to the known analytic value $1/4$ and similarly for the fits on other correlators. Indeed the MERA here reproduces exact $\langle{\sigma _x^{[r]} \sigma _x^{[r + d]} } \rangle$ correlators for the Ising model at a distance up to $d=10^9$ sites within $0.6\%$ accuracy. Critcal exponents for the Potts are also reproduced very accurately.  
  } \label{fig:MERAalg:IsingPottsCorr}
\end{center}
\end{figure}

\begin{figure}[!tbhp]
\begin{center}
\includegraphics[width=10cm]{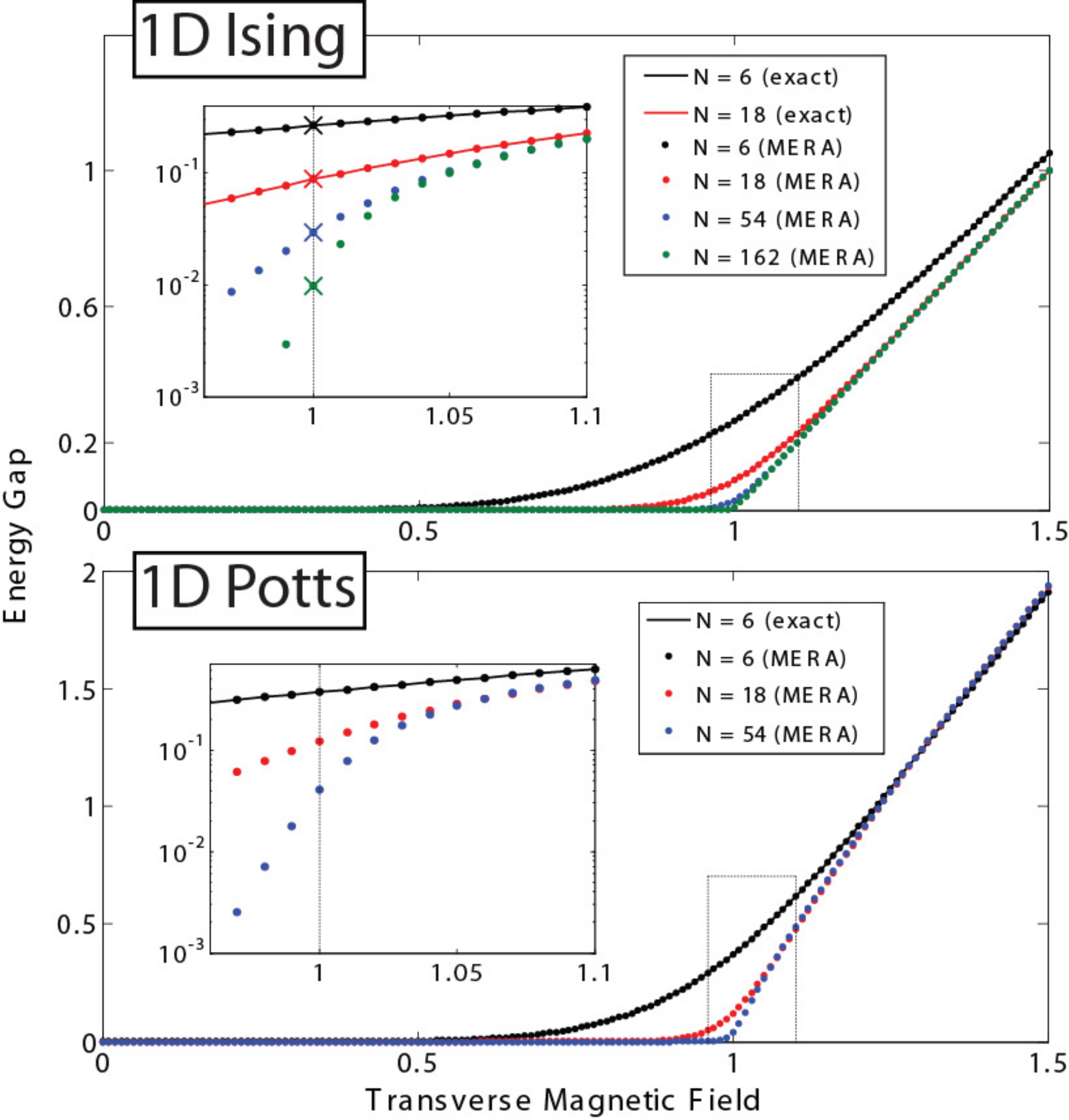}
\caption{ \emph{Top:} A $\chi=8$ MERA is used to compute the energy gap $\Delta E$ (the energy difference between the ground and $1^{\textrm{st}}$ excited state) of the Ising chains as a function of the transverse magnetic field. The gap computed with MERA for $N=6,18$ sites is in good agreement with that computed through exact diagonalization of the system. \emph{Inset:} Crosses show analytic values of energy-gaps at the critical point for $N=\{6,18,54,162 \}$. Even for the largest system considered, $N=162$, the gap computed with MERA $\Delta E_{\textrm{MERA}}=9.67\times 10^{-3}$ compares well with the exact value $\Delta E_{\textrm{ex}}=9.69\times 10^{-3}$. \emph{Bottom:} Equivalent data for the Potts Model where simulations have been performed with a $\chi=14$ MERA to account for the increased computational difficulty of this model.} \label{fig:MERAalg:IsingPottsGap}
\end{center}
\end{figure}

The previous results refer to local observables. Let us now consider correlators. A scale-invariant MERA, useful for the representation of critical systems, gives polynomial correlators at all length scales, as shown in Fig.~\ref{fig:MERAalg:IsingPottsCorr} for the critical Ising and Potts models. Note that Fig. \ref{fig:MERAalg:IsingPottsCorr} displays the correlators that are most convenient to compute (as per Fig.~\ref{fig:MERAalg:CorrEasy}). These occur at distances $d=3^q$ for $q=1,2,3\ldots$ and are evaluated with cost $O(\chi^8)$. Evaluation of arbitrary correlators is possible (see Fig.~\ref{fig:MERAalg:CorrDifficult}) but its cost is several orders of $\chi$ more expensive. The precision with which correlators are obtained is remarkable. A $\chi=22$ MERA for the Ising model gives $\langle{\sigma _x^{[r]} \sigma _x^{[r + d]} } \rangle$ correlators, at distances up to $d=10^9$ sites, accurate to within $0.6\%$ of exact correlators. Critical exponents $\eta$ are obtained through a fit of the form $C(r,r+d) \propto d^{ - \eta }$ with $C$ as correlators of $x,y$ or $z$ magnetization. For the Ising model, the exponents for $x,y$ and $z$ magnetizations are obtained with less than $0.02\%$ error each. For the Potts model exponents are obtained with less than $0.04\%$ and $0.08\%$ error for $x$ and $z$ magnetization respectively. 

Finally, we also demonstrate the ability of MERA to investigate low-energy excited states by computing the energy gap in the Ising and Potts models. Fig.~\ref{fig:MERAalg:IsingPottsGap} shows that the gap grows linearly with the magnetic field $\lambda$ and independent of $N$ in the disordered phase $\lambda >> \lambda_c$, whilst at criticality it closes as $1/N$. Even a relatively cheap $\chi=8$ MERA reproduces the known critical energy gaps to within $0.2\%$ for systems as large as $N=162$ sites. The expected value of arbitrary local observables (besides the energy) can also be easily evaluated for the excited state.

\section{Conclusions}

In this Chapter we have provided a rather self-contained description of an algorithm to explore low energy properties of lattice models (Sect. \ref{sect:MERAalg:optim}-\ref{sect:MERAalg:algorithm}), and benchmark calculations addressing 1$D$ quantum spin chains (Sect. \ref{sect:MERAalg:benchmark}).

Many of the features of the MERA algorithm highlighted by the present results were also observed in systems of free fermions and free bosons in Chapters \ref{chap:FreeFerm} and \ref{chap:FreeBoson}. These include (i) the ability to consider arbitrarily large systems, (ii) the ability to compute the low-energy subspace of a Hamiltonian, (iii) the ability to disentangle non-critical systems completely, (iv) the ability to find a scale-invariant representation of critical systems and finally (v) the reproduction of accurate polynomial correlators for critical systems. However, the algorithms of Chapters \ref{chap:FreeFerm} and \ref{chap:FreeBoson} exploit the formalism of Gaussian states that is characteristic of free fermions and bosons and cannot be easily generalized to interacting systems. Instead, the algorithms discussed in this Chapter can be used to address arbitrary lattice models with local Hamiltonians.

An alternative method to optimise the MERA is with a time evolution algorithm as described in Ref. \cite{rizzi08}. The time evolution algorithm has a clear advantage: it can be used both to compute the ground state of a local Hamiltonian (by simulating an evolution in imaginary time) and to study lattice dynamics (by simulating an evolution in real time). We find, however, that the algorithm described in the present Chapter is a better choice when it comes to computing ground states. On the one hand, the time-evolution algorithm has a time step $\delta t$ that needs to be sequentially reduced in order to diminish the error in the Suzuki-Trotter decomposition of the (euclidean) time evolution operator. In the present algorithm, convergence is faster and there is no need to fine tune a time step $\delta t$. In addition, the present algorithm allows to compute not only the ground state but also low energy excited states. It is unclear how to use the time evolution algorithm to achieve the same.

The benchmark calculations presented in this manuscript refer to 1$D$ systems. For such systems, however, DMRG \cite{white92,white93} already offers an extraordinarily successful approach. The strength of entanglement renormalization and the MERA relies on the fact that the present algorithms can also address large 2$D$ lattices, as is discussed Chapter \ref{chap:2DMera}.

\chapter{Entanglement renormalization and quantum criticality}
\label{chap:1DCrit}

\section{Introduction}
The study of quantum critical phenomena through real-space renormalization group (RG) techniques \cite{wilson75,white92,white93} has traditionally been obstructed by the accumulation, over successive RG transformations, of short-range entanglement across block boundaries. Entanglement renormalization was proposed as a technique to address this problem. By removing short-range entanglement at each iteration of the RG transformation, not only can arbitrarily large lattice systems be considered, but the scale invariance characteristic of critical phenomena is also seen to be restored. 

In this Chapter we explain how to use the MERA to investigate scale invariant systems. It has recently been shown that the \emph{scale invariant} MERA can represent the infra-red limit of topologically ordered phases \cite{aguado08,konig09}, here we focus instead on its use at quantum criticality. We present the following results: (i) given a critical Hamiltonian, an adaptation of the algorithm of the previous Chapter to compute a scale invariant MERA for its ground state; then, starting from a scale invariant MERA, (ii) a procedure to identify the scaling operators/dimensions of the theory and (iii) a closed expression for two-point and three-point correlators; (iv) a connection between the MERA and conformal field theory, which can be used to readily identify the continuum limit of a critical lattice model; finally (v) benchmark calculations for the Ising and Potts models. 

We note that result (ii) was already discussed by Giovannetti, Montangero and Fazio in Ref. \cite{giovannetti08} using the \emph{binary} MERA of Ref. \cite{vidal08}. Our derivations, as with those of the previous Chapter, are conducted using the \emph{ternary} MERA, in terms of which results (iii)-(iv) acquire a simple form.

We start by considering a finite 1$D$ lattice $\mathcal{L}$ made of $N$ sites, each one described by a vector space $\mathbb{V}$ of dimension $\chi$. The (ternary) MERA is a tensor network that serves as an ansatz for pure states $\ket{\Psi}\in \mathbb{V}^{\otimes N}$ of the lattice, see Fig. \ref{fig:1DCrit:TwoSiteCFT}. Its tensors, the \emph{disentanglers} and \emph{isometries}, are organized in $T\approx \log_3 N$ layers, each one implementing a RG transformation. Such transformations produce a sequence of lattices,
\begin{equation}
\mathcal{L}_0  ~\rightarrow  ~\mathcal{L}_1 ~\rightarrow ~\cdots ~\rightarrow ~\mathcal{L}_T, ~~~~~~~~~\mathcal{L}_{0} \equiv \mathcal{L},
\end{equation}
where lattice $\mathcal{L}_{\tau+1}$ is a coarse-graining of lattice $\mathcal{L}_{\tau}$, and the top lattice $\mathcal{L}_T$ is sufficiently small to allow exact numerical computations. Let $o$ denote a local observable supported on two contiguous sites of $\mathcal{L}$, and let $\rho_{T}$ be the density matrix that describes the state of the system on two contiguous sites of $\mathcal{L}_{T}$. Then the \emph{ascending} and \emph{descending} superoperators $\mathcal{A}_{\tau}$ and $\mathcal{D}_{\tau}$, 
\begin{equation}
	o_{\tau} = \mathcal{A}_{\tau}(o_{\tau-1}),~~~~~~~~~~ \rho_{\tau-1} = \mathcal{D}_{\tau}(\rho_{\tau}),
\end{equation}
generate a sequence of operators and density matrices 
\begin{eqnarray}
	&&o_0  \stackrel{\mathcal{A}_1}{\rightarrow}  o_1 \stackrel{\mathcal{A}_2}{\rightarrow} ~\cdots~ \stackrel{\mathcal{A}_T}{\rightarrow} o_T, ~~~~~~~~~ o_0 \equiv o, \label{eq:1DCrit:A}\\
	&&\rho_0  \stackrel{\mathcal{D}_1}{\leftarrow}  \rho_1 \stackrel{\mathcal{D}_2}{\leftarrow} ~\cdots~ \stackrel{\mathcal{D}_T}{\leftarrow} \rho_T, ~~~~~~~~~ \rho_0 \equiv \rho, \label{eq:1DCrit:D}
\end{eqnarray}
where $o_{\tau}$ and $\rho_{\tau}$ are supported on two contiguous sites of the lattice $\mathcal{L}_{\tau}$. Eq. (\ref{eq:1DCrit:A}) allows us to monitor how the local operator $o$ transforms under successive RG transformations, whereas its expected value $\left\langle o \right\rangle = \tr(\rho o)$ can be evaluated by computing $\rho$ in Eq. (\ref{eq:1DCrit:D}).


\section{RG fixed point}
\begin{figure}[!tbhp]
\begin{center}
\includegraphics[width=12cm]{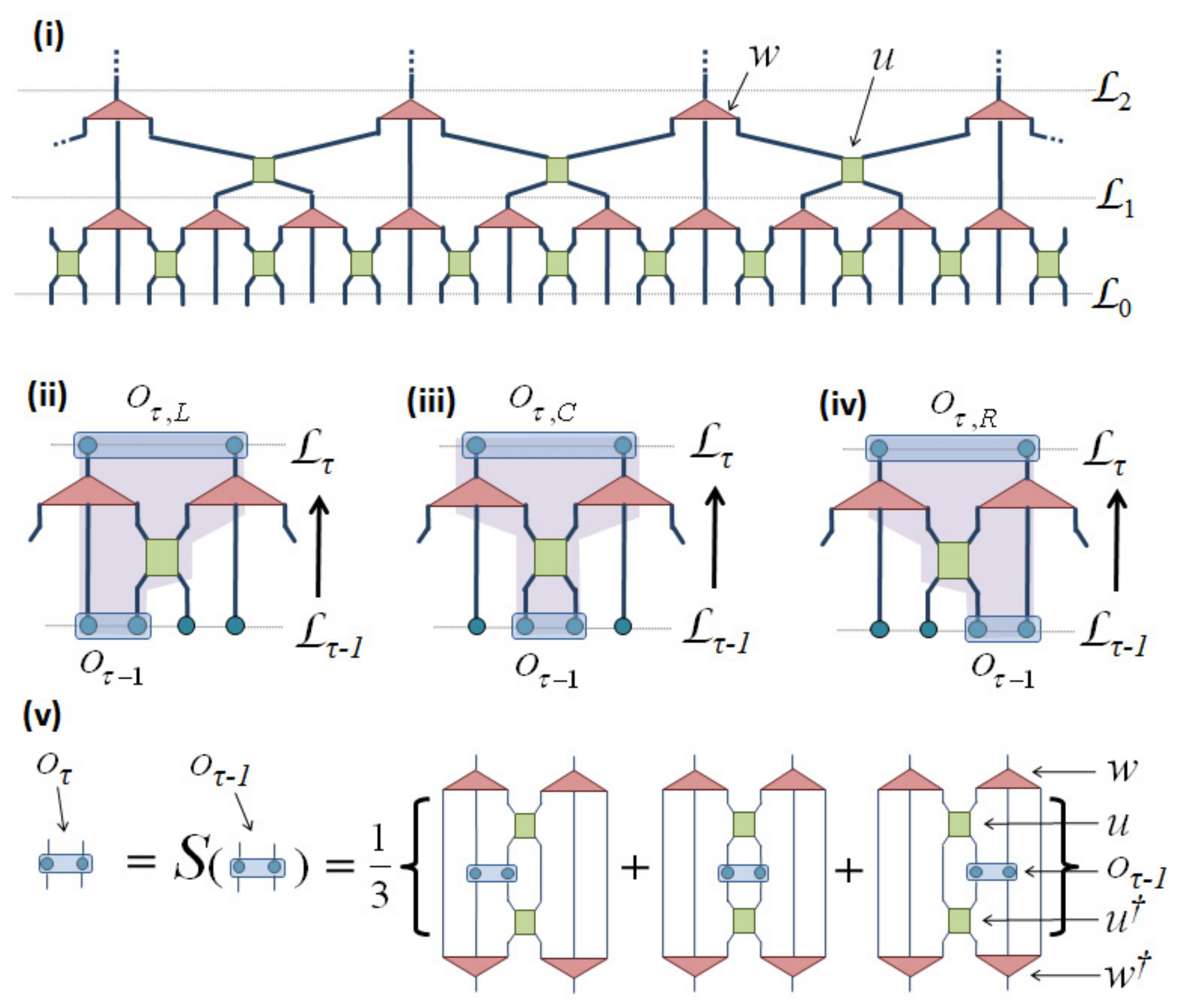}
\caption{($i$) Two lowest rows of disentanglers $u$ and isometries $w$ of the ternary MERA. They map the original infinite lattice $\mathcal{L}_0\equiv \mathcal{L}$ into increasingly coarse-grained lattices $\mathcal{L}_{1}$ and $\mathcal{L}_2$. Notice that three sites of $\mathcal{L}_{\tau-1}$ become one site of $\mathcal{L}_{\tau}$, hence the use of base $3$ logarithm throughout the Chapter. ($ii$)-($iv$) Under the coarse-graining transformation defined by the MERA, two-site operators supported on three different pairs of sites of $\mathcal{L}_{\tau-1}$ become supported on the same pair of sites of $\mathcal{L}_{\tau}$. ($v$) Accordingly, the scaling superoperator $\mathcal{S}$ is the average of three contributions, each of which (and thus also their average) is unital and contractive thanks to the isometric character of $u$ and $w$ \cite{vidal08}.} 
\label{fig:1DCrit:TwoSiteCFT}
\end{center}
\end{figure}

The scale invariant MERA corresponds to the limit of infinitely many layers, $T\rightarrow \infty$, and to choosing the disentanglers and isometries in all layers to be copies of a unique pair $u$ and $w$ \cite{vidal07,vidal08}. In this case we refer to the ascending superoperator $\mathcal{A}_{\tau}$, which no longer depends on $\tau$, as the \emph{scaling superoperator} $\mathcal{S}$ (see Fig. \ref{fig:1DCrit:TwoSiteCFT}), and to its dual $\mathcal{D}_{\tau}$ as $\mathcal{S}^{*}$. Notice that $\mathcal{S}$ is a fixed-point RG map. Then, as customary in RG analysis \cite{cardy96,francesco97}, the scaling operators $\phi_{\alpha}$ and scaling dimensions $\Delta_{\alpha}$ of the theory,
\begin{equation}
	\mathcal{S}(\phi_{\alpha}) = \lambda_{\alpha} \phi_{\alpha},~~~~~~~\Delta_{\alpha} \equiv -\log_3 \lambda_{\alpha},
	\label{eq:1DCrit:scaling}
\end{equation}
are obtained by diagonalizing this map,
\begin{equation}
	\mathcal{S}(\bullet) = \sum_{\alpha} \lambda_{\alpha} \phi_{\alpha} \tr(\hat{\phi}_{\alpha} \bullet),~~~~~~\tr(\hat{\phi}_{\alpha} \phi_{\beta}) = \delta_{\alpha\beta},
	\label{eq:1DCrit:spectral}
\end{equation}
where $\hat{\phi}_{\alpha}$ are the eigenvectors of the dual $S^{*}$, $S^{*}(\hat{\phi}_{\alpha}) = \lambda_{\alpha} \hat{\phi}_{\alpha}$. Eq. \ref{eq:1DCrit:spectral} was first discussed in Ref. \cite{giovannetti08} by Giovannetti, Montangero and Fazio. It formalizes a previous observation (see Eq. 5 of Ref. \cite{vidal08}) that the scale invariant MERA displays polynomial correlations. By construction, $\mathcal{S}$ is \emph{unital}, $\mathcal{S}(\mathbb{I}) = \mathbb{I}$, so that the identity operator $\mathbb{I}$ in $\mathbb{V}^{\otimes 2}$ is a scaling operator with eigenvalue $\lambda_{\mathbb{I}}=1$; and \emph{contractive}, meaning $|\lambda_{\alpha}|\leq 1$ \cite{bratteli79}. Here we will assume, as it is the case in the examples below, that only the identity operator $\mathbb{I}$ has eigenvalue $\lambda = 1$. Then the operator $\hat{\rho} \equiv \hat{\mathbb{I}}$ is a density matrix that corresponds to the \emph{unique} fixed point of $\mathcal{S}^{*}$, $\mathcal{S}^{*}(\hat{\rho})=\hat{\rho}$, and since
\begin{equation}
 \lim_{T\rightarrow \infty} \big(\underbrace{\mathcal{S}^{*}\circ \cdots \circ \mathcal{S}^{*}}_{T \mbox{ \scriptsize{times}}}\big) (\rho_T) = \hat{\rho}
	\label{eq:1DCrit:rho}
\end{equation}
for any starting $\rho_T$, it follows that $\hat{\rho}$ is the state of any pair of contiguous sites of $\mathcal{L}$ (consistent with scale invariance, $\hat{\rho}$ is also the state of any pair of contiguous sites of $\mathcal{L}_{\tau}$ for any finite $\tau$). The computation of the expected value of the local observable $o$ is then straightforward,
\begin{equation}
\left\langle o \right\rangle = \tr(\hat{\rho} o),
\end{equation}
which for the scaling operators reduces to	$\left\langle \phi_{\alpha} \right\rangle = \delta_{\alpha \mathbb{I}}$.

\section{Correlators}
\begin{figure}[!tbhp]
\begin{center}
\includegraphics[width=12cm]{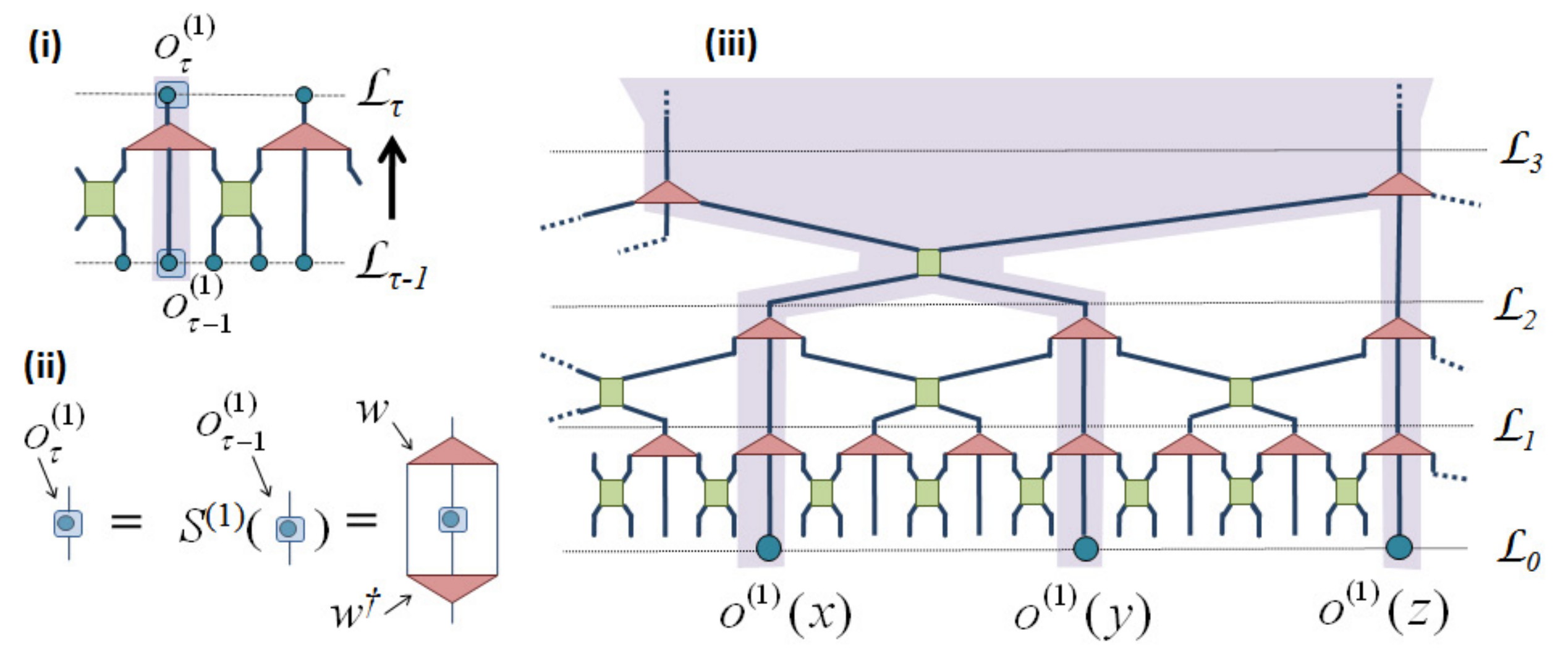}
\caption{($i$) One-site operators on special sites are coarse-grained into one-site operators. ($ii$) Scaling superoperator for one-site operators. ($iii$) In computing correlators on specific sites $x$ and $y$ (or $x$, $y$ and $z$), one-site operators are coarse-grained individually according to $\mathcal{S}^{(1)}$ until they become nearest neighbors (which in this case occurs at lattice $\mathcal{L}_{2}$, $q=2$).} 
\label{fig:1DCrit:OneSiteCFT}
\end{center}
\end{figure}

Let us now diagonalize the one-site scaling superoperator $\mathcal{S}^{(1)}$ of Fig. \ref{fig:1DCrit:OneSiteCFT}, 
\begin{equation}
	\mathcal{S}^{(1)}(\bullet) = \sum_{\alpha} \lambda^{(1)}_{\alpha} \phi^{(1)}_{\alpha} \tr(\hat{\phi}^{(1)}_{\alpha}\bullet), \label{eq:1DCrit:spectral1}
\end{equation}
where the scaling dimensions $\Delta^{(1)}_{\alpha}\equiv -\log_3 \lambda^{(1)}_{\alpha}$ coincide with $\Delta_{\alpha}$ \footnote{Our numerics show that the lowest $n_{\Delta}$ scaling dimensions fulfill $\Delta^{(1)}_{\alpha}\approx \Delta_{\alpha} \approx \Delta_{\alpha}^{\mbox{\tiny CFT}}$, where $n_{\Delta}$ grows with $\chi$.}. The correlator for two scaling operators $\phi_{\alpha}^{(1)}$ and $\phi_{\beta}^{(1)}$ placed on contiguous sites reads
\begin{equation}
	C_{\alpha\beta} \equiv \left\langle \phi_{\alpha}^{(1)}(1) \phi_{\beta}^{(1)}(0)\right\rangle 
	= \tr \big( (\phi^{(1)}_{\alpha}\otimes \phi^{(1)}_{\beta}) \hat{\rho} \big).
	\label{eq:1DCrit:C2}
\end{equation}
Suppose now that $\phi_{\alpha}^{(1)}$ and $\phi_{\beta}^{(1)}$ are placed in two special sites $x,y$ as in Fig. \ref{fig:1DCrit:OneSiteCFT}, where $r_{xy}\equiv x-y$ is such that $|r_{xy}| = 3^q$ for $q=1,2,\cdots$. Then after $q = \log_3 |r_{xy}|$ iterations of the RG transformation, $\phi_{\alpha}^{(1)}$ and $\phi_{\beta}^{(1)}$ become first neighbors again. Notice that each iteration contributes a factor $\lambda^{(1)}_{\alpha}\lambda^{(1)}_{\beta}$. Using the identity $a^{\log b}=b^{\log a}$ we find 
\begin{equation}
(\lambda^{(1)}_{\alpha}\lambda^{(1)}_{\beta})^{\log_3 |r_{xy}|} = |r_{xy}|^{\log_3 (\lambda^{(1)}_{\alpha}\lambda^{(1)}_{\beta})} = |r_{xy}|^{-\Delta^{(1)}_{\alpha}-\Delta^{(1)}_{\beta}}
\nonumber
\end{equation}
and obtain a closed expression for two-point correlators,
\begin{equation}
	\left\langle \phi_{\alpha}^{(1)}(x) \phi_{\beta}^{(1)}(y)\right\rangle 
	= \frac{C_{\alpha\beta}}{|r_{xy}|^{\Delta^{(1)}_{\alpha}+\Delta^{(1)}_{\beta}}}.
	\label{eq:1DCrit:two-point}
\end{equation}
For three-point correlators we define the constants
\begin{eqnarray}
	\Omega_{\alpha\beta}^{~\gamma} &\equiv& \Delta^{(1)}_{\alpha}+\Delta^{(1)}_{\beta}-\Delta^{(1)}_{\gamma}\\
	C_{\alpha\beta\gamma} &\equiv& 2^{\Omega_{\gamma\alpha}^{~\beta}}\tr \big( (\phi^{(1)}_{\alpha}\otimes \phi^{(1)}_{\beta}\otimes \phi^{(1)}_{\gamma}) \hat{\rho}^{(3)} \big) 
\label{eq:1DCrit:C3}
\end{eqnarray}
where the trace corresponds to the correlator on three consecutive sites and $\hat{\rho}^{(3)}$ is obtained from $\hat{\rho}$. For $|r_{xy}| = |r_{yz}| = |r_{xz}|/2 = 3^q$, analogous manipulations lead to
\begin{eqnarray}
	\left\langle \phi_{\alpha}^{(1)}(x) \phi_{\beta}^{(1)}(y) \phi_{\beta}^{(1)}(z)\right\rangle 
	= \frac{C_{\alpha\beta\gamma}}{
	|r_{xy}|{}^{\Omega_{\alpha\beta}^{~\gamma}}
	|r_{yz}|{}^{\Omega_{\beta\gamma}^{~\alpha}}
	|r_{zx}|{}^{\Omega_{\gamma\alpha}^{~\beta}}
	} \label{eq:1DCrit:three-point}
\end{eqnarray}

\section{Conformal field theory}
The continuous limit of a quantum criticial lattice system (scale invariant case) corresponds to a conformal field theory (CFT) \cite{cardy96,francesco97}. A CFT contains an infinite set of quasi-primary fields $\phi^{\mbox{\tiny CFT}}_{\alpha}$, with scaling dimensions $\Delta_{\alpha}^{\mbox{\tiny CFT}}$. The correlators involving two or three quasi-primary fields have expressions analogous to Eqs. \ref{eq:1DCrit:two-point} and \ref{eq:1DCrit:three-point}, and the (symmetric) coefficients $C_{\alpha\beta\gamma}^{\mbox{\tiny CFT}}$ for three-point correlators coincide with those in the so-called \emph{operator product expansion} (OPE). Moreover, quasi-primary fields are organized in \emph{conformal towers} corresponding to irreducible representations of the Virasoro algebra. Each tower contains one \emph{primary field} $\phi^{p}$ at the top, with conformal dimensions $(t,\bar{t})$ (such that its scaling dimension is $\Delta^{p} \equiv t + \bar{t}\,$), and its infinitely many descendants, which are quasi-primary fields with scaling dimension $\Delta = \Delta^{p} + n$ for some integer $n\geq 1$. 

A CFT is completely specified by its symmetries once the following conformal data has been provided: (i) the \emph{central charge} $c$, (ii) a complete list of \emph{primary fields} with their \emph{conformal dimensions} and (iii) the OPE for these primary fields. For instance, the Ising CFT in 1+1 dimensions has central charge $c=1/2$, three primary fields \emph{identity} $\mathbb{I}$, \emph{spin} $\sigma$ and \emph{energy} $\epsilon$ with conformal dimensions $(0,0)$, $(\frac{1}{16},\frac{1}{16})$ and $(\frac{1}{2},\frac{1}{2})$, and OPE coefficients
\begin{eqnarray}
	C^{\mbox{\tiny CFT}}_{\alpha\beta \mathbb{I}}\!= \!\delta_{\alpha\beta}, ~
  C^{\mbox{\tiny CFT}}_{\sigma\sigma\epsilon} \! = \frac{1}{2}, ~~
	C^{\mbox{\tiny CFT}}_{\sigma\sigma\sigma}\! = \!C^{\mbox{\tiny CFT}}_{\epsilon\epsilon\epsilon}\! =\! C^{\mbox{\tiny CFT}}_{\epsilon\epsilon\sigma}\! = 0.
	\label{eq:1DCrit:OPE_Ising}
\end{eqnarray}

The present analysis readily suggests a correspondence between the scaling operators $\phi_{\alpha}$ of the scale invariant MERA, defined on a lattice, and the quasi-primary fields $\phi_{\alpha}^{\mbox{\tiny CFT}}$ of a CFT, defined in the continuum. Together with the algorithm described below, this correspondence grants us numerical access, given a critical Hamiltonian $H$ on the lattice, to most of the conformal data of the underlying CFT, namely to scaling dimensions and OPE coefficients. The central charge $c$ can also be obtained from the von Neumann entropy $S(\rho)\equiv-\tr(\rho\log_2\rho)$, which for a block of $L$ sites scales, up to some additive constant, as $S = \frac{c}{3} \log_2 L$ \cite{vidal03,calabrese04}. We then have $S(\hat{\rho}) - S(\hat{\rho}^{(1)}) =  \frac{c}{3} (\log_2 2-\log_2 1) = \frac{c}{3}$, or simply
\begin{equation}
	c = 3 \left(S(\hat{\rho}) - S(\hat{\rho}^{(1)})\right).
	\label{eq:1DCrit:central}
\end{equation}


\section{Algorithm for scale invariant MERA}
Given a critical Hamiltonian $H$ for an infinite lattice, we obtain a scale invariant MERA for its ground state $\ket{\Psi}$ by adapting the general strategy discussed in Chapter \ref{chap:MERAalg}. Recall that tensors (disentanglers $u$ and isometries $w$) are optimized so as to minimize the energy $E \equiv \bra{\Psi} H \ket{\Psi}$. After linearization this reads 
\begin{equation}
	E = \tr(u \Upsilon_u) + k_1 = \tr(w \Upsilon_w) + k_2,
\end{equation}
where $\Upsilon_u$ and $\Upsilon_{w}$ are known as \emph{environments} and $k_1,k_2$ are two irrelevant constants. In the translation invariant case the environment for, say, an isometry $w$ at layer $\tau$ of the MERA, $\Upsilon_w = f(u_{\tau},w_{\tau},\rho_{\tau}, h_{\tau-1})$, is a function of the disentangler $u_{\tau}$ and isometry $w_{\tau}$ of that layer, a two-site density matrix $\rho_{\tau}$ and a two-site Hamiltonian term $h_{\tau-1}$. In the present case, we replace the above with the unique pair $(u,w)$, the fixed-point density matrix $\hat{\rho}$, and an average Hamiltonian $\bar{h} \equiv \sum_{\tau} h_{\tau}/3^{\tau}$, where the weights $1/3^{\tau}$ account for the relative number of tensors in different layers of the MERA. Then, starting from some initial pair $(u,w)$ and the critical Hamiltonian $H$ made of two-body terms $h$, the following steps are repeated until convergence:
\begin{enumerate}
	\item Given the latest $(u,w)$, compute $(\hat{\rho},\bar{h})$.
	\item Given $(u,w,\hat{\rho},\bar{h})$, update the pair $(u,w)$.
\end{enumerate}
In step A1, the scaling superoperator $\mathcal{S}$ is built as indicated in Fig. \ref{fig:1DCrit:TwoSiteCFT}. We compute the fixed-point density matrix $\hat{\rho}$ by sparse diagonalization of $\mathcal{S}$, and the average Hamiltonian $\bar{h}$ by using $h_{\tau} = \mathcal{S}(h_{\tau-1})$, $h_0\equiv h$ \footnote{In practice we only compute the first $k$ terms ($k \approx 2,3$) of the expansion $\bar{h} = h_0 + h_1/3 + h_2/9 + \cdots$. This average is only needed when $H$ contains operators that are irrelevant in the RG sense.}. Step A2 is decomposed into a sequence of alternating optimizations for $u$ and $w$ as in the generic algorithm of Chapter \ref{chap:MERAalg}, where each tensor is updated by computing a singular value decomposition of its environment.

\begin{figure}[!tb]
\begin{center}
\includegraphics[width=10cm]{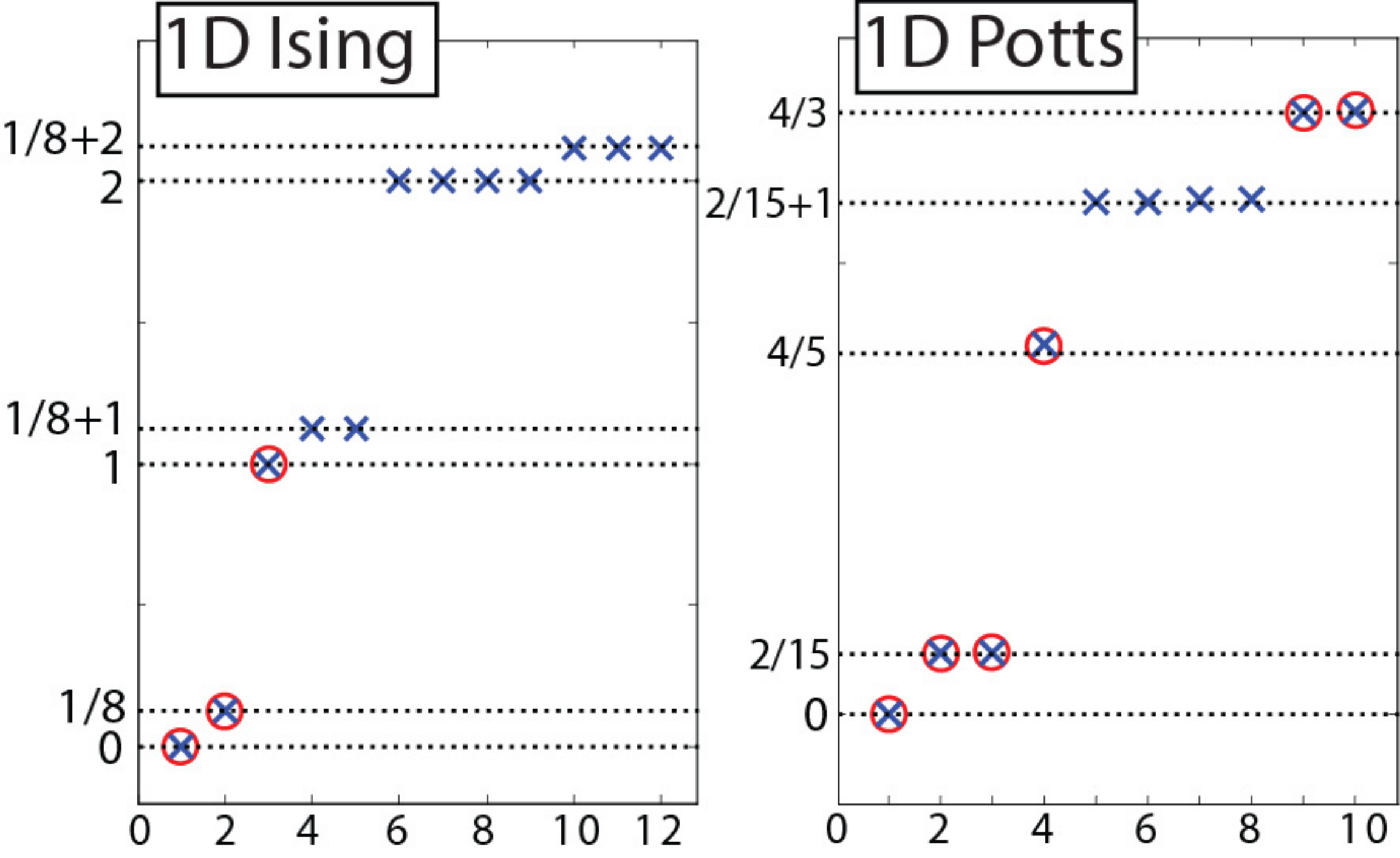}
\caption{Scaling dimensions $\Delta_{\alpha}$ obtained from the spectrum of the scaling superoperator $\mathcal{S}$. Circles indicate primary fields. \emph{Left:} For the Ising model we can identify the scaling dimensions of the three primary fields, the so-called identity $\mathbb{I}$, spin $\sigma$ and energy $\epsilon$, together with several of their descendants. \emph{Right:} The spectrum of $\mathcal{S}$ for the 3-level Potts model shows some of its primary fields, including its primary fields with multiplicity two, namely the spins $\sigma_1$ and $\sigma_2$ and the pair $Z_1$ and $Z_2$ \cite{francesco97}. 
} 
\label{fig:1DCrit:CFTscaling}
\end{center}
\end{figure}

\section{Benchmark results}
We illustrate the above ideas and the performance of the algorithm by considering the Ising and 3-level Potts quantum critical models in 1$D$,
\begin{eqnarray}
H_{{\rm{Ising}}}  &=& \sum_r \left(\lambda \sigma^{[r]}_z  + \sigma^{[r]}_x \sigma^{[r+1]}_x \right) \\
H_{{\rm{Potts}}}  &=& \sum_r \left(\lambda M^{[r]}_{z}  +  M^{[r]}_{x,1} M^{[r+1]}_{x,2} + M^{[r]}_{x,2} M^{[r+1]}_{x,1}\right) \nonumber
 \label{hams}
\end{eqnarray}
where $\sigma_z$ and $\sigma_x$ are Pauli matrices, and
\begin{eqnarray}
	M_{z} = \left( \begin{array}{ccc} 
		2 & 0 & 0\\
		0 & -1 & 0\\
		0 & 0 & -1
	\end{array} \right), M_{x,1}	= \left( \begin{array}{ccc} 
		0 & 1 & 0\\
		0 & 0 & 1\\
		1 & 0 & 0
	\end{array} \right),
\end{eqnarray}
$M_{x,2} = (M_{x,1})^2$. Notice that sites have a vector space of dimension $d=2$ or $d=3$. In order to use a scale invariant MERA with $\chi > d$, we allow the disentanglers and isometries of the first few (typically one to five) layers to be different from $u$ and $w$. We iterate steps A1-A2 about 1000 times. With a cost per iteration that scales as $\chi^8$ and using a 3 GHz dual core desktop with 8 Gb of RAM, simulations for $\chi=4,8,16,22$ take of the order of minutes, hours, days and weeks respectively. The following results correspond to $\chi = 22$.

From Eq. \ref{eq:1DCrit:central} we obtain an estimate for the central charge, namely $c_{{\rm{Ising}}} = .5007$ and $c_{{\rm{Potts}}} = .806$, to be compared with the exact results $0.5$ and $0.8$. Fig. \ref{fig:1DCrit:CFTscaling} shows the smallest scaling dimensions $\Delta_{\alpha}$ of the scaling superoperator $\mathcal{S}$ \footnote{Our numerics show that the lowest $n_{\Delta}$ scaling dimensions fulfill $\Delta^{(1)}_{\alpha}\approx \Delta_{\alpha} \approx \Delta_{\alpha}^{\mbox{\tiny CFT}}$, where $n_{\Delta}$ grows with $\chi$.}. We obtain remarkable agreement with those expected from CFT, as shown in 
Table~\ref{tab:fdims}.
\begin{table}
\begin{center}
\begin{tabular}{|c|c|c|c|}
  \hline
  Ising & $\Delta^{\mbox{\tiny CFT}}$    &  $\Delta$ \scriptsize{(MERA $\chi=22$)}& rel. error \\ \hline
  $\sigma$ & 1/8 = 0.125 & 0.124997 & 0.002$\%$\\
  $\epsilon$ & 1 &  1.0001 & 0.01$\%$\\
  \hline
  Potts &  $\Delta^{\mbox{\tiny CFT}}$  & $\Delta$  \scriptsize{(MERA $\chi=22$)}& rel. error\\ \hline
  $\sigma_1$ & 2/15 = 0.1$\hat{3}$ & 0.1339 & 0.4$\%$ \\
  $\sigma_2$ & 2/15 = 0.1$\hat{3}$ & 0.1339 & 0.4$\%$\\
  $\epsilon$ & 4/5 = 0.8 &  0.8204 & 2.5$\%$\\
  $Z_1$ & 4/3 = 1.$\hat{3}$ &  1.3346 & 0.1$\%$\\
  $Z_2$ & 4/3 = 1.$\hat{3}$ &  1.3351 & 0.1$\%$\\
  \hline
  \end{tabular} \nonumber
  \end{center}
\caption{\label{tab:fdims}Comparison of scaling dimensions of primary fields of the Ising and Potts models calculated using MERA ({$\Delta$\scriptsize{(MERA $\chi=22$)}}) with exact results known from CFT ({$\Delta^{\mbox{\tiny CFT}}$}).} 
\end{table}
Recall that all the critical exponents of the model can be obtained from the scaling dimensions of primary fields. For instance, for the Ising model the exponents $\nu$ and $\eta$ are $\nu = 2\Delta_{\sigma}$ and $\eta = \frac{1}{2-\Delta_{\epsilon}}$, whereas the \emph{scaling laws} express the critical exponents $\alpha,\beta,\gamma, \delta$ in terms of $\nu$ and $\eta$ \cite{francesco97}.
Further, the OPE coefficients for primary fields of, say, the critical Ising model are computed as follows. The matrix $C_{\alpha\beta}$ in Eq. \ref{eq:1DCrit:C2} is diagonal for the scaling operators corresponding to $\mathbb{I}$, $\sigma$ and $\epsilon$, which we normalize so that $C_{\alpha\beta} = \delta_{\alpha\beta}$. With this normalization, we then compute the coefficients $C_{\alpha\beta\gamma}$ using Eq. \ref{eq:1DCrit:C3}. We reproduce all the values of Eq. \ref{eq:1DCrit:OPE_Ising} with errors bounded by $3\times 10^{-4}$.


\section{Discussion}
In this Chapter we have explained how to compute the ground state of a critical Hamiltonian using the scale invariant MERA and how to extract from it the properties that characterize the system at a quantum critical point. Our results, which build upon those of Ref. \cite{vidal07,evenbly07a,evenbly07b,vidal08,aguado08,giovannetti08,evenbly08a}, also unveil a concise connection between the scale invariant MERA and CFT. This correspondence adds significantly to the conceptual foundations of entanglement renormalization. The scale invariant MERA can be regarded as approximately realizing an infinite dimensional representation of the Virasoro algebra \cite{cardy96,francesco97}. The finite value of $\chi$ effectively implies that only a finite number of the quasi-primary fields of the theory can be included in the description. Fields with small scaling dimension, such as primary fields, are retained foremost. 
As a result, given a Hamiltonian on an infinite lattice, we can numerically evaluate the scaling dimensions and OPE of the primary fields of the CFT that describes the continuum limit of the model. This approach differs in a fundamental way from, and offer an alternative to, the long-established techniques of Refs. \cite{cardy84,cardy86}, based instead on finite size scaling. We conclude by noting that most of our considerations rely on scale invariance alone and can be applied to study also critical ground states in 2$D$ systems, as is demonstrated next in Chapter \ref{chap:2DMera}.

\chapter{Entanglement renormalization in two spatial dimensions}
\label{chap:2DMera}

\section{Introduction}
Entanglement renormalization has been proposed as a real-space renormalization group (RG) method \cite{wilson75} to study extended quantum systems on a lattice. A highlight of the approach is the removal, before the coarse-graining step, of short-range entanglement by means of unitary transformations called \emph{disentanglers}. This prevents the accumulation of short-range entanglement over successive RG transformations. Such accumulation is the reason why the density matrix renormalization group (DMRG) \cite{white92,white93}-- an extremely powerful technique for lattices in one spatial dimension -- breaks down in two dimensions, where it can only address small systems. 

The use of disentanglers leads to a real-space RG transformation that can in principle be iterated indefinitely, enabling the study of very large systems in a quasi-exact way. This RG transformation also leads to the so-called \emph{multi-scale entanglement renormalization ansatz} (MERA) \cite{vidal08} to describe the ground state of the system -- or, more generally, a low energy sector of its Hilbert space. In a translation invariant lattice made of $N$ sites, the cost of simulations grows only as $\log N$ \cite{evenbly08a}. In the presence of scale invariance, this additional symmetry is naturally incorporated into the MERA and a very concise description, independent of the size of the lattice, is obtained in the infrared limit of a topological phase \cite{aguado08,konig09} or at a quantum critical point \cite{vidal07, vidal08, evenbly07a, evenbly07b, giovannetti08, pfeifer08, montangero09}.
 
The basic principles of entanglement renormalization are the same in any number of spatial dimensions, however, numerical work with 2$D$ lattices incurs a much larger computational cost. In Chapters \ref{chap:FreeFerm} and \ref{chap:FreeBoson} we explored the use of ER in 2$D$ lattices of free fermions and free bosons. Entanglement Renormalization has also been tested for an Ising model in a square lattice of small linear size $L\leq 8$ \cite{cincio08}. It must be emphasized, however, that the approach of Chapters \ref{chap:FreeFerm} and \ref{chap:FreeBoson} relies on the gaussian character of free particles and can not be generalised to the interacting case, whereas the results of Ref. \cite{cincio08} were obtained by exploiting a significant reduction in computational cost that occurs only for small 2$D$ lattices. 

In this Chapter we present an implementation of the MERA that allows us to consider, with modest computational resources, 2$D$ systems of arbitrary size, including infinite systems. In this way we demonstrate the scalability of entanglement renormalization in two spatial dimensions and decisively contribute to establishing the MERA as a competitive approach to systematically address 2$D$ lattice models. The key of the present scheme is a carefully planned organization of the tensors in the MERA, leading to simulation costs that grow as $O(\chi^{16})$, where $\chi$ is the dimension of the vector space of an effective site. This is drastically smaller than the cost $O(\chi^{28})$ of the 4-to-1 MERA scheme (see Fig. \ref{fig:MERAintro:2DAltSchemes}) that was used in Chapters \ref{chap:FreeFerm} and \ref{chap:FreeBoson} and in Ref. \cite{cincio08}. We also demonstrate the performance of the scheme by analysing the 2$D$ quantum Ising model, for which we obtain accurate estimates of the ground state energy and magnetizations, as well as two-point correlators (shown to scale polynomially at criticality), the energy gap, and the critical magnetic field and beta exponent. Finally, we discuss how the use of disentanglers affects the simulation costs, by comparing the MERA with a \emph{tree tensor network} (TTN) \cite{tagliacozzo09}.

\begin{figure}[!tbhp]
\begin{center}
\includegraphics[width=12cm]{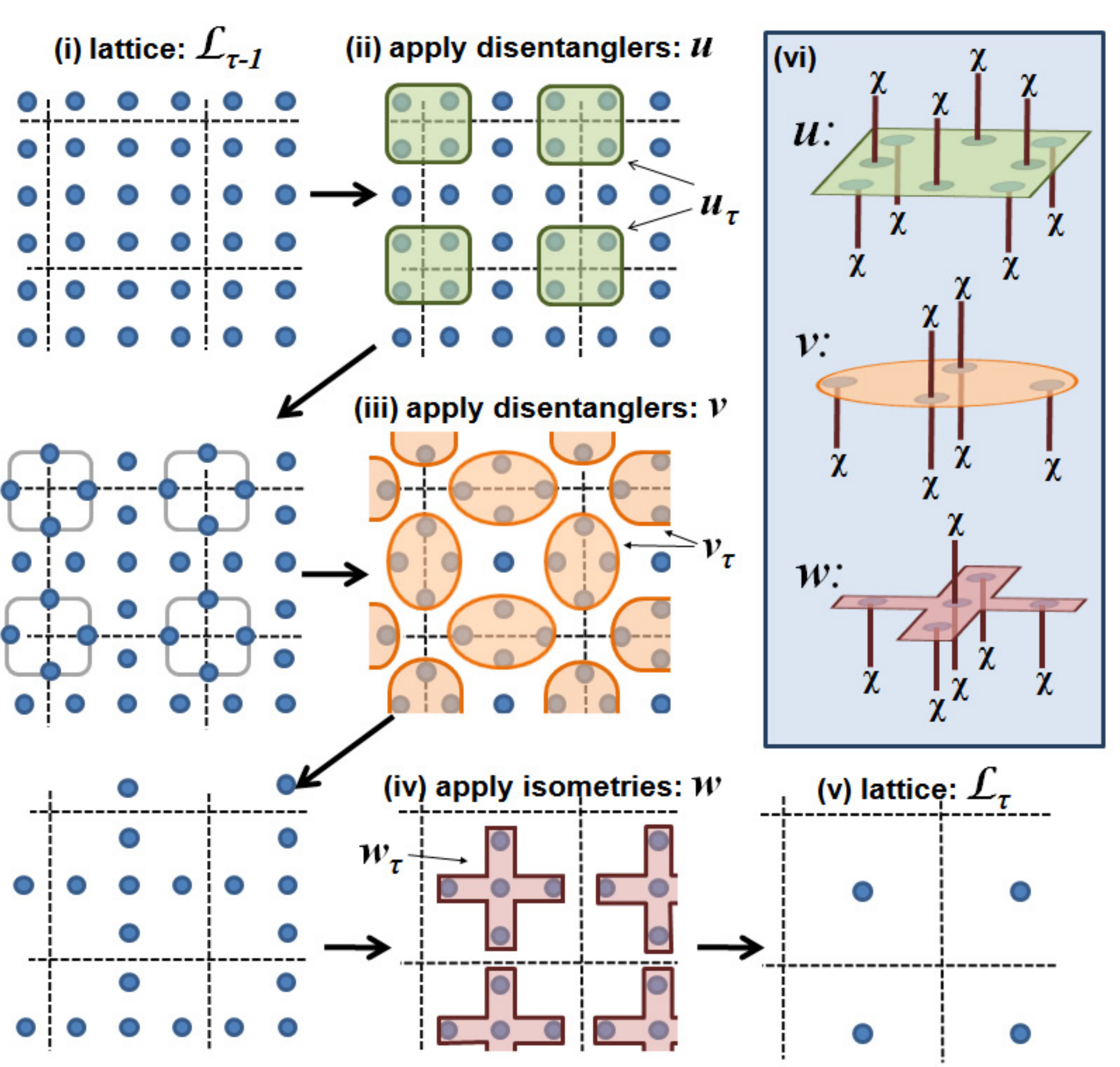}
\caption{ Entanglement renormalization scheme for a square lattice. A block of $3\times 3$ sites of lattice $\mathcal{L}_{\tau-1}$ (i) is mapped onto one site of $\mathcal{L}_{\tau}$ (v). The RG transformation involves (ii) applying disentanglers $u$ between the corners of adjacent blocks followed by (iii) disentanglers $v$ which act across the sides of adjacent blocks and (iv) isometries $w$ which act within a block. Tensors $u,v$ and $w$ have a varying number of incoming and outgoing indices (vi) according to Eq. \ref{eq:2DMera:tensors}.} \label{fig:2DMera:Scheme} 
\end{center}
\end{figure}
 
\section{The MERA on a 2D lattice}
Let us consider a square lattice $\mathcal{L}_0$ made of $N=L\times L$ sites, each one described by a Hilbert space $\mathbb{V}$ of finite dimension $d$. The proposed 2$D$ MERA is characterized by the coarse-graining transformation of Fig. (\ref{fig:2DMera:Scheme}), where blocks of $3 \times 3$ sites of lattice $\mathcal{L}_{0}$ are mapped onto single sites of a coarser lattice $\mathcal{L}_1$. This is achieved in three steps: first disentanglers $u$ are applied on the four sites located at the corners of four adjacent blocks; then disentanglers $v$ are applied at the boundary between two adjacent blocks, transforming four sites into two; finally, isometries $w$ are used to map a block into a single effective site. In this way, tensors $u,v$ and $w$ \footnote{The tensors of the MERA are called disentanglers $u$ or isometries $w$ depending on whether they are in charge of eliminating short-range entanglement or of mapping a block of sites into a single site \cite{vidal07,vidal08,evenbly08a}. This distinction is somewhat arbitrary: one can consider tensors that fulfill the two roles simultaneously, such as tensor $v$ in Fig. \ref{fig:2DMera:Scheme}, that we still call disentangler. All these tensors must be isometric, that is $u^{\dagger}u=I$, $v^{\dagger}v=I$, $w^{\dagger}w=I$. The hermitian conjugation ($\dagger$) in Eq. (\ref{eq:2DMera:tensors}) appears for consistency with previous references. },  
\begin{equation}
	u^{\dagger}\!: \mathbb{V}^{\otimes 4} \rightarrow \mathbb{V}^{\otimes 4},~~~
	v^{\dagger}\!: \mathbb{V}^{\otimes 4} \rightarrow \mathbb{V}^{\otimes 2},~~~
  w^{\dagger}\!: \mathbb{V}^{\otimes 5} \rightarrow \mathbb{V}, \label{eq:2DMera:tensors}
\end{equation}
transform the state $\ket{\Psi_0} \in \mathbb{V}^{\otimes N}$ of the lattice $\mathcal{L}_0$ in which we are interested (typically the ground state of a local Hamiltonian $H_0$) into a state $\ket{\Psi_1}\in \mathbb{V}^{\otimes N/9}$ of the effective lattice $\mathcal{L}_1$ through the sequence 
\begin{equation}
	\ket{\Psi_0} \stackrel{u}{\rightarrow} \ket{\Psi_0'} \stackrel{v}{\rightarrow} \ket{\Psi_0''} \stackrel{w}{\rightarrow} \ket{\Psi_1}.
\end{equation}
To understand the role of these tensors, it is useful to think of the state $\ket{\Psi_0}$ as possessing three different kinds of entanglement: short-range entanglement residing at the corners of four adjacent blocks, short-range entanglement residing near the boundary shared by two blocks, and long-range entanglement. Then the disentanglers $u$ and $v$ are used to reduce the amount of short-range entanglement residing near the corners and boundaries of the blocks. In other words, in states $\ket{\Psi_0'}$ and $\ket{\Psi_0''}$ increasing amounts of short-range entanglement from $\ket{\Psi_0}$ have been removed. This fact facilitates significantly the job of the isometry $w$, namely to compress into an effective site of $\mathcal{L}_1$ those degrees of freedom in a block that still remain entangled (now mostly through long-range entanglement) with degrees of freedom outside the block. Thus, the resulting state $\ket{\Psi_1}$ still contains the long-range entanglement of $\ket{\Psi_0}$, but most of its short-range entanglement is gone. We complete the above construction by noticing that a $d$-dimensional space $\mathbb{V}$ is often too small to accommodate all the relevant degrees of freedom left on a block. Accordingly, we shall describe the effective sites of $\mathcal{L}_1$ with a space of larger dimension $\chi$. This dimension $\chi$ determines both the accuracy and cost of the simulations. 

The transformation of Fig. \ref{fig:2DMera:Scheme} can now be applied to lattice $\mathcal{L}_1$, producing a coarser lattice $\mathcal{L}_2$. More generally, if $\mathcal{L}_0$ is finite, $O(\log N)$ iterations will produce a sequence of lattices $\{\mathcal{L}_0, \mathcal{L}_1, \mathcal{L}_2,\cdots , \mathcal{L}_{\mbox{\tiny top}} \}$ where the top lattice $\mathcal{L}_{\mbox{\tiny top}}$ contains only a small number of sites and can be addressed with exact numerical techniques. Thus, given a Hamiltonian $H_0$ on $\mathcal{L}_{0}$, we can use the above RG transformation to obtain a sequence of Hamiltonians $\{H_0, H_1, H_2, \cdots, H_{\mbox{\tiny top}}\}$, then diagonalize $H_{\mbox{\tiny top}}$ to find its ground state $\ket{\Psi_{\mbox{\tiny top}}}$, and finally recover the ground state $\ket{\Psi_0}$ of $H_0$ by reversing all the RG transformations:
\begin{equation}
	\ket{\Psi_{\mbox{\tiny top}}} \rightarrow \cdots \rightarrow \ket{\Psi_2} 
\rightarrow \ket{\Psi_1} \rightarrow \ket{\Psi_0}.
\label{eq:2DMera:sequence}
\end{equation}
This is precisely how the MERA is defined. Specifically, the MERA for $\ket{\Psi_0}$ is a tensor network containing (i) a top tensor, that describes $\ket{\Psi}_{\mbox{\tiny top}}$, and (ii) $O(\log N)$ layers of tensors (disentanglers and isometries), where each layer is used to invert one step of the coarse-graining transformation of Fig. \ref{fig:2DMera:Scheme} according to the sequence (\ref{eq:2DMera:sequence}).

\begin{figure}[!tbhp]
\begin{center}
\includegraphics[width=12cm]{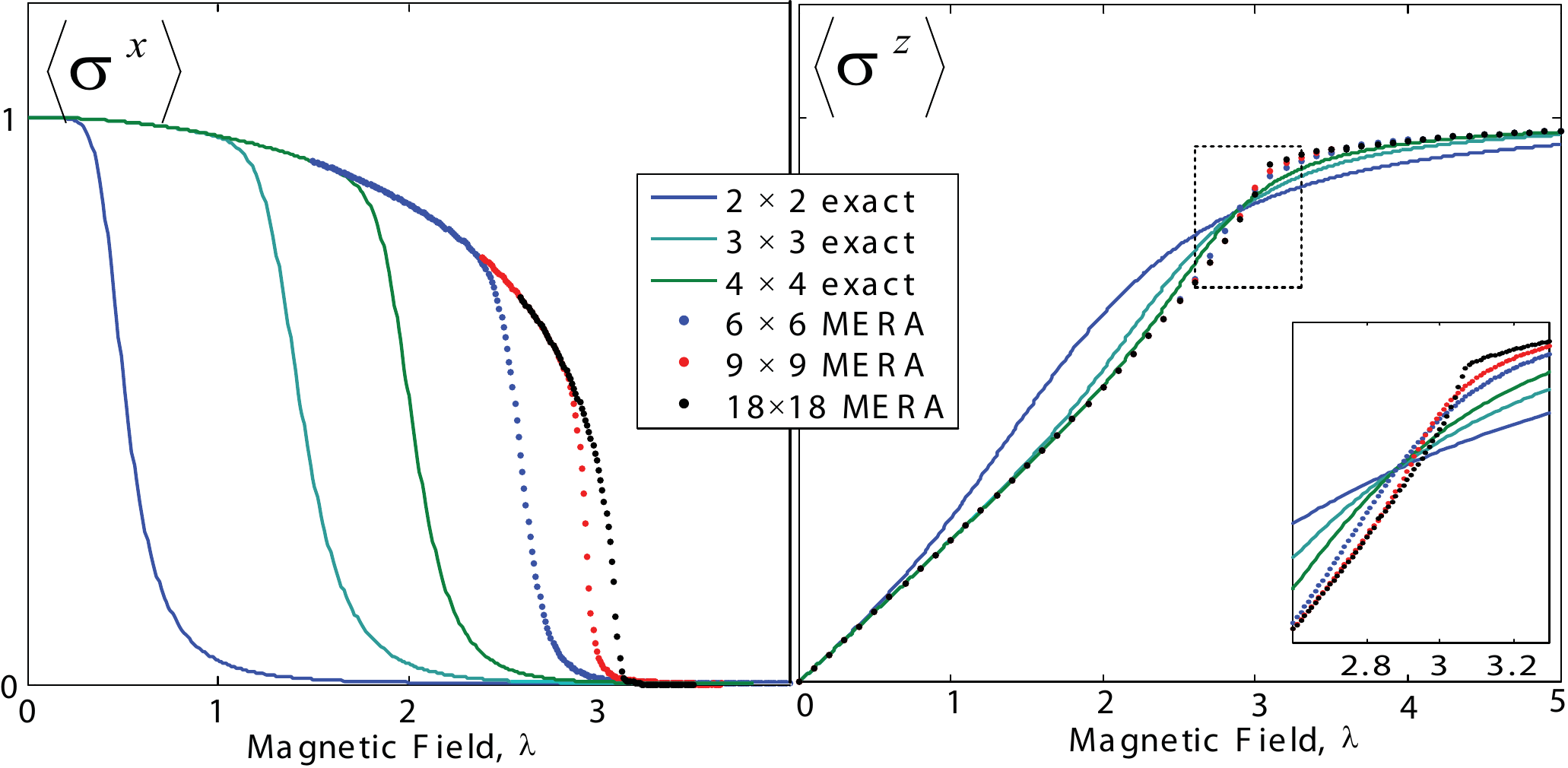}
\caption{ Spontaneous and transverse magnetizations $\left\langle {\sigma _x } \right\rangle$ and $\left\langle {\sigma _z } \right\rangle$ as a function of the applied magnetic field $\lambda$ and for different lattice sizes $L$. Results for small systems correspond to exact diagonalization whilst results for larger systems were obtained with a $\chi=6$ MERA. As $L$ increases, the magnetizations are seen to converge toward their thermodynamic limit values. Results for $L=54$ could not be visually distinguished from results for $L=18$ and have been omitted in the plot. As it is characteristic of a second order phase transition, for large $L$ both magnetizations develop a discontinuity in their derivative, with $\left\langle {\sigma _x } \right\rangle$ (the order parameter) suddenly dropping to zero at the quantum critical point (see Fig. \ref{fig:2DMera:MagRefine}).}\label{fig:2DMera:MagScale} 
\end{center}
\end{figure}

The technical details on how to numerically optimize the disentanglers and isometries of the MERA to approximate the ground state $\ket{\Psi_0}$ of $H_0$ are analogous to those discussed in Chapter \ref{chap:MERAalg} for a 1$D$ lattice and will not be repeated here. Instead, we focus on the key aspect that makes the present 2$D$ scheme much more efficient than the previous 4-to-1 scheme. For this purpose, we consider an operator $O_0$ whose support is contained within a block of $2\times 2$ sites of lattice $\mathcal{L}_0$. Direct inspection shows that, no matter where this block is placed with respect to the disentanglers and isometries of Fig. \ref{fig:2DMera:Scheme}, the support of the resulting coarse-grained operator $O_1$ is also contained within a block of $2\times 2$ sites of $\mathcal{L}_1$, and the same holds for any subsequent coarse-graining. This is in sharp contrast with the 2$D$ scheme of Refs. \cite{evenbly07a,evenbly07b,cincio08}, where the minimal stable support of local observables (or 'width' of past \emph{causal cones}) corresponded to blocks of $3\times 3$ sites. In the present case, much smaller objects (operators acting on $4$ sites instead of $9$ sites) are manipulated during the calculations, resulting in the announced dramatic drop in simulation costs.

\begin{figure}[!tbhp]
\begin{center}
\includegraphics[width=12cm]{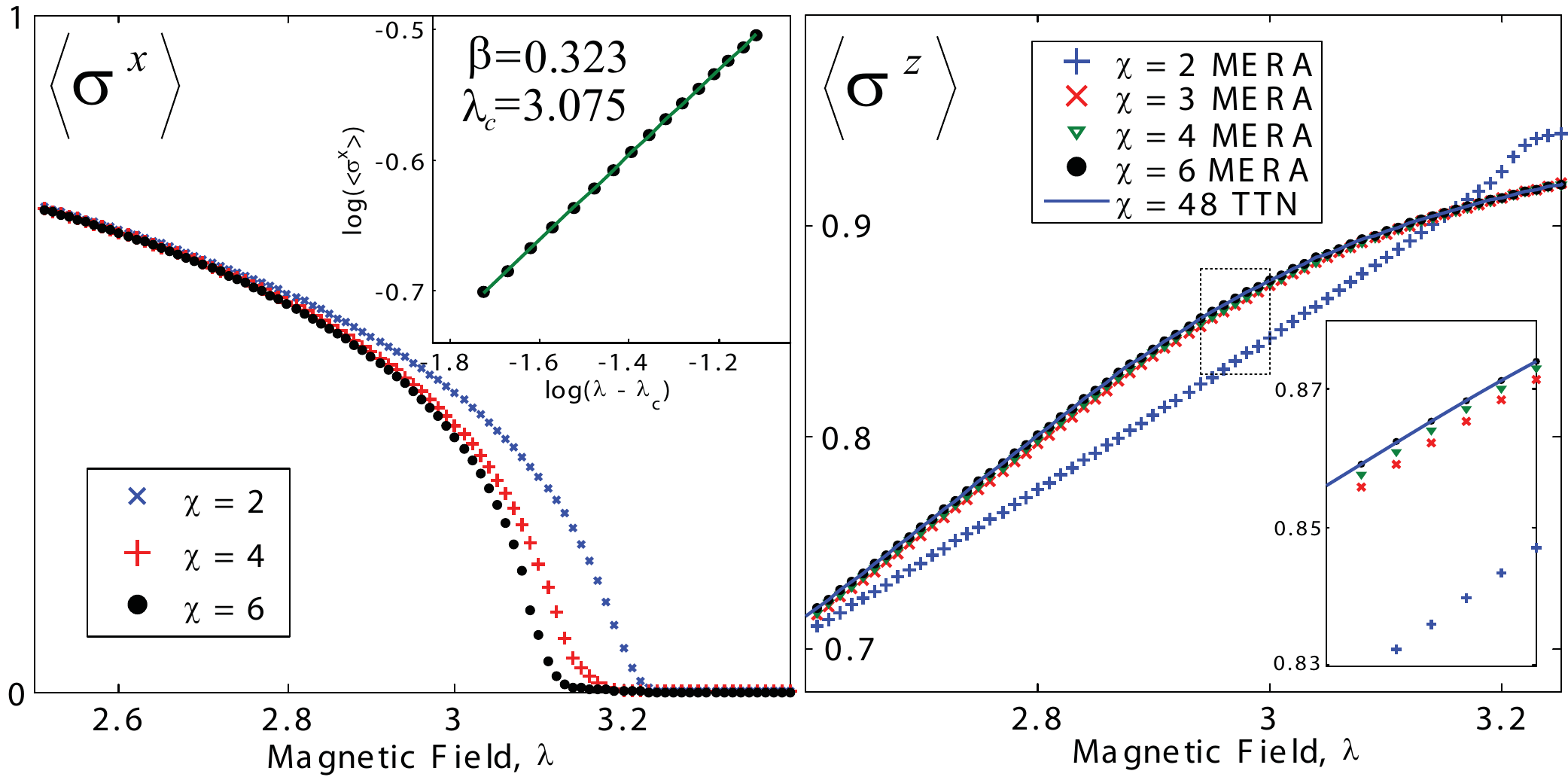}
\caption{ Magnetizations $\left\langle {\sigma _x } \right\rangle$ and $\left\langle {\sigma _z } \right\rangle$ as a function of the applied magnetic field $\lambda$ for different values of the refinement parameter $\chi$. \emph{Left:} 
Spontaneous magnetization $\left\langle {\sigma _x } \right\rangle$ for $L=54$. Data fits of the form $\left\langle {\sigma _x } \right\rangle  \sim \left( {\lambda  - \lambda _c } \right)^{\beta_c }$ near the critical point give a critical magnetic field $\lambda _c  = \left\{ {3.13,3.09,3.075} \right\}$ and critical exponent $\beta _c  = \{ 0.320,0.321,0.323\}$ for $\chi  = \left\{ {2,4,6} \right\}$. Current Monte Carlo estimates are $\lambda _c  = 3.044$ and $\beta _c  = 0.326$ \cite{rieger99,blote02}. Thus accuracy increases with $\chi$. \emph{Right:} Transverse magnetization $\left\langle {\sigma _z } \right\rangle$ for $L=6$. TTN results for large $\chi$ are taken as the exact solution (see Fig. \ref{fig:2DMera:2DEnergyErrorNew}). Whilst a $\chi=2$ MERA produces significantly different values, results for $\chi=3$ are already very similar and those for $\chi=6$ MERA agree with the TTN solution on at least 3 significant digits.} \label{fig:2DMera:MagRefine}
\end{center}
\end{figure}


\section{Results for the 2D Ising model}
We have tested the proposed scheme by investigating low energy properties of the quantum Ising model with transverse magnetic field,
\begin{equation}
H_{\mbox{\tiny Ising}} = \sum\limits_{\left\langle {r,r'} \right\rangle } {\sigma _x^{[r]} \sigma _x^{[r']}  + \lambda \sum\limits_r {\sigma _z^{[r]} } },
\end{equation}
on a square lattice with periodic boundary conditions (local dimension $d=2$). 
First of all, we consider a sequence of lattices with increasing linear size $L=\{6,9,18,54\}$. For each of them, a MERA approximation to the ground state of $H_{\mbox{\tiny Ising}}$ for different values $\lambda \in [0, 5]$ of the transverse magnetic field is obtained using $\chi=6$. Computing the ground state for $L=54$ and critical transverse magnetic field takes $\sim$ 4 days on a 3GHz dual-core desktop PC with 8Gb RAM when starting from a randomly initialized MERA \footnote{Calculations for $\chi=6$ are achieved by using a disentangler $u$ with $\chi=4$ on selected indices. The computation time is reduced to a few hours per point by re-using a MERA previously converged (for a similar magnetic field) as the starting point of a simulation.}. Fig. \ref{fig:2DMera:MagScale} displays the expected value of the parallel and transverse magnetizations, both of which show characteristic signs of a second order phase transition as $L$ increases. We emphasize that since the simulation costs grow only as the logarithm of $L$, it is straightforward to increase the system size until e.g. finite size effects become negligible on local observables.

Fig. \ref{fig:2DMera:MagRefine} shows how the parallel and transverse magnetizations change with increasing $\chi$, for $L=54$. Since the cost of the simulations grows as $O(\chi^{16})$, only small values of $\chi$ can be considered in practice. However, with $\chi=6$ one already obtains estimates for the location of the critical point and the critical exponent $\beta$ that already fall within $1\%$ of the best Monte Carlo results \cite{rieger99,blote02}.

\begin{figure}[!tbhp]
\begin{center}
\includegraphics[width=12cm]{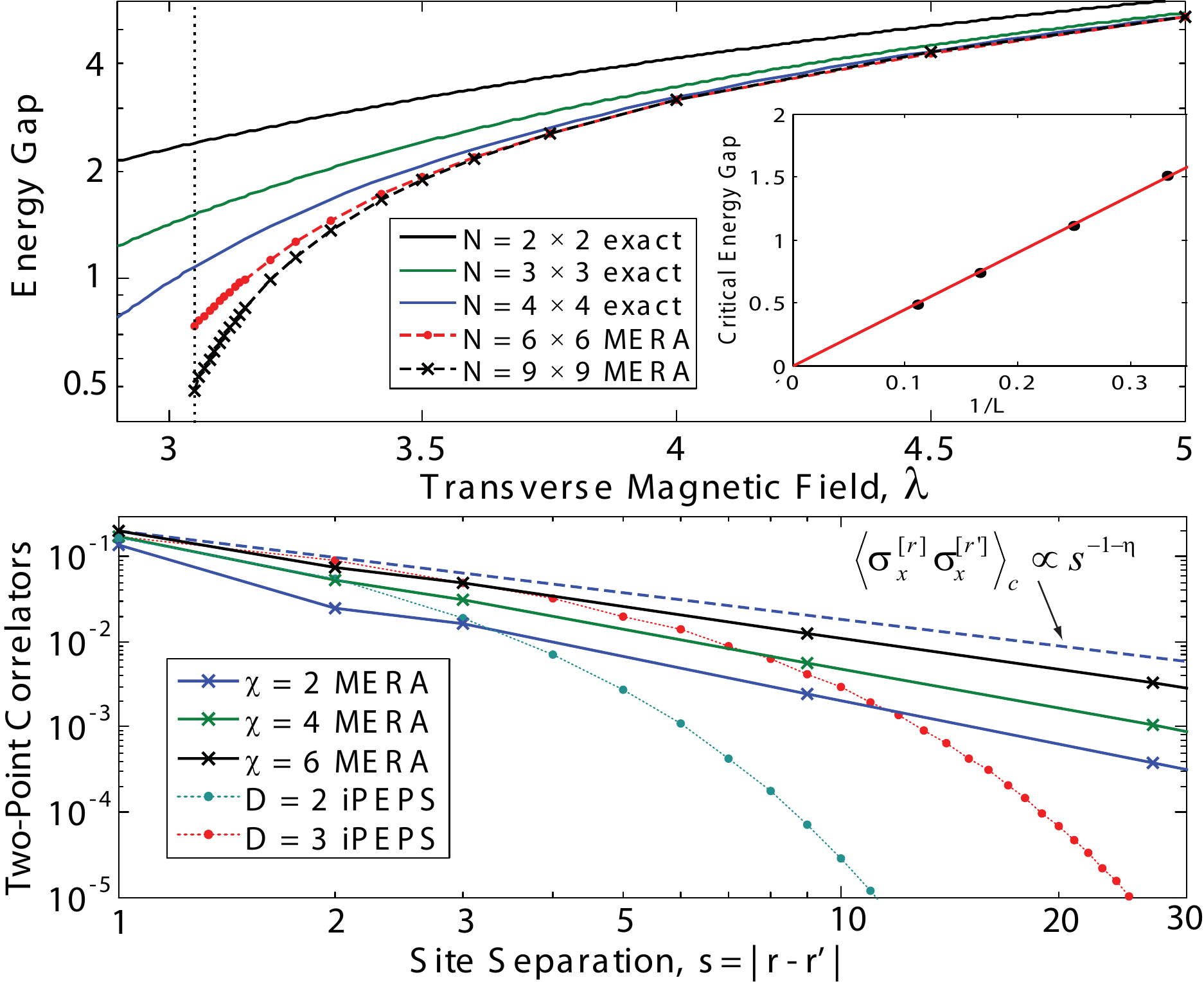}
\caption{ \emph{Top:} The energy gap as a function of the transverse magnetic field $\lambda$, computed by exact diagonalization for small system sizes $L=\left\{2,3,4\right\}$ and with a $\chi=6$ MERA for $L=\left\{6,9\right\}$. The gap scales as $1/L$ at the critical magnetic field. \emph{Bottom:} Two-point correlators $\langle {\sigma_x^{[r]} \sigma_x^{[r']} } \rangle_{c}$ at criticality and for different values of $\chi$. The scale invariant MERA produces correlators that decay polynomially with the distance $s \equiv |r-r'|$. As $\chi$ increases their asymptotic scaling approaches $1/s^{ 1 + \eta }$ with $\eta = 0.03 \pm 0.01$ \cite{pelissetto02}. Correlators have been computed at distances $s=3^k$ for $k=0,1,2,\ldots$, where they can be evaluated with cost $O(\chi^{16})$. For comparison, we have included correlators obtained with a $D=2$ and $D=3$ iPEPS \cite{jordan08}. The latter are very accurate for $s=1,2$ but decay exponentially after a few sites. } \label{fig:2DMera:GapCorr}
\end{center}
\end{figure}

By using the MERA to represent a two-dimensional subspace and minimizing the expectation value of $H_{\mbox{\tiny Ising}}$, we obtain the system's energy gap $\Delta E$. Fig. \ref{fig:2DMera:GapCorr} shows $\Delta E$ as a function of the transverse magnetic field and system size. Notice that at the critical point the gap closes with the system size as $1/L$ (dynamic exponent $z=1$). Two-point correlators can also be extracted. Fig. \ref{fig:2DMera:GapCorr} shows the correlator $\langle \sigma_{x}^{[r]} \sigma_{x}^{[r']} \rangle_{c} \equiv \langle \sigma_{x}^{[r]} \sigma_{x}^{[r']} \rangle - \langle \sigma_{x}^{[r]} \rangle\langle \sigma_{x}^{[r']}\rangle$ along a row or column of the lattice, obtained using the scale invariant algorithm of Chapter \ref{chap:1DCrit}, which directly addresses an infinite lattice at the critical point. 

\begin{figure}[!t]
\begin{center}
\includegraphics[width=10cm]{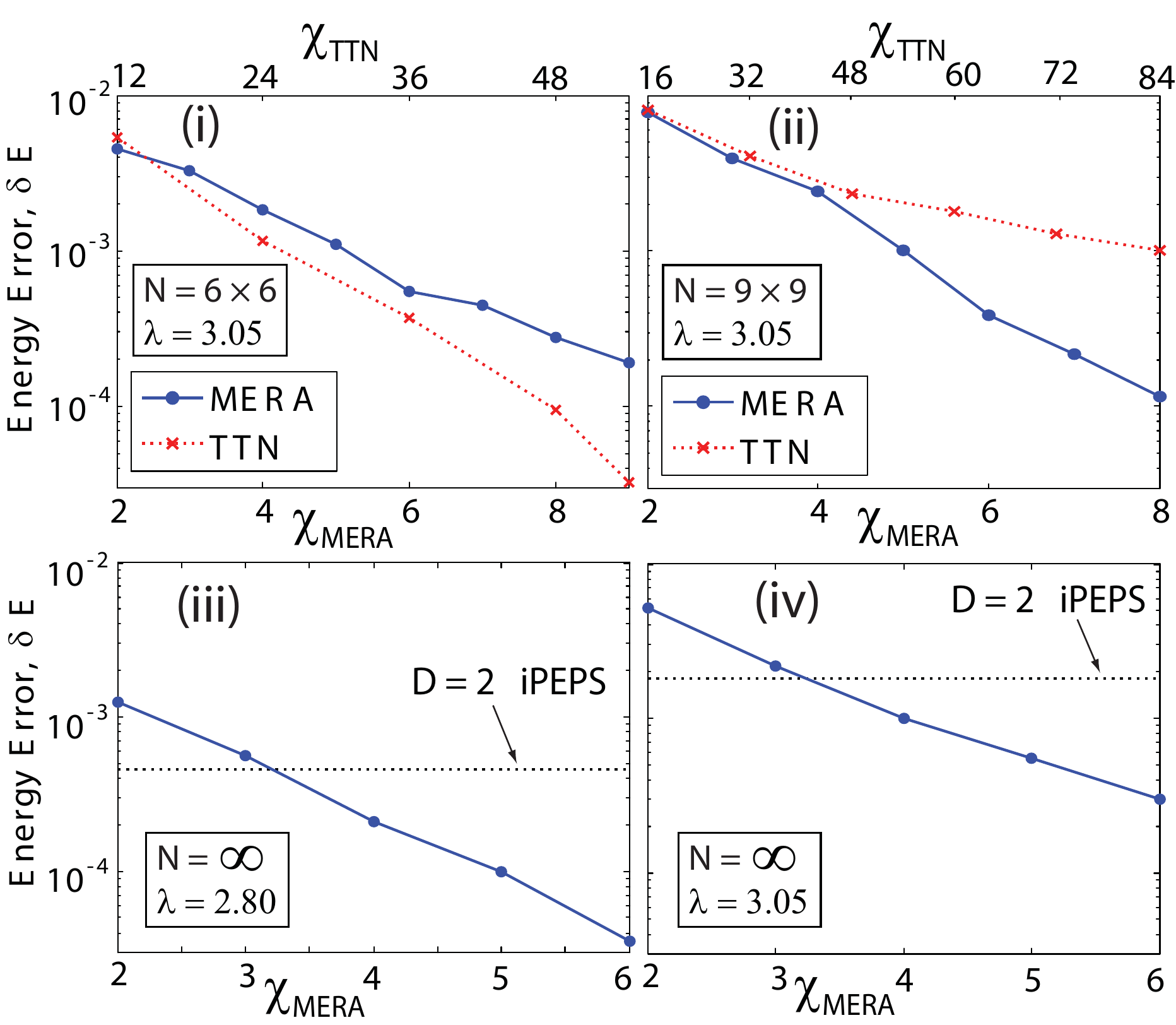}
\caption{ Energy error as a function of the refinement parameter $\chi$ for finite systems of different sizes and for infinite systems. In absence of an exact solution for ground state energies, the errors are defined relative to the results obtained with (i) a $\chi=60$ TTN, (ii) a $\chi=9$ MERA, (iii,iv) a $D=3$ iPEPS \cite{jordan08}. For finite systems (i,ii), the MERA is compared against the TTN. The double $x$-axes for $\chi_{\mbox{\tiny MERA}}$ and $\chi_{\mbox{\tiny TTN}}$ have been adjusted so that they roughly correspond to the same computational cost. For $L=6$ the TTN is more efficient whilst for $L=9$ the MERA already gives significantly better results. Comparison between MERA and iPEPS results for (iii) an infinite system off criticality and (iv) an infinite system at criticality shows very similar accuracy between $\chi=3$ MERA and $D=2$ iPEPS, whereas $D=3$ iPEPS gives a lower (better) energy than $\chi=6$ MERA.} \label{fig:2DMera:2DEnergyErrorNew} 
\end{center}
\end{figure}

\section{Role of disentanglers}
In order to highlight the importance of disentanglers, we have also performed simulations with a tree tensor network (TTN) \cite{tagliacozzo09}. This corresponds to a more orthodox real-space RG approach where the block of $3\times 3$ sites in Fig. \ref{fig:2DMera:Scheme} is directly mapped into an effective site without the use of disentanglers. Recall that a 2$D$ ground state typically displays a boundary law, $S_l \approx l$, for the entanglement entropy $S_l$ of a block of $l\times l$ sites. To reproduce this boundary law with a TTN, one needs to increase the dimension $\chi$ at each step of the coarse-graining. Specifically, $\chi_{\mbox{\tiny{TTN}}}$ must grow doubly exponentially with the linear size $L$ of the lattice. On the other hand, the cost of manipulating a 2$D$ TTN grows only as a small power of $\chi_{\mbox{\tiny{TTN}}}$. As a result, much larger values of $\chi$ can be used with a TTN, leading to a very competitive approach for small lattice sizes \cite{tagliacozzo09}. Fig \ref{fig:2DMera:2DEnergyErrorNew} (i and ii) compares the performance of the MERA and the TTN in lattices of size $6 \times 6$ and $9 \times 9$. It shows that a TTN is more efficient than the MERA in computing the ground state of the $6\times 6$ lattice; however, this trend is already reversed in the  $9\times 9$ lattice, where the cumulative benefit of using disentanglers clearly outweighs the large cost they incur. Disentanglers, by acting on the boundary of a block, readily reproduce the entropic boundary law (for any value of $\chi$) and allow us to consider arbitrarily large systems. Fig. \ref{fig:2DMera:2DEnergyErrorNew} (iii and iv) shows results for an infinite lattice near and at criticality.

To summarize, we have proposed an entanglement renormalization scheme for the square lattice and demonstrated its scalability by addressing the quantum Ising model on systems of linear size $L=\{6,9,18,54\}$, with cost $O(\chi^{16} \log L)$, and on an infinite system at criticality, with cost $O(\chi^{16})$. The key of the present approach is the use of two types of disentanglers that remove short-range entanglement residing near the corners and near the boundaries of the blocks while leading to narrow causal cones of $2\times 2$ sites. Similar schemes can be built for other lattice geometries, as is demonstrated for a Kagome lattice in the next Chapter.

\chapter[The spin-$\frac{1}{2}$ Kagome lattice Heisenberg model]{The ground state of the spin-$\frac{1}{2}$ Kagome lattice Heisenberg model with Entanglement Renormalization}
\label{chap:KagMera}

\section{Introduction}
Low dimensional spin-$\frac{1}{2}$ quantum systems have long been the focus of intense research efforts, largely fueled by the search for exotic states of matter. An important example of a geometrically frustrated quantum antiferromagnet \cite{lhuillier05} is the spin-$\frac{1}{2}$ kagome-lattice Heisenberg model (KLHM). Despite a long history of study, the nature of its ground state remains an open question. Leading proposals include valence bond crystal (VBC) \cite{marston91, syromyatnikov02, nikolic03, budnik04, singh07a, singh08a} and spin liquid (SL)  \cite{sachdev92, leung93, wang06, mila98, hastings00, hermele05, ran07, hermele08, jiang08} ground states. Interest has been further stimulated by recent experimental work on Herbertsmithite ZnCu$_3$(OH)$_6$Cl$_2$, a possible physical realization of the model \cite{shores05}.

Progress in our understanding of the KLHM has been hindered, as with many other models of frustrated antiferromagnets, by the inapplicability of quantum Monte Carlo methods due to the negative sign problem. Nevertheless, systems with up to 36 sites have been addressed with exact diagonalization \cite{leung93,laeuchli09}, whereas the density matrix renormalization group (DMRG) has been used to explore lattices of order $N\approx 100$ sites \cite{jiang08}. Unfortunately, it is very difficult to infer the nature of the ground state of an infinite system from these results. The reason is that these lattices are still relatively small given the 36-site unit cell of the leading VBC proposal, or the algebraic decay of correlations in some SL proposals. In larger systems, support for a SL ground state has also been obtained with a SL ansatz \cite{hermele05, ran07, hermele08}, whereas evidence for a VBC has been obtained for an infinite lattice with a series expansion around an arbitrary dimer covering \cite{singh07a,singh08a}, but both approaches are clearly biased.

\begin{figure}[!htbp]
\begin{center}
\includegraphics[width=12cm]{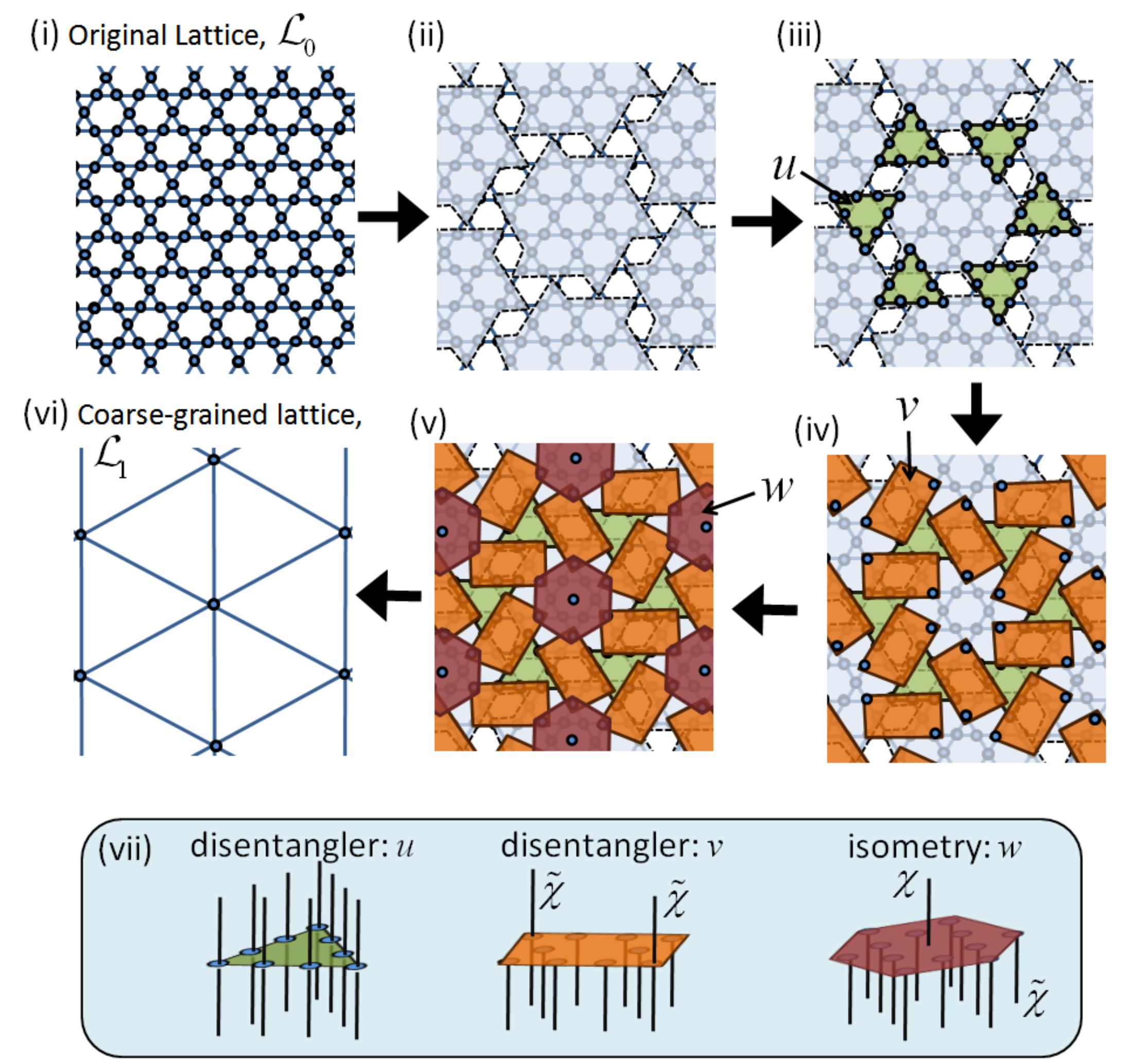}
\caption{ Coarse-graining transformation that maps (i) a kagome lattice $\mathcal L_0$ of $N$ sites into (vi) a coarser lattice $\mathcal L_1$ of $N/36$ sites. (ii) The lattice is first partitioned into blocks of 36 sites. (iii) Disentanglers $u$ are applied across the corners of three blocks, followed by (iv) disentanglers $v$ applied across the sides of two neighboring blocks. (v) Isometries $w$ map blocks to an effective site of the coarse-grained lattice. (vii) Tensors $u,v$ and $w$ have a varying number of incoming and outgoing indices according to Eq. \ref{eq:tensors1}. } \label{fig:KagScheme}
\end{center}
\end{figure}

In this Chapter we report new, independent numerical evidence in favor of a VBC ground state for the KLHM model obtained by a numerical study with entanglement renormalization. Entanglement renormalization is a real space RG approach which, through the proper removal of short-range entanglement, is capable of providing an approximation to ground states of large 2$D$ lattices, as has been demonstrated in Chapters \ref{chap:FreeFerm}, \ref{chap:FreeBoson} and \ref{chap:2DMera}, by means of a multi-scale entanglement renormalization ansatz (MERA). After describing a MERA scheme for the Kagome lattice with periodic boundary conditions, we address lattices of $N=\{36,144,\infty\}$ sites. Our simulations converge to a VBC state compatible with that first proposed by Marston and Zeng \cite{marston91} and revisited by Nikolic and Senthil \cite{nikolic03}, and by Singh and Huse \cite{singh07a, singh08a}. For an infinite lattice we obtain an energy per site E=-0.4322. This energy corresponds to an explicit (MERA) wave-function and therefore provides us with a strict upper bound for the true ground state energy. Importantly, its value is lower than the energy of any existing SL ansatz on a sufficiently large lattice, which we interpret as strong evidence for a VBC ground state in the thermodynamic limit. These results also demonstrate of the utility of entanglement renormalization to study 2$D$ lattice models that are beyond the reach of quantum Monte Carlo techniques.

\section{ER on the kagome lattice}
The present approach is based on the coarse-graining transformation of Fig. \ref{fig:KagScheme}, which is applied to a kagome lattice $\mathcal{L}_0$ made of $N$ sites. It maps blocks of 36 sites of $\mathcal{L}_0$ onto single sites of a coarser lattice $\mathcal{L}_1$ made of $N/36$ sites. A Hamiltonian $H_0$ defined on lattice $\mathcal{L}_0$ becomes an effective Hamiltonian $H_1$ on lattice $\mathcal{L}_1$. Analogously, the ground state $\ket{\Psi_0}$ of $H_0$ is transformed into the ground state $\ket{\Psi_1}$ of $H_1$. The transformation decomposes into three steps. Firstly disentanglers $u$, unitary tensors that act on $9$ sites, are applied across the corners of three neighboring blocks. Then disentanglers $v$ are applied across the sides of two neighboring blocks; these tensors reduce ten sites (each described by a vector space $\mathbb{C}_2$ of dimension $2$) into two effective sites (each described by a vector space $\mathbb{C}_{\tilde{\chi}}$ of dimension $\tilde{\chi}$). Finally isometries $w$ map the remaining sites of each block into a single effective site of $\mathcal L_1$. Thus the tensors $u$, $v$ and $w$,
\begin{eqnarray}
	u^{\dagger}\!&:& {\mathbb{C}_2}^{\otimes 9} \rightarrow {\mathbb{C}_2}^{\otimes 9}, ~~~~~~~~~~~~ u^{\dagger} u = I_{2^9},\nonumber\\
	v^{\dagger}\!&:& {\mathbb{C}_2}^{\otimes 10} \rightarrow {\mathbb{C}_{\tilde{\chi}}}^{\otimes 2}, ~~~~~~~~~~~ v^{\dagger} v = I_{\tilde{\chi}^2},\nonumber\\
  w^{\dagger}\!&:& {\mathbb{C}_2}^{\otimes 6} \otimes {\mathbb{C}_{\tilde{\chi}}}^{\otimes 6} \rightarrow {\mathbb{C}_{\chi}}, ~~~~w^{\dagger} w = I_{\chi}, \label{eq:tensors1}
\end{eqnarray}
transform the ground state $\ket{\Psi_0}$ of lattice $\mathcal{L}_0$ into the ground state $\ket{\Psi_1}$ of lattice $\mathcal{L}_1$ through the sequence 
\begin{equation}
	\ket{\Psi_0} \stackrel{u}{\rightarrow} \ket{\Psi_0'} \stackrel{v}{\rightarrow} \ket{\Psi_0''} \stackrel{w}{\rightarrow} \ket{\Psi_1}.
	\label{eq:Psi}
\end{equation}
The \emph{disentanglers} $u$ and $v$ aim at removing short-range entanglement across the boundaries of the blocks; therefore states $\ket{\Psi_0'}$ and $\ket{\Psi_0''}$ possess decreasing amounts of short range entanglement. If state $\ket{\Psi_0}$ only has short-range entanglement to begin with, then it is conceivable that the state $\ket{\Psi_1}$ has no entanglement left at all. For a finite lattice ($N=144$) we consider a state $\ket{\Psi_0}$ that after the coarse graining transformation give rise to an entangled state $\ket{\Psi_1}$ on $N/36=4$ sites. For an infinite lattice we will instead make an important assumption, namely that $\ket{\Psi_1}$ is a product (non-entangled) state.
How short-ranged must the entanglement in $\ket{\Psi_0}$ be for this assumption to be valid? By reversing the transformation on a product state $\ket{\Psi_1}$, it can be seen that each site in $\ket{\Psi_0}$ is still entangled with \emph{at least} $84$ neighboring sites.

\begin{figure}[!tbhp]
\begin{center}
\includegraphics[width=10cm]{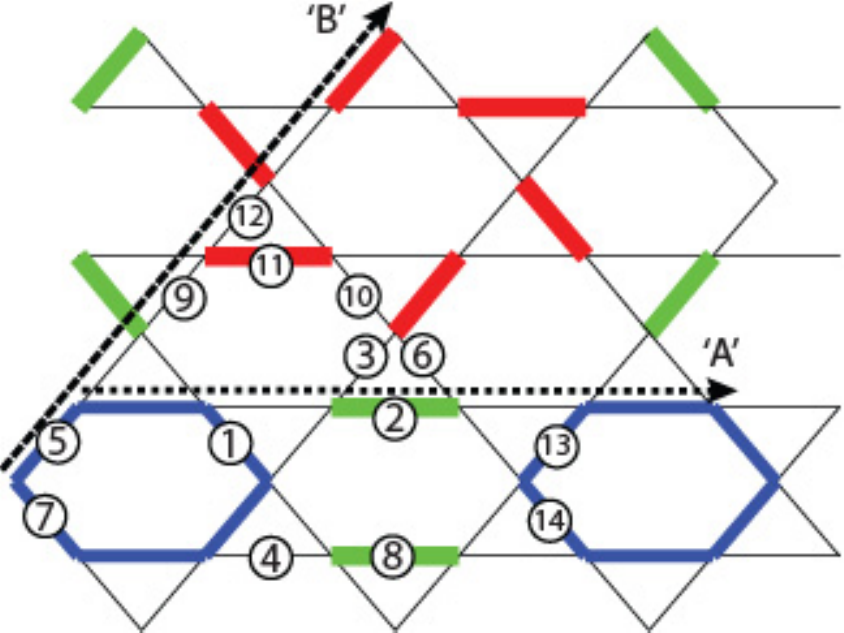}
\caption{ The 36-site unit cell for the honeycomb VBC, strong bonds are drawn with thick lines. Three different types of strong bonds can be identified; the six bonds belonging to the pinwheels (red), six bonds belonging to each `perfect hexagon' (blue) and the parallel bonds between perfect hexagons (green). Dotted arrows indicate the axis where spin-spin correlators have been computed. Bond-bond correlators have been computed between the reference bond (1) and the other numbered bonds.} \label{fig:UnitCell}
\end{center}
\end{figure}

\section{Results and discussion}
The disentanglers and isometries $(u,v,w)$ were initialized randomly and then optimized so as to minimize the expected value of the KLHM Hamiltonian,
\begin{equation}
H_0 = J\sum\limits_{\left\langle {i,j} \right\rangle } {{\rm{S}}_i  \cdot {\rm{S}}_j },
\label{eq:H}
\end{equation}
\begin{table}[!b]
\begin{center}
\caption{Ground state energies as a function of $\tilde\chi$.}
\begin{tabular}{cccc}
\hline\hline
 $\;\;\;\tilde\chi\;\;\;$  &   $\;\;N=\infty\;\;\;\;$   & $N=36$          & $N=36$        \\
         &                & (rand init)   & ($\ket{\textrm{h-VBC}}$ init)\\
   2     &    -0.42145    & -0.42164        & -0.42143      \\
   4     &    -0.42952    & -0.42816        & -0.42715      \\
   8     &    -0.43081    & -0.43199        & -0.43148      \\
   12    &    -0.43114    & -0.43371        & -0.43298      \\
   16    &    -0.43135    & -0.43490        & -0.43420      \\
   20    &    -0.43162    & -0.43611        & -0.43541      \\
   26    &    -0.43193    &                 &               \\
   32    &    -0.43221    &                 &               \\
\hline\hline
\end{tabular}
\label{table:Energy}
\end{center}
\end{table}
by following the algorithms described Chapter \ref{chap:MERAalg}, with cost $O(2^{12}\tilde \chi^6 \chi^2)$ \footnote[1]{The computational cost of simulations can be signifcantly reduced by incorporating symmetries of the system into the MERA tensor network \cite{sukhi09}. In this work, a U(1) symmetry is incorporated into the MERA to allow large $\tilde \chi$ simulations that would otherwise be unaffordable.}. Specifically, for lattices with $N=36$ and $144$ sites, the resulting (one-site and four-site) Hamiltonian $H_1$ is diagonalized exactly. Instead, for $N=\infty$, we use the finite correlation range algorithm (Sect. \ref{sect:MERAalg:constant}). All computations led to highly dimerized wave-functions of the VBC type. In order to explain the results, consider the \emph{exact} `honeycomb' VBC state, denoted $\ket{\textrm{h-VBC}}$, whose 36-site unit cell is shown in Fig. \ref{fig:UnitCell}. Each unit cell contains two `perfect hexagons' (resonating bonds around a hexagon) and a `pinwheel'. Three different types of strong bonds can be identified: those of the pinwheels (red), parallel bonds (green) and perfect hexagons (blue). The pinwheel and parallel bonds are singlets (energy per bond $=-0.75$) while the perfect hexagons are in the ground state of a periodic Heisenberg chain of 6 sites (energy per bond $=-0.4671$). The rest of links have zero energy. We call a `honeycomb' VBC a state that has strong bonds according to the above pattern, even though the rest of bonds (weak bonds) need not have zero energy. The `honeycomb' VBC was originally proposed by Marston and Zeng \cite{marston91} (see also \cite{nikolic03,singh07a,singh08a}). Our simulations with $N=144$ and $N=\infty$ produce a VBC of this type as the best MERA approximation to the ground state.

The energies obtained for an infinite lattice are shown in Table \ref{table:Energy}. For each value of $\tilde{\chi}$, the MERA is an explicit wave-function and therefore provides an upper bound to the exact ground state energy. Energies computed for the $N=144$ lattice matched those of the infinite lattice to within $0.02\%$ and have been omitted. These $N=\infty$ energies also match closely those obtained by series expansion in Ref. \cite{singh07a}, and are lower than those obtained in Ref. \cite{jiang08} with DMRG ($E = -0.43160$ for $N=108$) and in Ref. \cite{ran07} with fermonic mean-field theory and Gutzwiller projection ($E = -0.42863$ for $N=432$). We further notice that, where finite size effects are still relevant, such as in the $N=108$ case, they tend to decrease the ground state energy.

\begin{figure}[!tbhp]
\begin{center}
\includegraphics[width=10cm]{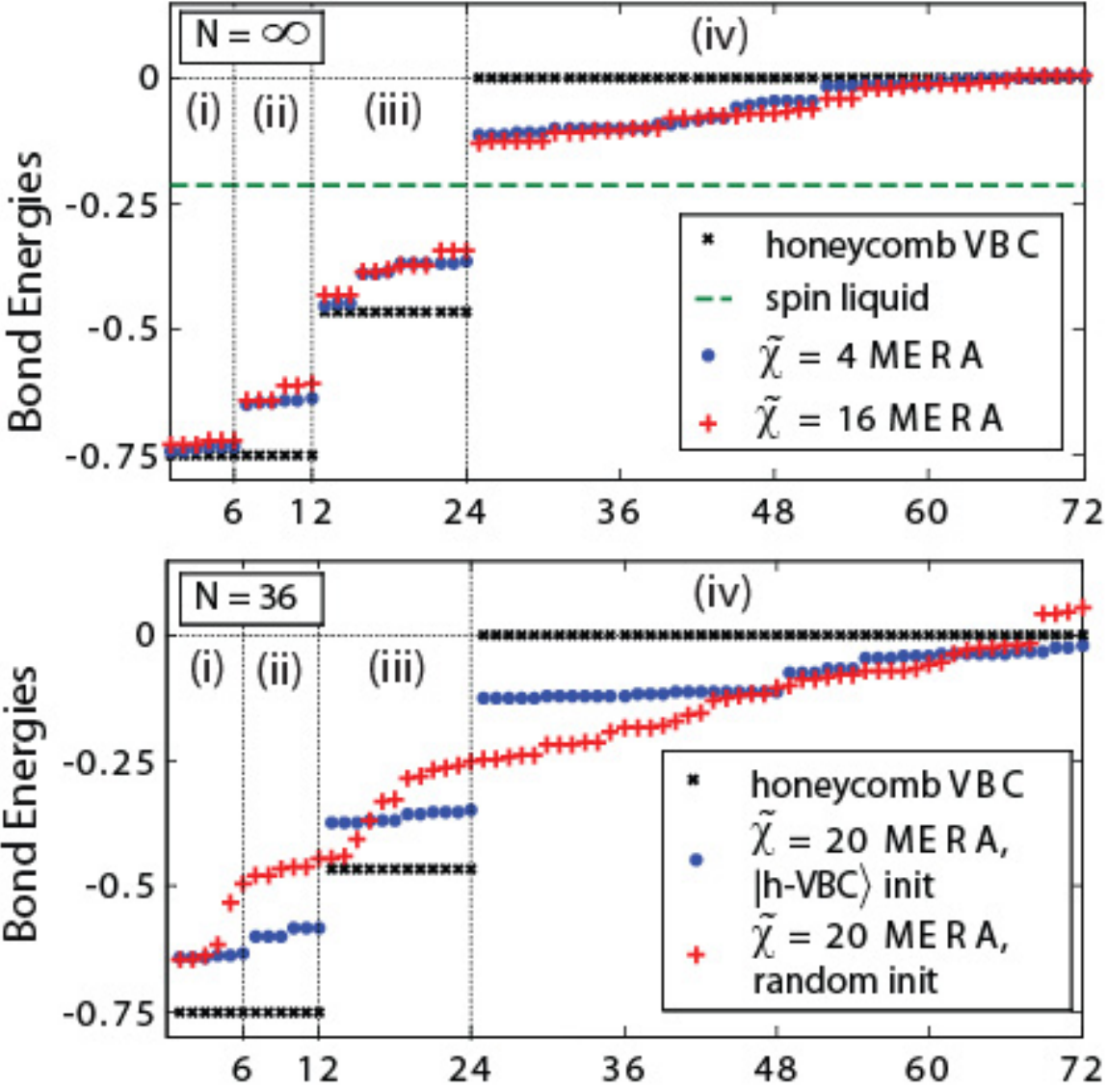}
\caption{(top) Bond energies for the 36-site unit cell of infinite MERA wave-functions, for two different values of $\tilde\chi$, as compared to those of an exact honeycomb VBC, $\ket{\textrm{h-VBC}}$, and those of a spin liquid, which by definition has all equal strength bonds. The MERA wave-functions clearly match the proposed honeycomb VBC; we identify (i) the six strong `pinwheel' bonds (red bonds), the six `parallel' bonds (green bonds) and (iii) the 12 `perfect hexagon' bonds (blue bonds). The (iv) remaining 48 bonds are the weak bonds of the unit cell. (bottom) Bond energies for the 36-site lattice. Here a randomly initialized MERA converges to a dimerized state that does not match the honeycomb VBC pattern, but gives lower overall energy than a honeycomb VBC initialized MERA of the same $\tilde \chi$.} \label{fig:BondEnergy}
\end{center}
\end{figure}

Fig. \ref{fig:BondEnergy} shows the distribution of bond energies obtained for the $N=\infty$ lattice. With $\tilde{\chi}=4$, one observes an energy increase per site over $\ket{\textrm{h-VBC}}$ of $\approx 0.08$ in the parallel (green) bonds and also in some of the hexagon (blue) bonds, with the weak bonds having lower energy in return. As $\tilde{\chi}$ is increased, the energy of the `strong' bonds becomes slightly larger and that of the `weak' bonds continues to decrease. However, the dimerization clearly survives: the bond energies are not seen to converge to a uniform distribution as required for a SL. 

\begin{figure}[!tbhp]
\begin{center}
\includegraphics[width=10cm]{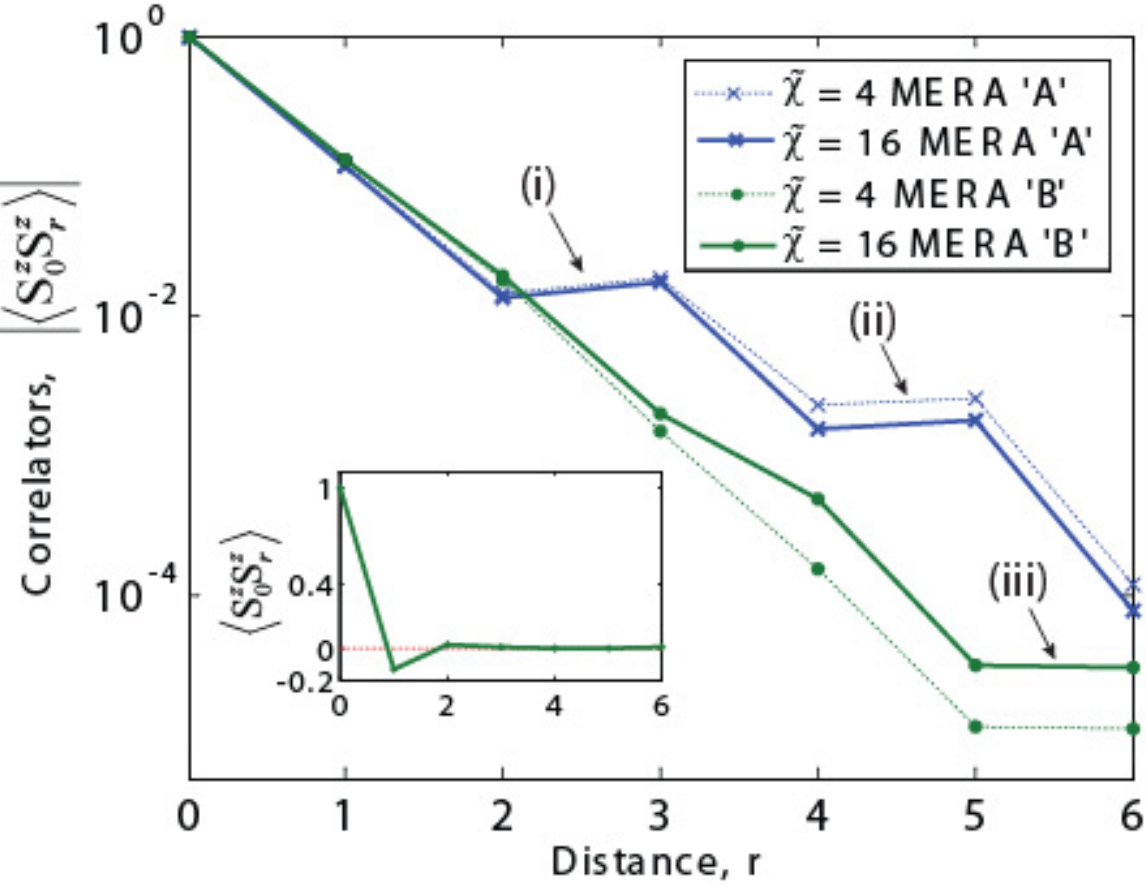}
\caption{Spin-Spin correlators along arrows `A' and `B' of Fig. \ref{fig:UnitCell} for infinite lattice MERA of $\tilde\chi=4$ and $\tilde\chi=16$. Although along both lattice directions considered the correlators decay exponentially, the decay along arrow `A' (a line joining two perfect hexagons) is seen to be slower than along arrow `B' (a line joining perfect hexagon to pinwheel). The plateaus marked (i), (ii) and (iii) show the correlation is the same with both spins of a strong bond.  } \label{fig:SpinSpinCorr}
\end{center}
\end{figure}

Fig. \ref{fig:SpinSpinCorr} shows spin-spin correlators evaluated along two different lattice axis $A$ and $B$ (cf. Fig. \ref{fig:UnitCell}) for $N=\infty$. These correlators decay exponentially with well defined `plateaus', where the correlation is the same with both spins of a strong bond. Correlations along the line joining a perfect hexagon and a pinwheel are seen to decay faster than along the line joining two perfect hexagons, consistent with the observation from Fig. \ref{fig:BondEnergy} that the perfect hexagon bonds remain almost exact singlets even for high values of $\tilde \chi$. Table \ref{table:BondCorr} shows bond-bond connected and disconnected correlators, $C_{1,\alpha}$ and $D_{1,\alpha }$, 
\begin{eqnarray}
C_{1,\alpha }  &\equiv& \left\langle {\left( {\vec S \cdot \vec S} \right)_1 \left( {\vec S \cdot \vec S} \right)_\alpha  } \right\rangle \\
D_{1,\alpha }  &\equiv&  C_{1,\alpha }  - \left\langle {\left( {\vec S \cdot \vec S} \right)_1 } \right\rangle \left\langle {\left( {\vec S \cdot \vec S} \right)_\alpha  } \right\rangle, \label{eq:Corr}
\end{eqnarray}
between a reference bond `1' and a surrounding bond $\alpha=1,\cdots, 14$ (cf. Fig.\ref {fig:UnitCell}). While disconnected correlators decay exponentially with distance, some connected correlators remain significant at arbitrary distances, demonstrating the long-range order of the VBC state. 

\begin{table}[!thb]
\begin{center}
\caption{Bond-Bond correlators for  $\tilde\chi = 16$ ($N=\infty$).}
\begin{tabular}{cccc}
\hline\hline
Bond    & $C_{1,\alpha}$ & $D_{1,\alpha}$ & $\left\langle {\left( {\vec S \cdot \vec S} \right)_\alpha  } \right\rangle$ \\ 
   1      &  0.38639  &   0.22815         & -0.39780       \\
   2      &  0.29490  &   0.03496         & -0.65342       \\
   3      &  0.00422  &  -0.00115         & -0.01349       \\
   4      &  0.09503  &   0.04644         & -0.12211       \\
   5      &  0.22488  &   0.06609         & -0.39918       \\
   6      & -0.00609  &  -0.01426         & -0.02054       \\
   7      & -0.10372  &  -0.03376         & -0.34561       \\
   8      &  0.26083  &   0.01252         & -0.62421       \\
   9      &  0.00336  &   0.00368         &  0.00082       \\
   10     & -0.00001  &  -0.00017         & -0.00042       \\
   11     &  0.29113  &   0.00196         & -0.72693       \\
   12     &  0.00090  &  -0.00036         & -0.00321       \\
   13     &  0.16632  &   0.00026         & -0.41744       \\
   14     &  0.15674  &  -0.00037         & -0.39496       \\
\hline

\hline\hline
\end{tabular}
\label{table:BondCorr}
\end{center}
\end{table}

Let us discuss the results for a lattice with $N=36$ sites. When initialized with random tensors, the MERA produced VBC type configurations which typically did not match the honeycomb VBC, although simulations initialized in the state $\ket{\textrm{h-VBC}}$ retained the honeycomb VBC pattern (see Fig. \ref{fig:BondEnergy}). Here the randomly initialized VBC produced a lower energy ($0.5\%$ above the exact diagonalization result $E=-0.438377$ of \cite{leung93}) than the honeycomb VBC type solution for an equivalent value of $\tilde{\chi}$ (cf. Table \ref{table:Energy}). These results strongly suggest that finite size effects in the $N=36$ site lattice lead to a significant departure from the physics of the infinite system. 

\section{Defective valence bond crystals}
\begin{figure}[!tbhp]
\begin{center}
\includegraphics[width=14cm]{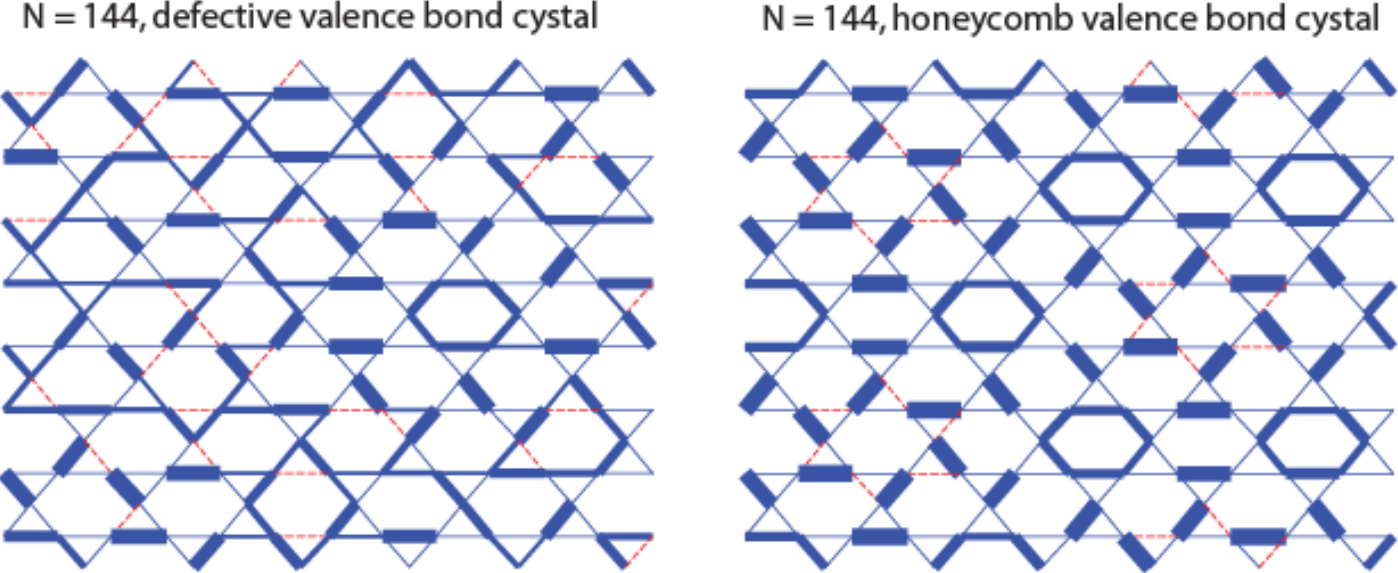}
\caption{ Converged $\tilde\chi=8$ MERA solutions for the $N=144$ site KLHM. The thickness of a line between two lattice sites is proportional to the absolute bond energy, $|E|$. Blue solid lines are used for negative energy bonds, $E<0$, and red dashed lines for positive energy bonds, $E>0$. (Top) An example of a `defective' VBC. Local structures that match the `pinwheels' and `perfect hexagons' of the honeycomb VBC are observed, but long range order is absent. (Bottom) By restricting the rate of lattice `cooling', as per Eq.\ref{eq:cooling}, one more reliably obtains the honeycomb VBC from a randomly initialized MERA. On average a defective VBC had $0.05\%$ greater energy than the honeycomb VBC of the same value of $\chi$. } \label{fig:BondPic}
\end{center}
\end{figure}

In this section we detail a modification of the MERA optimization algorithm of Sect. \ref{sect:MERAalg:algorithm} that was found necessary to ensure proper convergence of some simulations of the KLHM. All simulations of the KLHM we found to converge to highly dimerized VBC type states. For lattices of $N=\{144,\infty\}$ sites the lowest energy state that was found, or best approximation to the ground-state, had a pattern of strong bonds matching the `honeycomb' VBC state. However, not all randomly initialized simulations converged to this state; in some instances the MERA wave-function converged to a `defective' VBC state, an example of which is shown Fig. \ref{fig:BondPic}. The observed defective VBC's ranged from states which matched the honeycomb VBC with only a few misplaced strong bonds to states with an almost completely disordered placement of strong bonds. A defective VBC was observed to have on average $0.05\%$ higher energy per site than the corresponding honeycomb VBC type MERA of the same $\tilde \chi$. The exact difference in energy depended on the amount of defects; typically, states that deviated more from the honeycomb VBC had higher energy than states which deviated less. In the instances that the method did not converge to the honeycomb VBC one may conclude that the optimization of the MERA became trapped in a local minimum.

Becoming trapped in a local minimum is not surprising given that the optimization of the MERA is based upon updates of individual tensors. During the initial iterations of the optimization, locally stable structures, such as singlets between neighboring spins, form. But they do not necessarily form in places compatible with a global minimization of the energy. Once the method has converged to a particular VBC state, the transformation required to bring the state to a different VBC, with a new pattern of strong bonds, requires simultaneous shifts of many strong bonds. This is unlikely to occur with an optimization based on energy-lowering updates of individual tensors.

There are many ways to decrease the risk of becoming trapped in a local minimum due to the formation of local singlets. In the present work this is achieved by restricting how much a tensor is allowed to change in one single update; this may be thought of as decreasing the rate at which the MERA is `cooled-down' from an initial (high-energy) state to the (low-energy) approximation to the ground-state. The goal of this restriction is to prevent locally stable structures from forming too early in the optimization. 

In order to discuss how this restriction may be implemented, we first refresh some details of the variational MERA optimization, see Sect. \ref{sect:MERAalg:algorithm}. In the standard algorithm, a tensor $w$ in the MERA is replaced with a better (that is leading to lower energy) tensor $w'$ as follows. First one computes the linearized environment of $w$, denoted $\Upsilon _w$. Then, given the singular value decomposition of the environment, $\Upsilon _w = U S V^\dag$, one chooses the updated tensor to be $w'=V U^\dag$. Here we wish to restrict the amount $\varepsilon$ the tensor can change, $\left\| {w - w'} \right\| < \varepsilon$. This is achieved by using a modified environment $\tilde \Upsilon _w$ defined as 
\begin{equation}
\tilde \Upsilon _w  \equiv \Upsilon _w  + \lambda w 
\label{eq:cooling}
\end{equation}
for some $\lambda>0$. As before, given the singular value decomposition of the modified environment $\tilde \Upsilon _w = U S V^\dag$, the updated tensor $w'$ is chosen to be $w'=V U^\dag$. The parameter $\lambda$ acts as a soft constraint to ensure the tensor change $\varepsilon$ is kept small with each update, a larger value of $\lambda$ giving a smaller change $\varepsilon$. In the limit of $\lambda\rightarrow\infty$, the tensor $w$ would remain constant, $\varepsilon=0$, in the update. Simulations of the KLHM with a suitable value of $\lambda$, say $\lambda>10$ for the initial optimization, were much more likely to converge to a defect free honeycomb VBC than were unrestrained ($\lambda=0$) optimizations.

\section{Conclusions}
To summarize, we have used entanglement renormalization techniques to obtain new numerical evidence indicating that the ground state of the KLHM is of the honeycomb VBC type. In order to assess the robustness of this result, we briefly discuss some of the limitations of the present approach.

Firstly, the coarse-graining transformation of Fig. \ref{fig:KagScheme}, which maps 36 sites into one site, was designed to ensure compatibility with the 36-site unit cell of honeycomb VBC type solutions. While our approach did not preclude solutions with a smaller, compatible unit cell (such as a 12-site unit cell or a fully translation invariant solution), we cannot rule out the possibility that a state with an incompatible unit cell might have a lower energy. 

Secondly, the infinite lattice was investigated by restricting the range of entanglement in the ansatz to blocks of 84 spins, imposed through an unentangled state $\ket{\Psi_1}$ in Eq. \ref{eq:Psi}. This restriction was only implemented after preliminary simulations with $\tilde{\chi}=12$ had produced identical energies irrespectively of whether $\ket{\Psi_1}$ was allowed to be entangled. However, it could still be that entanglement in $\ket{\Psi_1}$ would make a big difference for larger values of $\tilde{\chi}$. We find this scenario quite unlikely, but could not test it due to computational limitations.

Finally, the MERA is an essentially unbiased method provided that the candidates to be the ground state of the system have all a relatively small amount of entanglement. But when deciding between a VBC (which mostly has short-range entanglement) and e.g. the algebraic SL of Refs. \cite{ran07,hermele08} (significantly more entangled at all length scales), it might well be that the MERA is biased toward the low entanglement solution. Therefore our results do not conclusively exclude a SL ground state. We emphasized, however, that the ground state energies obtained with the MERA are lower than the SL energies of Refs. \cite{ran07, hermele08, jiang08}.

Future work includes the computation of singlet and triplet excitation gaps, and studying the effect of adding a magnetic field and anisotropic terms to the Hamiltonian. Such additions to the Hamiltonian can be considered without modifying the present approach.

\chapter{Conclusion and Future Outlook}
\label{chap:Conclusion}

\section{Summary}
The content and major results of this thesis are summarised below. Firstly, Chapter \ref{chap:MERAintro} reviewed the foundations of entanglement renormalization and the MERA as first introduced in Refs. \cite{vidal07,vidal08}. Chapters \ref{chap:FreeFerm} and \ref{chap:FreeBoson} then explored the performance of entanglement renormalization in the simple setting of free fermions and free bosons respectivly. Here it was demonstrated that a real-space renormalization group transformation based upon entanglement renormalization could induce a sustainable coarse-graining transformation in variety of $D=1,2$ dimensional free fermion and free boson models. The substainability of the RG transformation with entanglement renormalization allowed investigation of arbitrarily large $D=1,2$ dimensional lattice systems without loss of accuracy, as was shown through comparison with exact solutions. An exception was found for critical $2D$ free-fermion models with a $1D$ fermi surface, whose entropy is known to scale as $S_L \sim L\textrm{log}(L)$, where ER could no longer produce a be sustainable RG transformation. It was concluded that a generalized MERA would be required to analyse such systems.

Chapter \ref{chap:MERAalg} described how to compute expected values of observables from a MERA and also presented optimization algorithms to compute the ground-state MERA for an arbitrary local Hamiltonian. A highlight of algorithms was their computational cost which, for a chain of $N$ sites, scales as $O(N)$ for non-translation invariant systems, $O(\textrm{log} (N))$ for translation invariant systems and scales independant of $N$ for scale-invariant systems. This Chapter also presented benchmark results for several $1D$ spin models, mainly for Ising and Potts models, including accurate computation of ground energy, local observables, correlators and energy gaps. Chapter \ref{chap:1DCrit} analysed the connection between entanglement renormalization and conformal field theory (CFT). It was demonstrated explicitly for several 1$D$ spin models that this connection could be exploited to extract most of the conformal data of the CFT that described the models in the continuum limit. This included accurate calculation of the scaling dimensions, the central charge and OPE coefficients for Ising and Potts models.

In Chapter \ref{chap:2DMera} the optimisation algorithms presented in Chapter \ref{chap:MERAalg} were generalised to the specific problem of analysing quantum many-body problems on $D=2$ dimensional lattices. Benchmark calculations with the quantum Ising model demonstrated the ability of the MERA to give accurate ground state properties on arbitrarily large (or infinite) lattice systems for a non-integrable model in $D=2$ spatial dimensions. The benchmark results included accurate calculations of critical exponents, expected values of local observables, two-points correlators and energy gaps. Finally, Chapter \ref{chap:KagMera} applied ER to study the spin-$\frac{1}{2}$ kagome lattice Heisenberg model (KLHM), a long-standing problem in condensed matter physics. The results of this Chapter, which is the first demonstration of ER to study an open problem in condensed matter physics and a problem beyond the reach of quantum Monte-Carlo, offers convincing evidence that the ground-state of the infinite KLHM is a honeycomb valence bond crystal. In particular, the energy of the MERA approximation to the ground-state of the infinite system, which serves as an exact upper bound to the true ground energy, is significantly better than any such bound previously produced.  

\section{Future Work and Outlook}

Entanglement renormalization offers a non-perturbative means to investigate quantum many-body systems, as such ER has a diverse range of applications that remain to been explored or further developed. 

Firstly one could consider extensions to the research of Chapter \ref{chap:1DCrit} (see also \cite{giovannetti08,pfeifer08,montangero09}) which made a connection between the MERA and conformal field theory (CFT). The procedure to compute the local scaling operators (and their corresponding scaling dimensions) of a critical theory, as described Chapter \ref{chap:1DCrit}, could be generalised to obtain a class of non-local scaling operators. Thus it might be possible to obtain a complete set of the scaling operators which characterise a CFT. It also remains to generalise the ER algorithm to the study of types of conformally invariant systems other than the 1$D$ infinite chains previously considered. These may include systems with a boundary, homogeneous systems with a defect (or several defects), an interface between two different critical systems, or a Y-intersection of three spin chains. 
  
There is much potential for the use of ER to study 2$D$ quantum systems. In particular ER may have a big impact in the areas of frustated magnets, systems of interacting fermions and topologically ordered systems, as these are problems for which previously there has often not been any satisfactory investigative techniques. The results in Chapter \ref{chap:KagMera} for the kagome lattice Heisenberg model are a demonstration of the potential of the method for frustrated magnets. Entanglement renormalization could be similarly applied to investigate many other models of magnetism that remain not well understood; these may include $J1-J2$ models, ring exchange models and quantum dimer models. In particular, it is possible that ER could be used to study the existance of a spin-liquid phase in a 2$D$ magnetic system, a problem that is of much interest to the condensed matter community. Recent developments have led to a generalisation of the ER optimisation algorithm to allow the investigation of systems of interacting fermions \cite{corboz09a,corboz09b,pineda09,barthel09}. At present most of the studies of fermions with ER have been tests to benchmark the algorithm, and there remains a wide range of interesting fermionic models that could potentially be investigated with ER. In regards to topological order, previous studies also have shown that ER can be used to characterise the infra-red fixed point of topologically ordered systems \cite{aguado08,konig09}. Further research could utilise ER to investigate, for instance, the stability of a topological phase under perturbation.

Prior to the further application of ER to study 2$D$ lattice systems it may be necessary to devise ways to improve the computational efficiency of the algorithm, which typically scales as a large power of the bond dimension $\chi$. Significant reductions in computational cost can come from incorporation of global or local symmetries into the tensor network; research into the incorporation of global symmetries is already well progressed \cite{sukhi09}. It may also be possible to reduce costs by introducing approximations into contraction of the MERA tensor-network or to incorporate monte-carlo type sampling into the algorithm.

In the long term, it is possible that entanglement renormalization, as a tool for analysing many-body systems, may find applications beyond the more immediate applications in condensed matter physics. Applications could arise wherever quantum many-body effects are important; for instance in areas such as string-theory, quantum gravity and quantum chromodynamics. Entanglement renormalization allows the evaluation of the continuum limit of a lattice model, hence may be useful for analysing problems that would be difficult to analyze in the continuum directly. Recent work on the holographic principle \cite{swingle09} might be a first step in this direction.


\appendix
\chapter{The ground-state of free fermion models in two dimensions}
\label{chap:FermGround}

\section{Particle conserving free-fermions}
In this appendix we derive an expression for the ground-state correlation matrix of free-fermion models on a 2$D$ square lattice (a thorough derivation of ground-state correlation matrices for 1$D$ free-fermion models can be found e.g. \cite{latorre04}). The ground-state covariance matrix is the starting point for the analysis of free fermions with entanglement renormalization considered in Chapter \ref{chap:FreeFerm}. We begin with by considering nearest neighbour free-fermion models that have only particle conserving terms on a 2$D$ lattice of $N\times N$ sites
\begin{align}
\hat H = &\sum\limits_{r_1,r_2 =  - (N - 1)/2}^{(N - 1)/2} {\frac{{1 }}{2}\left( {a_{r_1 + 1,r_2}^\dag  a_{r_1,r_2}  + a_{r_1,r_2}^\dag  a_{r_1 + 1,r_2} } \right)}\nonumber\\ 
&+\frac{{1 }}{2}\left( {a_{r_1,r_2 + 1}^\dag  a_{r_1,r_2}  + a_{r_1,r_2}^\dag  a_{r_1,r_2 + 1} } \right) - \lambda a_{r_1,r_2}^\dag  a_{r_1,r_2}\label{a1e1}.
\end{align}
We now perform a 2$D$ Fourier transform of the fermionic operators $a$ into a new set of operators $b$ defined by  
\begin{align}
 b_{k_1 ,k_2}  &= \frac{1}{N}\sum\limits_{r_1,r_2 = -(N-1)/2}^{(N - 1)/2} {a_{r_1,r_2} e^{ - 2\pi i r_1 k_1 /N} e^{ - 2\pi i r_2 k_2 /N} }  \nonumber\\ 
 b_{k_1 ,k_2}^\dag   &= \frac{1}{N}\sum\limits_{r_1,r_2 = -(N-1)/2}^{(N - 1)/2} {a_{r_1,r_2}^\dag  e^{2\pi i r_1 k_1 /N} e^{2\pi i r_2 k_2 /N} }\label{a1e2}.   
\end{align}
The inverse transformation is similarly defined
\begin{align}
 a_{r_1,r_2}  &= \frac{1}{N}\sum\limits_{k_1,k_2 = -(N-1)/2}^{(N - 1)/2} {b_{k_1 ,k_2} e^{2\pi ik_1 r_1/N} e^{2\pi ik_2 r_2/N} }  \nonumber\\ 
 a_{r_1,r_2}^\dag   &= \frac{1}{N}\sum\limits_{k_1,k_2 = -(N-1)/2}^{(N - 1)/2} {b_{k_1 ,k_2}^\dag  e^{ - 2\pi ik_1 r_1 /N} e^{ - 2\pi ik_2 r_2/N} }.\label{a1e3}  
\end{align}
By exploiting the orthogonality properties of the (unitary) fourier transform, it is seen that the hopping terms in Eq. \ref{a1e1} become on-site interaction terms in the fourier basis
\begin{align}
 \sum\limits_{r_1,r_2} {a_{r_1 + 1,r_2}^\dag  a_{r_1,r_2} }  &= \sum\limits_{k_1,k_2} {e^{ - 2\pi i{k_1}/N} b_{k_1,k_2}^\dag  b_{k_1,k_2} }  \nonumber\\ 
 \sum\limits_{r_1,r_2} {a_{r_1,r_2 + 1}^\dag  a_{r_1,r_2} }  &= \sum\limits_{k_1,k_2} {e^{ - 2\pi i{k_2}/N} b_{k_1,k_2}^\dag  b_{k_1,k_2} }.\label{a1e4}   
\end{align}
The on-site interaction term in Eq. \ref{a1e1}, which may be interpreted as a chemical potential, remains invariant under the fourier transform
\begin{equation}
\sum\limits_{r_1,r_2} {a_{r_1,r_2}^\dag  a_{r_1,r_2} }  = \sum\limits_{k_1,k_2} {b_{k_1,k_2}^\dag  b_{k_1,k_2} }. \label{a1e5}
\end{equation}
On subsitution of Eqns. \ref{a1e4} and \ref{a1e5}, the original Hamiltonian of Eq. \ref{a1e1} is recast into diagonal form
\begin{equation}
\hat H = \sum\limits_{k_1,k_2 =  - (N - 1)/2}^{(N - 1)/2} {\Lambda _{k_1,k_2} b_{k_1,k_2}^\dag  } b_{k_1,k_2} \label{a1e6} 
\end{equation}
with disperion relation $\Lambda$ given as
\begin{equation}
\Lambda _{k_1,k_2}  =  - \cos (2\pi k_1/N) - \cos (2\pi k_2/N) - \lambda. \label{a1e7}
\end{equation}
The ground-state of Hamiltonian \ref{a1e6} is defined as the state with all negative energy modes, $\Lambda_{k_1,k_2}<0$, occupied (the fermi sea) and the rest of the fermionic modes empty. It follows that the ground-state correlation matrix, in the fourier basis, is given by
\begin{equation}
\left\langle {b_{k_1,k_2}^\dag  b_{k_1',k_2'} } \right\rangle _\textrm{GS}  = \left\{ {\begin{array}{*{20}c}
   {\delta _{k_1,k_1'} \delta _{k_2,k_2'} } & {\Lambda _{k_1,k_2}  < 0}  \\
   0 & {{\rm{else}}}  \\
\end{array}} \right. .\label{a1e8}
\end{equation}
The inverse transformation relations of Eq.\ref{a1e3} can be used to transform the correlation matrix into the original (spatial) basis
\begin{align}
 \left\langle {a_{0,0}^\dag  a_{r_1,r_2}} \right\rangle _{\textrm{GS}}  &= \frac{1}{{N^2 }}\sum\limits_{k_1,k_2,k_1',k_2'} \left( \left\langle {b_{k_1,k_2}^\dag  b_{k_1',k_2'} } \right\rangle _\textrm{GS} e^{  2\pi ik_1'r_1/N} e^{  2\pi ik_2'r_2/N}\right)   \nonumber\\ 
  &= \frac{1}{{N^2 }}\sum\limits_{k_1,k_2 \in \mathcal{F}} {e^{  2\pi ik_1 r_1/N} e^{  2\pi ik_2 r_2/N} }\label{a1e9}   
\end{align}
where $\mathcal{F}$ denotes the \emph{fermi sea}. In order to take the thermodynamic limit of Eq.\ref{a1e9}, we first define new variables
\begin{equation}
\phi_1  = \frac{{2\pi k_1}}{N},\qquad \phi_2  = \frac{{2\pi k_2}}{N} \label{a1e10}
\end{equation}
where, by definition, $\phi_1, \phi_2 \in [-\pi,\pi]$. The thermodynamic limit, $N\rightarrow \infty$, is now taken 
\begin{equation}
\left\langle {a_{0,0}^\dag  a_{r_1,r_2} } \right\rangle _{\rm{0}}  = \frac{1}{{4\pi ^2 }}\int\limits_\mathcal{F}  {e^{ir_1\phi_1 } e^{ir_2\phi_2 } d\phi_1 d\phi_2 } \label{a1e11}
\end{equation}
where the integral is taken over those modes within the fermi sea $\mathcal{F}$. Explicitly, the integral should be evaluated over the region $\Lambda \ge 0$, where $\Lambda _{\phi_1,\phi_2}  =  - \cos (\phi_1) - \cos (\phi_2) - \lambda$. In most cases the intergal in Eq.\ref{a1e11} cannot be evaluated analytically and one must resort to numerical integration to obtain approximate values for the correlators. However, for the case of half-filling ($\lambda =0$), the fermi-sea has a particularly simple form and the integral may be carried out analytically as follows. Firstly, a change of variables is made
\begin{equation}
\varphi_1  = \phi_1  + \phi_2 ,\quad \varphi_2  = \phi_1  - \phi_2. \label{a1e12}
\end{equation}
Expressed in the new variables, the integral of Eq.\ref{a1e11} can be evaluated
\begin{align}
 \left\langle {a_{0,0}^\dag  a_{r_1,r_2} } \right\rangle _{\textrm{GS}}  &= \frac{1}{{8\pi ^2 }}\int\limits_{\varphi_1  =  - \pi }^\pi  {\int\limits_{\varphi_2  =  - \pi }^\pi  {e^{ir_1(\varphi_1  + \varphi_2 )/2} e^{ir_2(\varphi_1  - \varphi_2 )/2} d\varphi_2 d\varphi_1 } }  \nonumber\\ 
  &= \frac{1}{{8\pi ^2 }}\int\limits_{\varphi_1  =  - \pi }^\pi  {e^{i\varphi_1 (r_1 + r_2)/2} d\varphi_1 } \int\limits_{\varphi_2  =  - \pi }^\pi  {e^{i\varphi_2 (r_1 - r_2)/2} d\varphi_2 }  \nonumber \\ 
  &= \left\{ {\begin{array}{*{20}c}
   {1/2} \hfill & {r_1 = r_2 = 0} \hfill  \\
   0 \hfill & {r_1 =  \pm r_2} \hfill  \\
   {2f(r_1 + r_2)f(r_1 - r_2)} \hfill & {{\rm{else}}} \hfill  \\
\end{array}} \right.  \label{a1e13}
\end{align}
with the function $f$ defined as
\begin{equation}
f(x) \equiv \frac{{\sin (x\pi /2)}}{x \pi}.\label{a1e14}
\end{equation}

\section{Non particle conserving free-fermions}
We now consider more general quadratic models of fermions than was initially considered with Hamiltonian \ref{a1e1}. Specifically, we consider Hamiltonians with non particle conserving terms of the form $a a$ and $a^\dag a^\dag$. The generic nearest neighbour Hamiltonian on an $N\times N$ lattice is written
\begin{align}
\hat H &= \sum\limits_{r_1,r_2 =  - (N - 1)/2}^{(N - 1)/2} \frac{1}{2}\left( a_{r_1 + 1,r_2}^\dag  a_{r_1,r_2}+ a_{r_1,r_2}^\dag a_{r_1 + 1,r_2} + a_{r_1,r_2 + 1}^\dag  a_{r_1,r_2} + a_{r_1,r_2}^\dag  a_{r_1,r_2 + 1}  \right) \nonumber\\
&\quad +\frac{\gamma }{2}\left( a_{r_1,r_2}^\dag  a_{r_1 + 1,r_2}^\dag + a_{r_1 + 1,r_2} a_{r_1,r_2}  + a_{r_1,r_2}^\dag  a_{r_1,r_2 + 1}^\dag + a_{r_1,r_2 + 1} a_{r_1,r_2}  \right) - \lambda a_{r_1,r_2}^\dag  a_{r_1,r_2}.\label{a1e15} 
\end{align}
This model of free-fermions depends on a parameter $\lambda$, with represents the chemical potential, and another parameter $\gamma$, which represents the anisotrophy. The free-fermion model of Eq. \ref{a1e1} corresponds to setting the anisotrophy term to zero in the present model. Similar to the derivation of last section, we shall first proceed with a fourier transform of the fermionic operators $a$. However, as we shall see, the Hamiltonian of Eq. \ref{a1e15} also requires an additional Bogoliubov transformation to bring it into diagonal form. Applying the Fourier transform of the fermionic operators defined Eq. \ref{a1e2}, the non particle conserving terms of Eq.\ref{a1e15} transform as
\begin{align}
 \sum\limits_{r_1,r_2} {a_{r_1,r_2}^\dag  a_{r_1 + 1,r_2}^\dag  } & = \sum\limits_{k_1,k_2} {e^{2\pi i{k_1}/N} b_{k_1,k_2}^\dag  b_{ - k_1, - k_2}^\dag  }\nonumber  \\ 
 \sum\limits_{r_1,r_2} {a_{r_1 + 1,r_2} a_{r_1,r_2} } & = \sum\limits_{k_1,k_2} {e^{ - 2\pi i{k_2}/N} b_{ - k_1, - k_2} b_{k_1,k_2} }. \label{a1e16}  
\end{align}
while the other terms in the Hamiltonian transform in the same manner as Eqns. \ref{a1e4} and \ref{a1e5}. The Hamiltonian expressed in the Fourier basis $b$ is given
\begin{equation}
\hat H = \sum\limits_{k_1,k_2 =  - (N - 1)/2}^{(N - 1)/2} {\Lambda _{k_1,k_2} b_{k_1,k_2}^\dag  b_{k_1,k_2}  } + i\Phi_{k_1,k_2} \left( {b_{k_1,k_2}^\dag  b_{ - k_1, - k_2}^\dag   + b_{k_1, k_2} b_{ - k_1,-k_2} } \right)\label{a1e17}
\end{equation}
with $\Lambda$ defined the same as Eq.\ref{a1e7} and $\Phi$ defined
\begin{equation}
\Phi \equiv \frac{\gamma}{2}\left( \sin \frac{2 \pi k_1}{N} + \frac{2\pi k_2}{N}\right). \label{a1e17b}
\end{equation}
While the Fourier transform has eliminated most of the off-diagonal couplings from the Hamiltonian, couplings of the form $b_{k_1, k_2} b_{-k_1, -k_2}$ have been introduced. We now require a Bogoliubov type transformation to fully diagonalise the Hamiltonian of Eq.\ref{a1e17}. The Bogoliubov transformation, which acts independantly for every value $(k_1,k_2)$, involves taking linear combinations of the annihilation/creation operators $b^\dag , b$ to form new fermionic operators $c^\dag , c$ and is defined
\begin{equation}
c_{k_1, k_2} = \cos \left( {\theta _{k_1,k_2} /2} \right) b_{k_1, k_2}  - i\sin \left( {\theta _{k_1,k_2} /2} \right) b_{-k_1, -k_2}^\dag \label{a1e18}
\end{equation}
with $\theta_{k_1,k_2}$ as a yet undefined function of $(k_1, k_2)$. In order to determine the function $\theta_{k_1,k_2}$ it is first assumed that that Hamiltonian is diagonal in the Bogoliubov basis $c$
\begin{equation}
\hat H = \sum\limits_{k_1,k_2 =  - (N - 1)/2}^{(N - 1)/2} {\Omega _{k_1,k_2} c_{k_1,k_2}^\dag  c_{k_1,k_2}  }\label{a1e19}
\end{equation}
and then one works backwards to match the coefficients of Eq. \ref{a1e19} with those of Eq. \ref{a1e17}. Using Eq. \ref{a1e18} to expand the $c_{k_1,k_2}^\dag  c_{k_1,k_2}$ term gives
\begin{align}
c_{k_1, k_2}^\dag c_{k_1, k_2} & = \left( \cos \left( {\theta _{k_1,k_2} /2} \right) b_{k_1,k_2}^\dag + i\sin \left( {\theta _{k_1,k_2} /2} \right) b_{-k_1,-k_2}\right)\times \nonumber\\
& \quad \left( \cos \left( {\theta _{k_1,k_2} /2} \right) b_{k_1,k_2} - i\sin \left( {\theta _{k_1,k_2} /2} \right) b_{-k_1,-k_2}^\dag \right)\nonumber\\
& = \cos^2 \left( {\theta _{k_1,k_2} /2} \right) b_{k_1,k_2}^\dag b_{k_1,k_2} - i\sin \left( {\theta _{k_1,k_2} /2} \right) \cos \left( {\theta _{k_1,k_2} /2} \right) b_{k_1,k_2}^\dag b_{-k_1,-k_2}^\dag \nonumber\\
& + \sin^2 \left( {\theta _{k_1,k_2} /2} \right) b_{-k_1,-k_2} b_{-k_1,-k_2}^\dag + i\sin \left( {\theta _{k_1,k_2} /2} \right) \cos \left( {\theta _{k_1,k_2} /2} \right) b_{-k_1,-k_2} b_{k_1,k_2}. \label{a1e20}
\end{align}
On subsitution of this expression into the Hamiltonian of Eq. \ref{a1e19} we get
\begin{align}
\hat H = &\sum\limits_{k_1,k_2 =  - (N - 1)/2}^{(N - 1)/2} \Omega _{k_1,k_2} \left( \cos \left( {\theta _{k_1,k_2}} \right) b_{k_1,k_2}^\dag  b_{k_1,k_2}\right. \nonumber\\
& \left. + \frac{i}{2} \sin \left( {\theta _{k_1,k_2}} \right) \left( b_{k_1,k_2} b_{-k_1,-k_2} - b_{k_1,k_2}^\dag  b_{-k_1,-k_2}^\dag \right) \right) \label{a1e21}
\end{align}
where we have made use of the trigonometric identities
\begin{align}
\cos(x) &= \cos^2 (x/2) -\sin^2 (x/2),\nonumber\\
\sin(x) &= 2\sin (x/2) \cos (x/2).\label{a1e22}
\end{align}
By matching the coefficients of Eq. \ref{a1e17} with Eq. \ref{a1e21} is seen that the correct choice for the transform weights is
\begin{align}
  \cos \left( {\theta _{k_1,k_2} } \right) &\equiv \Lambda _{k_1,k_2} / \Omega _{k_1,k_2} \nonumber \\ 
  \sin \left( {\theta _{k_1,k_2} } \right) &\equiv \Phi _{k_1,k_2} / \Omega _{k_1,k_2} \label{a1e23}   
\end{align}
where the dispersion relation $\Omega$ may be written as
\begin{equation}
\Omega _{k_1,k_2}  = \sqrt {\left( {\Lambda _{k_1,k_2} } \right)^2  + \left( {\Phi _{k_1,k_2} } \right)^2 }. \label{a1e24}
\end{equation}
The Hamiltonian of Eq. \ref{a1e15} has now been tranformed into a diagonal form 
\begin{equation}
\hat H = \sum\limits_{k_1,k_2 =  - (N - 1)/2}^{(N - 1)/2} {\Omega _{k_1,k_2} c_{k_1,k_2}^\dag  c_{k_1,k_2} }. \label{a1e25}
\end{equation}
Note that, due to the action of the Bogoliubov transformation which mixes creation and annihilation operators, the dispersion in Eq. \ref{a1e25} is positive defined, unlike that of Eq. \ref{a1e6} from the previous section. Now that the correct transforms to diagonalise the free-fermion Hamiltonian has been identified, we turn our attention to finding the ground-state correlators. The ground-state correlation matrix is defined in the $c$ basis as
\begin{equation}
\left\langle {c_{k_1,k_2} c_{k_1,k_2}^\dag  } \right\rangle _\textrm{GS}  = \delta_{k_1 k_2}  \qquad \forall k_1,k_2.\label{a1e26}
\end{equation}
From this definition we can derive expressions for the ground-state correlators $\left\langle {a_{00}^\dag  a_{r_1 r_2} } \right\rangle _\textrm{GS}$ in the original (spatial) basis
\begin{align}
 \left\langle {a_{00}^\dag  a_{r_1 r_2} } \right\rangle _\textrm{GS}  &= \frac{1}{{N^2 }}\sum\limits_{k_1,k_2,k_1',k_2' =  - (N - 1)/2}^{(N - 1)/2} {\left\langle {b_{k_1,k_2}^\dag  b_{k_1',k_2'} } \right\rangle } e^{2\pi i(k_1'r_1+k_2'r_2)}  \nonumber\\ 
  &= \frac{1}{{N^2 }}\sum\limits_{k_1,k_2,k_1',k_2'} {\left\langle {\left( {i\sin(\theta_{k_1,k_2} /2) c_{ - k_1, - k_2}  + \cos(\theta_{k_1,k_2} /2) c_{k_1,k_2}^\dag  } \right)} \right.} \nonumber \\ 
 &\quad \left. \left( \cos(\theta_{k_1',k_2'} /2) c_{k_1',k_2'}  - i\sin(\theta_{k_1',k_2'} /2) c_{ - k_1', - k_2'}^\dag   \right) \right\rangle e^{2\pi i(k_1'r_1+k_2'r_2)}   \nonumber\\ 
  &= \frac{1}{{N^2 }}\sum\limits_{k_1,k_2} { \sin^2(\theta_{k_1,k_2} /2) \left\langle {c_{ - k_1, - k_2} c_{ - k_1, - k_2}^\dag  } \right\rangle } e^{2\pi ik_1 r_1} e^{2\pi ik_2 r_2}  \nonumber\\ 
  &= \frac{1}{{N^2 }}\sum\limits_{k_1,k_2} {\left( {\frac{{1 - \cos (\theta _{k_1,k_2} )}}{2}} \right)} e^{2\pi ik_1 r_1} e^{2\pi ik_2 r_2}.\label{a1e27}  
\end{align}
The other correlation matrix $\left\langle {a_{00}^\dag  a_{r_1 r_2}^\dag } \right\rangle _\textrm{GS}$ is similarly derived
\begin{align}
 \left\langle {a_{00}^\dag  a_{r_1 r_2}^\dag  } \right\rangle _\textrm{GS}  &= \frac{1}{{N^2 }}\sum\limits_{k_1,k_2,k_1',k_2'} {\left\langle {b_{k_1,k_2}^\dag  b_{k_1',k_2'} } \right\rangle } e^{ - 2\pi i(k_1'r_1 + k_2'r_2)}  \nonumber\\ 
  &= \frac{1}{{N^2 }}\sum\limits_{k_1,k_2,k_1',k_2'} {\left\langle {\left( {i\sin(\theta_{k_1,k_2} /2) c_{ - k_1, - k_2}  + \cos(\theta_{k_1,k_2} /2) c_{k_1,k_2}^\dag  } \right)} \right.}  \nonumber\\ 
 &\quad \left. {\left( {i\sin(\theta_{k_1',k_2'} /2) c_{ - k_1', - k_2'}  + \cos(\theta_{k_1',k_2'} /2) c_{k_1',k_2'}^\dag  } \right)} \right\rangle e^{ - 2\pi i(k_1'r_1 + k_2'r_2)} \nonumber \\ 
  &= \frac{1}{{N^2 }}\sum\limits_{k_1,k_2,k_1',k_2'} {i\sin(\theta_{k_1,k_2} /2) \cos(\theta_{k_1,k_2} /2) \left\langle {c_{ - k_1, - k_2} c_{k_1',k_2'}^\dag  } \right\rangle } e^{ - 2\pi i(k_1'r_1 + k_2'r_2)} \nonumber \\ 
  &= \frac{1}{{N^2 }}\sum\limits_{k_1,k_2} {i\sin(\theta_{k_1,k_2} /2) \cos(\theta_{k_1,k_2} /2) } e^{2\pi i(k_1 r_1 + k_2 r_2)} \nonumber \\ 
  &= \frac{1}{{N^2 }}\sum\limits_{k_1,k_2} {\left( {\frac{{i\sin (\theta _{k_1,k_2} )}}{2}} \right)} e^{2\pi i(k_1 r_1 + k_2 r_2)}.  \label{a1e28}
\end{align}
Finally, for clarity, we restate the Fourier/Bogoliubov transform weights
\begin{align}
 \cos (\theta _{k_1,k_2} ) &= \frac{{\Lambda _{k_1,k_2} }}{{\sqrt {(\Lambda _{k_1,k_2} )^2  + (\Phi _{k_1,k_2} )^2 } }} \nonumber\\ 
 \sin (\theta _{k_1,k_2} ) &= \frac{{\Phi _{k_1,k_2} }}{{\sqrt {(\Lambda _{k_1,k_2} )^2  + (\Phi _{k_1,k_2} )^2 } }}\label{a1e29}
\end{align} 
with $\Lambda$ defined Eq. \ref{a1e7} and $\Phi$ defined Eq. \ref{a1e17b}. As with the previous section one could take the thermodynamic limit to obtain integral equations for \ref{a1e27} and \ref{a1e28}. However, as the resultant integrals do not have an analytic solution, it is convenient to leave these equations as summations.

\chapter{The ground-state of harmonic lattice systems} \label{chap:BosonGround}

In this appendix we diagonalise a nearest-neighbour harmonic lattice in order to derive an expression for the ground-state covariance matrices of the system. Similar derivations may also be found e.g. \cite{audenaert02,plenio05,skrovseth05}. The ground-state covariance matrix is the starting point for the real-space RG analysis considered in Sect. \ref{Sec:Boson:GroundRG} of Chapter \ref{chap:FreeBoson}. We also offer a procedure for regularising the case of zero-mass, which would otherwise be divergent. As introduced in Eq. \ref{eq:Boson:s1e1}, we focus on Harmonic systems with Hamiltonian
\begin{equation}
\hat H = \sum\limits_{r  =  - (N - 1)/2}^{(N - 1)/2} \left( {\hat p_r^2  + m^2 \omega ^2 \hat q_r^2  + 2\tilde K\left( {\hat q_{r + 1}  - \hat q_r } \right)^2 }\right). \label{a2e1}
\end{equation}
Recall that operators $\hat p_r$ and $\hat q_r$ are the usual canonical coordinates with commutation $[\hat p_r , \hat q_{r'}]=i\hbar \delta_{rr'}$. By a Fourier transform of the canonical coordinates
\begin{align}
 \check p_\kappa   = \frac{1}{{\sqrt N }}\sum\limits_{r = 1}^N {\hat p_r e^{ - 2\pi ir\kappa /N} }  \nonumber\\ 
 \check q_\kappa   = \frac{1}{{\sqrt N }}\sum\limits_{r = 1}^N {\hat q_r e^{ - 2\pi ir\kappa /N} } .  \label{a2e2}
\end{align}
the Hamiltonian of Eq. \ref{a2e1} is bought into diagonal form
\begin{equation}
\hat H = \sum\limits_{\kappa  =  - (N - 1)/2}^{(N - 1)/2} \left( {\check p_\kappa ^2 + \left( m^2\omega^2 + 8\tilde K\sin ^2 ( \pi \kappa / N ) \right) \check q_\kappa ^2 }\right) .\label{a2e3} 
\end{equation}
It is seen that Eq. \ref{a2e3} represents a system of $N$ uncoupled oscillators, each with an effective mass dependant on $\kappa$. It follows that the ground-state in this basis is the product of the single oscillator ground-states
\begin{align}
 \left\langle {\check p_{\kappa _1 } \check p_{\kappa _2 } } \right\rangle _{{\rm{GS}}}  &= \frac{1}{2}\delta _{\kappa _1 ,\kappa _2 } \sqrt {m^2 \omega ^2  + 8\tilde K\sin ^2 (\pi \kappa _1 /N)}  \nonumber\\ 
 \left\langle {\check q_{\kappa _1 } \check q_{\kappa _2 } } \right\rangle _{{\rm{GS}}}  &= \frac{1}{2}\delta _{\kappa _1 ,\kappa _2 } \frac{1}{\sqrt {m^2 \omega ^2  + 8\tilde K\sin ^2 (\pi \kappa _1 /N)} } .\label{a2e4}
\end{align}
The covariance matrices in the original (spatial) basis are derived by using the inverse of the Fourier transforms of Eq. \ref{a2e2} to obtain
\begin{align}
 \left\langle {\hat p_0 \hat p_r } \right\rangle_{\textrm{GS}}  &= \frac{1}{2N}\sum\limits_{\kappa  = \frac{1-N}{2}}^{\frac{(N - 1)}{2}} \cos \left( \frac{2\pi r\kappa }{N} \right)\sqrt {m^2 \omega ^2  + 8\tilde K\sin ^2 \left( {{\frac{\pi \kappa }{N}}} \right)}   \nonumber\\ 
 \left\langle {\hat q_0 \hat q_r } \right\rangle _{{\textrm{GS}}}  &= \frac{1}{2N}\sum\limits_{\kappa  = {\frac{1-N}{2}}}^{{\frac{(N - 1)}{2}}} {\frac{{\cos \left( {{\frac{2\pi r\kappa }{N}}} \right)}}{{\sqrt {m^2 \omega ^2  + 8\tilde K\sin ^2 \left( {{\frac{\pi \kappa }{N}}} \right)} }}}.\label{a2e5}
\end{align}
As we are interested in the systems of infinite extent, it is possible to take the thermodynamic ($N\rightarrow\infty$) limit of Eq. \ref{a2e5}, in which the sums will be replaced by an integrals. However, in all but a particular case, to be addressed shortly, the resulting integral equations cannot be solved analytically. It is often more convenient, if one desires correlators from the infinite system between sites at most $R$ sites distant, to use the finite $N$ equations \ref{a2e5} with $N\gg R$ to in order to compute approximate correlators for the infinite system. For the $m=0$ case it is possible to evaluate the integral equations exactly; taking the thermodynamic limit of Eq. \ref{a2e5} with $k = {\frac{2\pi \kappa }{N}}$ gives
\begin{align}
 \left\langle {\hat p_0 \hat p_r } \right\rangle_{\textrm{GS}} &= \frac{{\sqrt {2\tilde K} }}{2\pi }\int\limits_{k =  - \pi }^\pi  {\cos \left( {kr} \right)\left| {\sin \left( {{\frac{k}{2}}} \right)} \right|dk} \label{a2e6}\\ 
\left\langle {\hat q_0 \hat q_r } \right\rangle _{{\textrm{GS}}}  &= \frac{1}{{8\pi \sqrt {2\tilde K} }}\int\limits_{k =  - \pi }^\pi  {\frac{{\cos \left( {kr} \right)}}{{\left| {\sin \left( {{\frac{k}{2}}} \right)} \right|}}dk}.\label{a2e7}  
\end{align}
The integral for the $p$-quadrature evaluates as
\begin{equation}
\left\langle {\hat p_0 \hat p_r } \right\rangle_{\textrm{GS}}  = \frac{{\sqrt {2\tilde K} }}{\pi }\left[ {\frac{1}{{2r + 1}} - \frac{1}{{2r - 1}}} \right] \label{a2e8}
\end{equation}
Unfortunately, the integral for the $q$-quadrature in Eq. \ref{a2e7} is \emph{divergent}, $(\gamma_q)_{0r}=\infty \quad$, for all $r$. One can proceed by regularizing the integrals with a small momentum cut-off $\varepsilon$; that is only modes with $\left| k \right| > \varepsilon$ are integrated. The limit as $\varepsilon\rightarrow 0$ is then investigated. To evaluate the $q$-quadrature integral of Eq. \ref{a2e7} first the substitution $x = e^{ik/2}$ is made 
\begin{equation}
\left\langle {\hat q_0 \hat q_r } \right\rangle _{{\textrm{GS}}; \varepsilon}   = \frac{1}{{\pi \sqrt {2\tilde K} }}\left[ {\int\limits_{k = \varepsilon }^\pi  {\frac{{x^{2r} }}{{x^2  - 1}}dx - \int\limits_{k =  - \pi }^{ - \varepsilon } {\frac{{x^{2r} }}{{x^2  - 1}}dx} } } \right].\label{a2e9}
\end{equation}
The integrand in Eq. \ref{a2e9} is known to have a finite series expansion
\begin{equation}
\frac{{x^{2r} }}{{x^2  - 1}} = \sum\limits_{s = 1}^r {x^{2s - 2} }  + \frac{1}{{x^2  - 1}}.\label{a2e10}
\end{equation}
Using Eq. \ref{a2e10} one may split the integral of Eq. \ref{a2e9} into a convergent quantity $f(r)$ (which contains the correlations) and a divergent constant $\Omega_\varepsilon$ (which is \emph{independent} of $r$). Thus the integral of Eq. \ref{a2e9} is evaluated to give
\begin{equation}
\left\langle {\hat q_0 \hat q_r } \right\rangle _{{\textrm{GS}}; \varepsilon}  = \Omega _\varepsilon   - f(r)  + O\left( {\varepsilon ^2 r^2 } \right)\label{a2e11}
\end{equation}
with spatial correlators $f(r)$ defined
\begin{equation}
f(r)\equiv \frac{1}{{2\pi \sqrt {2\tilde K} }}\sum\limits_{s = 1}^r {\left[ {\frac{1}{{s - \frac{1}{2}}}} \right]}\label{a2e12}
\end{equation}
and the constant $\Omega_\varepsilon$ (which divergent in $\varepsilon$) defined
\begin{equation}
\Omega _\varepsilon   \equiv \frac{1}{{4\pi \sqrt {2\tilde K} }}\log \left( {\cot \left( {{\frac{\varepsilon}{4}}} \right)} \right).\label{a2e13}
\end{equation}
By using the regularized expression for correlators of Eq. \ref{a2e11}, quantities such as the \emph{difference between correlators} are seen to remain finite in the limit of $\varepsilon$ taken to zero and may be evaluated exactly
\begin{equation}
\mathop {\lim }\limits_{\varepsilon  \to 0} \left( \left\langle {\hat q_0 \hat q_r } \right\rangle _{{\textrm{GS}}; \varepsilon}  - \left\langle {\hat q_0 \hat q_{r'} } \right\rangle _{{\textrm{GS}}; \varepsilon} \right) = f\left( {r'} \right) - f\left( r \right).\label{a2e14}
\end{equation}
Similarly it can be shown that, although the entanglement entropy of a block of $L$ modes diverges in the massless case, the \emph{difference} in entropy between two blocks is also convergent in the limit as $\varepsilon$ is taken to zero. Thus, for instance, in the zero-mass system we may still make sense of how the entropy of a block of $L$ modes changes along the RG flow as was demonstrated Fig. \ref{fig:Boson:BoseEntPlotFlat}.

\bibliographystyle{alpha}
\bibliography{mybib}

\newcommand{\etalchar}[1]{$^{#1}$}
\begin{thebibliography}{AEPW02}

\bibitem[AEPW02]{audenaert02}
K.~Audenaert, J.~Eisert, M.~B. Plenio, and R.~F. Werner.
\newblock Entanglement properties of the harmonic chain.
\newblock {\em Phys. Rev. A}, 66(4):042327, Oct 2002.

\bibitem[AK10]{alford09}
Mark~G Alford and Andrei Kryjevski.
\newblock Mitigating the sign problem for non-relativistic fermions on the
  lattice.
\newblock {\em Journal of Physics G: Nuclear and Particle Physics},
  37(2):025002, 2010.

\bibitem[And75]{anderson75}
James~B. Anderson.
\newblock A random-walk simulation of the schrodinger equation: H3+.
\newblock {\em The Journal of Chemical Physics}, 63(4):1499--1503, 1975.

\bibitem[AV08]{aguado08}
Miguel Aguado and Guifr\'{e} Vidal.
\newblock Entanglement renormalization and topological order.
\newblock {\em Physical Review Letters}, 100(7):070404, 2008.

\bibitem[BA04]{budnik04}
Ran Budnik and Assa Auerbach.
\newblock Low-energy singlets in the heisenberg antiferromagnet on the kagome
  lattice.
\newblock {\em Phys. Rev. Lett.}, 93(18):187205, Oct 2004.

\bibitem[Bax82]{baxter82}
R.~J. Baxter.
\newblock {\em Exactly solved models in statistical mechanics}.
\newblock Academic Press, 1982.

\bibitem[BCS06]{barthel06}
T.~Barthel, M.-C. Chung, and U.~Schollw\"{o}ck.
\newblock Entanglement scaling in critical two-dimensional fermionic and
  bosonic systems.
\newblock {\em Physical Review A (Atomic, Molecular, and Optical Physics)},
  74(2):022329, 2006.

\bibitem[BD02]{blote02}
Henk W.~J. Bl\"ote and Youjin Deng.
\newblock Cluster monte carlo simulation of the transverse ising model.
\newblock {\em Phys. Rev. E}, 66(6):066110, Dec 2002.

\bibitem[BG85]{burkhardt85}
T.~W. Burkhardt and I.~Guim.
\newblock Finite-size scaling of the quantum ising chain with periodic, free,
  and antiperiodic boundary conditions.
\newblock {\em Journal of Physics A: Mathematical and General}, 18(1):L33--L38,
  1985.

\bibitem[BHR03]{bergkvist03}
Sara Bergkvist, Patrik Henelius, and Anders Rosengren.
\newblock Reduction of the sign problem using the meron-cluster approach.
\newblock {\em Phys. Rev. E}, 68(1):016122, Jul 2003.

\bibitem[BPE09]{barthel09}
Thomas Barthel, Carlos Pineda, and Jens Eisert.
\newblock Contraction of fermionic operator circuits and the simulation of
  strongly correlated fermions.
\newblock {\em Phys. Rev. A}, 80(4):042333, Oct 2009.

\bibitem[BR79]{bratteli79}
O.~Bratteli and D.~W. Robinson.
\newblock {\em Operator Algebras and Quantum Statistical Mechanics}.
\newblock Springer, New York, 1979.

\bibitem[Car84]{cardy84}
J~L Cardy.
\newblock Conformal invariance and universality in finite-size scaling.
\newblock {\em Journal of Physics A: Mathematical and General},
  17(7):L385--L387, 1984.

\bibitem[Car86]{cardy86}
John~L. Cardy.
\newblock {Operator Content of Two-Dimensional Conformally Invariant Theories}.
\newblock {\em Nucl. Phys.}, B270:186--204, 1986.

\bibitem[Car96]{cardy96}
J.~Cardy.
\newblock {\em Scaling and Renormalization in Statistical Physics}.
\newblock Cambridge University Press, Cambridge, 1996.

\bibitem[CC04]{calabrese04}
Pasquale Calabrese and John Cardy.
\newblock Entanglement entropy and quantum field theory.
\newblock {\em Journal of Statistical Mechanics: Theory and Experiment},
  2004(06):P06002, 2004.

\bibitem[CD04]{corney04}
J.~F. Corney and P.~D. Drummond.
\newblock Gaussian quantum monte carlo methods for fermions and bosons.
\newblock {\em Phys. Rev. Lett.}, 93(26):260401, Dec 2004.

\bibitem[CDR08]{cincio08}
Lukasz Cincio, Jacek Dziarmaga, and Marek~M. Rams.
\newblock Multiscale entanglement renormalization ansatz in two dimensions:
  Quantum ising model.
\newblock {\em Physical Review Letters}, 100(24):240603, 2008.

\bibitem[Cep95]{ceperley95}
D.~M. Ceperley.
\newblock Path integrals in the theory of condensed helium.
\newblock {\em Rev. Mod. Phys.}, 67(2):279--355, Apr 1995.

\bibitem[CEPi06]{cramer06}
M.~Cramer, J.~Eisert, M.~B. Plenio, and J.~Drei\ss ig.
\newblock Entanglement-area law for general bosonic harmonic lattice systems.
\newblock {\em Physical Review A (Atomic, Molecular, and Optical Physics)},
  73(1):012309, 2006.

\bibitem[CEVV10]{corboz09a}
Philippe Corboz, Glen Evenbly, Frank Verstraete, and Guifr\'e Vidal.
\newblock Simulation of interacting fermions with entanglement renormalization.
\newblock {\em Phys. Rev. A}, 81(1):010303, Jan 2010.

\bibitem[CV09]{corboz09b}
Philippe Corboz and Guifr\'e Vidal.
\newblock Fermionic multiscale entanglement renormalization ansatz.
\newblock {\em Phys. Rev. B}, 80(16):165129, Oct 2009.

\bibitem[Del04]{delamotte04}
Bertrand Delamotte.
\newblock A hint of renormalization.
\newblock {\em American Journal of Physics}, 72(2):170--184, 2004.

\bibitem[DEO08]{dawson08}
C.~M. Dawson, J.~Eisert, and T.~J. Osborne.
\newblock Unifying variational methods for simulating quantum many-body
  systems.
\newblock {\em Physical Review Letters}, 100(13):130501, 2008.

\bibitem[EV09a]{evenbly08a}
G.~Evenbly and G.~Vidal.
\newblock Algorithms for entanglement renormalization.
\newblock {\em Physical Review B (Condensed Matter and Materials Physics)},
  79(14):144108, 2009.

\bibitem[EV09b]{evenbly08b}
G.~Evenbly and G.~Vidal.
\newblock Entanglement renormalization in two spatial dimensions.
\newblock {\em Physical Review Letters}, 102(18):180406, 2009.

\bibitem[EV10a]{evenbly07b}
G~Evenbly and G~Vidal.
\newblock Entanglement renormalization in free bosonic systems: real-space
  versus momentum-space renormalization group transforms.
\newblock {\em New Journal of Physics}, 12(2):025007, 2010.

\bibitem[EV10b]{evenbly07a}
G.~Evenbly and G.~Vidal.
\newblock Entanglement renormalization in noninteracting fermionic systems.
\newblock {\em Phys. Rev. B}, 81(23):235102, Jun 2010.

\bibitem[EV10c]{evenbly09a}
G.~Evenbly and G.~Vidal.
\newblock Frustrated antiferromagnets with entanglement renormalization: Ground
  state of the spin-$12$ heisenberg model on a kagome lattice.
\newblock {\em Phys. Rev. Lett.}, 104(18):187203, May 2010.

\bibitem[Fis98]{fisher98}
Michael~E. Fisher.
\newblock Renormalization group theory: Its basis and formulation in
  statistical physics.
\newblock {\em Rev. Mod. Phys.}, 70(2):653--681, Apr 1998.

\bibitem[GK06]{gioev06}
Dimitri Gioev and Israel Klich.
\newblock Entanglement entropy of fermions in any dimension and the widom
  conjecture.
\newblock {\em Physical Review Letters}, 96(10):100503, 2006.

\bibitem[GLSW09]{gu09}
Zheng-Cheng Gu, Michael Levin, Brian Swingle, and Xiao-Gang Wen.
\newblock Tensor-product representations for string-net condensed states.
\newblock {\em Physical Review B (Condensed Matter and Materials Physics)},
  79(8):085118, 2009.

\bibitem[GLW08]{gu08}
Zheng-Cheng Gu, Michael Levin, and Xiao-Gang Wen.
\newblock Tensor-entanglement renormalization group approach as a unified
  method for symmetry breaking and topological phase transitions.
\newblock {\em Physical Review B (Condensed Matter and Materials Physics)},
  78(20):205116, 2008.

\bibitem[GMF08]{giovannetti08}
V.~Giovannetti, S.~Montangero, and Rosario Fazio.
\newblock Quantum multiscale entanglement renormalization ansatz channels.
\newblock {\em Physical Review Letters}, 101(18):180503, 2008.

\bibitem[Has00]{hastings00}
M.~B. Hastings.
\newblock Dirac structure, rvb, and goldstone modes in the kagom\'e
  antiferromagnet.
\newblock {\em Phys. Rev. B}, 63(1):014413, Dec 2000.

\bibitem[HRLW08]{hermele08}
Michael Hermele, Ying Ran, Patrick~A. Lee, and Xiao-Gang Wen.
\newblock Properties of an algebraic spin liquid on the kagome lattice.
\newblock {\em Physical Review B (Condensed Matter and Materials Physics)},
  77(22):224413, 2008.

\bibitem[HS00]{henelius00}
Patrik Henelius and Anders~W. Sandvik.
\newblock Sign problem in monte carlo simulations of frustrated quantum spin
  systems.
\newblock {\em Phys. Rev. B}, 62(2):1102--1113, Jul 2000.

\bibitem[HSF05]{hermele05}
Michael Hermele, T.~Senthil, and Matthew P.~A. Fisher.
\newblock Algebraic spin liquid as the mother of many competing orders.
\newblock {\em Phys. Rev. B}, 72(10):104404, Sep 2005.

\bibitem[Hub63]{hubbard63}
J.~Hubbard.
\newblock Electron correlations in narrow energy bands.
\newblock {\em Proceedings of the Royal Society of London. Series A,
  Mathematical and Physical Sciences}, 276(1365):238--257, 1963.

\bibitem[JOV{\etalchar{+}}08]{jordan08}
J.~Jordan, R.~Or\'{u}s, G.~Vidal, F.~Verstraete, and J.~I. Cirac.
\newblock Classical simulation of infinite-size quantum lattice systems in two
  spatial dimensions.
\newblock {\em Physical Review Letters}, 101(25):250602, 2008.

\bibitem[JWS08]{jiang08}
H.~C. Jiang, Z.~Y. Weng, and D.~N. Sheng.
\newblock Density matrix renormalization group numerical study of the kagome
  antiferromagnet.
\newblock {\em Physical Review Letters}, 101(11):117203, 2008.

\bibitem[KGH{\etalchar{+}}67]{kadanoff67}
Leo~P. Kadanoff, Wolfgang G\"otze, David Hamblen, Robert Hecht, E.~A.~S. Lewis,
  V.~V. Palciauskas, Martin Rayl, J.~Swift, David Aspnes, and Joseph Kane.
\newblock Static phenomena near critical points: Theory and experiment.
\newblock {\em Rev. Mod. Phys.}, 39(2):395--431, Apr 1967.

\bibitem[KRV09]{konig09}
Robert K\"{o}nig, Ben~W. Reichardt, and Guifr\'{e} Vidal.
\newblock Exact entanglement renormalization for string-net models.
\newblock {\em Physical Review B (Condensed Matter and Materials Physics)},
  79(19):195123, 2009.

\bibitem[LDY{\etalchar{+}}06]{li06}
Weifei Li, Letian Ding, Rong Yu, Tommaso Roscilde, and Stephan Haas.
\newblock Scaling behavior of entanglement in two- and three-dimensional
  free-fermion systems.
\newblock {\em Physical Review B (Condensed Matter and Materials Physics)},
  74(7):073103, 2006.

\bibitem[LE93]{leung93}
P.~W. Leung and Veit Elser.
\newblock Numerical studies of a 36-site kagome antiferromagnet.
\newblock {\em Phys. Rev. B}, 47(9):5459--5462, Mar 1993.

\bibitem[Lhu05]{lhuillier05}
Claire Lhuillier.
\newblock Frustrated quantum magnets.
\newblock {arXiv}:cond-mat/0502464v1, 2005.

\bibitem[LL09]{laeuchli09}
Andreas Laeuchli and Claire Lhuillier.
\newblock Dynamical correlations of the kagome s=1/2 heisenberg quantum
  antiferromagnet.
\newblock {arXiv}:0801.2449v1, 2009.

\bibitem[LRV04]{latorre04}
J.~I. Latorre, E.~Rico, and G.~Vidal.
\newblock Ground state entanglement in quantum spin chains.
\newblock {\em Quantum Inf. Comput.}, 4(1):48--92, 2004.

\bibitem[LSM61]{lieb61}
Elliott Lieb, Theodore Schultz, and Daniel Mattis.
\newblock Two soluble models of an antiferromagnetic chain.
\newblock {\em Annals of Physics}, 16(3):407 -- 466, 1961.

\bibitem[LW05]{levin05}
Michael~A. Levin and Xiao-Gang Wen.
\newblock String-net condensation: A physical mechanism for topological phases.
\newblock {\em Phys. Rev. B}, 71(4):045110, Jan 2005.

\bibitem[McM65]{mcmillan65}
W.~L. McMillan.
\newblock Ground state of liquid he4.
\newblock {\em Phys. Rev.}, 138(2A):A442--A451, Apr 1965.

\bibitem[Mil98]{mila98}
F.~Mila.
\newblock Low-energy sector of the $s =1/2$ kagome antiferromagnet.
\newblock {\em Phys. Rev. Lett.}, 81(11):2356--2359, Sep 1998.

\bibitem[MRGF09]{montangero09}
S.~Montangero, M.~Rizzi, V.~Giovannetti, and Rosario Fazio.
\newblock Critical exponents with a multiscale entanglement renormalization
  ansatz channel.
\newblock {\em Physical Review B (Condensed Matter and Materials Physics)},
  80(11):113103, 2009.

\bibitem[MS93]{mandl93}
Franz Mandl and Graham Shaw.
\newblock {\em Quantum field theory}.
\newblock J. Wiley, 1993.

\bibitem[MW94]{morningstar94}
Colin~J. Morningstar and Marvin Weinstein.
\newblock Contractor renormalization group method: A new computational
  technique for lattice systems.
\newblock {\em Phys. Rev. Lett.}, 73(14):1873--1877, Oct 1994.

\bibitem[MW96]{morningstar96}
Colin~J. Morningstar and Marvin Weinstein.
\newblock Contractor renormalization group technology and exact hamiltonian
  real-space renormalization group transformations.
\newblock {\em Phys. Rev. D}, 54(6):4131--4151, Sep 1996.

\bibitem[MZ91]{marston91}
J.~B. Marston and C.~Zeng.
\newblock Spin-peierls and spin-liquid phases of kagom[e-acute] quantum
  antiferromagnets.
\newblock {\em 35th annual conference on magnetism and magnetic materials},
  69(8):5962--5964, 1991.

\bibitem[NS03]{nikolic03}
P.~Nikolic and T.~Senthil.
\newblock Physics of low-energy singlet states of the kagome lattice quantum
  heisenberg antiferromagnet.
\newblock {\em Phys. Rev. B}, 68(21):214415, Dec 2003.

\bibitem[PBE10]{pineda09}
Carlos Pineda, Thomas Barthel, and Jens Eisert.
\newblock Unitary circuits for strongly correlated fermions.
\newblock {\em Phys. Rev. A}, 81(5):050303, May 2010.

\bibitem[PDFS97]{francesco97}
P.~Mathieu P.~Di~Francesco and D.~Senechal.
\newblock {\em Conformal Field Theory}.
\newblock Springer, New York, 1997.

\bibitem[PEDC05]{plenio05}
M.~B. Plenio, J.~Eisert, J.~Drei\ss{}ig, and M.~Cramer.
\newblock Entropy, entanglement, and area: Analytical results for harmonic
  lattice systems.
\newblock {\em Phys. Rev. Lett.}, 94(6):060503, Feb 2005.

\bibitem[PEV09]{pfeifer08}
Robert N.~C. Pfeifer, Glen Evenbly, and Guifr\'{e} Vidal.
\newblock Entanglement renormalization, scale invariance, and quantum
  criticality.
\newblock {\em Physical Review A (Atomic, Molecular, and Optical Physics)},
  79(4):040301, 2009.

\bibitem[Pfe70]{pfeuty70}
Pierre Pfeuty.
\newblock The one-dimensional ising model with a transverse field.
\newblock {\em Annals of Physics}, 57(1):79 -- 90, 1970.

\bibitem[PV02]{pelissetto02}
Andrea Pelissetto and Ettore Vicari.
\newblock {Critical phenomena and renormalization-group theory}.
\newblock {\em Phys. Rept.}, 368:549--727, 2002.

\bibitem[PVC06]{porras06}
D.~Porras, F.~Verstraete, and J.~I. Cirac.
\newblock Renormalization algorithm for the calculation of spectra of
  interacting quantum systems.
\newblock {\em Physical Review B (Condensed Matter and Materials Physics)},
  73(1):014410, 2006.

\bibitem[RHLW07]{ran07}
Ying Ran, Michael Hermele, Patrick~A. Lee, and Xiao-Gang Wen.
\newblock Projected-wave-function study of the spin-1/2 heisenberg model on the
  kagom[e-acute] lattice.
\newblock {\em Physical Review Letters}, 98(11):117205, 2007.

\bibitem[RK99]{rieger99}
H.~Rieger and N.~Kawashima.
\newblock Application of a continuous time cluster algorithm to the
  two-dimensional random quantum ising ferromagnet.
\newblock {\em Eur. Phys. J. B}, 9(2):233--236, 1999.

\bibitem[RMV08]{rizzi08}
Matteo Rizzi, Simone Montangero, and Guifre Vidal.
\newblock Simulation of time evolution with multiscale entanglement
  renormalization ansatz.
\newblock {\em Physical Review A (Atomic, Molecular, and Optical Physics)},
  77(5):052328, 2008.

\bibitem[RS10]{richter09}
{Richter, J.} and {Schulenburg, J.}
\newblock The spin-1/2 j1-j2 heisenberg antiferromagnet on the square lattice:
  Exact diagonalization for n = 40 spins.
\newblock {\em Eur. Phys. J. B}, 73(1):117--124, 2010.

\bibitem[Sac92]{sachdev92}
Subir Sachdev.
\newblock Kagome-lattice and triangular-lattice heisenberg antiferromagnets:
  Ordering from quantum fluctuations and quantum-disordered ground states with
  unconfined bosonic spinons.
\newblock {\em Phys. Rev. B}, 45(21):12377--12396, Jun 1992.

\bibitem[Sch05]{scholl05}
U.~Schollw\"ock.
\newblock The density-matrix renormalization group.
\newblock {\em Rev. Mod. Phys.}, 77(1):259--315, Apr 2005.

\bibitem[SDV06]{shi06}
Y.-Y. Shi, L.-M. Duan, and G.~Vidal.
\newblock Classical simulation of quantum many-body systems with a tree tensor
  network.
\newblock {\em Physical Review A (Atomic, Molecular, and Optical Physics)},
  74(2):022320, 2006.

\bibitem[SH07]{singh07a}
Rajiv R.~P. Singh and David~A. Huse.
\newblock Ground state of the spin-1/2 kagome-lattice heisenberg
  antiferromagnet.
\newblock {\em Physical Review B (Condensed Matter and Materials Physics)},
  76(18):180407, 2007.

\bibitem[SH08]{singh08a}
Rajiv R.~P. Singh and David~A. Huse.
\newblock Triplet and singlet excitations in the valence bond crystal phase of
  the kagome lattice heisenberg model.
\newblock {\em Physical Review B (Condensed Matter and Materials Physics)},
  77(14):144415, 2008.

\bibitem[Sha94]{shankar94}
R.~Shankar.
\newblock Renormalization-group approach to interacting fermions.
\newblock {\em Rev. Mod. Phys.}, 66(1):129--192, Jan 1994.

\bibitem[Shi99]{shirkov99}
D.~V. Shirkov.
\newblock Evolution of the bogoluibov renormalization group.
\newblock {arXiv}:hep-th/9909024v1, 1999.

\bibitem[Skr05]{skrovseth05}
Stein~Olav Skr\o{}vseth.
\newblock Entanglement in bosonic systems.
\newblock {\em Phys. Rev. A}, 72(6):062305, Dec 2005.

\bibitem[SM02]{syromyatnikov02}
A.~V. Syromyatnikov and S.~V. Maleyev.
\newblock Hidden long-range order in kagom\'e heisenberg antiferromagnets.
\newblock {\em Phys. Rev. B}, 66(13):132408, Oct 2002.

\bibitem[SNBN05]{shores05}
Matthew~P. Shores, Emily~A. Nytko, Bart~M. Bartlett, and Daniel~G. Nocera.
\newblock A structurally perfect s = 1/2 kagomé antiferromagnet.
\newblock {\em Journal of the American Chemical Society}, 127:13462, 2005.

\bibitem[SP81]{solyom81}
J.~S\'olyom and P.~Pfeuty.
\newblock Renormalization-group study of the hamiltonian version of the potts
  model.
\newblock {\em Phys. Rev. B}, 24(1):218--229, Jul 1981.

\bibitem[SPV09]{sukhi09}
Sukhwinder Singh, Robert Pfeifer, and Guifre Vidal.
\newblock Tensor network decompositions in the presence of a global symmetry.
\newblock {arXiv}:0907.2994v1, 2009.

\bibitem[Sre07]{srednicki07}
Mark~Allen Srednicki.
\newblock {\em Quantum field theory}.
\newblock Cambridge University Press, Cambridge, 2007.

\bibitem[Swi09]{swingle09}
Brian Swingle.
\newblock Entanglement renormalization and holography.
\newblock {arXiv}:0905.1317v1, 2009.

\bibitem[TEV09]{tagliacozzo09}
L.~Tagliacozzo, G.~Evenbly, and G.~Vidal.
\newblock Simulation of two-dimensional quantum systems using a tree tensor
  network that exploits the entropic area law.
\newblock {\em Phys. Rev. B}, 80(23):235127, Dec 2009.

\bibitem[Vid03]{vidal03b}
Guifr\'e Vidal.
\newblock Efficient classical simulation of slightly entangled quantum
  computations.
\newblock {\em Phys. Rev. Lett.}, 91(14):147902, Oct 2003.

\bibitem[Vid04]{vidal04}
Guifr\'e Vidal.
\newblock Efficient simulation of one-dimensional quantum many-body systems.
\newblock {\em Phys. Rev. Lett.}, 93(4):040502, Jul 2004.

\bibitem[Vid07]{vidal07}
G.~Vidal.
\newblock Entanglement renormalization.
\newblock {\em Physical Review Letters}, 99(22):220405, 2007.

\bibitem[Vid08]{vidal08}
G.~Vidal.
\newblock Class of quantum many-body states that can be efficiently simulated.
\newblock {\em Physical Review Letters}, 101(11):110501, 2008.

\bibitem[VLRK03]{vidal03}
G.~Vidal, J.~I. Latorre, E.~Rico, and A.~Kitaev.
\newblock Entanglement in quantum critical phenomena.
\newblock {\em Phys. Rev. Lett.}, 90(22):227902, Jun 2003.

\bibitem[VPC04]{verstraete04}
F.~Verstraete, D.~Porras, and J.~I. Cirac.
\newblock Density matrix renormalization group and periodic boundary
  conditions: A quantum information perspective.
\newblock {\em Phys. Rev. Lett.}, 93(22):227205, Nov 2004.

\bibitem[Whi92]{white92}
Steven~R. White.
\newblock Density matrix formulation for quantum renormalization groups.
\newblock {\em Phys. Rev. Lett.}, 69(19):2863--2866, Nov 1992.

\bibitem[Whi93]{white93}
Steven~R. White.
\newblock Density-matrix algorithms for quantum renormalization groups.
\newblock {\em Phys. Rev. B}, 48(14):10345--10356, Oct 1993.

\bibitem[Wil75]{wilson75}
Kenneth~G. Wilson.
\newblock The renormalization group: Critical phenomena and the kondo problem.
\newblock {\em Rev. Mod. Phys.}, 47(4):773--840, Oct 1975.

\bibitem[Wol06]{wolf06}
Michael~M. Wolf.
\newblock Violation of the entropic area law for fermions.
\newblock {\em Physical Review Letters}, 96(1):010404, 2006.

\bibitem[WV06]{wang06}
Fa~Wang and Ashvin Vishwanath.
\newblock Spin-liquid states on the triangular and kagome lattices: A
  projective-symmetry-group analysis of schwinger boson states.
\newblock {\em Physical Review B (Condensed Matter and Materials Physics)},
  74(17):174423, 2006.

\end{thebibliography}
\cleardoublepage %

\end{document}